\documentclass{article}
\usepackage{arxiv}

\usepackage[utf8]{inputenc}
\usepackage[T1]{fontenc}
\usepackage{url}
\usepackage{amsmath, amssymb, amsthm, mathtools}
\usepackage{graphicx}
\usepackage{subcaption}
\usepackage{booktabs}
\usepackage{threeparttable}
\usepackage{tabularx}
\usepackage{microtype}
\usepackage{xcolor}
\usepackage{adjustbox}
\usepackage{enumitem}
\usepackage{algorithm}
\usepackage{algpseudocode}
\usepackage[round]{natbib}
\usepackage{hyperref}

\hypersetup{colorlinks=true, linkcolor=blue, citecolor=blue, urlcolor=blue}
\graphicspath{{Figures/}}

\title{Small Area Estimation under Spatial Regimes: Spatially Clustered Fay–Herriot Models for Agricultural Indicators}
\renewcommand{\shorttitle}{Small Area Estimation under Spatial Regimes: SCFH}

\author{
  Paolo Maranzano \\
  Department of Economics, Management and Statistics (DEMS) \\
  University of Milano-Bicocca, Italy \& \\
  Fondazione Eni Enrico Mattei (FEEM), Italy \\
  \texttt{paolo.maranzano@unimib.it} \\
  \And
  Raffaele Mattera \\
  Department of Mathematics and Physics \\
  University of Campania "Luigi Vanvitelli", Italy \\
  \texttt{raffaele.mattera@unicampania.it}
  \And
  Shonosuke Sugasawa \\
  Faculty of Economics \\
  Keio University, Japan \\
  \texttt{sugasawa@econ.keio.ac.jp}
}

\begin{document}
\maketitle

\begin{abstract}
Area-level small area estimation (SAE) models, such as the Fay--Herriot (FH) model, borrow strength across domains through covariates and random effects, but they can struggle when the relationship between the covariates and the outcome is spatially heterogeneous, that is, when it changes across the spatial domain of interest. We propose a spatially-clustered FH (SC-FH) framework that simultaneously (i) estimates cluster-specific regression coefficients and random effects variances and (ii) generates spatially coherent partitions of the geographical domain.
Estimation maximizes a penalized likelihood that augments the FH likelihood with a Potts-type spatial cohesion term over the areal adjacency graph, through an efficient strategy that alternates between sequential label updates and closed-form FH updates within clusters.
In simulation experiments run on the real geography of the application, the method recovers the latent regimes almost exactly whenever they are separated in the covariate--response space and improves prediction accuracy over the standard FH benchmark, with the spatial penalty acting as a stabilizer of both classification and estimation.
An empirical application to the average standard output of farms in the Po Valley (Northern Italy) identifies two spatially compact production regimes with significantly different cluster-wise coefficients, and shows that the clusterwise predictor improves on the direct estimates while avoiding the over-shrinkage of the pooled model.
\end{abstract}


\clearpage
\section{Introduction}
Small area estimation (SAE) aims to produce reliable domain-level estimates when direct (i.e., design-based) estimators are too noisy due to small, sparse, or unevenly distributed samples \citep{RaoMol2015}. Model-based SAE combines direct estimators (with known or estimated sampling variances) with auxiliary covariates through hierarchical or mixed-effects models, enabling \emph{borrowing strength} across areas and providing empirical best linear unbiased predictors (EBLUPs) and measures of uncertainty. The model introduced by \citet{FH79}, hereafter FH, is a classical area-level linear mixed model for SAE, where direct survey estimators are linked to auxiliary covariates through a regression model with area-specific random effects. This hierarchical structure allows information to be shared across areas, improving estimation accuracy when direct estimates are based on small or uneven samples.

In this paper, we focus on FH models for areal spatial data, where small areas are associated with fixed geographical units and spatial information is incorporated to account for structured heterogeneity across neighboring domains. 

\subsection{Spatial heterogeneity in small area estimation models}
A key limitation of the classical FH area-level model is the implicit assumption that the linking relationship between the direct estimator and the auxiliary covariates is spatially homogeneous (stationary), aside from an area-level random effect, which is also assumed to act globally. Also, while the classical FH model assumes independent random effects across areas, spatially-structured extensions incorporate neighborhood dependence to capture smooth spatial dependence in random effects, e.g., through conditional autoregressive or simultaneous autoregressive structures \citep[e.g.][]{chandra2013exploring,PratesiSalvati2008} and to improve prediction when nearby areas exhibit similar latent behavior.
The assumption of global dependence patterns, implicitly assuming that the latent process varies smoothly across the entire spatial domain, implies that a single regression relationship and a globally-parameterized, i.e., homoskedastic, random effect hold across all domains, so that ``borrowing strength'' is coherent throughout the whole area of interest.
However, in many official statistics, especially in socio-economic and environmental applications, empirical relationships vary across space due to unobserved regimes, structural breaks, policy differences and geographically-localized mechanisms \citep[e.g. see][]{anselin2024endogenous,mattsson2025modeling,otto2016detection}. It follows that smoothness-based spatial borrowing might be too restrictive when indicators exhibit \emph{regime-like} behavior: neighboring areas may share similar relationships within subregions, while relationships differ sharply across boundaries. In these settings, \textit{global models} can lead to over-smoothing at borders and can mask discontinuities that are substantively meaningful for policy and interpretation.

The above circumstances motivate models that encourage spatial clustering or heterogeneous spatial smoothing rather than global smoothing. Accordingly, the literature has evolved along two main directions: spatially non-stationary FH formulations and mixture-based FH models that capture latent heterogeneity through regime-specific linking structures.

Non-stationary FH \citep{ChandraEtAl2015} and mixture-based FH \citep{GardiniEtAl2025} can be viewed as two complementary strategies to model spatially varying or regime-dependent relationships: the former emphasizes continuous or locally adaptive variation in the linking model, whereas the latter emphasizes discrete latent regimes (possibly with covariate-dependent gating), both of which are highly relevant when the underlying spatial mechanism is better represented by heterogeneous regimes than by global stationarity.

\cite{ChandraEtAl2015} proposed a spatially non-stationary FH model where regression parameters vary geographically, yielding EBLUPs that adapt to local conditions rather than enforcing a single global slope. This contribution builds on earlier ideas of geographically weighted or locally adaptive SAE under non-stationarity \citep{BrunsdonFotheringham1998,ChandraEtAl2012}, and it complements the broader literature on spatial FH formulations that introduce structured spatial random effects (e.g., CAR/SAR-like dependence) to borrow strength across neighboring areas while retaining the FH area-level sampling framework \citep[e.g.,][]{PratesiSalvati2008,MarhuendaMolinaMorales2013,PorterEtAl2014}. In applied domains, these developments are discussed and contextualized, for instance, in methodological treatments of agricultural SAE where spatial heterogeneity and domain-specific regimes are empirically salient \citep{BertarelliEtAl2021}.

A complementary and increasingly influential direction replaces the notion of a single global linking model with Mixture and Mixture-of-Experts (MoE) structures, effectively allowing multiple data-generating regimes to coexist and to be inferred from the data.
Finite-mixture FH-type models introduce latent classes of areas, each with its own regression relationship and/or variance components, enabling cluster-specific borrowing of strength and greater robustness to multi-modality and unobserved heterogeneity \citep{ArticusBurgard2014}. MoE-FH models push this idea further by letting the mixing weights depend on covariates (or contextual information), so that regime membership becomes a probabilistic, covariate-driven mechanism rather than a purely latent partition. A recent work by \cite{GardiniEtAl2025} developed a MoE extension of FH motivated by the presence of rural against urban spatial regimes, showing how the MoE structure improves flexibility while preserving the interpretability of SAE models. From a broader perspective, mixture-based FH extensions are naturally related to robust mixed-model SAE, where departures from standard random effects assumptions are handled via more flexible distributions, by relaxing the restrictive homogeneity assumptions without sacrificing the borrow-strength principle \citep{FabriziTrivisano2010,Fabrizi2009}.

The demand for such flexible SAE tools is driven, to a large extent, by applications. Agricultural and agri-environmental statistics have been a natural domain for SAE since the county-level crop-area predictions of \citet{BatteseEtAl1988}, and the recent availability of administrative, environmental, and remote-sensing layers has multiplied the auxiliary information that area-level models can exploit. Recent examples include poverty mapping from satellite imagery and machine-learning features \citep{NewhouseEtAl2025}, spatio-temporal FH modeling of agricultural indicators for the Agrarian Sub-Regions of Northern Italy \citep{CarilloEtAl2024}, and the small area estimation of the agricultural carbon footprint in the Po Valley, where satellite-derived ammonia emissions enter the linking model and the additional uncertainty generated by the data-integration step is propagated by a parametric bootstrap \citep{PajnoEtAl2026}. In these applications the auxiliary information is increasingly rich, but the linking relationship is still typically assumed to be common to all domains --- precisely the assumption that the present paper relaxes.

\subsection{A third way to address spatial heterogeneity in small area estimation}
In this paper, we pursue an alternative route to both mixture-of-experts and spatially non-stationary Fay--Herriot models by adopting a spatially-clustered regression perspective. Rather than modeling continuous spatial variation or probabilistic regime switching, our approach explicitly identifies spatially coherent clusters of areas and performs cluster-specific borrowing of strength within the Fay--Herriot framework.

Clusterwise regression, also known as $k$-means regression \citep{desarbo1988maximum,spath1979algorithmus}, addresses the heterogeneity by assuming that observations belong to latent clusters, each with its own regression coefficients \citep[see also,][]{park2017algorithms,kuang2024performance}.
Recently, \citet{SugasawaMurakami2021} introduced a spatially-clustered regression (SCR) approach that promotes spatially coherent clusters, implying that areas are partitioned into an unknown small number of spatially coherent groups, and model parameters are allowed to differ across groups. The proposed approach is implemented by augmenting the likelihood with a Potts-type penalty \citep{potts1952some} over an adjacency graph.

A complementary stream of research in spatial econometrics extended classical spatial autoregressive models by allowing clusterwise varying coefficients and jointly estimating regression parameters and spatial clusters through penalized likelihood. For instance, \citet{MaranzanoEtAl2025} proposed a spatially-clustered spatial autoregression (SC-SAR) framework that generalizes SAR/SEM/SLM-type models, focusing on spatial heterogeneity in spillovers and covariate effects while enforcing spatial coherence of clusters via a neighborhood penalty.

The above contributions highlight two features that are particularly attractive for SAE modeling. First, the ability to represent non-smooth spatial heterogeneity through a parsimonious set of regimes; second, an estimation approach that is compatible with large sparse adjacency structures. Building on this idea, we develop a spatially-clustered Fay--Herriot (SC-FH) that targets piecewise-homogeneous spatial variation in regression effects: within each cluster, areas share the same coefficient vector and random effects variance, while different clusters capture distinct regimes. The Potts penalty encourages contiguous cluster membership for neighboring areas but does not enforce it strictly; boundaries are determined jointly by fit and spatial cohesion.

The SC-FH model sits in a middle ground between (i) independent-area FH models, which may under-utilize spatial information, and (ii) globally smoothed spatial SAE models, which may under-represent spatial regimes. The goal is to retain the SAE advantage of borrowing strength, while aligning the dependence structure with empirical spatial regimes by allowing cluster-specific regression relations and random effect variances.

Our development is also motivated by a concrete empirical setting, to which we return throughout the paper: the Po Valley in Northern Italy, one of the most productive and intensive agricultural regions in Europe and a recognized example of regime-like spatial heterogeneity. The fertile alluvial plain hosts a dense crop--livestock system whose economic performance, policy exposure, and environmental pressure have been extensively documented at the farm level \citep{BaldoniEtAl2017,CoderoniEtAl2018,CortignaniCoderoni2022,CoderoniVanino2022}, while the alpine and Apennine arcs that frame the plain host structurally different, more extensive farming systems; the divide is clearly visible in the Italian agricultural censuses and in the FADN accounts \citep{AltamoreEtAl2024,CardilloEtAl2023}. The livestock districts of the central--eastern plain concentrate ammonia emissions and secondary particulate formation, making the basin a European hotspot for agriculture-related air pollution \citep{Agrimonia2023,MarongiuEtAl2024,RodeschiniEtAl2024,ColomboEtAl2023}, and the exposure of the same territory to climatic stress has been dramatically exposed by the recent Po River droughts \citep{MontanariEtAl2023,CoderoniPagliacci2023}. Recently, survey-based direct estimates, farm-level economic indicators, and environmental layers for this territory were harmonized in the SCARFACE database \citep{MaranzanoScarface2026}, which provides both the geography on which our simulation experiments run and the data of the empirical application. In such a territory, a single linking model between farm output and production endowments is hardly credible --- the plain and the mountain arc operate under different technologies --- and the boundaries between regimes are themselves an object of inference: exactly the task the spatially-clustered FH model is designed for.

The rest of the paper is structured as follows. In Section~\ref{sec2} we present the spatially-clustered FH model, with cluster-specific regression coefficients and random effect variances estimated jointly with a spatially coherent partition of the areas: estimation is formulated as penalized Restricted Maximum Likelihood (REML) with a Potts-type cohesion term, solved by an efficient alternating optimization strategy that leverages closed-form FH updates within clusters. Section~\ref{sec3} tests the methodology through simulation experiments run on the real geography of the Po Valley, reflecting different degrees and drivers of spatial heterogeneity, and illustrates how the number of clusters and the spatial penalty affect cluster recovery, coefficient estimation, and prediction accuracy. Section~\ref{sec:compinf} collects the computational and inferential aspects of the procedure: algorithmic details, cluster-wise variance estimation, the data-driven selection of the number of clusters and of the spatial penalty, scaling and transformations, and bootstrap inference. This section is deliberately placed after the simulations; indeed the numerical validation of the estimation and inference routines is discussed there only briefly, referring back to simulation designs and results that the reader has already seen in Section~\ref{sec3}. Section~\ref{sec4} applies the model to the average standard output of farms in the Po Valley, and Section~\ref{sec5} concludes with final remarks and future research directions. Extended simulation and application results are collected in the online supplementary material.

\section{Spatially-clustered Fay--Herriot model}\label{sec2}
\subsection{Background: spatially-clustered regression with Potts-type penalties}
Let $\{(y_d,\mathbf{x}_d)\}_{d=1}^D$ denote a sample of $D$ georeferenced observations with $y_d(s_d)\in\mathbb{R}$ being the response variable for the $d$-th statistical unit at location $s_d$ and $\mathbf{x}_d(s_d)\in\mathbb{R}^P$ being the $P$-dimensional vector of covariates at location $s_d$. Clusterwise regression assumes the existence of $k=1,\dots,K$ latent clusters having size $D_k$ (with $D_k < D$) and such that for unit $d$ assigned to cluster $k$, the following relationship holds:
\begin{equation}
  y_{dk} = \mathbf{x}_{dk}^\top \boldsymbol{\theta}_k + \varepsilon_{dk}, \quad d=1,\dots,D_k \quad k=1,\dots,K
  \label{eq:cwr}
\end{equation}
with $\varepsilon_{dk} \sim N(0,\sigma^2_{\varepsilon k})$ being the cluster-specific errors, and $\boldsymbol{\theta}_k=[\theta_{1k}, \dots, \theta_{Pk}]'$ and $[\sigma^2_{\varepsilon 1}, \dots, \sigma^2_{\varepsilon K}]'$ the $P$-dimensional vector of unknown cluster-specific regression coefficients and the vector of cluster-specific error variances, respectively.

Given an initial clustering assignment of the units, we assume that the conditional density of $y_{dk}$ given $x_{dk}$, that is $f_d\left(y_{dk} \mid \mathbf{x}_{dk}; \boldsymbol{\theta}\right)$, is known and, therefore, the parameters within each cluster $k$ can be estimated by maximizing the following Gaussian within-cluster log-likelihood function
\begin{equation} \label{eq:func}
    \ell(\boldsymbol{\theta},\boldsymbol{\sigma}^2_\varepsilon,k) = \sum_{d=1}^{D} \log f_{dk} \left(y_d \mid \mathbf{x}_d, \boldsymbol{\theta}_{k},\boldsymbol{\sigma}^2_{\varepsilon k} \right)
\end{equation}
In practice the clustering structure is not known, thus a $k$-means-like algorithm can be employed to jointly estimate the cluster membership (i.e., maximizing the likelihood contribution of each unit at each potential clustered sub-model while keeping fixed all the other units) and the within-cluster coefficients iteratively \citep[e.g. see][]{desarbo1988maximum,spath1979algorithmus}.

Let $W=(w_{dj})$ be an adjacency (or proximity) weight matrix with elements $w_{dj}\in [0,1]$ encoding neighborhood relations among areas $d$ and $j$. When observations are georeferenced, it is often desirable that nearby units share the same cluster. To enforce spatial coherence of the latent partition, \cite{SugasawaMurakami2021} introduced a penalty that discourages label changes across edges of the adjacency graph by augmenting the log-likelihood in Equation (\ref{eq:func}) with a Potts-like penalty term \citep{potts1952some}:
\begin{equation}
  Q(\boldsymbol{\theta},\boldsymbol{\sigma}^2_\varepsilon,\boldsymbol{k})
  =
  \sum_{d=1}^D \log f_{dk_d}(y_d\mid \mathbf{x}_d;\boldsymbol{\theta}_{k_d},\sigma^2_{\varepsilon k_d}) + \phi \sum_{d<j} w_{dj}\, \mathbb{I}(k_d=k_j),
  \label{eq:potts_generic}
\end{equation}
where $\mathbb{I}(\cdot)$ is the indicator function and $k_d$ is the unknown cluster label representing the assignment of the $d$-th unit. The hyperparameter $\phi\ge 0$ controls the strength of spatial cohesion: large $\phi$ favors contiguous clusters and reduces isolated assignments, whereas $\phi=0$ reduces to non-spatial clusterwise regression. 


\subsection{The clusterwise Fay--Herriot model with spatial penalty}
Let us consider a finite population $U$ consisting of $N$ georeferenced units, which are distributed across $D$ disjoint small areas or domains indexed by $d=1,\ldots,D$. Let $y_d$ denote a direct estimate (e.g., Horvitz--Thompson) for a target unknown quantity of interest $\tau_d$ in area $d$; for instance, consider the task of estimating the true average value of a given variable of interest at the area-level.

In the standard FH model, the sampling model is
\begin{equation}
  y_d = \tau_d + e_d, \qquad e_d\sim N(0,\sigma^2_{ed}),
  \label{eq:sampling}
\end{equation}
with known sampling variances $\sigma^2_{e d}$, and the linking model is
\begin{equation}
  \tau_d = \mathbf{x}_d^\top \boldsymbol{\theta} + u_d, \qquad u_d\sim N(0,\sigma^2_u),
  \label{eq:linking}
\end{equation}
where $\mathbf{x}_d$ is a $P$-vector of covariates and $\boldsymbol{\theta}$ is a $(P+1) \times 1$ vector of unknown regression coefficients.
Combining \eqref{eq:sampling}--\eqref{eq:linking} yields the linear mixed model with area-level random effects:
\begin{equation}
  y_d = \mathbf{x}_d^\top \boldsymbol{\theta} + u_d + e_d.
  \label{eq:fh_lmm}
\end{equation}
The parameters $\boldsymbol{\theta}=[\theta_0,\theta_1,\ldots,\theta_P]'$ and $\sigma^2_u$ are typically estimated by Maximum Likelihood (ML) or Restricted Maximum Likelihood (REML). The empirical best linear unbiased predictor (EBLUP) of $\tau_d$ follows from best linear prediction with plug-in estimates \citep{RaoMol2015,MoralesSAE} and can be expressed as follows:
\begin{equation}\label{eq:EBLUP}
  \hat{\tau}_d = \frac{\hat{\sigma}^2_u}{\hat{\sigma}^2_u + \sigma^2_{ed}}y_d + \frac{\sigma^2_{ed}}{\hat{\sigma}^2_u + \sigma^2_{ed}}\mathbf{x}_d^\top\hat{\boldsymbol{\theta}},
\end{equation}
where $\hat{\boldsymbol{\theta}}$ and $\hat{\sigma}^2_u$ are consistent estimators of the regression coefficients and the random effects' variance, respectively.

To accommodate spatial heterogeneity in the regression relationship, we assume that each area $d$ belongs to one of the $K$ latent clusters, denoted by $k\in\{1,\ldots,K\}$.
Conditional on $k$, that is, assuming that unit $d$ belongs to the $k$-th cluster, we posit a cluster-specific FH model
\begin{equation}
  y_{dk} = \mathbf{x}_{dk}^\top \boldsymbol{\theta}_{k} + u_{dk} + e_{d},
  \qquad u_{dk} \sim N(0,\sigma^2_{u k}) \qquad e_d\sim N(0,\sigma^2_{e d}).
  \label{eq:scfh}
\end{equation}
Consequently, the parameters to be estimated become $\boldsymbol{\theta}=[\boldsymbol{\theta}_1,\ldots,\boldsymbol{\theta}_k,\ldots,\boldsymbol{\theta}_K]$ and $\boldsymbol{\sigma}^2_u = [\sigma^2_{u1},\ldots,\sigma^2_{uk},\ldots,\sigma^2_{uK}]$, with $\boldsymbol{\theta}_k=[\theta_{0k},\theta_{1k},\ldots,\theta_{Pk}]$. Compared to the standard FH model, Equation \eqref{eq:scfh} allows the regression coefficients (and, if desired, the random effect variance) to vary across clusters. This mirrors the clusterwise coefficient idea in spatially-clustered regression \citep{SugasawaMurakami2021} and in spatially-clustered spatial econometric models \citep{MaranzanoEtAl2025}, but tailored to the area-level SAE setting with known sampling variances.

Let $f_{dk}(y_d\mid \mathbf{x}_d;\boldsymbol{\theta}_k,\sigma^2_{u k})$ denote the marginal normal density implied by Equation \eqref{eq:scfh}. If cluster labels were observed, estimation would reduce to $K$ separate FH fits. However, the labels are unknown and must be inferred jointly with parameters. We estimate the unknown cluster-specific regression parameters $\{\boldsymbol{\theta}_k,\sigma^2_{u k}\}_{k=1}^K$ and the partition $\hat{\mathcal{P}}=(k_1,\ldots,k_D)$ of the $D$ small areas into $K$ groups with length $D_k$ for $k=1,\ldots,K$ by maximizing the following penalized log-likelihood
\begin{equation}
  Q(\boldsymbol{\theta},\boldsymbol{\sigma}^2_u,\hat{\mathcal{P}})
  =
  \sum_{d=1}^D \log f_{d k}(y_d\mid \mathbf{x}_d;\boldsymbol{\theta}_{k_d},\sigma^2_{u k_d}) + \phi \sum_{d<d'} w_{dd'}\, \mathbb{I}(k_d=k_{d'}),
  \label{eq:scfh_obj}
\end{equation}
where $d$ and $d'$ are the indices for two generic areas, and $w_{dd'}$ encodes area adjacency and $\phi\ge 0$ controls spatial cohesion. In particular, the penalty term couples the cluster labels across space and encourages spatially coherent regimes.

The objective \eqref{eq:scfh_obj} reduces to (non-spatial) clusterwise FH when $\phi=0$ and to a single global FH model when $K=1$.

\subsection{Estimation strategy: alternating optimization}
Direct maximization of Equation \eqref{eq:scfh_obj} is challenging because parameters and labels are intertwined. We adopt an alternating optimization scheme that iterates between (i) updating model parameters given labels, and (ii) updating labels given parameters, akin to EM-like procedures for spatially-clustered regression \citep{SugasawaMurakami2021} and clusterwise regression algorithms \citep{desarbo1988maximum}. The procedure operates as follows:
\begin{enumerate}
  \item \textbf{Parameter update (given labels $\mathcal{P}$).} Conditional on a current partition, for each cluster $k$, fit a FH model to the subset of areas $\{d: k_d=k\}$, obtaining updated $\hat{\boldsymbol{\theta}}_k$ and $\hat{\sigma}^2_{u k}$ by ML or REML and treating $\{y_d,\mathbf{x}_d,\sigma^2_{ed}\}$ for $d$ with $k_d=k$ as the data for that cluster. In the FH linear mixed formulation, $\hat{\boldsymbol{\theta}}_k$ can be updated by generalized least squares given $\hat{\sigma}^2_{u k}$, and $\hat{\sigma}^2_{u k}$ can be updated by Fisher scoring \citep{MoralesSAE}.
  In our implementation, $\hat{\sigma}^2_{u k}$ maximizes the profile restricted log-likelihood through one of the three numerically equivalent updates described in \citet[Sect.~16.5.4]{MoralesSAE} --- Fisher scoring on the REML score, its residual-based variant, or a fixed-point iteration on the estimated random effects --- embedded in a safeguarded scheme that projects each update on the admissible space $\sigma^2_{u k}\geq 0$ and enforces monotone ascent of the restricted likelihood via step halving. To prevent the boundary degeneracies discussed in Section~\ref{sec:boundary}, variance components are estimated by adjusted REML by default. Implementation details are collected in Appendix~\ref{app:computational}.
  \item \textbf{Label update (given parameters $\boldsymbol{\theta},\boldsymbol{\sigma}^2_u$).} Given the current values of the parameters, each area $d$ has a cluster-specific log-likelihood contribution $\ell_{dk}=\log f_{dk}(y_d\mid \mathbf{x}_d;\boldsymbol{\theta}_k,\sigma^2_{uk})$. Then, for each area $d$, we update $k_d$ by maximizing the local contribution of Equation \eqref{eq:scfh_obj} while holding other labels fixed:
  \[
    k_d \leftarrow \arg\max_{k\in\{1,\ldots,K\}}
      \left\{
        \log f_{d k}(y_d\mid \mathbf{x}_d;\boldsymbol{\theta}_{k},\sigma^2_{u k}) + \phi \sum_{d':\, w_{dd'}=1}\mathbb{I}(k=k_{d'})
      \right\}.
  \]
  This step balances model fit with agreement with neighboring labels.
  Labels are updated sequentially --- one area at a time, in the spirit of iterated conditional modes or ICM \citep{Besag1986} --- so that each update is conditional on the current labels of all other areas and the penalized objective \eqref{eq:scfh_obj} cannot decrease within a sweep. Compared with a simultaneous update of all labels, the sequential scheme rules out cyclic label configurations.
\end{enumerate}
The two steps are iterated until $Q$ reaches convergence or until labels stabilize. Operationally, the algorithm stops when the partition is unchanged between consecutive iterations or when the improvement of $Q$ falls below a fixed tolerance. As an additional safeguard, visited partitions are tracked and, if any configuration reappears (a cycle), the best-scoring visited partition is returned.



\section{Simulation experiments}\label{sec3}

\subsection{Design and evaluation metrics}\label{sec3:design}
\begin{figure}[!htb]
  \centering
  \includegraphics[width=0.99\linewidth]{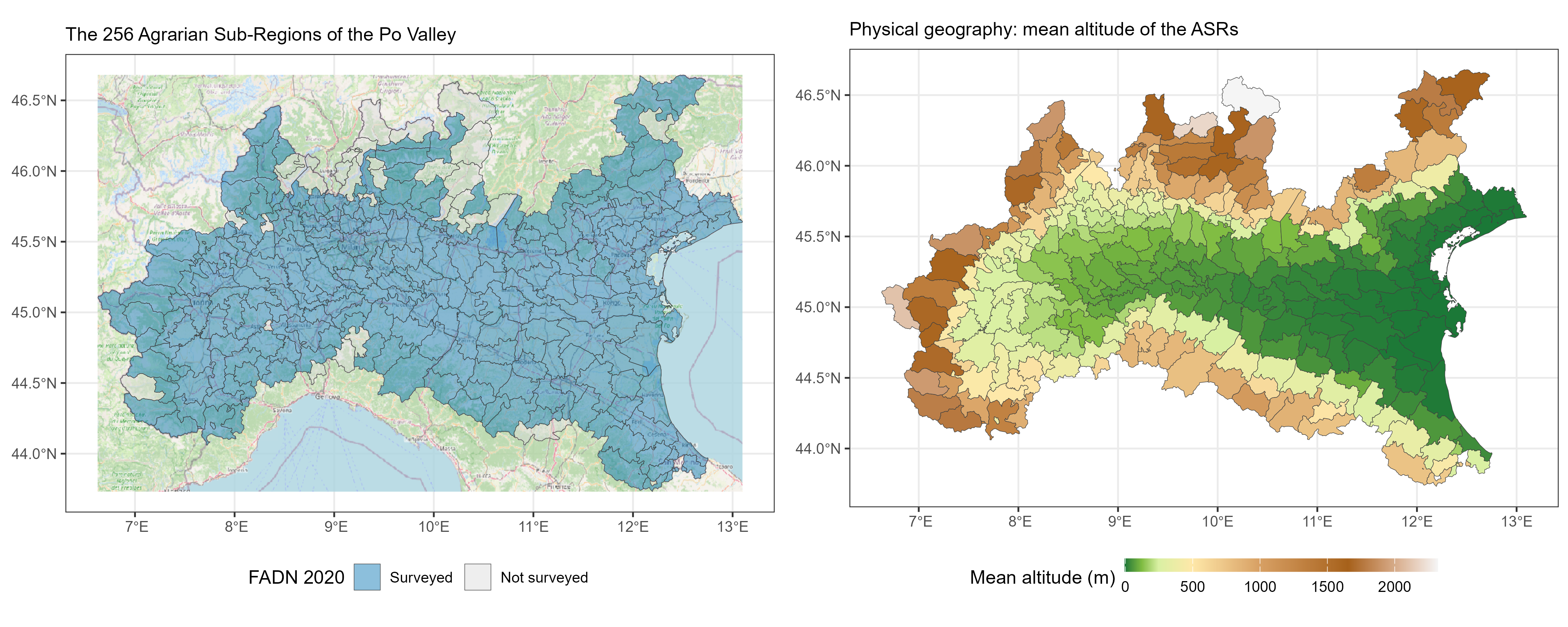}
  \caption{The Po Valley in Northern Italy. Left: the partition into $D=256$ Agrarian Sub-Regions, with the areas surveyed by FADN in 2020 highlighted. Right: mean altitude of the ASRs, showing the alluvial plain framed by the alpine and Apennine reliefs.}
  \label{fig:data_map}
\end{figure}

The simulation study investigates the finite-sample performance of the proposed spatially-clustered Fay--Herriot model under different sources of spatial heterogeneity and varying degrees of cluster separability. All the experiments run on the real geography of the empirical application, namely the $D=256$ Agrarian Sub-Regions (ASRs) partitioning the Po Valley in Northern Italy, shown in Figure~\ref{fig:data_map} together with their physical geography --- a wide alluvial plain framed by the alpine arc to the north and west and by the Apennine ridge to the south. This orographic structure makes the simulation designs geographically meaningful: the true regimes are defined on the actual map, either as longitudinal bands or as altitude classes, so they form contiguous blocks whose shape follows the real terrain.

We simulate data according to the following intercept-plus-one-covariate area-level model with known sampling variances, which are heterogeneous across clusters, mimicking realistic survey settings. In particular, the model can be written as follows
\begin{equation}\label{eq:sim_dgp}
\hat{\tau}^{*}_{dk} = \beta_{0k} + \beta_{1k} x^{*}_{dk} + u^*_{dk} + e^*_{dk},
\end{equation}
where $x^{*}_{dk} \sim N(0,\sigma_X^2)$ with $\sigma_X^2=4$, $u^*_{dk} \sim N(0,\sigma^2_{uk})$ denotes the area-level random effect, and $e^*_{dk} \sim N(0,\sigma^2_{ek})$ represents the sampling error with known variance.

Variance components are chosen to target a decomposition of the share of explained variability\footnote{We adopt the definition of marginal coefficient of determination for each component as in \cite{NakagawaSchielzeth2013} and \cite{NakagawaJohnsonSchielzeth2017}, that is, we compute the proportion of the total variance explained by the fixed effects ($R^2_{fe}$), by the random effects ($R^2_{re}$) and by the sampling errors ($R^2_{se}$).} $R^2_{fe}+R^2_{re}+R^2_{se}=1$ across fixed effects ($fe$), random effects ($re$), and sampling error ($se$), with ``clearly separated'' and ``poorly separated'' regimes corresponding to a higher or a lower fixed-effects signal. The spatial dependence of the simulated data arises from the regimes themselves: the true groups are defined on the actual Po Valley geography, so they form contiguous blocks and the response inherits a spatially structured mean surface, while the covariate, the random effects, and the sampling errors are drawn independently across areas. The full simulation procedure is summarized in Algorithm \ref{AlgSims}.

\begin{algorithm}
\caption{Simulation study for the Spatially-Clustered Fay-Herriot}
\label{AlgSims}
\begin{algorithmic}[]
\State Set $K=3$: number of true clusters
\State \,\,\,\,\,\,\,\,\,\, $\boldsymbol{\beta}_{0k} = [50,75,100]$ for $k=1,2,3$: group-specific intercept
\State \,\,\,\,\,\,\,\,\,\, $\boldsymbol{\beta}_{1k} = [-5,2,10]$ for $k=1,2,3$: group-specific regression coefficients
\State \,\,\,\,\,\,\,\,\,\, $\rho_{XY}=0.95$ (``clearly separated'') or $\rho_{XY}=0.50$ (``poorly separated''): correlation among the predictor and the direct estimate
\State \,\,\,\,\,\,\,\,\,\, $D=256$: number of small areas of interest
\State \,\,\,\,\,\,\,\,\,\, $R^2_{fe}, R^2_{re}, \text{ and } R^2_{se}$ according to the ``clearly separated'' and ``poorly separated''
\State \,\,\,\,\,\,\,\,\,\,\,\,\,\,\,\,\,\,\,\,\,\,\,\, regimes while respecting the condition $R^2_{fe}+R^2_{re}+R^2_{se}=1$, that is,
\State \,\,\,\,\,\,\,\,\,\,\,\,\,\,\,\,\,\,\,\,\,\,\,\,\,\,\,\,\,\,\,\,\,\, ``clearly separated'': $R^2_{fe}=0.90$, $R^2_{re}=0.07$ and $R^2_{se}=0.03$
  \State \,\,\,\,\,\,\,\,\,\,\,\,\,\,\,\,\,\,\,\,\,\,\,\,\,\,\,\,\,\,\,\,\,\, ``poorly separated'': $R^2_{fe}=0.25$, $R^2_{re}=0.30$ and $R^2_{se}=0.45$
    \State \,\,\,\,\,\,\,\,\,\, $\sigma_X^2=4$: variance of the predictor variable
    \State Compute $\sigma^2_{uk}$ and $\sigma^2_{ek}$ based on $R^2_{fe},R^2_{re},R^2_{se},\rho_{XY}\text{ and }\sigma_X^2$
    \State Load (or simulate) the geometries of the $D$ areas of interest
    
    \For{$i=1,\ldots,M=1000$}
    \State Partition the $D$ small areas into three groups with length $D_k \quad \forall k=1,2,3$ according
    \State \,\,\,\,\,\,\,\,\,\,\,\,\,\,\,\, one of the scenarios of interest
    \For{$k=1,2,3$}
    \State Simulate the $D_k\times1$ vector of covariates as $x^{*}_{dk} \sim N(0,\sigma_X^2)$
      \State Simulate the $D_k\times1$ vector of sampling errors as $e^*_{dk} \sim N(0,\sigma^2_{edk})$
      \State Simulate the $D_k\times1$ vector of random effects as $u^*_{dk} \sim N(0,\sigma^2_{uk})$
      \State Simulate the $D_k\times1$ vector of response variable as $\hat{\tau}^{*}_{dk} = \beta_{0k} + \beta_{1k} x^{*}_{dk} + u^*_{dk} + e^*_{dk}$  
        \EndFor
      \State Stack the vectors $\hat{\tau}^{*}_{dk}$, $x^{*}_{dk}$, $u^*_{dk}$, and $e^*_{dk}$ across the $K=3$ groups
      \State Run the SC-FH algorithm using $\left(\hat{\tau}^{*}_{dk},x^{*}_{dk},\sigma^2_{ed}\right)'$
      \State Store the output of interest for the generic iteration $i$
    \EndFor
\end{algorithmic}
\end{algorithm}

\noindent We assume $K=3$ clusters, with cluster-specific regression coefficients fixed at $\beta_{0k} = \{50,75,100\}$ and $\beta_{1k} = \{-5,2,10\}$ for $k=1,2,3$. Given the $K=3$ clusters, we analyze the following four distinct scenarios, differing in the spatial driver of clustering and in the degree of separation between clusters.
\begin{enumerate}[label=Scenario \arabic*]
    \item \textbf{Longitude-based clusters with clear separation.} Clusters are defined along the longitudinal dimension, generating spatially contiguous groups with clearly separated regression coefficients. Clear separation is induced by imposing a high goodness-of-fit of the fixed-effects component, set to $R^2_{fe}=0.90$, resulting in limited overlap between clusters in the response--covariate space.
    \item \textbf{Longitude-based clusters with poor separation.} The spatial structure is identical to Scenario~1, but clusters are poorly separated in the variable space. This is achieved by reducing the explanatory power of the fixed-effects component to $R^2_{fe}=0.25$, leading to substantial overlap across clusters and increased fuzziness in cluster assignment. 
    \item \textbf{Altitude-based clusters with clear separation.} Clusters are defined according to altitude, inducing a spatial partition that differs from the longitudinal ordering. As in Scenario~1, coefficients are clearly separated by imposing $R^2_{fe}=0.90$, yielding well-defined spatial regimes aligned with elevation patterns. 
    \item \textbf{Altitude-based clusters with poor separation.} This scenario combines altitude-driven clustering with poorly separated coefficients ($R^2_{fe}=0.25$). This represents the most challenging setting, featuring both a complex spatial structure and substantial overlap in the covariate space.
\end{enumerate}

\noindent Scenarios~1 and~2 reproduce settings in which spatial clustering is either aligned with a simple geographical ordering or characterized by limited empirical separability, yielding results that mirror, respectively, ideal and weak-signal cases. By contrast, Scenarios~3 and~4 are based on altitude-driven spatial clustering, which better reflects realistic spatial partitioning mechanisms. Moreover, they span two complementary regimes of practical relevance: a setting with clearly identifiable spatial clusters and a more challenging configuration with weak cluster separation. Together, these scenarios provide a comprehensive assessment of the proposed method under conditions that are both empirically plausible and methodologically demanding, while maintaining coherence with the results obtained across all simulation designs.

\begin{figure}[!htb]
  \centering
  \includegraphics[width=0.95\linewidth]{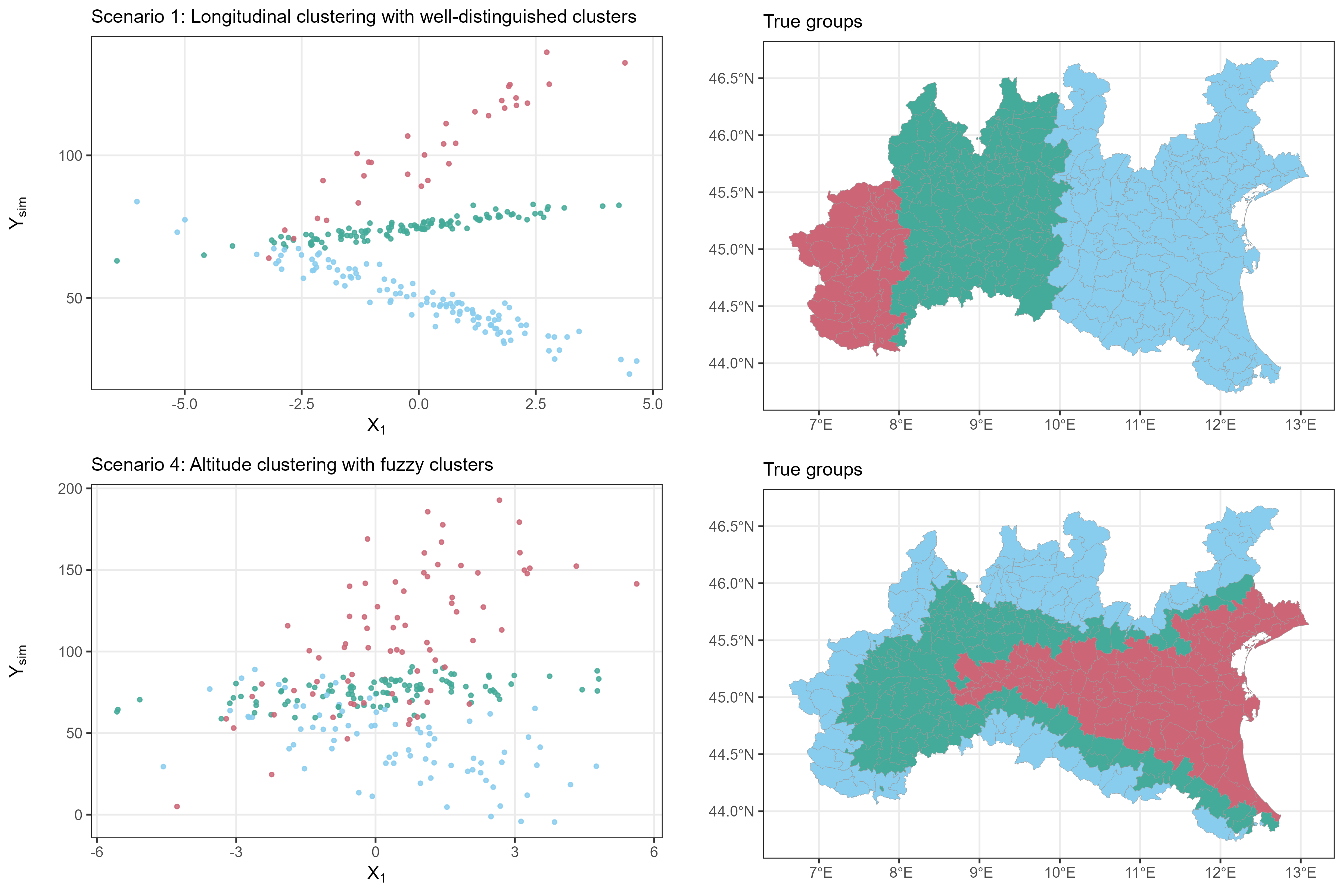}
  \caption{Examples of simulated data according to Scenario 1 (upper plots) and Scenario 4 (bottom plots).}
  \label{fig:Scenario14}
\end{figure}

Figure \ref{fig:Scenario14} illustrates examples of simulated data according to Scenario 1 (upper plots) and Scenario 4 (bottom plots), that is, longitude-based clustering under the clear-separation regime and altitude-based clustering under the poor-separation regime. In addition, Figures from \ref{fig:sim_scen1} to \ref{fig:sim_scen4} in Appendix \ref{AppendixA} display representative Monte Carlo replications for the four scenarios, highlighting the spatial distribution of the latent process under different clustering mechanisms.

For each scenario, we consider $M=1{,}000$ Monte Carlo replications and evaluate 45 combinations of hyperparameters, obtained by varying the number of clusters ($K=1,\dots,5$) and the spatial penalty parameter ($\phi \in [0,2]$ with step $0.25$). In all fits, the cluster-wise variance components are estimated by adjusted REML, labels are updated by the sequential scheme of Section~\ref{sec2}, and the initial partition is obtained by $k$-means on the auxiliary covariate with multiple restarts and a minimum-cluster-size admissibility filter. Simulation results are compared according to:
\begin{itemize}
  \item \textbf{Clustering accuracy:} measured, for each replication, by (i) the Adjusted Rand Index (ARI) between the estimated and the true partition, which is invariant to label switching by construction, and (ii) the share of correctly assigned areas after relocating the estimated labels to the true ones by a majority rule on the confusion matrix;
  \item \textbf{Parameter recovery:} Monte Carlo bias and standard deviation of $\widehat{\beta}_{1k}$ and $\widehat{\sigma}^2_{u k}$ over the $M$ replications, after majority-rule relocation of the cluster labels;
  \item \textbf{Prediction accuracy:} for each replication, the area-level root mean squared error of the EBLUP against the true conditional mean $\mu_d$; we report its Monte Carlo mean and standard deviation and its ratio to the pooled FH benchmark ($K=1$, the standard Fay--Herriot model) fitted on the same data, so that values below one quantify the gain from clustering.
\end{itemize}

\subsection{Results for the baseline scenarios}\label{sec3:results}
\begin{figure}[!htb]
  \centering
  \begin{subfigure}{0.93\linewidth}
    \includegraphics[width=\linewidth]{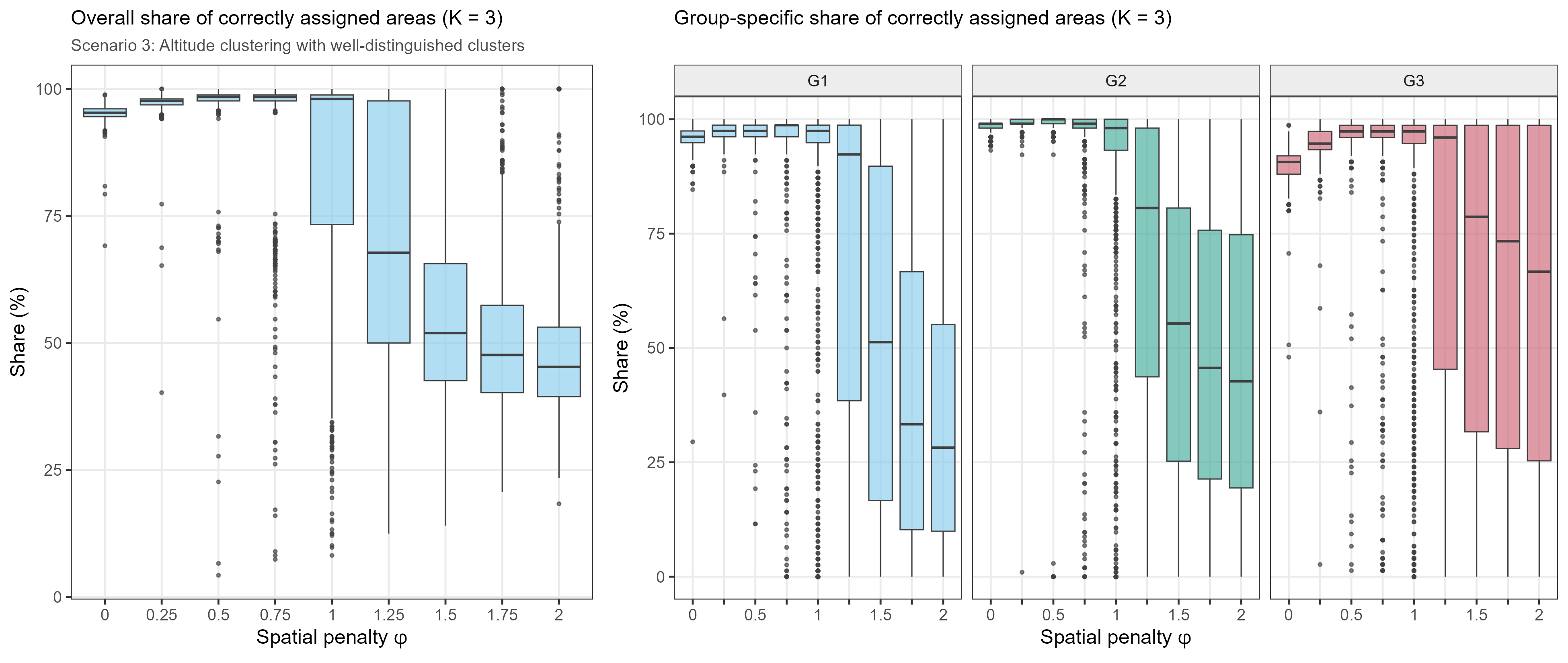}
    \caption{Scenario 3}
  \end{subfigure}\\[2pt]
  \begin{subfigure}{0.93\linewidth}
    \includegraphics[width=\linewidth]{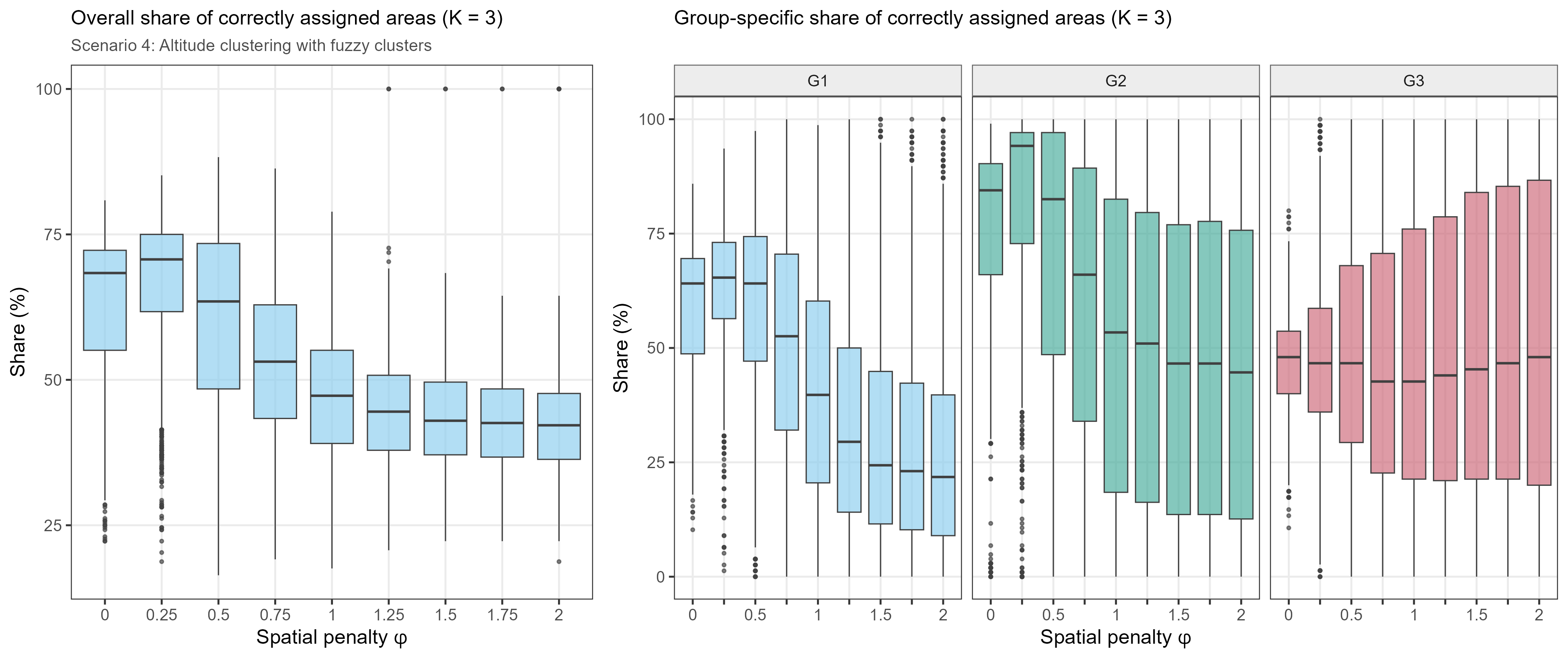}
    \caption{Scenario 4}
  \end{subfigure}
  \caption{Overall (left) and group-specific (right) share of correctly assigned areas at the true $K=3$, by spatial penalty $\phi$, across the $1{,}000$ MC replications. Top: Scenario~3; bottom: Scenario~4.}
  \label{fig:sim_s34_share}
\end{figure}

\begin{figure}[!htb]
  \centering
  \begin{subfigure}{0.93\linewidth}
    \includegraphics[width=\linewidth]{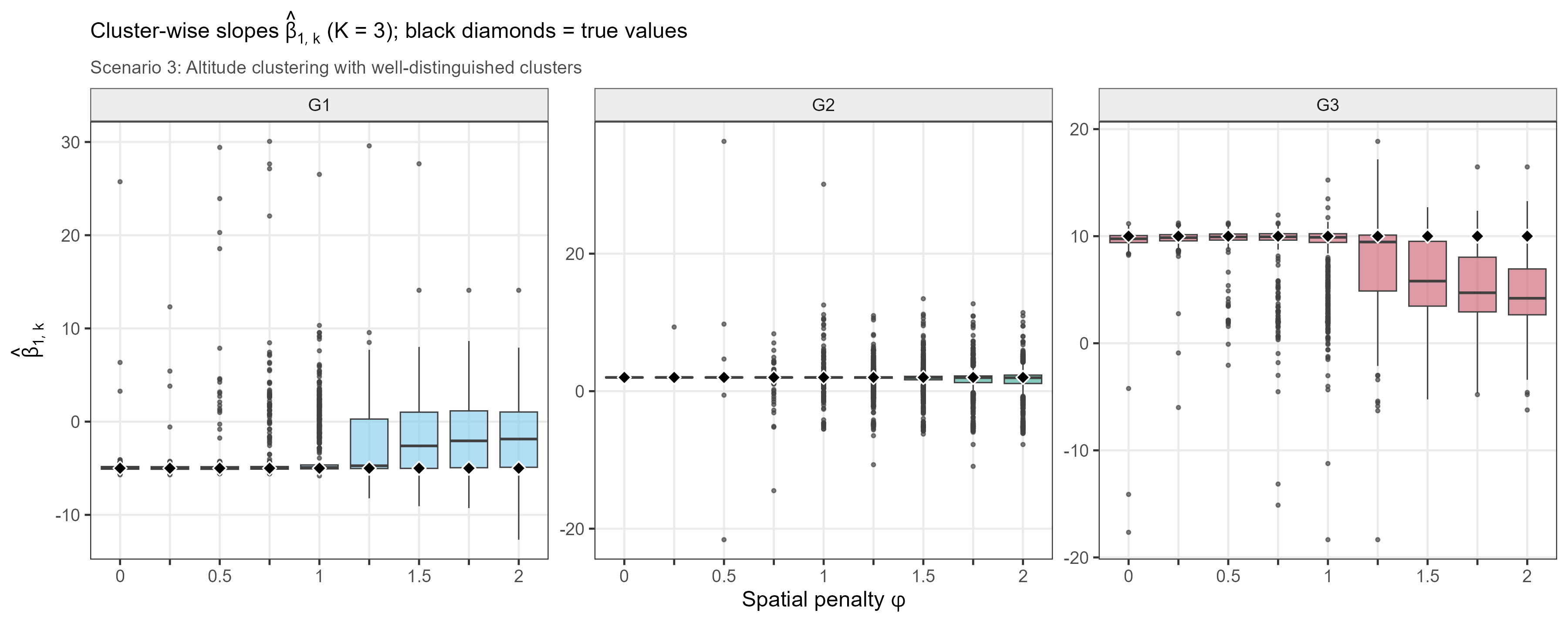}
    \caption{Scenario 3}
  \end{subfigure}\\[2pt]
  \begin{subfigure}{0.93\linewidth}
    \includegraphics[width=\linewidth]{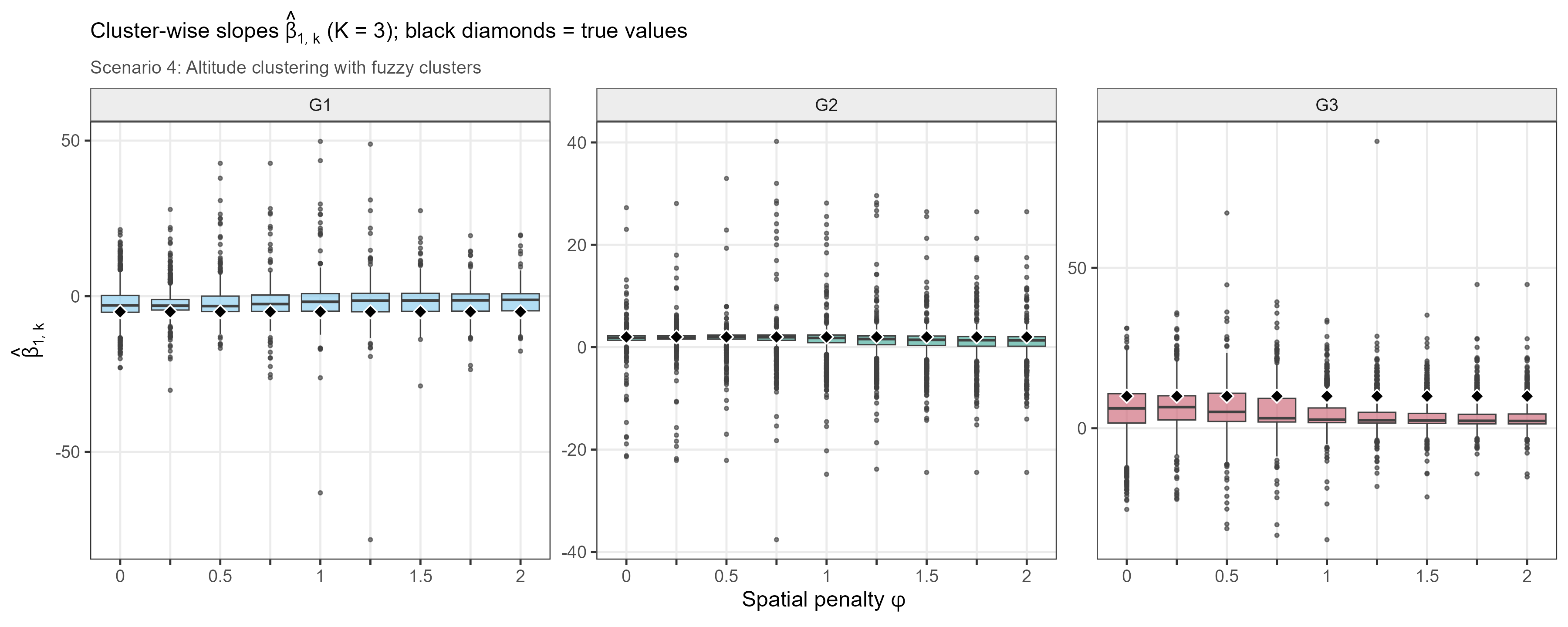}
    \caption{Scenario 4}
  \end{subfigure}
  \caption{Cluster-specific slope estimates at the true $K=3$, by spatial penalty $\phi$, across the $1{,}000$ MC replications. Black diamonds mark the true values $\beta_{1k}=(-5,2,10)$. Top: Scenario~3; bottom: Scenario~4.}
  \label{fig:sim_s34_beta}
\end{figure}

\begin{figure}[!htb]
  \centering
  \begin{subfigure}{0.93\linewidth}
    \includegraphics[width=\linewidth]{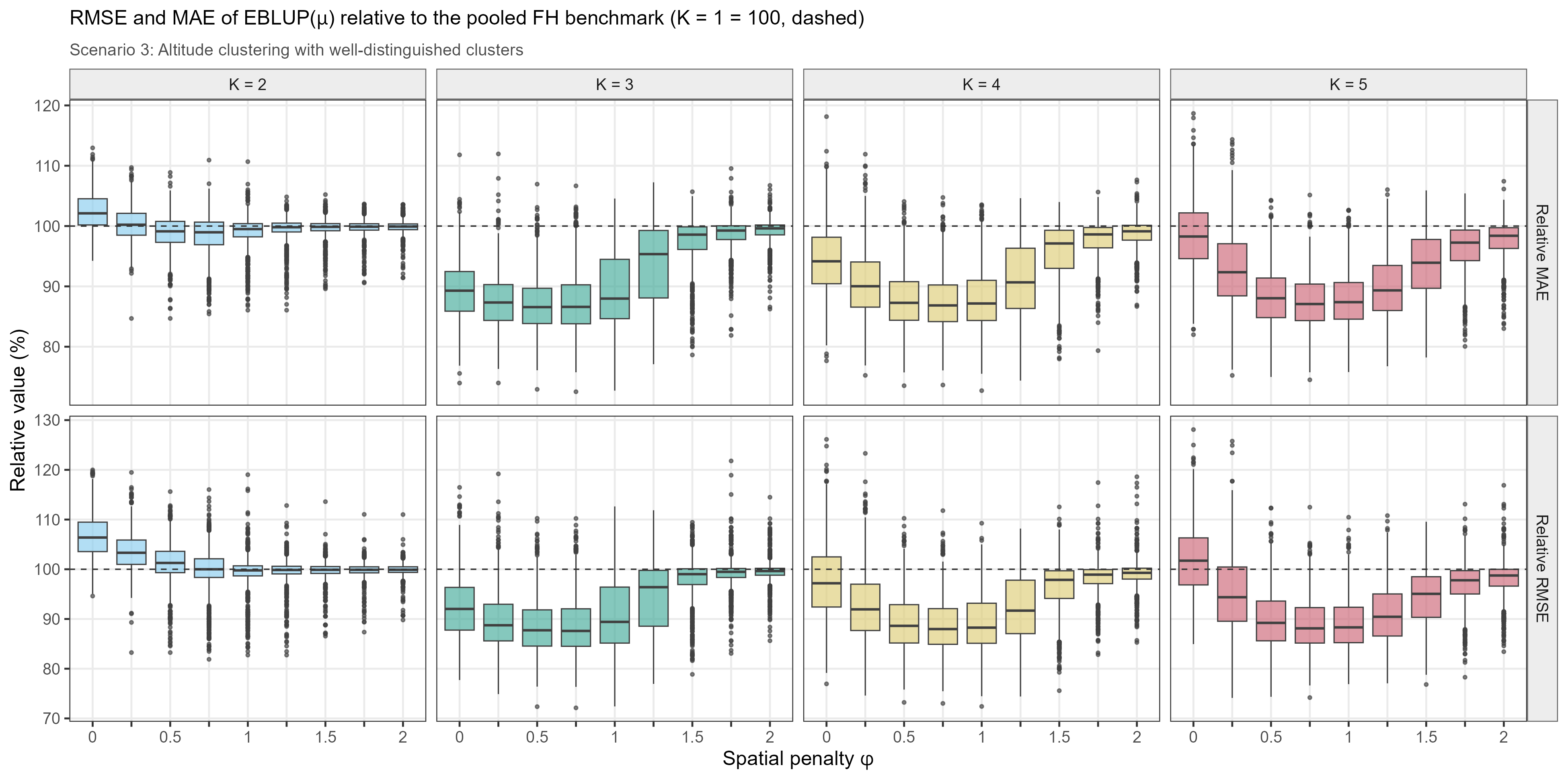}
    \caption{Scenario 3}
  \end{subfigure}\\[2pt]
  \begin{subfigure}{0.93\linewidth}
    \includegraphics[width=\linewidth]{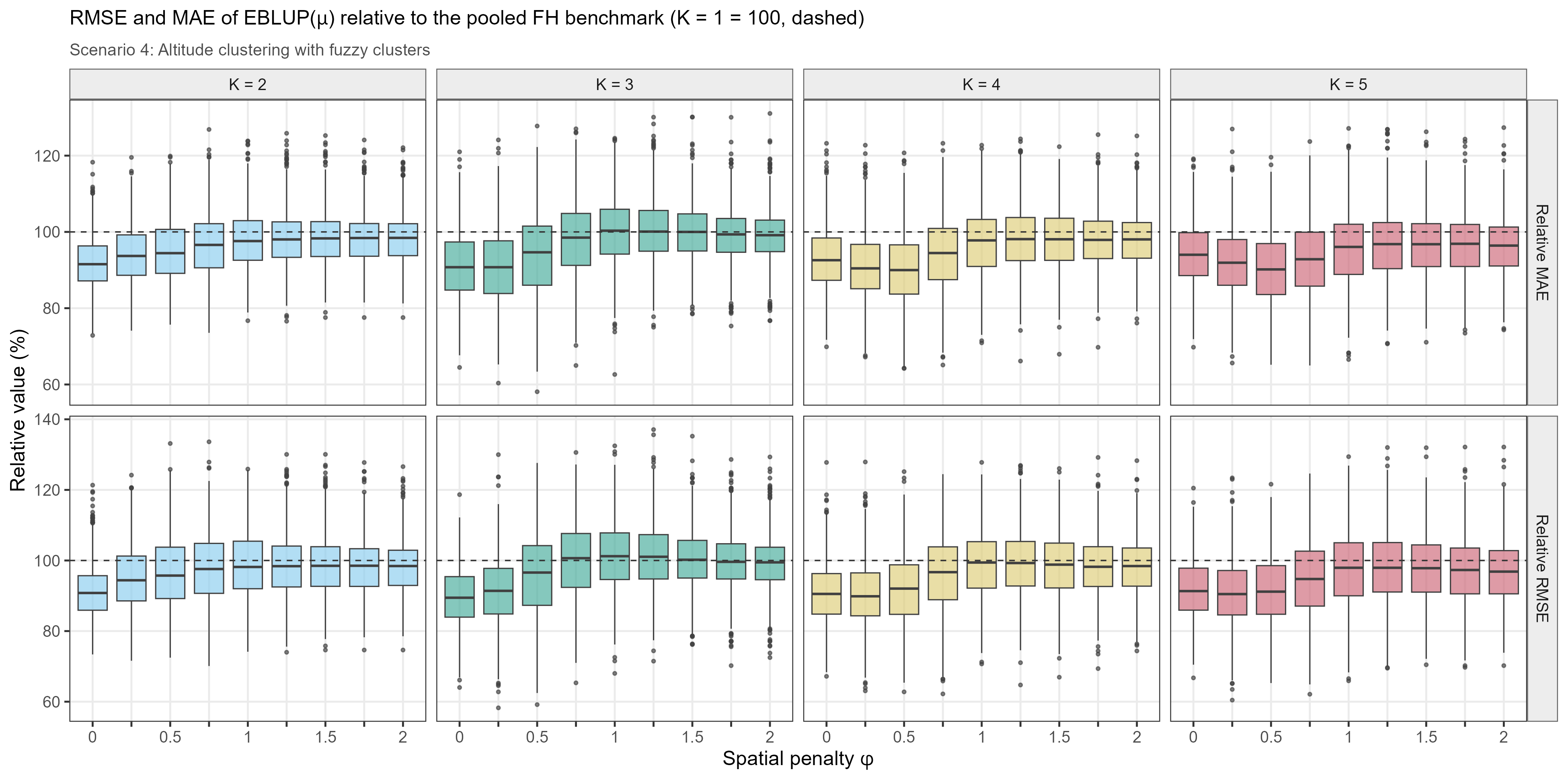}
    \caption{Scenario 4}
  \end{subfigure}
  \caption{RMSE of the EBLUP for the true signal, relative to the pooled FH benchmark ($K=1$, dashed line at $100\%$), by number of clusters $K$ and spatial penalty $\phi$, across the $1{,}000$ MC replications. Top: Scenario~3; bottom: Scenario~4.}
  \label{fig:sim_s34_pred}
\end{figure}

\begin{figure}[!htb]
  \centering
  \begin{subfigure}{0.93\linewidth}
    \includegraphics[width=\linewidth]{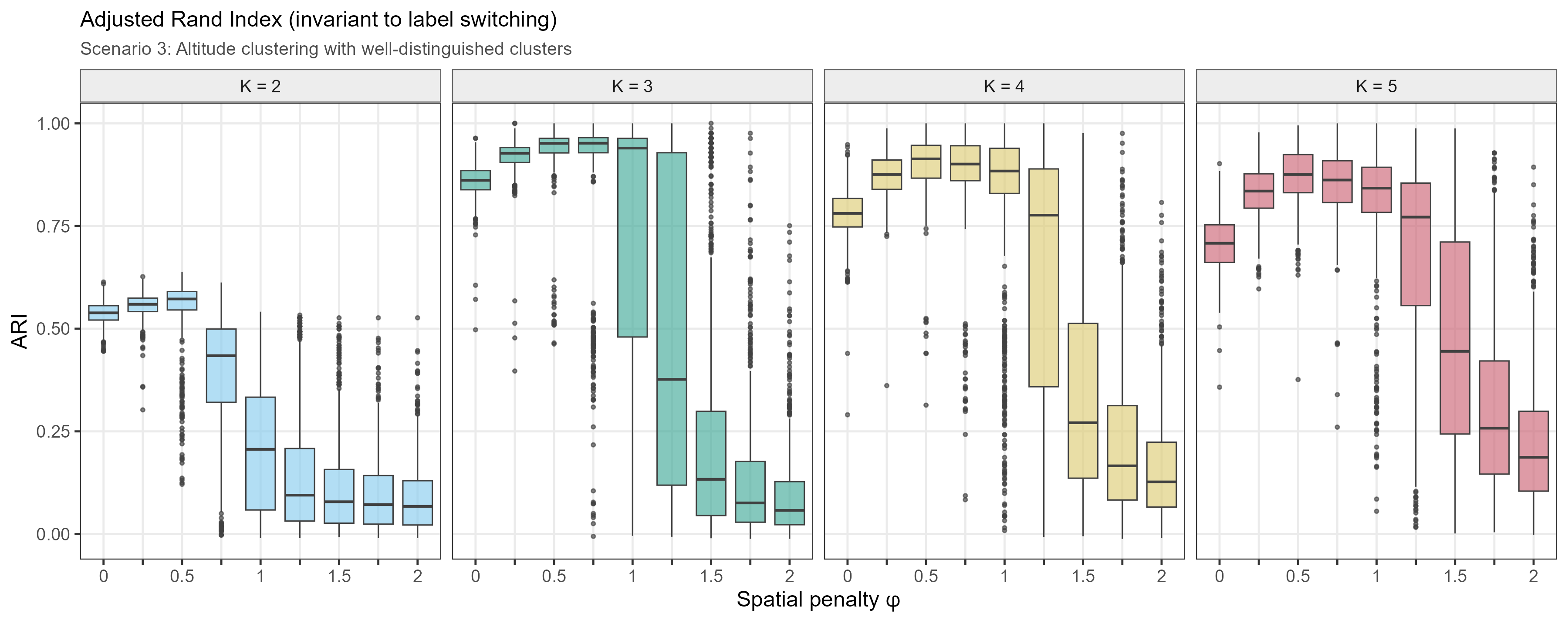}
    \caption{Scenario 3}
  \end{subfigure}\\[2pt]
  \begin{subfigure}{0.93\linewidth}
    \includegraphics[width=\linewidth]{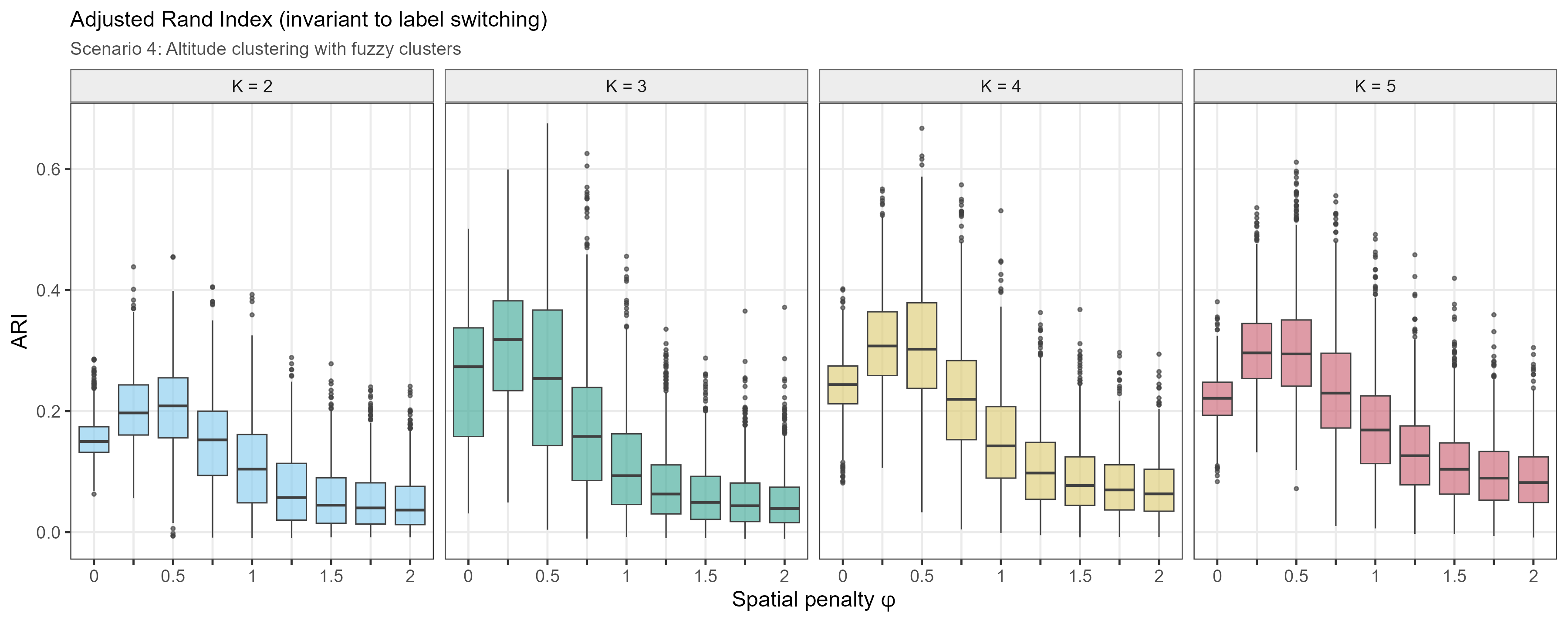}
    \caption{Scenario 4}
  \end{subfigure}
  \caption{Adjusted Rand Index between the estimated and the true partition, by number of clusters $K$ and spatial penalty $\phi$, across the $1{,}000$ MC replications. Top: Scenario~3; bottom: Scenario~4.}
  \label{fig:sim_s34_ari}
\end{figure}

For reasons of space, the discussion in the main text focuses on the simulation results
corresponding to Scenarios~3 and~4. Along with space constraints, this choice is also motivated by two arguments. First, we anticipate that the remaining Scenarios~1 and~2 deliver qualitatively consistent with those of Scenarios~3 and~4 evidence and are therefore reported in the Appendix. Second, we remark that Scenarios~3 and~4 are the most informative from a methodological standpoint, as they combine a realistic spatial partitioning mechanism with two complementary regimes of signal strength. Unlike longitude-based clustering, altitude-driven clusters generate spatial structures that are irregular, non-monotone, and not trivially ordered, closely resembling the type of latent spatial heterogeneity encountered in applied small area estimation problems. The results for Scenarios~3 and~4 are illustrated side by side in Figures~\ref{fig:sim_s34_share} (share of correct assignments), \ref{fig:sim_s34_beta} (recovery of the cluster-wise slopes), \ref{fig:sim_s34_pred} (prediction accuracy relative to the pooled FH benchmark) and \ref{fig:sim_s34_ari} (ARI), with Scenario~3 in the top panels and Scenario~4 in the bottom panels. The extended results for all the considered scenarios are provided in Sections~1 and ~2 of the Supplementary Material.

Scenario~3 shows the algorithm at its best along all three dimensions of interest. In terms of classification, when $K=3$ the average share of correctly assigned areas is $95.2\%$ (ARI $0.86$) already without spatial penalization, and rises to $97.4\%$ (ARI $0.94$) at $\phi=0.5$ (Figure~\ref{fig:sim_s34_share}); the group-specific panels show that no regime is systematically sacrificed. In terms of parameter recovery, at $\phi \leq 0.5$ the cluster-wise slopes are estimated with negligible bias --- the largest absolute bias across the three clusters is $0.32$ at $\phi=0$ and $0.18$ at $\phi=0.25$, against true values $\beta_{1k}=(-5,2,10)$ --- and the same holds for the random effect variances, whose Monte Carlo distributions are centered on the true values (see Figure~\ref{fig:sim_s34_beta} and supplementary tables).

Moderate penalization also acts as a variance-reduction device: at $\phi=0.25$ the Monte Carlo standard deviation of the slope estimators in the two extreme regimes is $30$--$40\%$ smaller than its unpenalized value ($0.74$ against $1.09$, and $0.78$ against $1.31$), because the penalty removes the sporadic misassignments that contaminate the cluster-wise fits.

Finally, the gains translate directly into signal reconstruction (Figure~\ref{fig:sim_s34_pred}). The RMSE of the clusterwise EBLUP against the true simulated signal is $92.6\%$ of the pooled FH benchmark at $\phi=0$ and falls to $88.4\%$ at $\phi=0.5$ (i.e., an average accuracy gain of about $12\%$), stable across replications and mirrored by the MAE.

The right-hand side of every panel documents the cost of over-penalization: beyond $\phi \approx 1.25$ contiguous regimes are merged, the extreme slopes shrink toward each other (bias up to $+3.3$ and $-5.1$ at $\phi=2$), the variance components inflate to absorb the unmodeled heterogeneity, and the predictive advantage over the benchmark disappears.

Scenario~4 exposes the natural limit of the approach. The endogenous partition is driven by the covariate--response relationship, so when the fixed-effects signal is weak ($R^2_{fe}=0.25$) the true regimes can no longer be reconstructed reliably. The ARI at the true $K$ peaks at a modest $0.31$ ($\phi=0.25$) and, in contrast with the spatially regular Scenario~2 discussed below, spatial penalization cannot substitute for the missing signal: beyond $\phi \approx 0.75$ it becomes actively harmful, as label smoothing consolidates wrong assignments instead of correcting them (see Figure~\ref{fig:sim_s34_ari}). The method nevertheless degrades gracefully rather than failing. Parameter recovery weakens but preserves the regime structure: the slope of the central regime remains essentially unbiased (bias $-0.18$ at $\phi=0.25$), while the two extreme slopes are attenuated toward the center by $2.6$--$3.1$ units, about $20\%$ of the true range, and the ordering of the three regimes is preserved in virtually every replication. Signal reconstruction also remains competitive: at $\phi \leq 0.25$ the clusterwise EBLUP still improves on the pooled benchmark by about $10\%$ (relative RMSE $89.6\%$ at $\phi=0$), because even a noisy partition adapts the fit locally while the EBLUP shrinkage absorbs part of the misclassification noise. However, we stress that Scenario~4 is the only design in which the relative RMSE crosses the $100\%$ line (for $\phi$ between $1$ and $1.5$), so under weak covariate signal the spatial penalty should be kept minimal.

\subsection{Additional scenarios}\label{sec3:additionalscenarios}
Two additional simulation designs stress the clustering mechanism in opposite directions: a \emph{perfect separation} design (Scenario~5), in which the cluster-specific slopes are close ($\beta_{1k}=(-2,1,4)$) and the regimes differ mainly in level, and a \emph{high variability} design (Scenario~6) with widely spread coefficients ($\beta_{0k}=(100,150,200)$, $\beta_{1k}=(-20,10,40)$). Under perfect separation the partition is recovered essentially exactly even without spatial penalization (ARI $0.99$ at $\phi=0$) and the prediction gain over the pooled benchmark is the largest of the whole study (relative RMSE $78.0\%$). This means that level-separated regimes are the most favorable configuration for the covariate-based initialization, and the penalty can only preserve (but not improve) an already perfect classification. Under high variability setup the unpenalized fit is markedly noisier (ARI $0.68$), because widely spread coefficients amplify the leverage of individual areas on the cluster-wise fits, and the spatial penalty yields its largest marginal benefit (at $\phi=0.5$ the ARI is restored to $0.94$ and the relative RMSE falls to $87.5\%$). Both designs otherwise reproduce the qualitative patterns of the baseline scenarios, including the over-penalization decay. Readers are referred to figures and tables in Section S3 of the Supplementary Material for scenario-specific results.

\subsection{A synthesis of the simulation experiments}\label{sec3:synthesis}
\begin{figure}[!htb]
  \centering
  \includegraphics[width=0.98\linewidth]{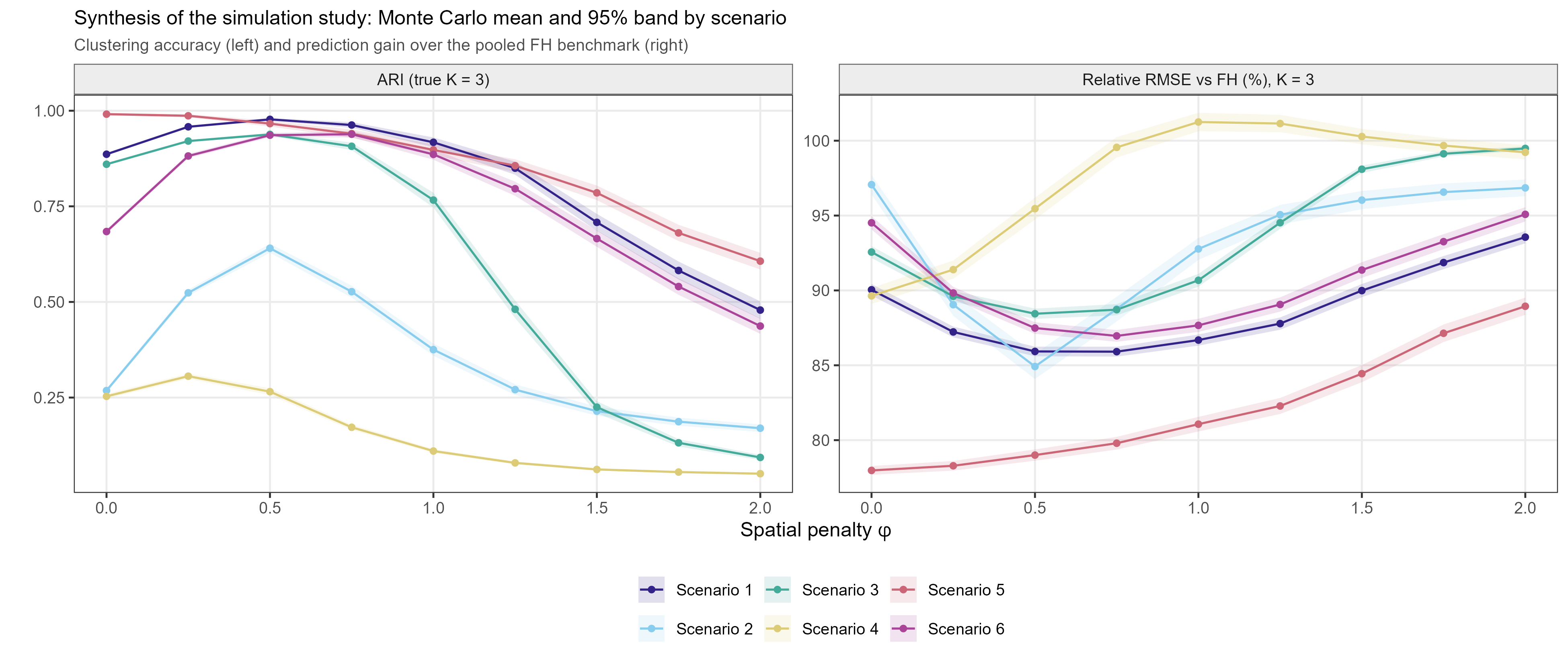}
  \caption{Synthesis of the simulation study: Monte Carlo mean (lines) and 95\% confidence bands (shaded) of the Adjusted Rand Index (left panel) and of the relative RMSE with respect to the pooled Fay--Herriot benchmark (right panel, in \%), at the true $K=3$, as functions of the spatial penalty $\phi$, for all simulation designs.}
  \label{fig:sims_summary}
\end{figure}

Taken together, the six designs deliver three take-home messages. First, whenever the regimes are separated in the covariate--response space (i.e., the clear-separation Scenarios~1 and~3 and the two additional designs of Section~\ref{sec3:additionalscenarios}) the partition is recovered almost exactly at moderate penalties and the cluster-wise parameters are estimated with negligible bias. Second, the spatial penalty is primarily a stabilizer: at moderate values it overrules isolated misassignments through the neighborhood consensus, sharply reducing the Monte Carlo dispersion of both the classification rates and the cluster-wise estimators, whereas pushing it too far merges genuine regimes and degrades every metric. Third, under a weak covariate signal the penalty complements but cannot replace the missing information: it rescues the classification only where the regimes are spatially coherent (i.e., the wide longitudinal bands of Scenario~2), not where they are fragmented (i.e., the altitude-driven Scenario~4). Figure~\ref{fig:sims_summary} and Table~\ref{tab:sims_summary} quantify this picture across all designs. The moderate range $\phi \in [0.25,0.75]$ drives the relative RMSE down to between $78\%$ and $88\%$ of the pooled Fay--Herriot benchmark (prediction gains of $12\%$ to $22\%$) while the ARI at the true $K$ reaches $0.94$--$0.99$; the clusterwise predictor beats the pooled benchmark at every penalty in five designs out of six, the sole exceptions being the strongly penalized fits of Scenario~4. Moreover, on the computational side, the adjusted REML estimator of Section~\ref{sec:boundary} kept all $270{,}000$ fits of the study strictly interior, without a single boundary solution.

\begin{table}[!htb]
  \centering
  \caption{Synthesis of the simulation study: Monte Carlo mean (standard deviation) of the ARI and of the relative RMSE (\%) at the true $K=3$ for $\phi \in \{0, 0.5, 1\}$, and share of replications (\%) in which BIC selects the true $K$ at $\phi=0.5$.}
  \label{tab:sims_summary}
  \resizebox{\linewidth}{!}{
\begin{tabular}{llllllll}
\toprule
Scenario & ARI\_0 & ARI\_0.5 & ARI\_1 & RelRMSE\_0 & RelRMSE\_0.5 & RelRMSE\_1 & BIC\_trueK\_phi05 \\
\midrule
Scenario 1 & 0.886 (0.053) & 0.977 (0.057) & 0.917 (0.205) & 90.044 (6.242) & 85.924 (5.150) & 86.684 (5.977) & 24.900 \\
Scenario 2 & 0.268 (0.079) & 0.640 (0.189) & 0.375 (0.258) & 97.066 (10.405) & 84.919 (13.516) & 92.772 (11.801) & 12.100 \\
Scenario 3 & 0.860 (0.040) & 0.938 (0.062) & 0.766 (0.313) & 92.562 (6.408) & 88.442 (5.548) & 90.677 (6.936) & 15.200 \\
Scenario 4 & 0.253 (0.100) & 0.265 (0.143) & 0.110 (0.082) & 89.637 (8.611) & 95.460 (11.834) & 101.249 (9.954) & 8.800 \\
Scenario 5 & 0.991 (0.058) & 0.966 (0.124) & 0.898 (0.229) & 77.978 (4.719) & 79.003 (6.075) & 81.067 (7.861) & 45.100 \\
Scenario 6 & 0.684 (0.091) & 0.936 (0.090) & 0.886 (0.223) & 94.517 (7.186) & 87.487 (6.208) & 87.666 (7.087) & 17.600 \\
\bottomrule
\end{tabular}
}
\end{table}

\subsection{Robustness to misspecification}\label{sec3:robust}
Finally, a separate battery of experiments evaluates the algorithm resilience when the data-generating process (DGP) violates the working model (i.e., Gaussian working likelihood, adjusted REML, $K=3$, $\phi=0.5$). We consider five DGP misspecifications that mimic empirically-relevant situations, each of them replicated $500$ times. The construction details and the full tables are collected in Appendix~\ref{app:misspec} and in Section~S4 of the Supplementary Material. While reading each design against the correctly specified control (i.e., GAUSS, identical to Scenario~3: ARI $0.938$, relative RMSE $0.884$), the simulations provide the following outcomes\footnote{We considered the Adjusted Rand Index against the true partition and the RMSE of the clusterwise EBLUP relative to the pooled benchmark}:
\begin{itemize}
  \item \textbf{Asymmetric sampling errors} (SKEW-E): positively skewed direct estimates, the typical SAE situation, violate the Gaussian sampling model. Classification is essentially unaffected (ARI $0.936$) and the predictive gain is preserved (relative RMSE $0.910$).
  \item \textbf{Asymmetric random effects} (SKEW-U): the same skewness moved to the latent field violates the Gaussian linking model, with a marginally larger but still negligible effect (ARI $0.926$, relative RMSE $0.861$).
  \item \textbf{Understated sampling variances} (VARDIR-UNDER): the true $\sigma^2_{ed}$ are twice the stated ones, so the model over-trusts the direct estimates; classification and prediction are again barely touched (ARI $0.926$, relative RMSE $0.859$).
  \item \textbf{Residual spatial correlation} (SAR-U): a simultaneous autoregressive field ($\rho=0.4$) on the true contiguity graph violates the independence of the random effects. Calibrated on the application data, it is the mildest departure of all, indistinguishable from the control (ARI $0.938$, relative RMSE $0.882$).
  \item \textbf{No discrete regimes} (SMOOTH): intercept and slope vary continuously in space, so the piecewise-constant working model is qualitatively wrong. Here, and only here, clustering brings no advantage (relative RMSE $0.997$) and BIC over-clusters ($\widehat{K}=4$ most frequently).
\end{itemize}
Several insights follow. First, the good properties documented on the baseline designs are not artifacts of a perfectly-specified DGP: under skewed errors or random effects, understated design variances, and moderate residual spatial correlation, the classification performance remains very high (i.e., ARI lies within $0.012$ from that of the Gaussian control) and the prediction gain survives. Second, the SAR-U setup seems to be the least harmful among all misspecification scenarios and its effects are negligible or very moderate. Third, when the truth is a smooth gradient rather than a set of regimes, the clusterwise predictor merely matches the pooled model, and an information criterion evaluated on the endogenously optimized partition yields to spurious clusters, consistently with the fact that a genuinely smooth surface produces no stability plateau in the stability step of the selection rule (see Section~\ref{sec:tuning}).

\section{Computational and inferential aspects}\label{sec:compinf}

\subsection{Algorithmic details}\label{sec:algdetails}
Several implementation choices complete the description of the estimation strategy of Section~\ref{sec2}.

\paragraph{Initialization.} The initial partition is obtained by $k$-means on the (PCA-compressed and standardized) auxiliary covariates only, with multiple random restarts and an admissibility filter that discards starting partitions whose smallest cluster cannot support a cluster-wise fit (minimum-cluster-size admissibility filter). For intercept-only models the spatial coordinates are used instead. Initializing on the covariates leaves spatial contiguity entirely to the Potts penalty, which can then be read as the price of spatial coherence rather than as a constraint built into the starting point.


\paragraph{Label updating.} The default update is sequential, in the spirit of the iterated conditional modes algorithm of \citet{Besag1986}: areas are visited in turn and each label maximizes its own penalized contribution given the current labels of all the others, so that the penalized objective cannot decrease within a sweep and cyclic configurations are ruled out. The alternative \emph{simultaneous} update --- that is, all labels refreshed at once given the previous configuration, as in the spatially-clustered regression of \citet{SugasawaMurakami2021} and in the clusterwise spatial econometric models of \citet{CerquetiEtAl2025_JABES} and \citet{MaranzanoEtAl2025} --- is retained as an option: it is faster per sweep, but it offers no monotonicity guarantee and can cycle, which the sequential scheme avoids by construction.

\paragraph{Empty and small clusters.} Along the iterations a cluster can shrink below the minimum size required by its cluster-wise fit ($P+2$ areas). Its parameters are then frozen at the last valid values and the cluster remains available to the assignment step, while clusters that never reached a valid fit are excluded from it. Configurations whose final partition still contains empty or under-sized clusters are flagged as inadmissible and excluded from the model selection step.

\paragraph{Scaling and transformations.}
When response and covariates are standardized to drive the clustering phase, the known sampling variances are rescaled accordingly (i.e., $\sigma^2_{ed}/\mathrm{var}(y)$), since they are tied to the scale of the direct estimates.

\noindent In many SAE applications, including ours, the direct estimates are positively skewed and the model is more plausibly specified on the logarithmic scale, with the sampling variances converted by the delta method (i.e., $\sigma^2_{ed,\log} \approx \sigma^2_{ed}/\hat{Z}_d^2$). In that case the SC-FH predictors are reported on both scales, that is, on the log scale as EBLUPs, and on the original scale through the lognormal bias correction $\widehat{Z}_d = \exp\{\hat{\theta}_d + g_{1d}/2\}$, where $g_{1d} = (1-\gamma_d)\,\hat{\sigma}^2_{u k_d}$ is the conditional variance of the log-scale linking parameter given the data in the FH case, so that the naive predictor $\exp\{\hat{\theta}_d\}$, which is downward biased for $Z_d$, is corrected at first order. We refer the readers to \citet{SludMaiti2006} for prediction and MSE estimation in transformed FH models and to \citet{SugasawaKubokawa2015} for general parametric transformations in SAE contexts.

\noindent Final cluster-wise estimates, EBLUPs, and information criteria are always recomputed on the original scale, which makes them comparable across $(K,\phi)$ configurations and with the pooled FH model.

\paragraph{Final refit.} Upon convergence, the cluster-wise models are re-estimated once on the final partition, holding the allocations fixed: all the reported coefficients, variance components, EBLUPs and information criteria come from this final refit --- performed on the original scale of the data --- and not from the last step of the alternating iterations.

\paragraph{Computational costs.}
Recall that each cluster $k=1,\ldots,K$ contains a number of areas equal to $D_k$. Each iteration involves the $D_k \times D_k$ matrix within each cluster $k$, with $D_k$ at most a few hundred in many area-level applications. Therefore, each iteration is $O(D_k^3)$ at worst and the cluster-wise fits are trivially parallelizable. Following the ICM paradigm, the label update is a single sweep over the $D$ areas with cost $O(D \cdot K \cdot \bar{d})$, where $\bar{d}$ is the average number of neighbors.

\subsection{Variance estimation: algorithms, boundary solutions, and software.}\label{sec:boundary}
Each iteration requires fitting $K$ distinguished FH models on cluster subsets. For area-level FH, these fits are usually lightweight, thus implementations can leverage existing routines \citep[e.g., Fisher scoring as in \texttt{sae},][]{Molina2015saeAR} based on either ML or REML. Specifically, we adapted the maximization of the REML building on the Fisher-scoring routines from the \texttt{sae} package in \texttt{R} \citep{Molina2015saeAR}, modified to incorporate spatially penalized cluster updates and potentially allowing for straightforward extensions, such as the spatial FH model with SAR random effects \citep{PratesiSalvati2008} or the spatio-temporal FH model \citep{MoralesSAE}.

The interplay between endogenous clustering and cluster-specific variance components deserves special attention. Within-cluster homogeneity is precisely what the partitioning step seeks: as areas are reallocated to maximize the (penalized) likelihood, the residual between-area variability within clusters shrinks, and the cluster-specific REML estimate $\hat{\sigma}^2_{u k}$ is increasingly attracted to the boundary of the parameter space, $\hat{\sigma}^2_{u k}=0$. The tendency is strongest when the spatial penalty is weak: for $\phi$ close to zero the partition is driven by fit alone, and the algorithm can assemble clusters whose areas are so well interpolated by the cluster-specific regression --- relative to their known sampling variances --- that the random effect vanishes altogether.

A boundary solution is not a mere numerical nuisance, as it carries two substantive consequences. First, in a cluster with $\hat{\sigma}^2_{u k}=0$ the EBLUP \eqref{eq:EBLUP} collapses to the synthetic regression predictor: the model no longer borrows strength from the direct estimates in that cluster. Second, and more insidiously, zero-variance clusters inflate the maximized log-likelihood, since the Gaussian densities concentrate on the areas with the smallest sampling variances --- in close analogy with the degenerate-likelihood problem of Gaussian mixtures.

To remove the degeneracy at its root, we estimate the cluster-specific variance components by adjusted REML in the sense of \citet{LiLahiri2010}: instead of maximizing the restricted likelihood $L_{R}(\sigma^2_{u k})$, we maximize the adjusted objective $\sigma^2_{u k} \cdot L_{R}(\sigma^2_{u k})$, whose maximizer is strictly positive by construction, remains consistent for $\sigma^2_{u k}>0$, and preserves the usual asymptotic properties \citep[refined adjustment factors are discussed in][]{YoshimoriLahiri2014}. Adjusted REML keeps every cluster in the interior of the parameter space, thereby preserving the borrowing-strength interpretation of the EBLUP and restoring the comparability of likelihood-based criteria across configurations.

The reliability of all the estimation procedures is assessed by a dedicated simulation experiment. On all six simulation designs of Section~\ref{sec3}, the full SC-FH procedure is run with each of the three REML updating schemes of \citet[Sect.~16.5.4, pp.~434--437]{MoralesSAE} --- Fisher scoring on the REML score, its residual-based variant, and the fixed-point iteration on the estimated random effects --- and with the adjusted REML estimator. Since the three REML updates maximize the same restricted likelihood, they must agree up to numerical tolerance wherever the solution is interior. The experiment verifies this equivalence within the full endogenous-clustering loop (i.e., identical partitions, coefficients, and variance components), quantifies the incidence of boundary solutions under unadjusted REML, and confirms that adjusted REML removes them by construction at a negligible cost in the interior of the parameter space. Over the $500$ replications, the three REML updates yield exactly the same partition as one another (ARI $=1$), with variance components agreeing to machine precision for the residual-based variant and to the tolerance of its fixed-point recursion for the third update; adjusted REML eliminates the boundary solutions entirely --- under standard REML they affect up to $96\%$ of the replications in the poor-separation designs at $\phi=0$ --- while returning essentially the same partition wherever standard REML is interior (ARI against M1 between $0.96$ and $1.00$ in the well-separated designs) and an equally or slightly more accurate one where it is not. The full comparison is reported in Appendix~\ref{app:remlrobust}.

\paragraph{Software and reproducibility.}
The estimation routines are implemented in \texttt{R}, building on the Fisher-scoring REML routines of the \texttt{sae} package \citep{Molina2015saeAR} and on the three equivalent REML updates described in \citet[Sect.~16.5.4]{MoralesSAE}, extended with projected monotone iterations, explicit boundary detection, adjusted REML variance estimation \citep{LiLahiri2010,YoshimoriLahiri2014}, and spatially penalized sequential cluster updates. As an internal validation, the pooled version of our routine reproduces the estimates of \texttt{sae::eblupFH} up to $10^{-8}$, and the three REML updates agree up to numerical tolerance on all fitted models.

\subsection{Tuning and model selection.}\label{sec:tuning}
Within the SC-FH framework, the number of clusters $K$ and the spatial regularization penalty $\phi$ are treated as hyperparameters that govern the model complexity. We consider a pairwise grid of candidate values for the hyperparameters (e.g., $K=1,\ldots,K_{\max}$ and $\phi=0,\ldots,\phi_{\max}$) and select $K$ and $\phi$ using an information criterion computed from the (unpenalized) log-likelihood evaluated at the fitted partition, with an effective degrees-of-freedom adjustment for clusterwise parameters. This is consistent with the practice in clusterwise spatial econometrics, where both $K$ and $\phi$ are selected by penalized likelihood criteria or stability diagnostics \citep{MaranzanoEtAl2025}. Alternatively, one could choose predictive criteria like cross-validated mean squared prediction error of $\tau_d$ (i.e., $\hat{\tau}_d$) or complementing the analysis by stability diagnostics and interpretability considerations.

Two caveats apply. First, information criteria should be computed only on non-degenerate fits: comparisons involving boundary configurations ($\hat{\sigma}^2_{u k}=0$) are vitiated by the likelihood inflation discussed in Section~\ref{sec:boundary}, which is why adjusted REML estimation is preferred to the other algorithms. Second, even with interior variance estimates, the partition itself is optimized on the data, so the maximized likelihood retains an optimism bias that a parameter count of the form $K(P+2)$ does not fully correct. Consequently, in-sample criteria tend to favor small $\phi$ and large $K$. The simulation study of Section~\ref{sec3:results} confirms this model-selection behavior. In Scenario~3, BIC selects $K=5$ in $81\%$ of the replications at $\phi=0$; the effect recedes as the penalty regularizes the partition, and BIC recovers the true $K=3$ with the highest frequency at $\phi \in [0.75, 1]$ (up to $54\%$ of the replications), while AIC and, to a lesser extent, KIC overselect at every penalty level (true-$K$ shares at most $30\%$ and $41\%$, respectively). Two features make this behavior manageable in practice. On the one hand, over-selection is benign for prediction: at moderate penalties, the relative RMSE at $K=4$ or $K=5$ stays within $1.5\%$ of its value at the true $K=3$ (in Scenario~3 at $\phi=0.5$, $89.3\%$ and $89.9\%$ against $88.4\%$), because supernumerary clusters are either small or near-duplicates of existing regimes. On the other hand, under-selection exerts a negative effect on prediction accuracy: with $K=2$ the misspecified partition makes the clusterwise predictor worse than the pooled benchmark ($101$--$107\%$ at $\phi \leq 0.5$). 

Building on the two caveats, we therefore adopt a two-step model selection rule such that $K$ is selected via ICs at a fixed moderate penalty, then $\phi$ is chosen within the moderate range by stability or predictive benchmarking, always retaining the pooled FH model as the $K=1$ reference. Let $\Phi_M$ denote a \emph{moderate-penalty band} within the candidate grid (e.g., $\phi \in [0.25, 1]$ in our empirical exercise):
\begin{enumerate}
  \item[(S1)] \emph{Number of clusters, by BIC within the band.} For each $\phi \in \Phi_M$, record the BIC-minimizing $K$; select the modal winner $\widehat{K}$ across the band, resolving ties toward the smaller $K$. Restricting the criterion to $\Phi_M$ excludes the region where the optimism bias of the endogenously optimized partition is strongest ($\phi \approx 0$), which is precisely where in-sample criteria overselect: in the simulations, BIC recovers the true $K$ most frequently at moderate-to-high penalties, and its residual over-selection is benign for prediction, whereas under-selection is the harmful direction.
  \item[(S2)] \emph{Spatial penalty, at the onset of the stability plateau.} At $K = \widehat{K}$, compute for each $\phi \in \Phi_M$ the stability index $S(\phi)$, defined as the average Adjusted Rand Index between the partition estimated at $\phi$ and the partitions estimated at the adjacent grid values. Because a stronger penalty makes the partition more rigid, $S(\phi)$ increases with $\phi$ almost by construction, and its maximizer would be systematically biased toward over-smoothing; the rule therefore selects the \emph{smallest} penalty on the stability plateau, $\widehat{\phi} = \min\{\phi \in \Phi_M : S(\phi) \geq \max_{\Phi_M} S - \delta\}$ for a small tolerance $\delta$ (we use $\delta = 0.05$): the minimal amount of spatial forcing that already delivers a reproducible partition --- a clustering-stability argument in the spirit of \citet{vonLuxburg2010}. As a validation, the same plateau can be inspected under \emph{data perturbation}, by refitting the procedure on parametric-bootstrap replicates at each candidate penalty and measuring the agreement of the refitted partitions with the estimated one; this second curve measures the reproducibility that ultimately matters and confirms (or refutes) the plateau identified under tuning perturbation.
\end{enumerate}
Notice that only \emph{admissible} configurations\footnote{A fit is admissible when each of its $K$ clusters is non-empty and large enough to support the final cluster-wise re-estimation, since a configuration with collapsed clusters has fewer effective regimes than its nominal $K$, its likelihood is not comparable with the others, and no parametric bootstrap can be generated from it.} enter the rule. Moreover, the pooled FH model ($K=1$) is always retained as the reference against which the selected configuration must be justified in terms of fit, interpretability, and area-level uncertainty.

\subsection{Bootstrap inference}\label{sec:bootinf}
A practical advantage of the proposed framework is that, conditional on the partition $\mathcal{P}$, it reduces to standard FH estimation within clusters, enabling familiar measures for uncertainty quantification such as mean squared error approximations and bootstrap resampling. To propagate label uncertainty, we recommend a parametric bootstrap that resamples the direct estimates $y_d$ from the fitted model, refits the clustering procedure, and computes empirical variability of the resulting SAE predictions $\hat{\tau}_d$; details on the parametric bootstrap procedure are provided in Algorithm \ref{AlgParBoot}.

\begin{algorithm}
\caption{Parametric bootstrap procedure for inference in the SC-FH model}\label{AlgParBoot}
\begin{algorithmic}[]
	\State Set $B$: number of bootstrap iterations
    \State \,\,\,\,\, \, $K$: the number of clusters
	\State \,\,\,\,\, \, $\phi$: the spatial penalty
	\State \,\,\,\,\, \, $b=1$
    \State Load the vectors  having the geometries for $D$ small areas of interest
    \State Load the geometries of the $D$ areas of interest
    \State Load a $D \times 1$ vector of direct estimates $y_{d}$ and the corresponding $D \times1 $ vector of area-specific sampling error variances $\sigma^2_{ed}$
    \State Load a $D \times p$ matrix of predictors $\mathbf{x}_d$   
    \State Run the SC-FH algorithm with fixed $K$ and $\phi$ using the original data $\left(y_{d},\mathbf{x}_d,\sigma^2_{ed}\right)'$ to obtain the estimates of the regression coefficients $\hat{\boldsymbol{\theta}}_k=[\hat{\theta}_{1k}, \dots, \hat{\theta}_{Pk}]'$ and of the random effects' variances $\hat{\sigma}_{uk}^2$ for $k=1,\ldots,K$, the partition $\hat{\mathcal{P}}$ of the $D$ small areas into $K$ groups with length $D_k$ for $k=1,\ldots,K$ and the corresponding $D \times 1$ vector of EBLUPs $\hat{\boldsymbol{\mu}}$ having elements $\hat{\tau}_d$ 
    \While{$b \leq B$}
        \For{$k=1,\ldots,K$}
            \State Simulate the $D_k\times1$ vector of bootstrap sampling errors as $e^{*(b)}_{dk} \sim N(0,\sigma^2_{ed})$
            \State Simulate the $D_k\times1$ vector of bootstrap random effects as $u^{*(b)}_{dk} \sim N(0,\hat{\sigma}^2_{uk})$
            \State Generate the $D_k\times1$ vector of true bootstrap quantity as $\tau^{{*(b)}}_{dk} = \mathbf{x}_{dk}^\top \hat{\boldsymbol{\theta}} + u^{*(b)}_{dk}$
            \State Generate the $D_k\times1$ vector of bootstrap direct estimates as $y^{{*(b)}}_{dk} = \tau^{{*(b)}}_{dk} + e^{*(b)}_{dk}$
        \EndFor
        \State Stack the vectors $\hat{\tau}^{{*(b)}}_{dk}$ and $\mathbf{x}_{dk}^\top$ across the $K$ groups into $\hat{\tau}^{{*(b)}}_{d}$ and $\mathbf{x}_{d}^\top$, respectively
        \State Restore the original area ordering of $\hat{\tau}^{{*(b)}}_{d}$, $\mathbf{x}_{d}^\top$ and $\sigma^2_{ed}$, so that they are aligned with the rows of the adjacency matrix $W$
        \State Run the SC-FH algorithm using the ordered values $\left(y^{*(b)}_{d},x_{d},\sigma^2_{ed}\right)'$ to obtain $\hat{\boldsymbol{\theta}}_k^{(b)}$ and $\hat{\sigma}^{2(b)}_u$ for $k=1,\ldots,K$, the partition $\hat{\mathcal{P}}^{(b)}$ and the corresponding $D \times 1$ vector of EBLUPs $\hat{\boldsymbol{\tau}}^{(b)}$ with elements $\hat{\tau}_d^{*(b)}$
        \State Store the estimates for the generic iteration $b$: $A[[b]]\leftarrow \left(\hat{\boldsymbol{\theta}}^{(b)}, \hat{\boldsymbol{\sigma}}^{2(b)}_u,\hat{\mathcal{P}}^{(b)},\hat{\boldsymbol{\tau}}^{(b)}\right)$
        \State $b=b+1$     
    \EndWhile
    \State \Return The aligned bootstrap draws of the cluster-wise parameters and of the EBLUPs
    \State \Return Confidence intervals for the cluster-wise parameters (normal, basic, percentile, studentized, bias-corrected) and percentile tests for their between-cluster differences
    \State \Return Bootstrap estimate of the area-level mean square error of the EBLUPs and area-level prediction intervals
    $$ MSE^{*}(\hat{\tau}_d) = \frac{1}{B} \sum_{b=1}^{B}\left( \hat{\tau}_d^{*(b)} - \tau^{{*(b)}}_{d} \right)^2 \qquad \forall \hspace{0.1cm} d=1,\ldots,D $$
\end{algorithmic}
\end{algorithm}

The distinguishing feature of Algorithm~\ref{AlgParBoot} is that it re-runs the entire clustering procedure at every bootstrap draw, rather than conditioning on the estimated partition: the partition varies across draws, so its uncertainty is propagated into the bootstrap distribution of the cluster-wise estimates. Moreover, because each bootstrap sample is generated area by area --- the resampled direct estimate of area $d$ is drawn from its own covariates and from the parameters of the cluster it belongs to --- the spatial structure of the original data is preserved: the fixed-effects surface $\mathbf{x}_d^\top\hat{\boldsymbol{\theta}}_{k_d}$ inherits the contiguity of the estimated regimes, while every area retains its position in the adjacency graph, so that the spatial penalty in each refit acts on the correct neighbors. 

From a single bootstrap run, after relocating the labels of every refit onto the clusters of the original fit by the majority rule, the implementation returns the aligned draws of all cluster-wise parameters and EBLUPs and computes several families of confidence intervals --- normal, basic, percentile, studentized, and bias-corrected, in the sense of \citet{DavisonHinkley1997} --- together with equal-tailed percentile tests for the between-cluster differences of the coefficients and area-level prediction intervals for the EBLUPs based on the bootstrap prediction errors \citep{HallMaiti2006,ChatterjeeLahiriLi2008}.

The bootstrap is validated as an inferential tool by a dedicated simulation experiment, whose extended results, accompanied by a detailed discussion, are provided in Section S6 of the Supplementary Material. On the clear- and poor-separation altitude designs of Section~\ref{sec3} (Scenarios~3 and~4, fitted at $K=3$ and $\phi=0.5$), each of the $500$ Monte Carlo replications is paired with a full bootstrap of $B=200$ refits, and three properties are measured: the empirical coverage of the percentile confidence intervals for the cluster-wise regression coefficients, the ratio between the average bootstrap standard error and the Monte Carlo standard deviation of the same estimators, and the agreement between the bootstrap MSE of the EBLUPs and their Monte Carlo mean squared error.

The results mirror the operating characteristics of the point estimator (Section~\ref{sec3:results}). In Scenario~3 the percentile intervals are essentially calibrated --- average empirical coverage $0.95$ across the coefficient--group combinations at the $95\%$ nominal level and $0.89$ at the $90\%$ level --- and the bootstrap MSE of the EBLUPs tracks its Monte Carlo counterpart almost exactly (median area-level ratio $1.01$, overall relative accuracy $0.99$). In the weak-signal Scenario~4 the bootstrap inherits the attenuation of the point estimates: the intervals for the well-identified central regime remain calibrated ($0.95$--$0.98$ at the $95\%$ level), those for the two extreme regimes undercover (around $0.69$--$0.81$), and the bootstrap MSE understates the Monte Carlo MSE (overall relative accuracy $0.58$).

\section{Empirical application: the standard output of farms in the Po Valley}\label{sec4}

\subsection{Context, data and specification}
We illustrate the SC-FH approach within the SCARFACE database, which combines the Italian FADN survey with environmental and socio-economic layers to study the economic and environmental dimensions of farming in the Po Valley, Northern Italy \citep{MaranzanoScarface2026}. The target variable is the \emph{average standard output per farm}, in thousand euros. The standard output (SO) is the standard European measure of the economic size of a farm \citep{ECtypology2008}. Specifically, for each product it is the average monetary value of the gross agricultural output at farm-gate prices --- a regional coefficient in euros per hectare of crop or per head of livestock, averaged over a reference period --- and a farm's SO is the sum of these coefficients over all the crops it grows and the animals it raises, net of subsidies and taxes. It therefore summarizes both the scale and the product mix of the holding through standardized prices, so that products with a high per-unit value, notably livestock, weigh heavily on it. For each of the $D=224$ Agrarian Sub-Regions surveyed in 2020 (out of the $256$ composing the full Po Valley geography), here highlighted in the left panel of Figure~\ref{fig:data_map}, we observe the Horvitz--Thompson direct estimator and its design-based variance. The direct estimates are strongly right-skewed (median $118$, maximum above $2{,}000$ thousand euros per farm), so the model is estimated on the log scale with area predictors reported on both scales. The regressors of the working model are production endowments, namely the livestock endowment\footnote{The endowment spans four orders of magnitude across the surveyed ASRs, from a few dozen heads along the alpine and Apennine arcs to almost $800{,}000$ in the intensive districts of the plain: the logarithm compresses this range and prevents a handful of very large livestock districts from acting as high-leverage points in the cluster-wise fits, a concern amplified by the endogenous clustering, where few influential areas can steer the fit of a small cluster. The unit shift, in turn, keeps the transform well defined for livestock-free domains, so that the same specification carries over unchanged to the yearly refits used in the stability diagnostics below; and since all observed counts are at least in the tens, $\log(1+x)$ is numerically indistinguishable from $\log x$, so the coefficient retains its interpretation as an elasticity.}, entering as $\log(1 + \text{bovine plus swine heads})$, the share of agricultural workers, and the extension of arable land. These variables are selected among the SCARFACE auxiliary information through the screening reported in Section S7 of the Supplementary Material; in particular, the environmental state and pressure variables (e.g., ammonia emissions from manure management and agricultural-related airborne pollutants concentrations, like PM$_{2.5}$) are deliberately kept out of the regression and used to profile the estimated regimes ex post, so that the regimes are defined by the local production structure and their environmental footprint is described afterward.

\subsection{Selection, regimes and gains}
\begin{figure}[!htb]
  \centering
  \includegraphics[width=0.99\linewidth]{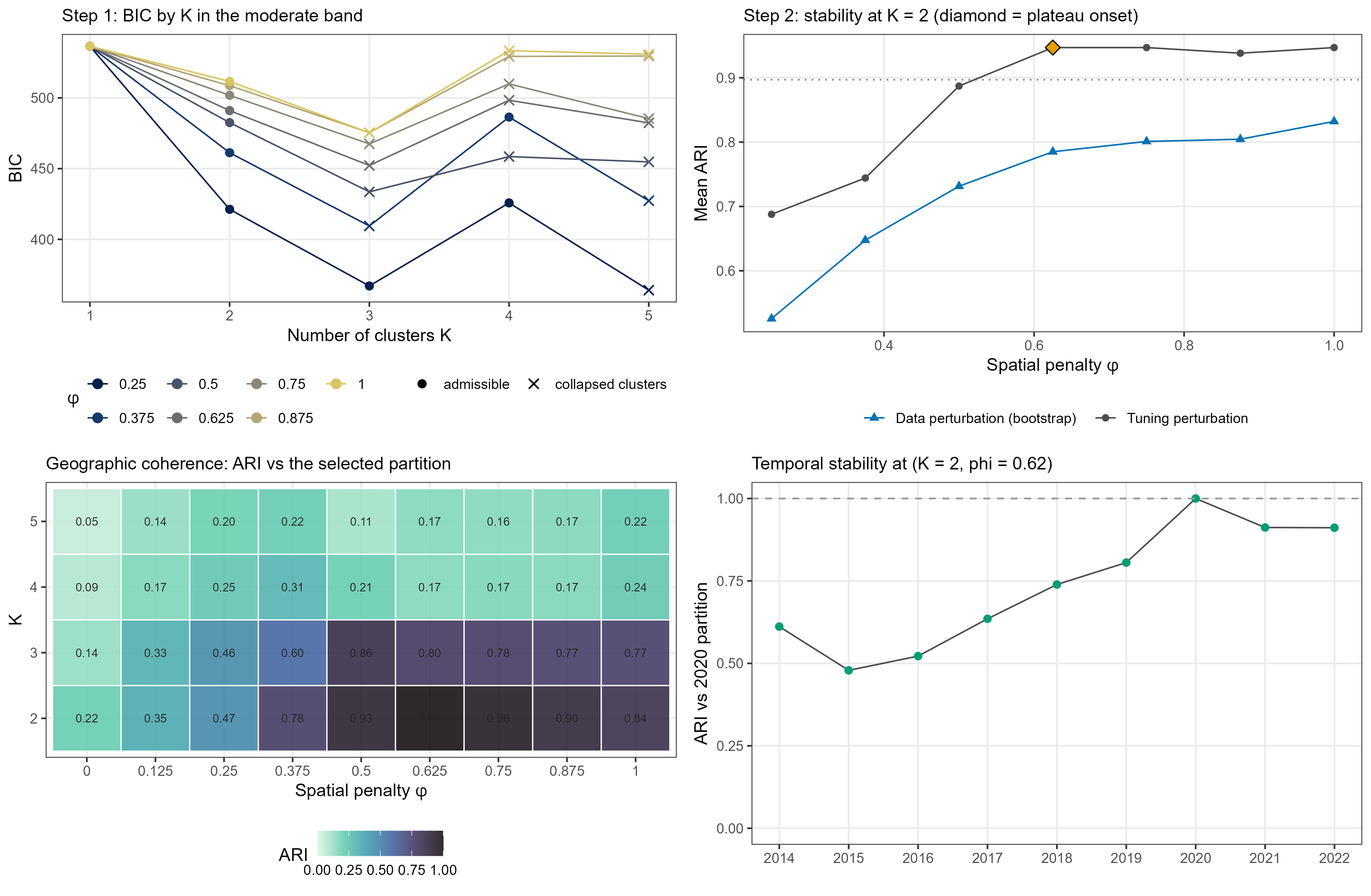}
  \caption{Selection and stability diagnostics. Top: BIC by $K$ in the moderate band, with collapsed-cluster configurations marked by crosses; tuning- and data-perturbation stability at $K=2$ with the plateau onset (diamond). Bottom: ARI between every grid partition and the selected one; ARI of the yearly refits against the 2020 partition.}
  \label{fig:app_stab}
\end{figure}

The model is fitted on the grid $K = 1,\ldots,5$ and $\phi \in [0,1]$ with step $0.125$, with adjusted REML throughout, and the configuration is chosen by the two-step rule of Section~\ref{sec:tuning}. The admissibility filter is not an idle precaution here, indeed $20$ of the $45$ configurations have collapsed clusters, and the raw BIC dips exactly there (crosses in the first panel of Figure~\ref{fig:app_stab}). Among the admissible fits, BIC selects $\widehat{K}=2$ throughout the moderate band, and the stability plateau starts at $\widehat{\phi}=0.625$ --- a choice confirmed by the data-perturbation curve, which flattens beyond that point (second panel of Figure~\ref{fig:app_stab}). The selected partition is robust: its ARI against the other admissible fits at $K=2$ in the moderate grid ranges between $0.78$ and $0.96$, and yearly refits of the same configuration reproduce it closely in recent years (ARI $0.91$ in 2021--2022 against the 2020 reference; third and fourth panels of Figure~\ref{fig:app_stab}).

\begin{figure}[!htb]
  \centering
  \includegraphics[width=0.99\linewidth]{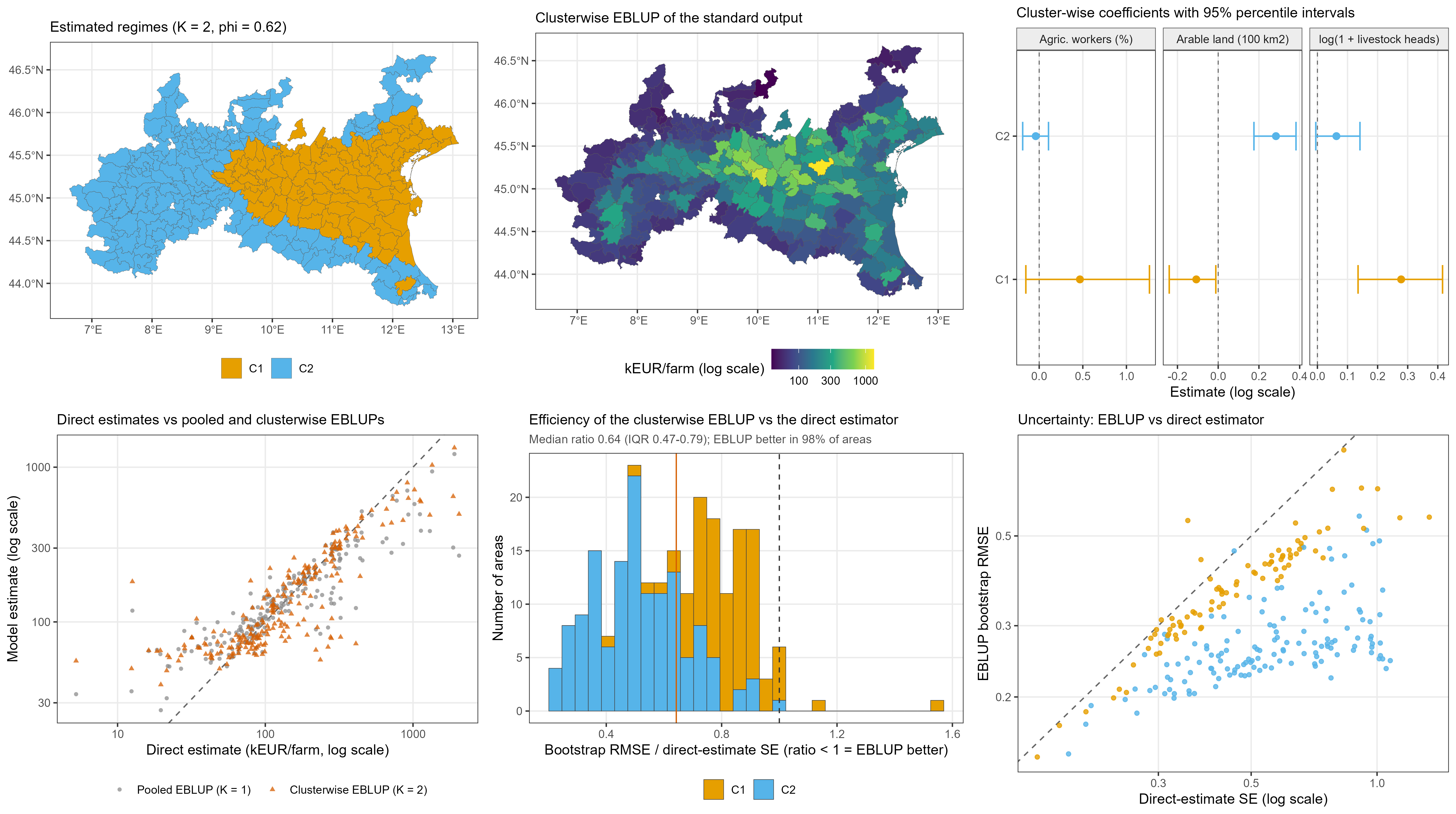}
  \caption{The selected SC-FH fit ($K=2$, $\phi=0.625$) for the average standard output per farm, 2020. Top: estimated regimes; clusterwise EBLUPs (original scale); cluster-wise coefficients with $95\%$ bootstrap percentile intervals. Bottom: direct estimates versus pooled and clusterwise EBLUPs; distribution of the area-level ratios between bootstrap RMSE and direct-estimate standard error; per-area uncertainty comparison.}
  \label{fig:app_main}
\end{figure}

According to Figure~\ref{fig:app_main} (upper panel) and Table~\ref{tab:app_main} the two regimes are spatially compact and provide interesting policy insights. Regime C1 collects $87$ areas of the central--eastern plain --- the intensive-livestock belt of eastern Lombardy, Emilia and the Veneto plain with a mean altitude of $100m$ --- with an average direct estimate of $388$ thousand euros per farm; its standard output responds to the livestock endowment. Regime C2 collects the remaining $137$ areas (the Piedmont plain, including the rice district, and the alpine and Apennine arcs; average altitude $615m$; average SO is $111$ thousand euros per farm), where the livestock slope is null and the arable-land extension drives the output. The refit-with-clustering bootstrap suggests that the between-regime differences of the livestock and arable-land slopes are significant (percentile $p$-values $0.02$ and $0.005$), while the agricultural-labor slope is positive in C1 but not significant at the selected penalty, and the intercepts do not differ significantly. The estimated elasticities admit a direct economic reading in terms of what the standard output measures. In C1 the farm economy turns on animal husbandry rather than on cropping: livestock products dominate the value of output, so a larger herd translates into a materially higher standard output, which is what the strong and significant livestock elasticity ($0.278$) captures, against an essentially null slope in C2. This same livestock intensity is what makes the regime an air-quality hotspot --- the livestock regime C1 records an average ammonia concentration of $8.9\mu g/m^3$ against $3.2\mu g/m^3$ in regiome C2, PM$_{2.5}$ of $19.4\mu g/m^3$ against $12.5\mu g/m^3$, manure-management ammonia emissions of $480$ against $93$ metric tons (see the ex-post environmental profile in Table~\ref{tab:app_main}) --- so that the economic and the environmental readings of C1 coincide. In C2, where orography limits both the stocking density and the availability of workable land, the binding endowment is instead arable land: holdings that command a larger arable surface reach a higher standard output ($0.284$), while herd size no longer discriminates between farms. In both regimes, by contrast, the share of agricultural workers carries no significant weight: once the productive endowments are accounted for, labor intensity per se does not explain the standard output, consistent with a measure built on the value of output rather than on employment.

The gains from the clusterwise specification are visible on both margins (Figure~\ref{fig:app_main}, bottom row). Against the direct estimates, the bootstrap RMSE of the clusterwise EBLUP is smaller than the design-based standard error in $98\%$ of the areas, with a median ratio of $0.64$ (interquartile range $0.47$--$0.79$); the gains are largest in regime C2, where the direct estimates are noisiest. Against the pooled model, the selected fit improves the BIC ($491$ versus $537$) and, more tellingly, the pooled EBLUP visibly over-shrinks the two tails of the distribution toward the overall mean, while the clusterwise EBLUP shrinks each area toward its own regime and preserves the between-regime differences. Bootstrap prediction intervals for every area are available from the same bootstrap run. The covariate screening, the between-regime tests, the alternative $K=3$ exhibit (whose third cluster is a small, spatially scattered group with poorly identified coefficients) and the full BIC grid are collected in Section S7 of the Supplementary Material.

\begin{table}[!htb]
  \centering
  \caption{The two regimes at a glance: cluster-wise coefficients with $95\%$ bootstrap percentile intervals, random effect variances, size, average direct estimate, ex-post environmental profile, and median efficiency ratio against the direct estimator.}
  \label{tab:app_main}
  \adjustbox{max width=0.8\linewidth}{
\begin{tabular}{lll}
\toprule
Quantity & C1 & C2 \\
\midrule
Intercept & 2.593 [1.090, 4.150] & 3.688 [2.954, 4.251] \\
log(1 + livestock heads) & 0.278 [0.135, 0.416] & 0.063 [-0.005, 0.141] \\
Agricultural workers (\%) & 0.467 [-0.153, 1.264] & -0.040 [-0.189, 0.107] \\
Arable land (100 km2) & -0.108 [-0.240, -0.012] & 0.284 [0.175, 0.382] \\
Random-effect variance (log scale) & 0.455 & 0.581 \\
Number of areas & 87 & 137 \\
Mean direct estimate (kEUR/farm) & 387.9 & 110.9 \\
NH3 concentration, CAMS (ug/m3) & 8.85 & 3.20 \\
PM2.5 concentration, EEA (ug/m3) & 19.40 & 12.52 \\
NH3 emissions, manure mgmt (t) & 480 & 93 \\
Mean temperature (C) & 14.58 & 11.84 \\
Mean altitude (m) & 100 & 615 \\
Median RMSE ratio vs direct & 0.830 & 0.501 \\
\bottomrule
\end{tabular}
}
\end{table}
\clearpage

\section{Conclusion}\label{sec5}
In this paper we proposed a clusterwise extension of the Fay--Herriot model that allows for cluster-specific relationships and addresses spatial heterogeneity in area-level models. The SC-FH framework is conceptually simple, computationally tractable, and interpretable: it offers a middle ground between two common strategies for spatial structure in SAE, that is, a single global regression with independent random effects, and globally smoothed spatial random effects. The algorithm jointly estimates cluster-specific SAE relationships and spatially-consistent clusters via penalized maximum likelihood estimation: the classic log-likelihood function of the FH model is augmented by a Potts-like penalty term on the contiguity graph of the spatial units, encouraging (but not forcing) cluster contiguity and spatial cohesion. Along the way, the paper addresses the computational and inferential questions that the endogenous partitioning raises in practice (i.e., the boundary degeneracy of the cluster-wise variance components and its adjusted-REML remedy, the optimism bias of in-sample criteria and the resulting two-step selection rule, and bootstrap inference that propagates the uncertainty of the partition). The proposed SC-FH framework thus contributes to the growing literature on spatially heterogeneous small area models by explicitly formalizing spatial regimes as contiguous latent clusters, making it particularly appealing in applied contexts where sharp spatial discontinuities and policy-relevant regional regimes are expected to coexist with within-regime homogeneity.

We demonstrate the practical relevance of the proposed framework through both simulations and an empirical application to the standard output of farms in the Po Valley (Northern Italy). The application identifies two spatially compact production regimes --- the intensive-livestock plain and a crop-and-mountain regime --- whose cluster-wise coefficients differ significantly under the refit-with-clustering bootstrap and align with the known structural and environmental heterogeneity of the region; the clusterwise EBLUP improves on the direct estimates in $98\%$ of the areas and avoids the over-shrinkage of the pooled model.
Beyond the empirical illustration, the Monte Carlo study provides systematic evidence on the operating characteristics of the framework. In particular, whenever the latent regimes are separated in the covariate--response space, SC-FH reconstructs them almost exactly at moderate spatial penalties, estimates the cluster-specific coefficients and random effect variances with negligible bias, and remarkably improves prediction accuracy over the standard Fay--Herriot benchmark. The spatial penalty acts as a stabilizer and it rescues the classification under weak separability whenever the regimes are spatially coherent. The weak-signal configuration marks the natural boundary of the approach, where the method still improves local fit but cannot properly recover the partition, and the penalty should be kept minimal.

Several extensions of the maximum likelihood framework developed here are worth pursuing. First, one could consider alternative random effects specifications: combining the clusterwise structure with spatially correlated random effects (e.g., SAR or CAR) could capture regime shifts and smooth residual dependence at the same time \citep{PratesiSalvati2008}. Second, the current hard assignment forces every area into exactly one regime, an arbitrary choice precisely at the borders between clusters, where the competing regimes are almost equally plausible; a \emph{fuzziness} parameter allowing partial, soft membership --- fuzzy rather than crisp spatial partitions, in the spirit of \citet{SugasawaMurakami2021} and of the fuzzy spatially-clustered estimator of \citet{CerquetiEtAl2025_ASTA} --- would let border areas borrow from several regimes at once, yielding a membership-weighted predictor that can improve accuracy over hard clustering exactly where the latter is least reliable. Third, data transformations such as the dual power transformation \citep{SugasawaKubokawa2015} could extend the variance-stabilization and non-Gaussian toolkit beyond the logarithmic case treated here. Finally, richer spatio-temporal specifications of the FH model \citep{MarhuendaMolinaMorales2013,MoralesSAE} would broaden applicability to panels of survey rounds, where the persistence of the regimes over time becomes itself an object of inference.

\clearpage
\appendix

\section{Computational details of the cluster-wise variance estimation}\label{app:computational}
This appendix details the estimation of the cluster-specific variance components used in the parameter-update step of the alternating algorithm of Section~\ref{sec2}. Throughout, we drop the cluster index $k$ and consider a generic FH fit on $m$ areas with design matrix $\mathbf{X}$ ($m \times (P+1)$), response vector $\mathbf{y}$, and known sampling variances $\sigma^2_{e1},\ldots,\sigma^2_{em}$; we write $A=\sigma^2_u$ for the variance component, $\mathbf{V}(A) = \mathrm{diag}(A + \sigma^2_{ed})$, and
\[
\mathbf{P}(A) \;=\; \mathbf{V}^{-1} - \mathbf{V}^{-1}\mathbf{X}\left(\mathbf{X}^\top \mathbf{V}^{-1}\mathbf{X}\right)^{-1}\mathbf{X}^\top \mathbf{V}^{-1}.
\]

\paragraph{Profile restricted likelihood.}
Up to an additive constant, the restricted log-likelihood profiled over $\boldsymbol{\theta}$ is
\begin{equation}\label{eq:remlobj}
R(A) \;=\; -\tfrac{1}{2}\log\lvert \mathbf{V}(A)\rvert \;-\; \tfrac{1}{2}\log\lvert \mathbf{X}^\top \mathbf{V}(A)^{-1}\mathbf{X}\rvert \;-\; \tfrac{1}{2}\, \mathbf{y}^\top \mathbf{P}(A)\, \mathbf{y},
\end{equation}
with score and Fisher information
\[
S(A) = -\tfrac{1}{2}\,\mathrm{tr}\,\mathbf{P} + \tfrac{1}{2}\,\mathbf{y}^\top \mathbf{P}^2 \mathbf{y},
\qquad
F(A) = \tfrac{1}{2}\,\mathrm{tr}\,\mathbf{P}^2,
\]
see \citet[Sect.~16.5.4, pp.~434--435]{MoralesSAE}.

\paragraph{Three equivalent updates.}
Following \citet[Sect.~16.5.4]{MoralesSAE}, the REML estimate can be computed by
\begin{enumerate}[label=Method (\roman*)]
    \item (p.~435), the Fisher-scoring recursion $A_{t+1} = A_t + S(A_t)/F(A_t)$;
    \item (pp.~435--436), the same recursion with the score evaluated through the GLS residuals, $S(A) = -\tfrac12 \mathrm{tr}\,\mathbf{P} + \tfrac12 \sum_d (y_d - \mathbf{x}_d^\top\tilde{\boldsymbol{\theta}}(A))^2/(A+\sigma^2_{ed})^2$, which coincides with Method 1 because $\mathbf{P}\mathbf{y} = \mathbf{V}^{-1}(\mathbf{y}-\mathbf{X}\tilde{\boldsymbol{\theta}}(A))$;
    \item (pp.~436--437), the fixed-point recursion $A_{t+1} = \sum_d \hat{u}_d(A_t)^2 / \{A_t\, \mathrm{tr}\,\mathbf{P}(A_t)\}$ with $\hat{u}_d(A) = A(A+\sigma^2_{ed})^{-1}(y_d - \mathbf{x}_d^\top\tilde{\boldsymbol{\theta}}(A))$, whose fixed point solves the score equation $S(A)=0$. 
\end{enumerate}
The three recursions target the same maximizer of Equation \eqref{eq:remlobj}.

Each candidate update is projected on the admissible interval $[0, A_{\max}]$, with $A_{\max}$ a generous data-driven bound, and is accepted only if it does not decrease $R(A)$; otherwise the step is halved (monotone ascent). If the recursion stalls, $R(A)$ is maximized directly by bounded golden-section search on $[0, A_{\max}]$. Finally, $\hat{A}=0$ is accepted only under the explicit boundary condition $R(0) \geq R(\hat{A}_{\mathrm{int}})$, where $\hat{A}_{\mathrm{int}}$ is the best interior candidate; the sign of $S(0)$ is reported as a diagnostic, since $S(0)\leq 0$ characterizes a genuine boundary maximum. These safeguards eliminate the divergence of the unconstrained Fisher-scoring recursion near the boundary, which otherwise oscillates and fails to converge precisely in the configurations where the clustering pressure pushes $A$ toward zero (Section~\ref{sec:boundary}).

\paragraph{Adjusted REML.}
Under the default adjusted REML option \citep{LiLahiri2010}, the estimate maximizes $\log A + R(A)$ on $(0, A_{\max}]$. Since $\log A \to -\infty$ as $A \to 0^+$, the maximizer is interior and strictly positive; no boundary decision is needed. The associated estimator is consistent for $A>0$ and carries a small-sample positive bias, the accepted price for positivity \citep{LiLahiri2010,YoshimoriLahiri2014}.

\clearpage
\section{Robustness of the variance-component estimation routines}\label{app:remlrobust}
This appendix documents the numerical equivalence and robustness of the four variance-component estimation routines available in our implementation: the three REML updates of \citet[Sect.~16.5.4]{MoralesSAE} --- Fisher scoring on the REML score (M1, p.~435), Fisher scoring with the residual-based expression of the score (M2, pp.~435--436), and the fixed-point recursion on the estimated random effects (M3, pp.~436--437) --- and the adjusted REML estimator of \citet{LiLahiri2010}. All four run inside the safeguarded scheme of Appendix~\ref{app:computational}.

For each of the six simulation designs (the four scenarios of Section~\ref{sec3} and the two robustness designs), the same simulated datasets are fitted with the four routines inside the full SC-FH algorithm at the true $K=3$ and $\phi \in \{0, 0.5\}$, over $500$ Monte Carlo replications, with $\phi = 0$ deliberately included because it is the configuration in which boundary solutions arise and the routines are stressed the most.

For each fit we record the ARI between the estimated partition and the truth, the ARI between the partition of each routine and that of M1 (agreement across routines), the median across replications of $\max_k |\hat{\sigma}^2_{u k} - \hat{\sigma}^2_{u k}(\mathrm{M1})|$, and the share of replications with at least one boundary cluster (Table~\ref{tab:remlrobust}).

\begin{table}[!htb]
\centering
\caption{Agreement and robustness of the estimation routines across the six simulation designs, at the true $K=3$ and $\phi \in \{0, 0.5\}$ over $500$ Monte Carlo replications: ARI against the truth, ARI against the M1 partition, median of the maximum absolute difference of the cluster-wise variance components from M1, and share of replications with a boundary cluster.}
\label{tab:remlrobust}
\adjustbox{max width=\linewidth, max totalheight=0.82\textheight}{
\begin{tabular}{lllllll}
\toprule
Scenario & Phi & Method & ARI\_truth & ARI\_vs\_M1 & MedDiffS2u & Boundary\_rate \\
\midrule
Scenario 1 & 0.000 & REML-M1 & 0.883 & 1.000 & 0.0e+00 & 0.002 \\
Scenario 1 & 0.000 & REML-M2 & 0.883 & 1.000 & 2.8e-14 & 0.002 \\
Scenario 1 & 0.000 & REML-M3 & 0.883 & 1.000 & 6.3e-05 & 0.002 \\
Scenario 1 & 0.000 & AdjREML & 0.886 & 0.991 & 4.4e+00 & 0.000 \\
Scenario 1 & 0.500 & REML-M1 & 0.976 & 1.000 & 0.0e+00 & 0.002 \\
Scenario 1 & 0.500 & REML-M2 & 0.976 & 1.000 & 2.5e-14 & 0.002 \\
Scenario 1 & 0.500 & REML-M3 & 0.976 & 1.000 & 7.1e-05 & 0.002 \\
Scenario 1 & 0.500 & AdjREML & 0.975 & 0.992 & 4.3e+00 & 0.000 \\
Scenario 2 & 0.000 & REML-M1 & 0.250 & 1.000 & 0.0e+00 & 0.960 \\
Scenario 2 & 0.000 & REML-M2 & 0.250 & 1.000 & 2.1e-14 & 0.960 \\
Scenario 2 & 0.000 & REML-M3 & 0.250 & 1.000 & 3.9e-04 & 0.960 \\
Scenario 2 & 0.000 & AdjREML & 0.266 & 0.803 & 5.8e+01 & 0.000 \\
Scenario 2 & 0.500 & REML-M1 & 0.627 & 1.000 & 0.0e+00 & 0.440 \\
Scenario 2 & 0.500 & REML-M2 & 0.627 & 1.000 & 7.1e-14 & 0.440 \\
Scenario 2 & 0.500 & REML-M3 & 0.627 & 1.000 & 1.7e-03 & 0.440 \\
Scenario 2 & 0.500 & AdjREML & 0.646 & 0.859 & 2.0e+02 & 0.000 \\
Scenario 3 & 0.000 & REML-M1 & 0.859 & 1.000 & 0.0e+00 & 0.000 \\
Scenario 3 & 0.000 & REML-M2 & 0.859 & 1.000 & 2.1e-14 & 0.000 \\
Scenario 3 & 0.000 & REML-M3 & 0.859 & 1.000 & 6.7e-05 & 0.000 \\
Scenario 3 & 0.000 & AdjREML & 0.859 & 0.996 & 1.7e+00 & 0.000 \\
Scenario 3 & 0.500 & REML-M1 & 0.934 & 1.000 & 0.0e+00 & 0.000 \\
Scenario 3 & 0.500 & REML-M2 & 0.934 & 1.000 & 2.1e-14 & 0.000 \\
Scenario 3 & 0.500 & REML-M3 & 0.934 & 1.000 & 6.7e-05 & 0.000 \\
Scenario 3 & 0.500 & AdjREML & 0.938 & 0.991 & 1.7e+00 & 0.000 \\
Scenario 4 & 0.000 & REML-M1 & 0.224 & 1.000 & 0.0e+00 & 0.892 \\
Scenario 4 & 0.000 & REML-M2 & 0.224 & 1.000 & 2.8e-13 & 0.892 \\
Scenario 4 & 0.000 & REML-M3 & 0.224 & 1.000 & 1.2e-03 & 0.892 \\
Scenario 4 & 0.000 & AdjREML & 0.252 & 0.802 & 1.7e+02 & 0.000 \\
Scenario 4 & 0.500 & REML-M1 & 0.255 & 1.000 & 0.0e+00 & 0.604 \\
Scenario 4 & 0.500 & REML-M2 & 0.255 & 1.000 & 1.7e-13 & 0.604 \\
Scenario 4 & 0.500 & REML-M3 & 0.255 & 1.000 & 1.7e-03 & 0.604 \\
Scenario 4 & 0.500 & AdjREML & 0.263 & 0.754 & 2.4e+02 & 0.000 \\
Scenario 5 & 0.000 & REML-M1 & 0.991 & 1.000 & 0.0e+00 & 0.000 \\
Scenario 5 & 0.000 & REML-M2 & 0.991 & 1.000 & 1.2e-14 & 0.000 \\
Scenario 5 & 0.000 & REML-M3 & 0.991 & 1.000 & 9.8e-06 & 0.000 \\
Scenario 5 & 0.000 & AdjREML & 0.992 & 0.995 & 6.7e-01 & 0.000 \\
Scenario 5 & 0.500 & REML-M1 & 0.962 & 1.000 & 0.0e+00 & 0.000 \\
Scenario 5 & 0.500 & REML-M2 & 0.962 & 1.000 & 1.2e-14 & 0.000 \\
Scenario 5 & 0.500 & REML-M3 & 0.962 & 1.000 & 9.6e-06 & 0.000 \\
Scenario 5 & 0.500 & AdjREML & 0.964 & 0.987 & 6.9e-01 & 0.000 \\
Scenario 6 & 0.000 & REML-M1 & 0.672 & 1.000 & 0.0e+00 & 0.040 \\
Scenario 6 & 0.000 & REML-M2 & 0.672 & 1.000 & 2.3e-13 & 0.040 \\
Scenario 6 & 0.000 & REML-M3 & 0.672 & 1.000 & 6.3e-04 & 0.040 \\
Scenario 6 & 0.000 & AdjREML & 0.683 & 0.964 & 8.6e+01 & 0.000 \\
Scenario 6 & 0.500 & REML-M1 & 0.930 & 1.000 & 0.0e+00 & 0.014 \\
Scenario 6 & 0.500 & REML-M2 & 0.930 & 1.000 & 2.3e-13 & 0.014 \\
Scenario 6 & 0.500 & REML-M3 & 0.930 & 1.000 & 1.2e-03 & 0.014 \\
Scenario 6 & 0.500 & AdjREML & 0.935 & 0.985 & 7.0e+01 & 0.000 \\
\bottomrule
\end{tabular}
}
\end{table}

The table confirms the two anticipated facts. First, the three REML updates are numerically interchangeable inside the clustering loop: the ARI between their partitions and M1's is exactly $1$ in every design and at every penalty, the median deviation of the variance components from M1 is at the level of machine precision for M2 ($10^{-14}$--$10^{-13}$) and at the tolerance of its fixed-point recursion for M3 ($10^{-5}$--$10^{-3}$), and the ARI against the truth coincides across the three to the third decimal. This is the expected behavior, since M1 and M2 are algebraically identical and the fixed point of M3 solves the same score equation; on the pooled model the routine also reproduces \texttt{sae::eblupFH} up to $10^{-8}$. Second, adjusted REML eliminates boundary solutions entirely (boundary rate $0.000$ everywhere), whereas under standard REML they affect up to $96\%$ and $89\%$ of the replications in the two poor-separation designs at $\phi=0$ ($44\%$ and $60\%$ at $\phi=0.5$) and up to $4\%$ elsewhere. Where standard REML is interior, that is, in the four well-separated designs, the adjusted REML returns essentially the same partition (ARI against M1 between $0.96$ and $1.00$) and the same classification accuracy. In the poor-separation designs the agreement with M1 necessarily drops (ARI $0.75$--$0.86$), since the two estimators differ precisely in the replications where REML sits on the boundary, and there the adjusted-REML partition is the equally or slightly more accurate one against the truth (e.g., ARI $0.25$ against $0.22$ in Scenario~4 at $\phi=0$). The systematic gap in the variance-component column for adjusted REML merely reflects its strictly positive estimates, the very property that removes the degeneracy of Section~\ref{sec:boundary}.

\clearpage
\section{Behavior under model misspecification}\label{app:misspec}
This appendix explores the behavior of the SC-FH algorithm when the data-generating process departs from the assumed model. The reference design is Scenario~3 (altitude-based clusters, clear separation); each design is replicated $500$ times and fitted with $K=3$, $\phi=0.5$, alongside the pooled FH benchmark. The designs are:
\begin{itemize}
  \item \textbf{GAUSS}: the correctly specified control, identical to Scenario~3, against which every other design is read;
  \item \textbf{SKEW-E}: sampling errors drawn from a standardized shifted lognormal (skewness $\approx 3.7$) rescaled to the stated variances $\sigma^2_{ed}$ --- the Gaussian working likelihood faces positively asymmetric direct estimates, the typical SAE situation;
  \item \textbf{SKEW-U}: area random effects from the same standardized lognormal, rescaled to $\sigma^2_{u k}$;
  \item \textbf{SMOOTH}: no discrete regimes --- intercept and slope vary continuously with longitude between the extreme cluster values; the fit selects $K \in \{1,\ldots,4\}$ by BIC at $\phi=0.5$, and the question is whether spurious clusters are created and at what predictive cost;
  \item \textbf{VARDIR-UNDER}: the true sampling variances are twice the stated ones, mimicking understated design-based variance estimates;
  \item \textbf{SAR-U}: the area random effects follow a simultaneous autoregressive process on the true contiguity graph ($\rho = 0.4$, marginal variances rescaled to $\sigma^2_{u k}$), so that the working assumption of independent random effects is violated by residual spatial correlation calibrated on the application data.
\end{itemize}
ARI and share of correct assignments (where a true partition exists), RMSE of the cluster-wise slopes, boundary rate, RMSE of the EBLUP for the true $\mu_d$ and its ratio to the pooled FH benchmark, and the modal BIC-selected number of clusters (SMOOTH only) are used as benchmark metrics.

\begin{table}[!htb]
\centering
\caption{SC-FH under model misspecification (reference design: Scenario 3; $500$ replications; $K=3$, $\phi=0.5$). Classification accuracy (ARI and share of correct assignments), RMSE of the cluster-wise slopes, boundary rate, EBLUP RMSE ratio against the pooled FH benchmark, and modal BIC-selected number of clusters (SMOOTH only).}
\label{tab:misspec}
\adjustbox{max width=\linewidth}{
\begin{tabular}{lllllll}
\toprule
DGP & ARI & Share\_correct & RMSE\_Beta1 & Boundary\_rate & RMSE\_ratio\_vs\_FH & G\_BIC\_mode \\
\midrule
GAUSS & 0.938 & 0.975 & 0.424 & 0.000 & 0.884 & -- \\
SKEW-E & 0.936 & 0.971 & 0.413 & 0.000 & 0.910 & -- \\
SKEW-U & 0.926 & 0.968 & 0.420 & 0.000 & 0.861 & -- \\
SMOOTH & -- & -- & -- & 0.000 & 0.997 & 4 \\
VARDIR-UNDER & 0.926 & 0.968 & 0.475 & 0.000 & 0.859 & -- \\
SAR-U & 0.938 & 0.975 & 0.398 & 0.000 & 0.882 & -- \\
\bottomrule
\end{tabular}
}
\end{table}

The algorithm proves to be remarkably robust (Table~\ref{tab:misspec}). Against the correctly specified control (GAUSS: ARI $0.938$, EBLUP RMSE ratio $0.884$), positive skewness in the sampling errors or in the random effects barely moves the classification (SKEW-E $0.936$, SKEW-U $0.926$) and leaves the predictive gain over the pooled benchmark intact (ratios $0.910$ and $0.861$); understated sampling variances behave similarly (VARDIR-UNDER, ARI $0.926$, ratio $0.859$). The spatially correlated random effects of the SAR-U design --- the most relevant departure for spatial applications --- are the mildest of all: the classification is indistinguishable from the Gaussian control (ARI $0.938$) and the prediction gain is preserved (ratio $0.882$), so that ignoring a moderate residual spatial correlation in the random effects costs essentially nothing. No boundary solution occurs in any design. The single instructive failure is SMOOTH, where by construction there is no discrete regime: the clusterwise predictor neither gains nor loses against the pooled benchmark (ratio $0.997$) and BIC most frequently selects $\widehat{K}=4$, a reminder that the discrete-regime model should not be forced on a genuinely smooth surface --- exactly the situation the stability-based selection rule of Section~\ref{sec:tuning} is designed to expose.

\clearpage
\section{Representative Monte Carlo replications for the four simulation scenarios} \label{AppendixA}
Each figure of this Appendix decomposes the data-generating process of one representative replication: fixed-effects surface (inheriting the contiguous regime structure), spatially independent random effects and covariate, simulated response, true groups, and response versus covariate by group. The representative replication is the one with complete predictions over the whole $(K,\phi)$ grid and the highest mean ARI at the true $K=3$.

\begin{figure}[!htb]
\centering
\includegraphics[width=0.95\textwidth]{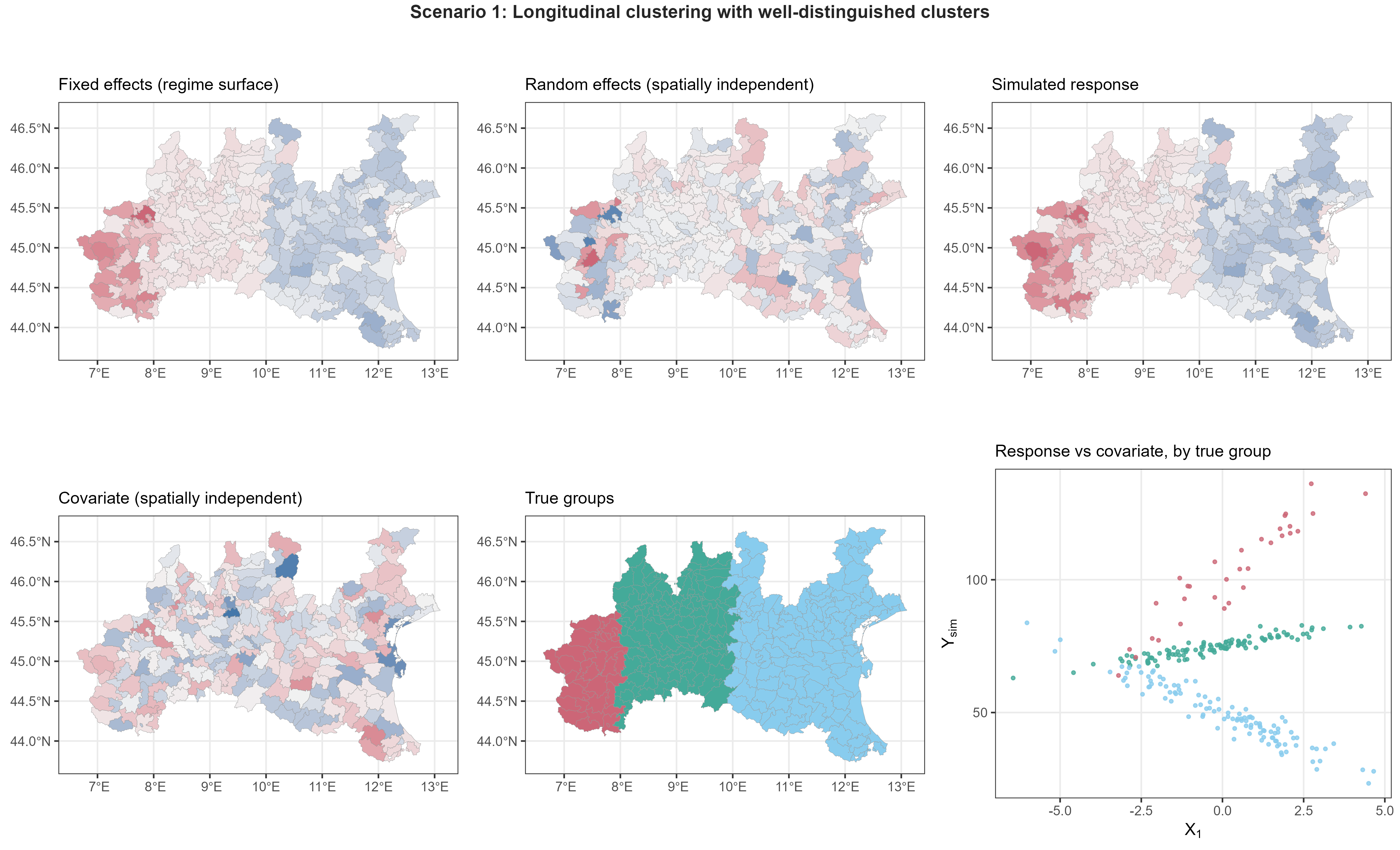}
\caption{Scenario 1 (longitude-based clusters with clearly separated coefficients): decomposition of the data-generating process for one representative replication.}
\label{fig:sim_scen1}
\end{figure}

\begin{figure}[!htb]
\centering
\includegraphics[width=0.95\textwidth]{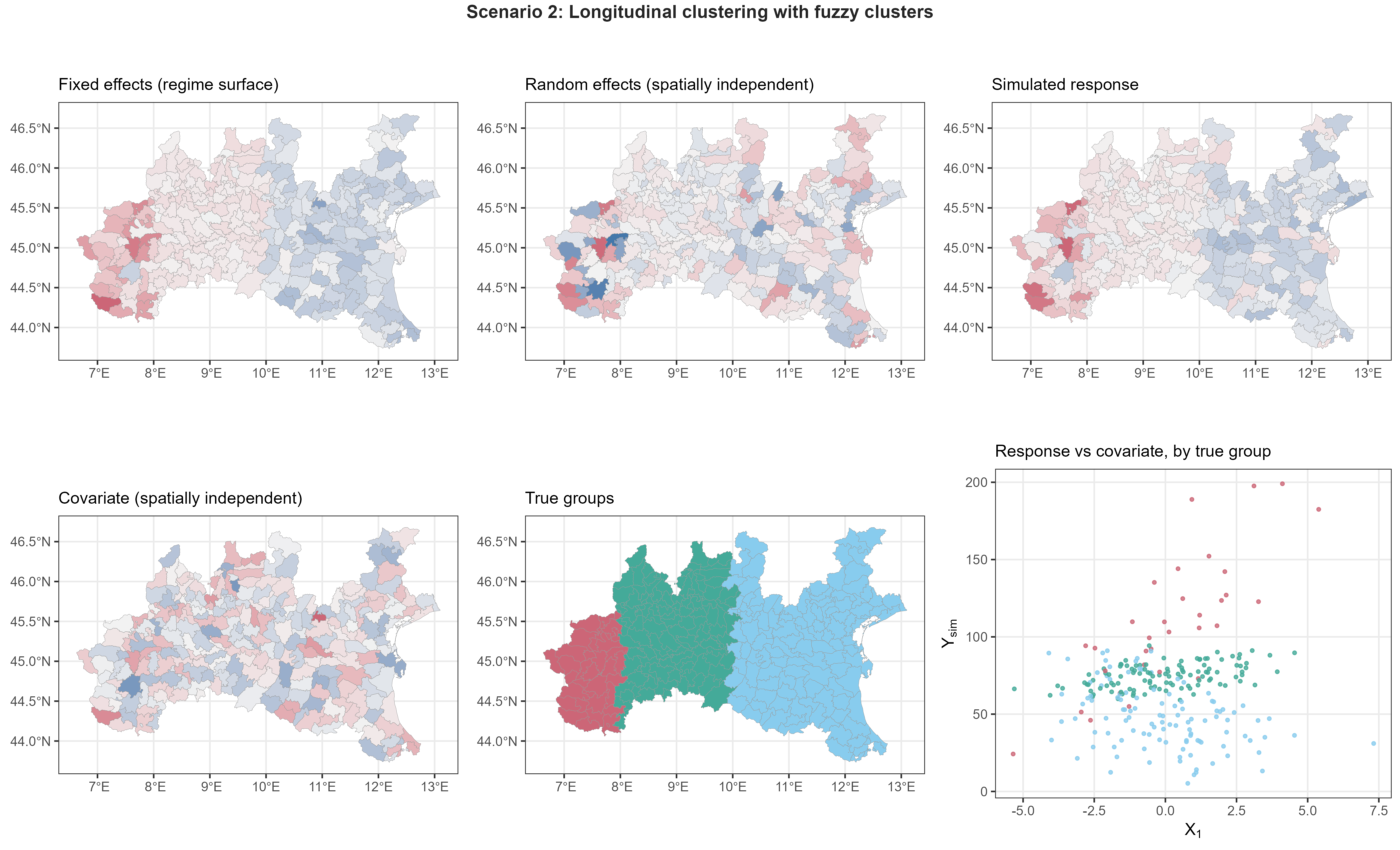}
\caption{Scenario 2 (longitude-based clusters with poorly separated coefficients): decomposition of the data-generating process for one representative replication.}
\label{fig:sim_scen2}
\end{figure}

\begin{figure}[!htb]
\centering
\includegraphics[width=0.95\textwidth]{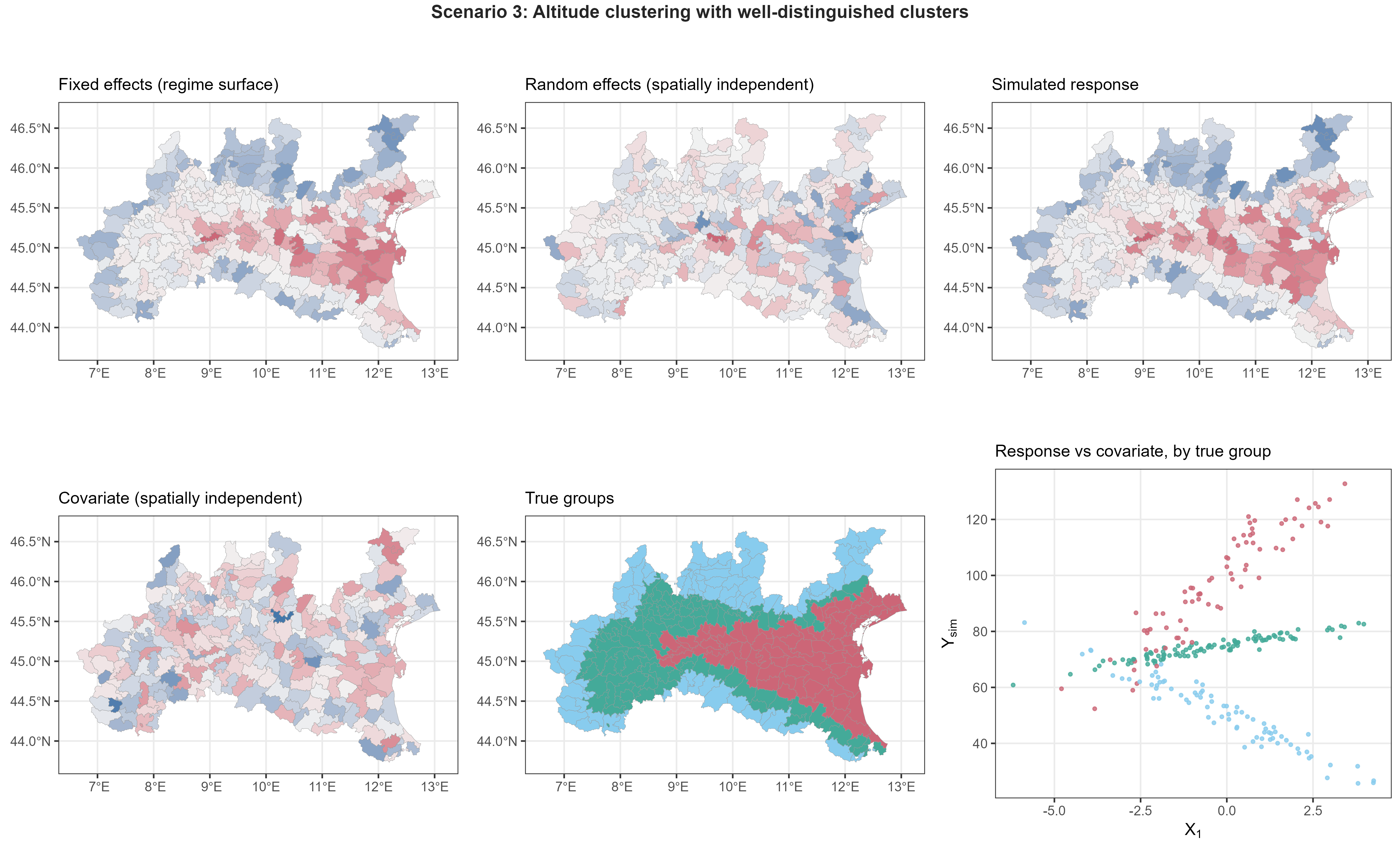}
\caption{Scenario 3 (altitude-based clusters with clearly separated coefficients): decomposition of the data-generating process for one representative replication.}
\label{fig:sim_scen3}
\end{figure}

\begin{figure}[!htb]
\centering
\includegraphics[width=0.95\textwidth]{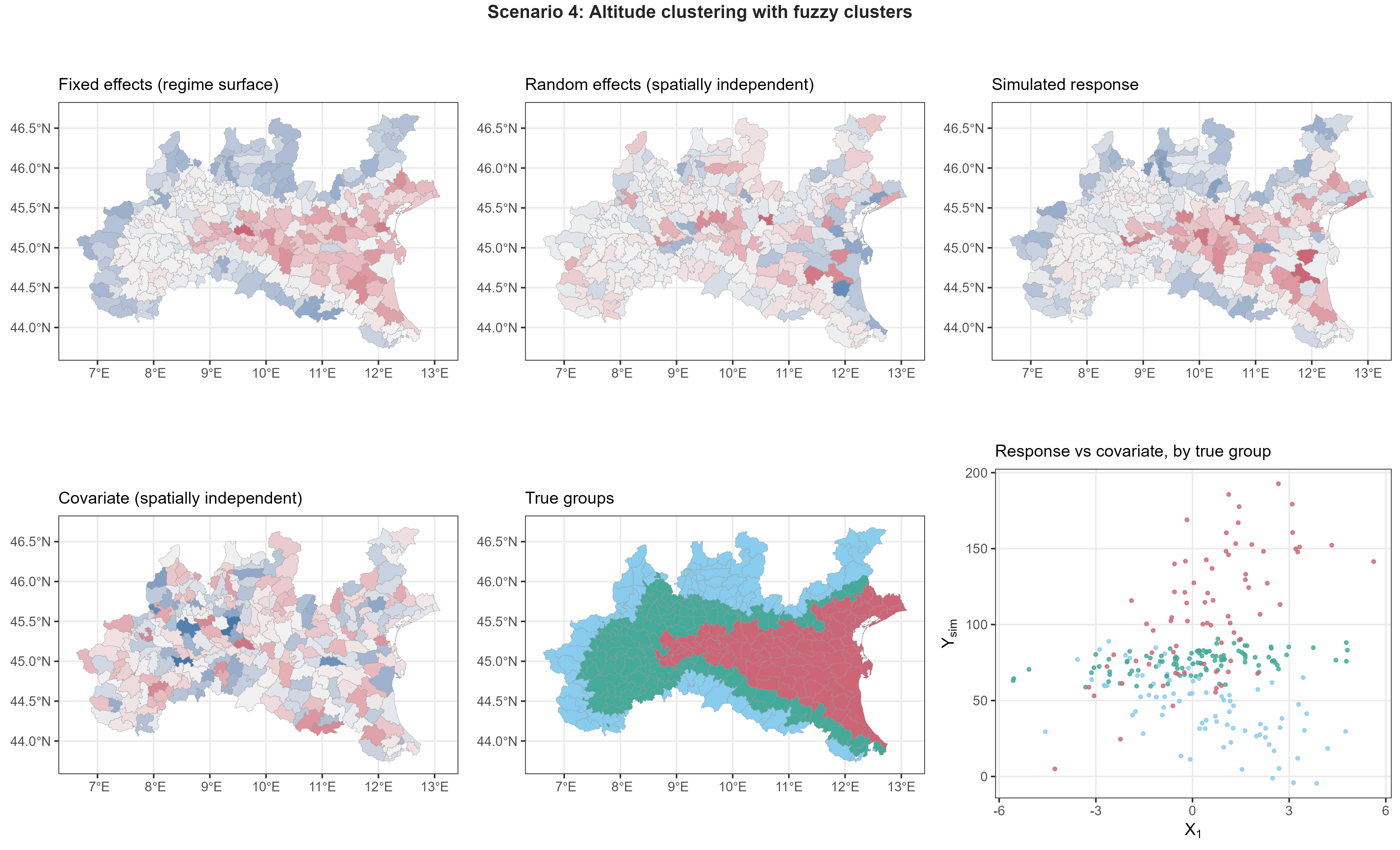}
\caption{Scenario 4 (altitude-based clusters with poorly separated coefficients): decomposition of the data-generating process for one representative replication.}
\label{fig:sim_scen4}
\end{figure}

\clearpage

\clearpage
\bibliographystyle{plainnat}
\bibliography{biblio}

@article{mattsson2025modeling,
  title={Modeling Spatial Regimes With Smooth Transitions},
  author={Mattsson, Ingrid and Lyhagen, Johan},
  journal={International Regional Science Review},
  volume={48},
  number={1},
  pages={38--61},
  year={2025},
  publisher={Sage Publications Sage CA: Los Angeles, CA}
}

@article{otto2016detection,
  title={Detection of spatial change points in the mean and covariances of multivariate simultaneous autoregressive models},
  author={Otto, Philipp and Schmid, Wolfgang},
  journal={Biometrical Journal},
  volume={58},
  number={5},
  pages={1113--1137},
  year={2016},
  publisher={Wiley Online Library}
}

@article{anselin2024endogenous,
  title={Endogenous spatial regimes},
  author={Anselin, Luc and Amaral, Pedro},
  journal={Journal of Geographical Systems},
  volume={26},
  number={2},
  pages={209--234},
  year={2024},
  publisher={Springer}
}

@article{chandra2013exploring,
  title={Exploring spatial dependence in area-level random effect model for disaggregate-level crop yield estimation},
  author={Chandra, Hukum},
  journal={Journal of Applied Statistics},
  volume={40},
  number={4},
  pages={823--842},
  year={2013},
  publisher={Taylor \& Francis}
}

@article{Molina2015saeAR,
  title={sae: An R Package for Small Area Estimation},
  author={Isabel Molina and Yolanda Marhuenda},
  journal={R J.},
  year={2015},
  volume={7},
  pages={81},
  url={https://api.semanticscholar.org/CorpusID:21804789}
}

@book{RaoMol2015,
	author = {Rao, J. N. K. and Molina, I.},
	publisher = {Wiley, New York},
	title = {Small Area Estimation},
	year = {2015}
}

@article{FH79,
author = {Robert E. Fay and Roger A. Herriot},
title = {Estimates of Income for Small Places: An Application of James-Stein Procedures to Census Data},
journal = {Journal of the American Statistical Association},
volume = {74},
number = {366a},
pages = {269--277},
year = {1979},
publisher = {ASA Website},
doi = {10.1080/01621459.1979.10482505},
URL = { 
         https://doi.org/10.1080/01621459.1979.10482505
},
eprint = {  https://doi.org/10.1080/01621459.1979.10482505}
}

@misc{ECtypology2008,
   author = {{European Commission}},
   title  = {Commission Regulation ({EC}) No.~1242/2008 of 8 {December} 2008 establishing a {Community} typology for agricultural holdings},
   year   = {2008},
   note   = {Official Journal of the European Union, L 335, 3--24}
}

@article{ColomboEtAl2023,
   author = {Colombo, Loris and Marongiu, Alessandro and Malvestiti, Giulia and Fossati, Giuseppe and Angelino, Elisabetta and Lazzarini, Matteo and Gurrieri, Gian Luca and Pillon, Silvia and Lanzani, Guido Giuseppe},
   title = {Assessing the impacts and feasibility of emissions reduction scenarios in the Po Valley},
   journal = {Frontiers in Environmental Science},
   volume = {11},
   ISSN = {2296-665X},
   DOI = {10.3389/fenvs.2023.1240816},
   url = {https://www.frontiersin.org/articles/10.3389/fenvs.2023.1240816},
   year = {2023},
   type = {Journal Article}
}

@book{MoralesSAE,
    author = {Morales, D. and Esteban, M.D. and Pérez, A. and Hobza, T.},
    title = {A Course on Small Area Estimation and Mixed Models},
    publisher = {Springer},
    year = {2021}
}

@inbook{CarilloEtAl2024,
   author = {Carillo, Felicetta and Maranzano, Paolo and Marcis, Laura and Pagliarella, Maria Chiara and Salvatore, Renato},
   title = {The spatio-temporal Fay-Herriot model using the state-space method: an application to Italian Lombard agrarian sub-regions},
   booktitle = {Book of Short Papers - 2nd Italian Conference on Economic Statistics (ICES 2024) - Statistical Analysis of Complex Economic Data: Recent Developments and Applications},
   publisher = {Casa Editrice Bonechi, Via Scipione Ammirato, 100 - 50136 Firenze (FI), info@bonechibooks.it, www.bonechi.it},
   address = {University of Florence - Department of Statistics, Computer Science, Applications “Giuseppe Parenti” (DiSIA)},
   pages = {66-69},
   ISBN = {978-88-476-2950-9},
   year = {2024},
   type = {Book Section}
}

@article{Agrimonia2023,
   author = {Fassò, Alessandro and Rodeschini, Jacopo and Moro, Alessandro Fusta and Shaboviq, Qendrim and Maranzano, Paolo and Cameletti, Michela and Finazzi, Francesco and Golini, Natalia and Ignaccolo, Rosaria and Otto, Philipp},
   title = {Agrimonia: a dataset on livestock, meteorology and air quality in the Lombardy region, Italy},
   journal = {Scientific Data},
   volume = {10},
   number = {1},
   pages = {143},
   ISSN = {2052-4463},
   DOI = {10.1038/s41597-023-02034-0},
   url = {https://doi.org/10.1038/s41597-023-02034-0},
   year = {2023},
   type = {Journal Article}
}

@article{RodeschiniEtAl2024,
   author = {Rodeschini, Jacopo and Fassò, Alessandro and Finazzi, Francesco and Fusta Moro, Alessandro},
   title = {Scenario analysis of livestock-related PM2.5 pollution based on a new heteroskedastic spatiotemporal model},
   journal = {Socio-Economic Planning Sciences},
   pages = {102053},
   ISSN = {0038-0121},
   DOI = {https://doi.org/10.1016/j.seps.2024.102053},
   url = {https://www.sciencedirect.com/science/article/pii/S0038012124002520},
   year = {2024},
   type = {Journal Article}
}

@article{MarongiuEtAl2024,
   author = {Marongiu, Alessandro and Collalto, Anna Gilia and Distefano, Gabriele Giuseppe and Angelino, Elisabetta},
   title = {Application of Machine Learning to Estimate Ammonia Atmospheric Emissions and Concentrations},
   journal = {Air},
   volume = {2},
   number = {1},
   pages = {38-60},
   ISSN = {2813-4168},
   url = {https://www.mdpi.com/2813-4168/2/1/3},
   year = {2024},
   type = {Journal Article}
}

@article{BaldoniEtAl2017,
   author = {Baldoni, Edoardo and Coderoni, Silvia and Esposti, Roberto},
   title = {The productivity and environment nexus with farm-level data. The Case of Carbon Footprint in Lombardy FADN farms},
   journal = {Bio-based and Applied Economics},
   volume = {6},
   number = {2},
   pages = {119-137},
   ISSN = {2280-6172},
   DOI = {https://doi.org/10.13128/BAE-19112 },
   year = {2017},
   type = {Journal Article}
}

@article{CoderoniEtAl2018,
   author = {Coderoni, Silvia and Esposti, Roberto},
   title = {CAP payments and agricultural GHG emissions in Italy. A farm-level assessment},
   journal = {Science of The Total Environment},
   volume = {627},
   pages = {427-437},
   ISSN = {0048-9697},
   DOI = {https://doi.org/10.1016/j.scitotenv.2018.01.197},
   url = {https://www.sciencedirect.com/science/article/pii/S0048969718302389},
   year = {2018},
   type = {Journal Article}
}

@article{CoderoniPagliacci2023,
   author = {Coderoni, Silvia and Pagliacci, Francesco},
   title = {The impact of climate change on land productivity. A micro-level assessment for Italian farms},
   journal = {Agricultural Systems},
   volume = {205},
   pages = {103565},
   ISSN = {0308-521X},
   DOI = {https://doi.org/10.1016/j.agsy.2022.103565},
   url = {https://www.sciencedirect.com/science/article/pii/S0308521X22002013},
   year = {2023},
   type = {Journal Article}
}

@article{CoderoniVanino2022,
   author = {Coderoni, Silvia and Vanino, Silvia},
   title = {The farm-by-farm relationship among carbon productivity and economic performance of agriculture},
   journal = {Science of The Total Environment},
   volume = {819},
   pages = {153103},
   ISSN = {0048-9697},
   DOI = {https://doi.org/10.1016/j.scitotenv.2022.153103},
   url = {https://www.sciencedirect.com/science/article/pii/S0048969722001930},
   year = {2022},
   type = {Journal Article}
}

@article{CortignaniCoderoni2022,
   author = {Cortignani, R. and Coderoni, S.},
   title = {The impacts of environmental and climate targets on agriculture: Policy options in Italy},
   journal = {Journal of Policy Modeling},
   volume = {44},
   number = {6},
   pages = {1095-1112},
   ISSN = {0161-8938},
   DOI = {https://doi.org/10.1016/j.jpolmod.2022.11.003},
   url = {https://www.sciencedirect.com/science/article/pii/S0161893822001089},
   year = {2022},
   type = {Journal Article}
}

@inproceedings{GardiniEtAl2025,
   author = {Gardini, Aldo and De Nicoló, Silvia and Fabrizi, Enrico},
   title = {Mapping Well-Being Through a Mixture-of-Experts Fay-Herriot Model},
   series = {Methodological and Applied Statistics and Demography I},
   publisher = {Springer Nature Switzerland},
   pages = {57-62},
   year = {2025},
   ISBN = {978-3-031-64346-0},
   type = {Conference Proceedings}
}

@article{NewhouseEtAl2025,
   author = {Newhouse, David and Ramakrishnan, Anusha and Swartz, Tom and Merfeld, Josh and Lahiri, Partha},
   title = {Small Area Estimation of Monetary Poverty in Mexico Using Satellite Imagery and Machine Learning},
   journal = {Oxford Bulletin of Economics and Statistics},
   year = {2025},
   ISSN = {0305-9049},
   DOI = {https://doi.org/10.1111/obes.12678},
   url = {https://onlinelibrary.wiley.com/doi/abs/10.1111/obes.12678},
   type = {Journal Article}
}

@article{CerquetiEtAl2025_ASTA,
   author = {Cerqueti, Roy and D’Urso, Pierpaolo and Mattera, Raffaele},
   title = {Fuzzy group fixed-effects estimation with spatial clustering},
   journal = {AStA Advances in Statistical Analysis},
   ISSN = {1863-818X},
   year = {2025},
   type = {Journal Article}
}

@article{CerquetiEtAl2025_JABES,
   author = {Cerqueti, Roy and Maranzano, Paolo and Mattera, Raffaele},
   title = {Spatially-Clustered Spatial Autoregressive Models with Application to Agricultural Market Concentration in Europe},
   journal = {Journal of Agricultural, Biological and Environmental Statistics},
   ISSN = {1537-2693},
   year = {2025},
   type = {Journal Article}
}

@article{SugasawaMurakami2021,
  title={Spatially clustered regression},
  author={Sugasawa, Shonosuke and Murakami, Daisuke},
  journal={Spatial Statistics},
  volume={44},
  pages={100525},
  year={2021},
  publisher={Elsevier}
}

@inproceedings{potts1952some,
  title={Some generalized order-disorder transformations},
  author={Potts, Renfrey Burnard},
  booktitle={Mathematical proceedings of the Cambridge Philosophical Society},
  volume={48},
  pages={106--109},
  year={1952},
DOI={https://doi.org/10.1017/S0305004100027419},
  organization={Cambridge University Press}
}

@article{MaranzanoEtAl2025,
   author = {Cerqueti, Roy and Maranzano, Paolo and Mattera, Raffaele},
   title = {Spatially-Clustered Spatial Autoregressive Models with Application to Agricultural Market Concentration in Europe},
   journal = {Journal of Agricultural, Biological and Environmental Statistics},
   ISSN = {1537-2693},
   DOI = {10.1007/s13253-024-00672-4},
   url = {https://doi.org/10.1007/s13253-024-00672-4},
   year = {2025},
   type = {Journal Article}
}

@article{desarbo1988maximum,
  title={A maximum likelihood methodology for clusterwise linear regression},
  author={DeSarbo, Wayne S and Cron, William L},
  journal={Journal of Classification},
  volume={5},
  pages={249--282},
  year={1988},
DOI={https://doi.org/10.1007/BF01897167},
  publisher={Springer}
}

@article{spath1979algorithmus,
  title={Algorithmus 39. Klassenweise lineare Regression},
  author={Sp{\"a}th, Helmuth},
  journal={Computing},
  volume={22},
  pages={367--373},
  year={1979},
DOI={https://doi.org/10.1007/BF02265317},
  publisher={Springer}
}

@article{ChandraEtAl2015,
   author = {Chandra, Hukum and Salvati, Nicola and Chambers, Ray},
   title = {A Spatially Nonstationary Fay-Herriot Model for Small Area Estimation},
   journal = {Journal of Survey Statistics and Methodology},
   volume = {3},
   number = {2},
   pages = {109-135},
   ISSN = {2325-0984},
   DOI = {10.1093/jssam/smu026},
   url = {https://doi.org/10.1093/jssam/smu026},
   year = {2015},
   type = {Journal Article}
}

@article{PratesiSalvati2008,
   author = {Pratesi, Monica and Salvati, Nicola},
   title = {Small area estimation: the EBLUP estimator based on spatially correlated random area effects},
   journal = {Statistical Methods and Applications},
   volume = {17},
   number = {1},
   pages = {113-141},
   ISSN = {1613-981X},
   DOI = {10.1007/s10260-007-0061-9},
   url = {https://doi.org/10.1007/s10260-007-0061-9},
   year = {2008},
   type = {Journal Article}
}

@inbook{BertarelliEtAl2021,
   author = {Bertarelli, Gaia and Spagnolo, Francesco Schirripa and Salvati, Nicola and Pratesi, Monica},
   title = {Small area estimation of agricultural data},
   booktitle = {Spatial Econometric Methods in Agricultural Economics Using R},
   publisher = {CRC Press},
   pages = {202-233},
   year = {2021},
   type = {Book Section}
}

@article{SugasawaKubokawa2015,
   author = {Sugasawa, Shonosuke and Kubokawa, Tatsuya},
   title = {Parametric transformed Fay–Herriot model for small area estimation},
   journal = {Journal of Multivariate Analysis},
   volume = {139},
   pages = {295-311},
   ISSN = {0047-259X},
   DOI = {https://doi.org/10.1016/j.jmva.2015.04.001},
   url = {https://www.sciencedirect.com/science/article/pii/S0047259X15000895},
   year = {2015},
   type = {Journal Article}
}

@techreport{ArticusBurgard2014,
   author = {Articus, Charlotte and Burgard, Jan Pablo},
   title = {A Finite Mixture Fay Herriot-type model for estimating regional rental prices in Germany},
   institution = {Research Papers in Economics},
   year = {2014},
   type = {Report}
}

@article{ChandraEtAl2012,
   author = {Chandra, Hukum and Salvati, Nicola and Chambers, Ray and Tzavidis, Nikos},
   title = {Small area estimation under spatial nonstationarity},
   journal = {Computational Statistics \& Data Analysis},
   volume = {56},
   number = {10},
   pages = {2875-2888},
   ISSN = {0167-9473},
   DOI = {https://doi.org/10.1016/j.csda.2012.02.006},
   url = {https://www.sciencedirect.com/science/article/pii/S0167947312000734},
   year = {2012},
   type = {Journal Article}
}

@article{kuang2024performance,
  title={Performance Characterization of Clusterwise Linear Regression Algorithms},
  author={Kuang, Ye Chow and Ooi, Melanie},
  journal={Wiley Interdisciplinary Reviews: Computational Statistics},
  volume={16},
  number={5},
  pages={e70004},
  year={2024},
  publisher={Wiley Online Library}
}

@article{park2017algorithms,
  title={Algorithms for generalized clusterwise linear regression},
  author={Park, Young Woong and Jiang, Yan and Klabjan, Diego and Williams, Loren},
  journal={INFORMS Journal on Computing},
  volume={29},
  number={2},
  pages={301--317},
  year={2017},
  publisher={INFORMS}
}

@article{Fabrizi2009,
  title={A comparison of adjusted Bayes Estimators of an ensemble of small area parameters},
  author={Fabrizi, Enrico},
  journal={Statistica},
  volume={69},
  number={4},
  pages={269--284},
  year={2009}
}

@article{FabriziTrivisano2010,
   author = {Fabrizi, Enrico and Trivisano, Carlo},
   title = {Robust linear mixed models for Small Area Estimation},
   journal = {Journal of Statistical Planning and Inference},
   volume = {140},
   number = {2},
   pages = {433-443},
   ISSN = {0378-3758},
   DOI = {https://doi.org/10.1016/j.jspi.2009.07.022},
   url = {https://www.sciencedirect.com/science/article/pii/S0378375809002304},
   year = {2010},
   type = {Journal Article}
}

@article{PorterEtAl2014,
   author = {Porter, Aaron T. and Holan, Scott H. and Wikle, Christopher K. and Cressie, Noel},
   title = {Spatial Fay–Herriot models for small area estimation with functional covariates},
   journal = {Spatial Statistics},
   volume = {10},
   pages = {27-42},
   ISSN = {2211-6753},
   DOI = {https://doi.org/10.1016/j.spasta.2014.07.001},
   url = {https://www.sciencedirect.com/science/article/pii/S2211675314000451},
   year = {2014},
   type = {Journal Article}
}

@article{BrunsdonFotheringham1998,
   author = {Brunsdon, C. and Fotheringham, S. and Charlton, M.},
   title = {Geographically Weighted Regression},
   journal = {Journal of the Royal Statistical Society: Series D (The Statistician)},
   volume = {47},
   number = {3},
   pages = {431-443},
   ISSN = {0039-0526},
   DOI = {https://doi.org/10.1111/1467-9884.00145},
   url = {https://rss.onlinelibrary.wiley.com/doi/abs/10.1111/1467-9884.00145},
   year = {1998},
   type = {Journal Article}
}

@article{MarhuendaMolinaMorales2013,
   author = {Marhuenda, Yolanda and Molina, Isabel and Morales, Domingo},
   title = {Small area estimation with spatio-temporal Fay–Herriot models},
   journal = {Computational Statistics \& Data Analysis},
   volume = {58},
   pages = {308-325},
   ISSN = {0167-9473},
   DOI = {https://doi.org/10.1016/j.csda.2012.09.002},
   url = {https://www.sciencedirect.com/science/article/pii/S0167947312003222},
   year = {2013},
   type = {Journal Article}
}

@article{NakagawaJohnsonSchielzeth2017,
   author = {Nakagawa, Shinichi and Johnson, Paul C. D. and Schielzeth, Holger},
   title = {The coefficient of determination R2 and intra-class correlation coefficient from generalized linear mixed-effects models revisited and expanded},
   journal = {Journal of The Royal Society Interface},
   volume = {14},
   number = {134},
   ISSN = {1742-5689},
   DOI = {10.1098/rsif.2017.0213},
   url = {https://doi.org/10.1098/rsif.2017.0213},
   year = {2017},
   type = {Journal Article}
}

@article{NakagawaSchielzeth2013,
   author = {Nakagawa, Shinichi and Schielzeth, Holger},
   title = {A general and simple method for obtaining R2 from generalized linear mixed‐effects models},
   journal = {Methods in ecology and evolution},
   volume = {4},
   number = {2},
   pages = {133-142},
   ISSN = {2041-210X},
   year = {2013},
   type = {Journal Article}
}

@article{LiLahiri2010,
   author = {Li, Huilin and Lahiri, Partha},
   title = {An adjusted maximum likelihood method for solving small area estimation problems},
   journal = {Journal of Multivariate Analysis},
   volume = {101},
   number = {4},
   pages = {882-892},
   DOI = {10.1016/j.jmva.2009.10.009},
   year = {2010},
   type = {Journal Article}
}

@article{YoshimoriLahiri2014,
   author = {Yoshimori, Masayo and Lahiri, Partha},
   title = {A new adjusted maximum likelihood method for the Fay-Herriot small area model},
   journal = {Journal of Multivariate Analysis},
   volume = {124},
   pages = {281-294},
   DOI = {10.1016/j.jmva.2013.10.012},
   year = {2014},
   type = {Journal Article}
}

@article{SludMaiti2006,
   author = {Slud, Eric V. and Maiti, Tapabrata},
   title = {Mean-squared error estimation in transformed Fay-Herriot models},
   journal = {Journal of the Royal Statistical Society Series B: Statistical Methodology},
   volume = {68},
   number = {2},
   pages = {239-257},
   DOI = {10.1111/j.1467-9868.2006.00542.x},
   year = {2006},
   type = {Journal Article}
}

@article{vonLuxburg2010,
  author  = {von Luxburg, Ulrike},
  title   = {Clustering stability: an overview},
  journal = {Foundations and Trends in Machine Learning},
  year    = {2010},
  volume  = {2},
  number  = {3},
  pages   = {235--274},
  doi     = {10.1561/2200000008}
}

@misc{MaranzanoScarface2026,
  author        = {Maranzano, Paolo and Colombo, Pietro and Carillo, Felicetta and Borgoni, Riccardo and Pajno, Riccardo and Borrotti, Matteo and Ferrero, Luca and Bolzacchini, Ezio},
  title         = {{SCARFACE}: a harmonized spatio-temporal dataset integrating socio-economic, environmental, and agricultural indicators for the {P}o {V}alley ({I}taly), 2011--2024},
  year          = {2026},
  eprint        = {2604.25940},
  archivePrefix = {arXiv},
  primaryClass  = {stat.AP},
  url           = {https://arxiv.org/abs/2604.25940}
}

@article{Besag1986,
  author  = {Besag, Julian},
  title   = {On the statistical analysis of dirty pictures},
  journal = {Journal of the Royal Statistical Society: Series B},
  year    = {1986},
  volume  = {48},
  number  = {3},
  pages   = {259--302}
}

@book{DavisonHinkley1997,
  author    = {Davison, Anthony C. and Hinkley, David V.},
  title     = {Bootstrap Methods and their Application},
  publisher = {Cambridge University Press},
  address   = {Cambridge},
  year      = {1997}
}

@article{HallMaiti2006,
  author  = {Hall, Peter and Maiti, Tapabrata},
  title   = {On parametric bootstrap methods for small area prediction},
  journal = {Journal of the Royal Statistical Society: Series B},
  year    = {2006},
  volume  = {68},
  number  = {2},
  pages   = {221--238}
}

@article{ChatterjeeLahiriLi2008,
  author  = {Chatterjee, Snigdhansu and Lahiri, Partha and Li, Huilin},
  title   = {Parametric bootstrap approximation to the distribution of {EBLUP} and related prediction intervals in linear mixed models},
  journal = {Annals of Statistics},
  year    = {2008},
  volume  = {36},
  number  = {3},
  pages   = {1221--1245}
}

@misc{PajnoEtAl2026,
  author        = {Pajno, Riccardo and Carillo, Felicetta and Maranzano, Paolo and Schmid, Timo and Borgoni, Riccardo},
  title         = {On the Use of Satellite Information to Estimate Agricultural Carbon Footprint in a Small Area Framework},
  year          = {2026},
  eprint        = {2604.25342},
  archivePrefix = {arXiv},
  primaryClass  = {stat.AP},
  url           = {https://arxiv.org/abs/2604.25342}
}

@article{BatteseEtAl1988,
  author  = {Battese, George E. and Harter, Rachel M. and Fuller, Wayne A.},
  title   = {An Error-Components Model for Prediction of County Crop Areas Using Survey and Satellite Data},
  journal = {Journal of the American Statistical Association},
  year    = {1988},
  volume  = {83},
  number  = {401},
  pages   = {28--36},
  doi     = {10.1080/01621459.1988.10478561},
  url     = {https://doi.org/10.1080/01621459.1988.10478561}
}

@article{AltamoreEtAl2024,
  author  = {Altamore, Luca and Chinnici, Pietro and Bacarella, Simona and Chironi, Stefania and Ingrassia, Marzia},
  title   = {Current Framework of Italian Agriculture and Changes between the 2010 and 2020 Censuses},
  journal = {Agriculture},
  year    = {2024},
  volume  = {14},
  number  = {9},
  pages   = {1603},
  doi     = {10.3390/agriculture14091603},
  url     = {https://doi.org/10.3390/agriculture14091603}
}

@article{CardilloEtAl2023,
  author  = {Cardillo, Claudia and others},
  title   = {The Farm's Orientation towards Sustainability: An Assessment Using {FADN} Data in Italy},
  journal = {Land},
  year    = {2023},
  volume  = {12},
  number  = {2},
  pages   = {301},
  doi     = {10.3390/land12020301},
  url     = {https://doi.org/10.3390/land12020301}
}

@article{MontanariEtAl2023,
  author  = {Montanari, Alberto and Nguyen, Hung and Rubinetti, Sara and Ceola, Serena and Galelli, Stefano and Rubino, Angelo and Zanchettin, Davide},
  title   = {Why the 2022 {P}o {R}iver drought is the worst in the past two centuries},
  journal = {Science Advances},
  year    = {2023},
  volume  = {9},
  number  = {32},
  pages   = {eadg8304},
  doi     = {10.1126/sciadv.adg8304},
  url     = {https://www.science.org/doi/abs/10.1126/sciadv.adg8304}
}

\clearpage
\setcounter{section}{0}
\setcounter{figure}{0}
\setcounter{table}{0}
\setcounter{equation}{0}
\renewcommand{\thesection}{S\arabic{section}}
\renewcommand{\thefigure}{S\arabic{figure}}
\renewcommand{\thetable}{S\arabic{table}}
\renewcommand{\theequation}{S\arabic{equation}}
\begin{center}\Large\textbf{Supplementary Material}\end{center}
\bigskip

\noindent This document collects the extended results of all the simulation experiments accompanying the main paper: the full per-scenario results for the four baseline scenarios (Section~\ref{sm:baseline}), the two robustness designs (Section~\ref{sm:robust}), the misspecification study (Section~\ref{sm:misspec}), the comparison of the REML updating schemes (Section~\ref{sm:reml}), and the validation of the parametric bootstrap (Section~\ref{sm:boot}). All figures and tables are auto-generated by the analysis scripts. The scenario results (Sections~\ref{sm:baseline} and~\ref{sm:robust}) are based on $1{,}000$ Monte Carlo runs; the misspecification and REML studies (Sections~\ref{sm:misspec} and~\ref{sm:reml}) on $500$ runs each; and the bootstrap validation (Section~\ref{sm:boot}) on $500$ replications with $B=200$ bootstrap refits each.

\section{Overview of the simulation designs}\label{sm:designs}
All the experiments share the area-level data-generating process described in the main paper (Section~3), on the full Po Valley geography. Six scenarios are considered:
\begin{description}
  \item[Scenario 1] longitude-based clusters with clearly separated coefficients ($R^2_{fe}=0.90$);
  \item[Scenario 2] longitude-based clusters with poorly separated coefficients ($R^2_{fe}=0.25$);
  \item[Scenario 3] altitude-based clusters with clearly separated coefficients ($R^2_{fe}=0.90$);
  \item[Scenario 4] altitude-based clusters with poorly separated coefficients ($R^2_{fe}=0.25$);
  \item[Scenario 5] \emph{perfect separation}: cluster-specific slopes close to each other ($\beta_{1k}=(-2,1,4)$), regimes differing mainly in level;
  \item[Scenario 6] \emph{high variability}: widely spread coefficients ($\beta_{0k}=(100,150,200)$, $\beta_{1k}=(-20,10,40)$).
\end{description}
Each scenario is fitted over the full hyperparameter grid $K=1,\ldots,5$ and $\phi \in \{0, 0.25, \ldots, 2\}$, with adjusted REML variance estimation and sequential label updates. For each scenario we report, in order: the classification accuracy at the true $K=3$ (overall and by group) and over the whole grid (share and Adjusted Rand Index); the recovery of the cluster-wise slopes and random-effect variances; the prediction accuracy of the EBLUP, in absolute terms and relative to the pooled Fay--Herriot benchmark ($K=1$); the behavior of the information criteria; the mean profiles of the key metrics; and the incidence of boundary solutions. Two auto-generated tables per scenario summarize parameter recovery and overall performance.

\clearpage
\section{Extended results for the baseline scenarios}\label{sm:baseline}

\subsection{Scenario 1: longitude-based clusters, clear separation}

\begin{figure}[!htb]
  \centering
  \includegraphics[width=0.95\linewidth]{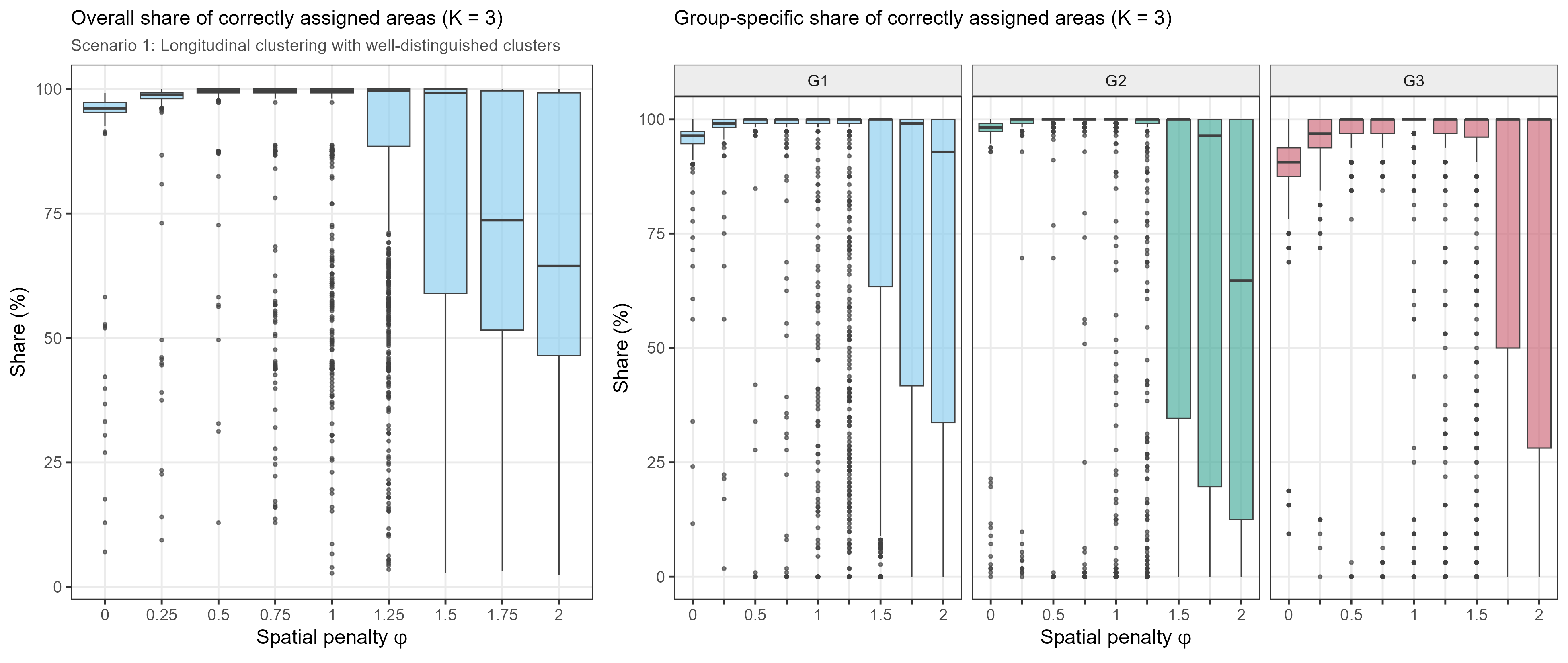}
  \caption{Scenario 1: overall (left) and group-specific (right) share of correctly assigned areas at the true $K=3$, by spatial penalty $\phi$.}
\end{figure}

\begin{figure}[!htb]
  \centering
  \includegraphics[width=0.9\linewidth]{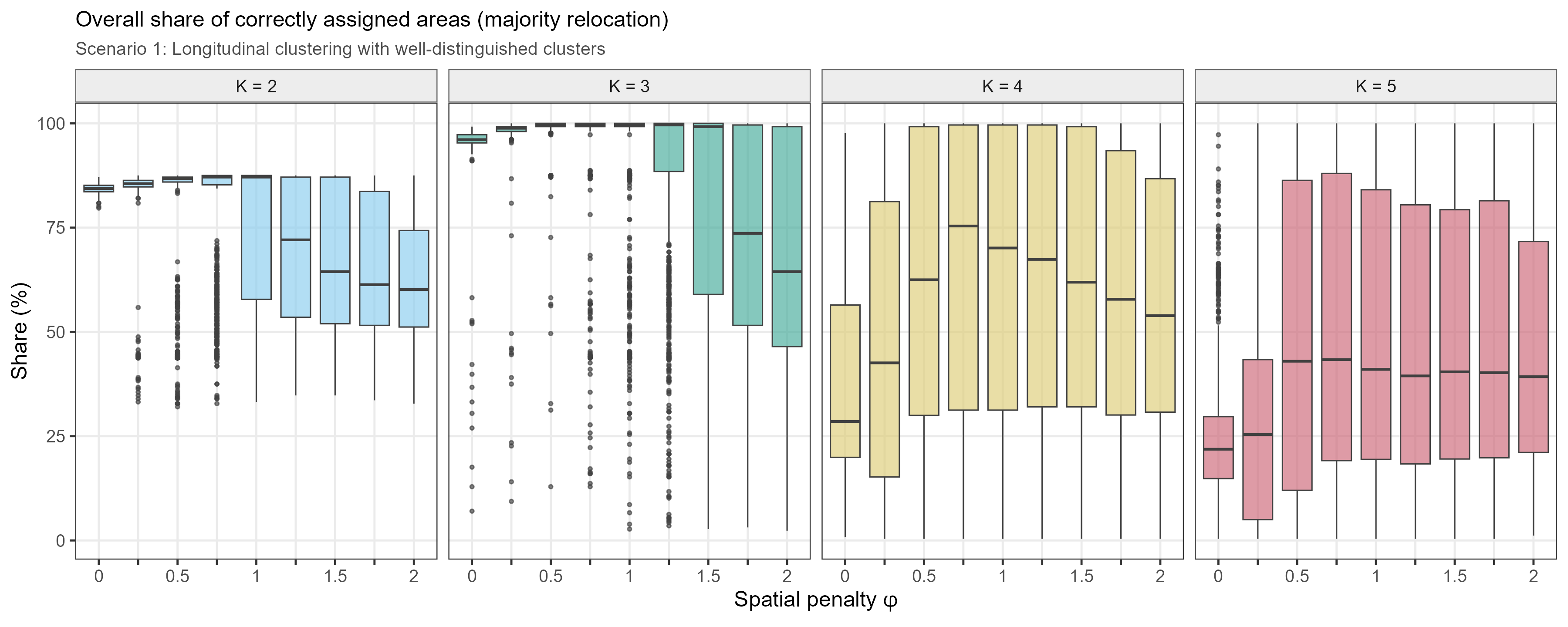}
  \caption{Scenario 1: overall share of correctly assigned areas by number of clusters $K$ and spatial penalty $\phi$.}
\end{figure}

\begin{figure}[!htb]
  \centering
  \includegraphics[width=0.9\linewidth]{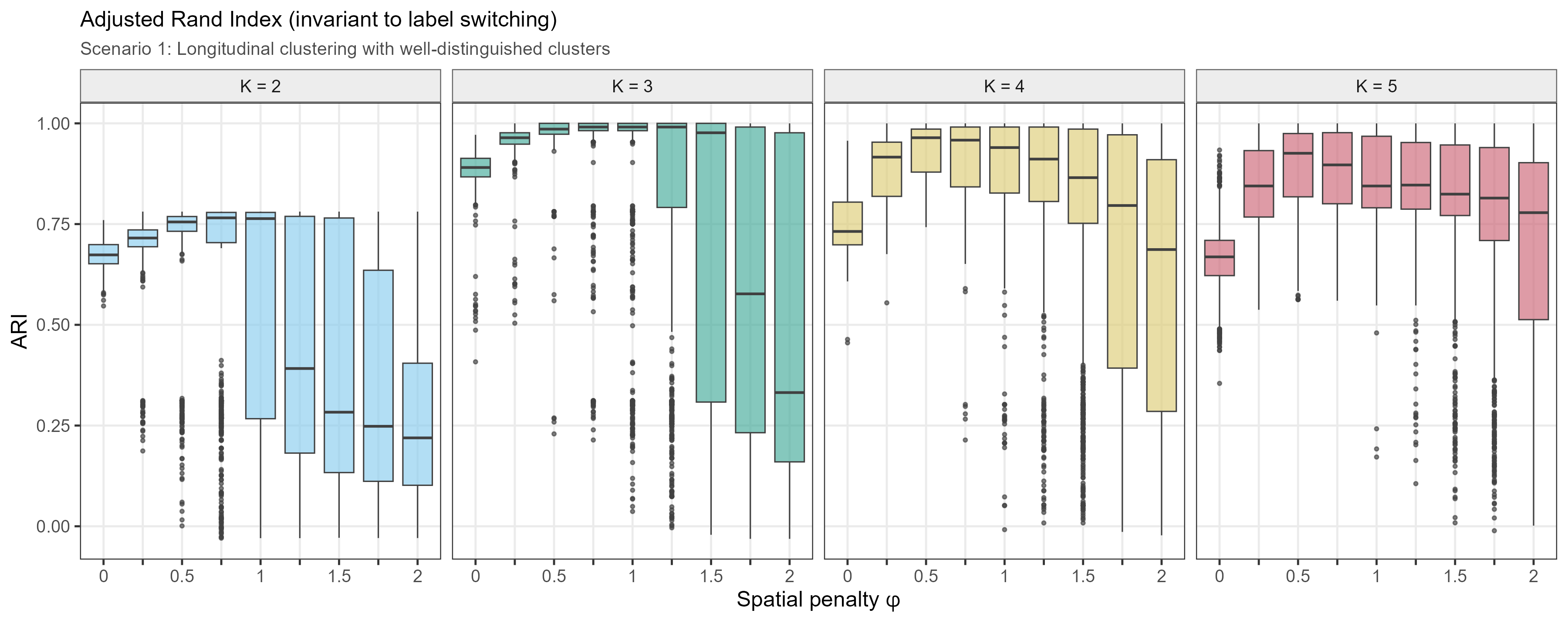}
  \caption{Scenario 1: Adjusted Rand Index between estimated and true partition, by $K$ and $\phi$.}
\end{figure}

\begin{figure}[!htb]
  \centering
  \includegraphics[width=0.95\linewidth]{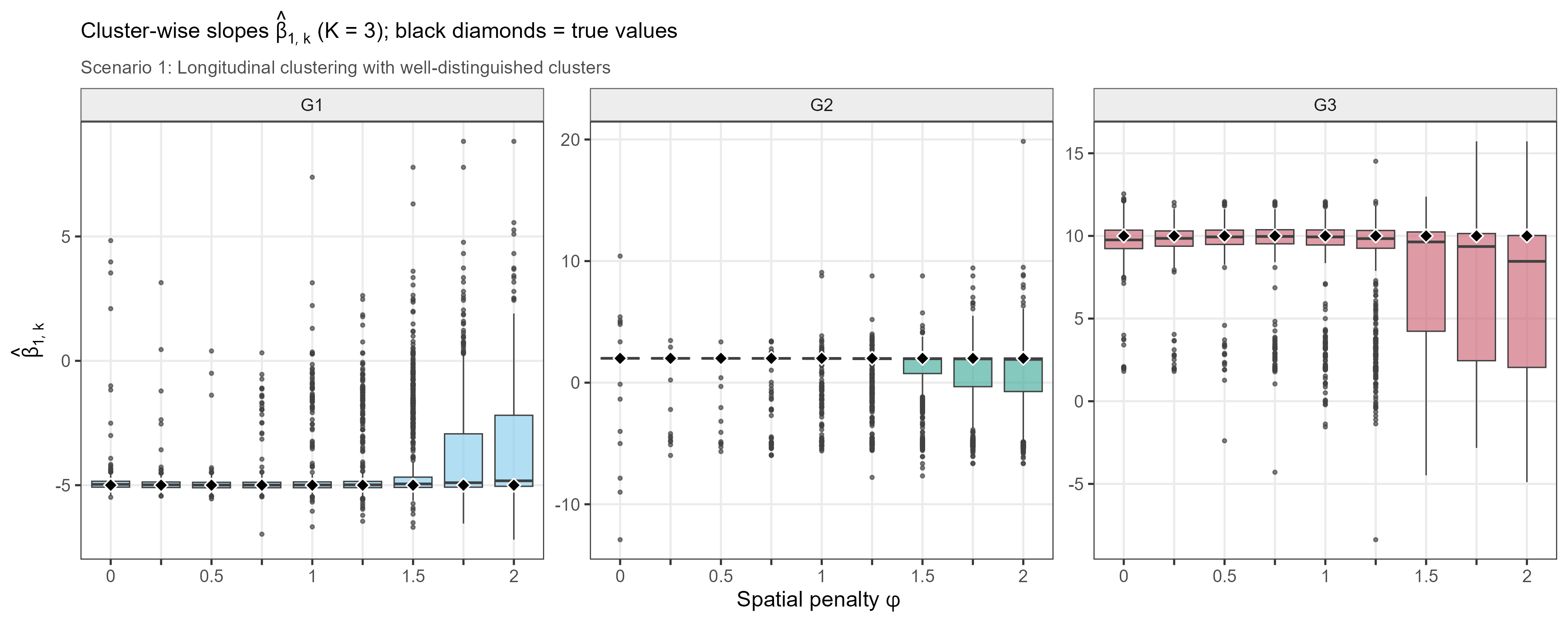}
  \caption{Scenario 1: cluster-specific slope estimates at the true $K=3$, by $\phi$. Black diamonds mark the true values.}
\end{figure}

\begin{figure}[!htb]
  \centering
  \includegraphics[width=0.95\linewidth]{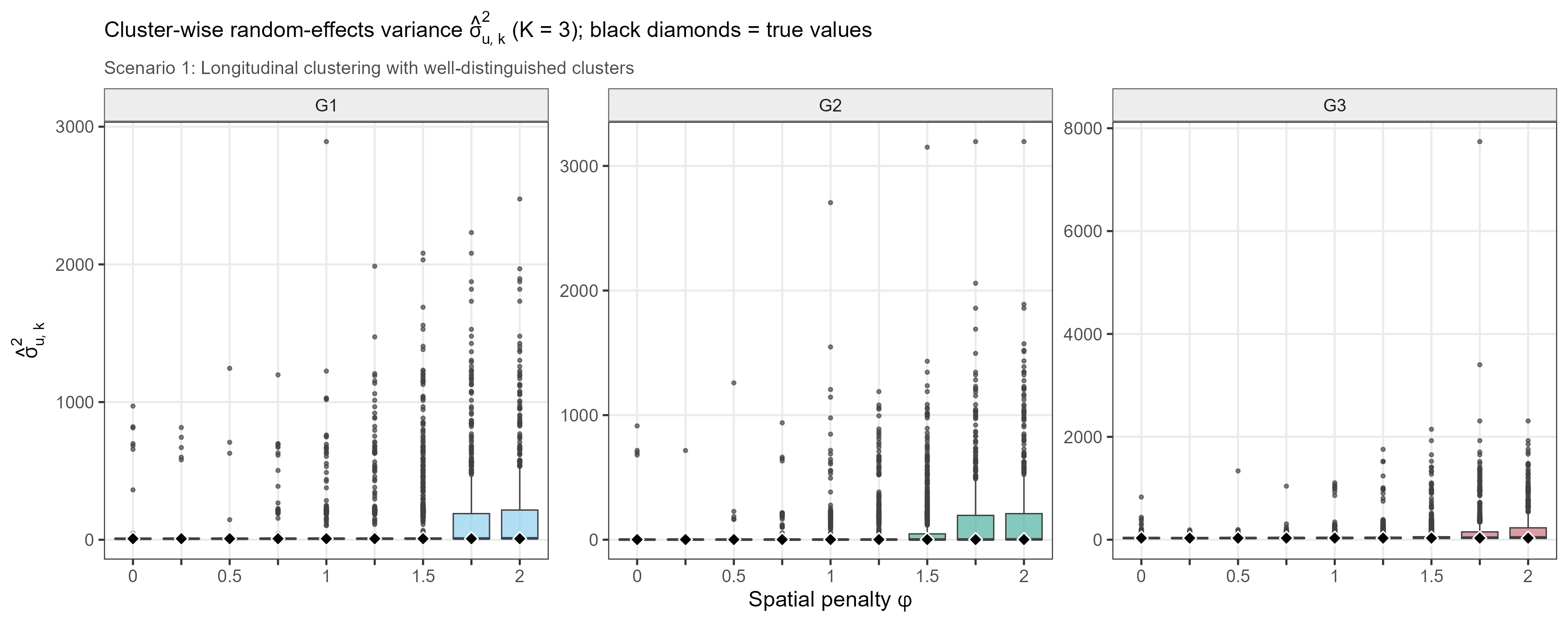}
  \caption{Scenario 1: cluster-specific random-effect variance estimates at the true $K=3$, by $\phi$. Black diamonds mark the true values.}
\end{figure}

\clearpage

\begin{figure}[!htb]
  \centering
  \includegraphics[width=0.95\linewidth]{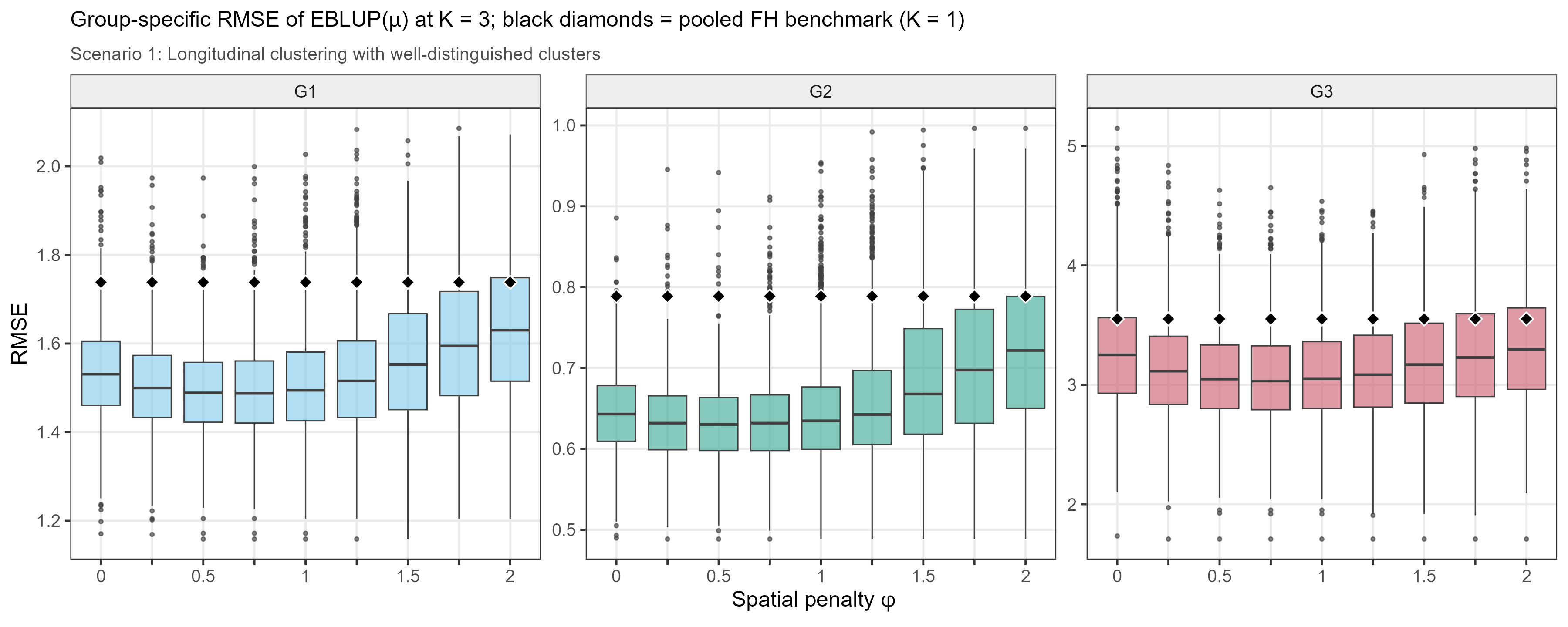}
  \caption{Scenario 1: RMSE of the EBLUP by true group at $K=3$. Black diamonds mark the mean RMSE of the pooled FH benchmark ($K=1$).}
\end{figure}

\begin{figure}[!htb]
  \centering
  \includegraphics[width=0.9\linewidth]{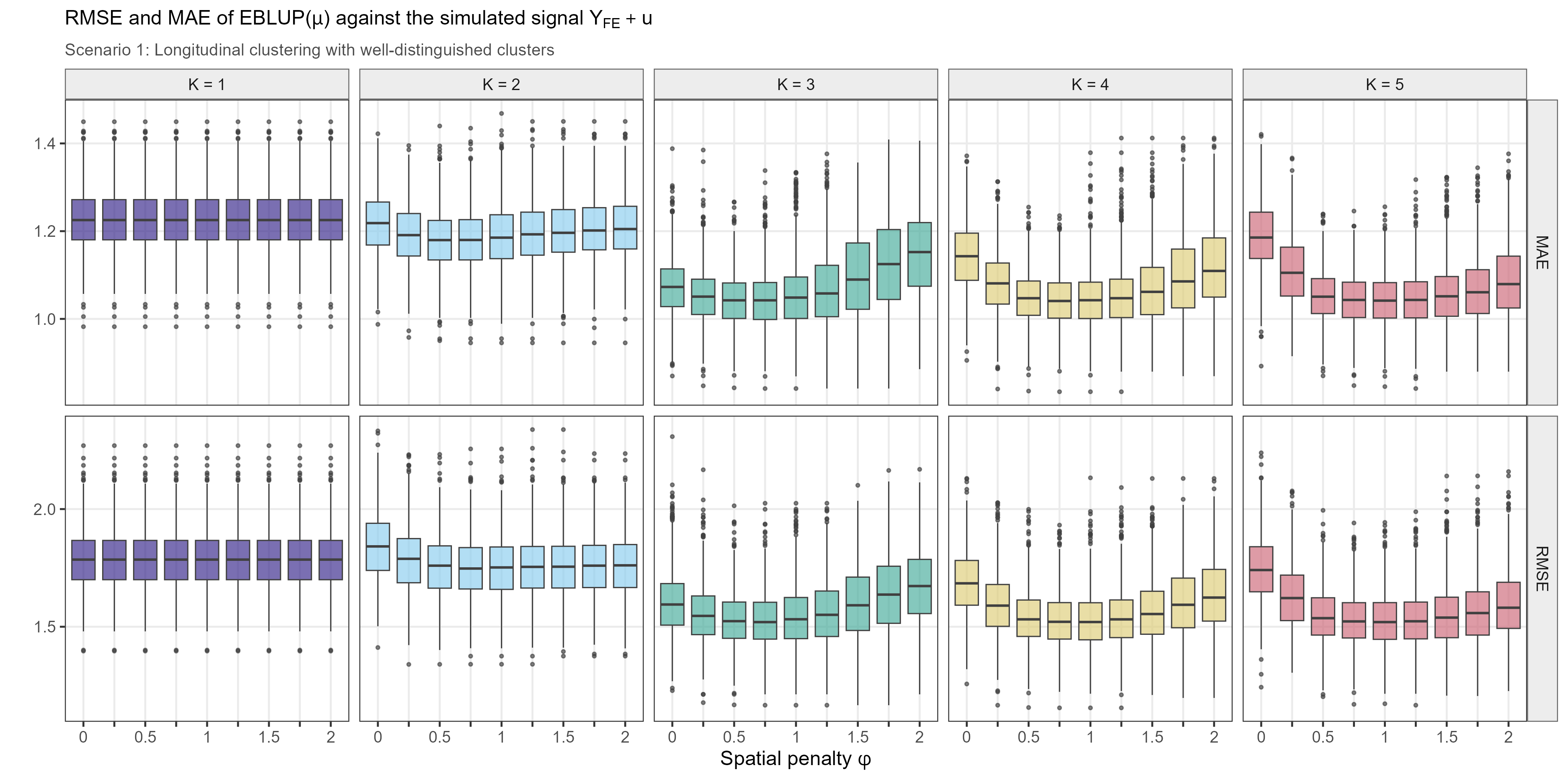}
  \caption{Scenario 1: RMSE of the EBLUP by $K$ and $\phi$.}
\end{figure}

\begin{figure}[!htb]
  \centering
  \includegraphics[width=0.9\linewidth]{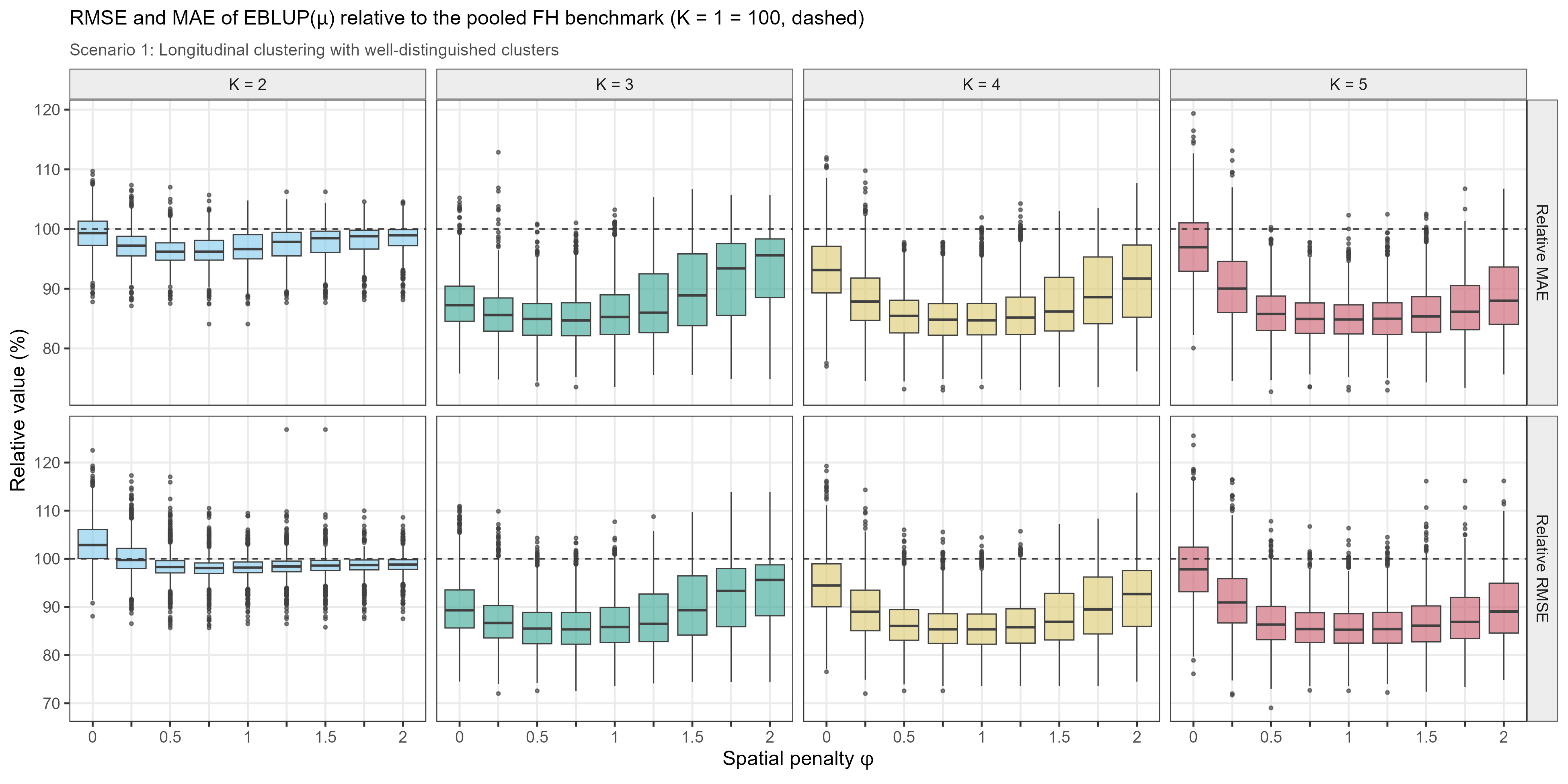}
  \caption{Scenario 1: RMSE of the EBLUP relative to the pooled FH benchmark (\%, dashed line at 100), by $K$ and $\phi$.}
\end{figure}

\begin{figure}[!htb]
  \centering
  \includegraphics[width=0.95\linewidth]{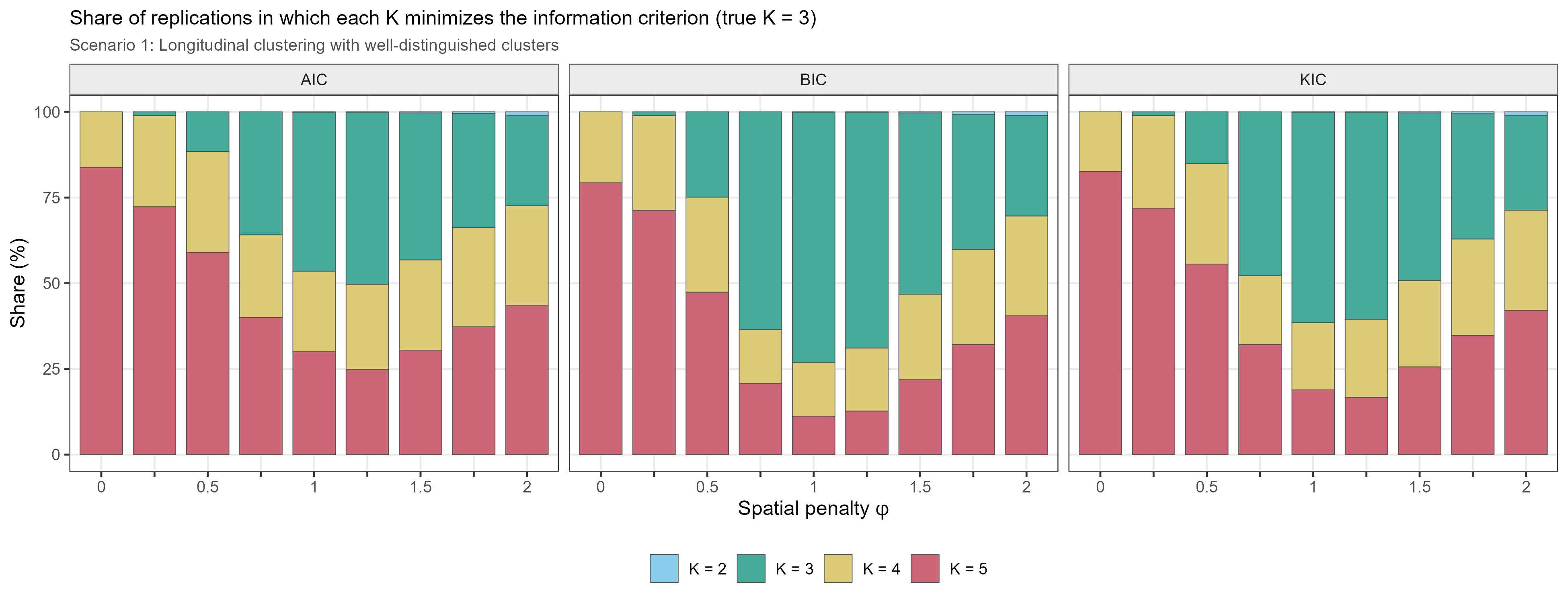}
  \caption{Scenario 1: share of replications in which each $K$ minimizes AIC, BIC, and KIC, by $\phi$.}
\end{figure}

\begin{figure}[!htb]
  \centering
  \includegraphics[width=0.95\linewidth]{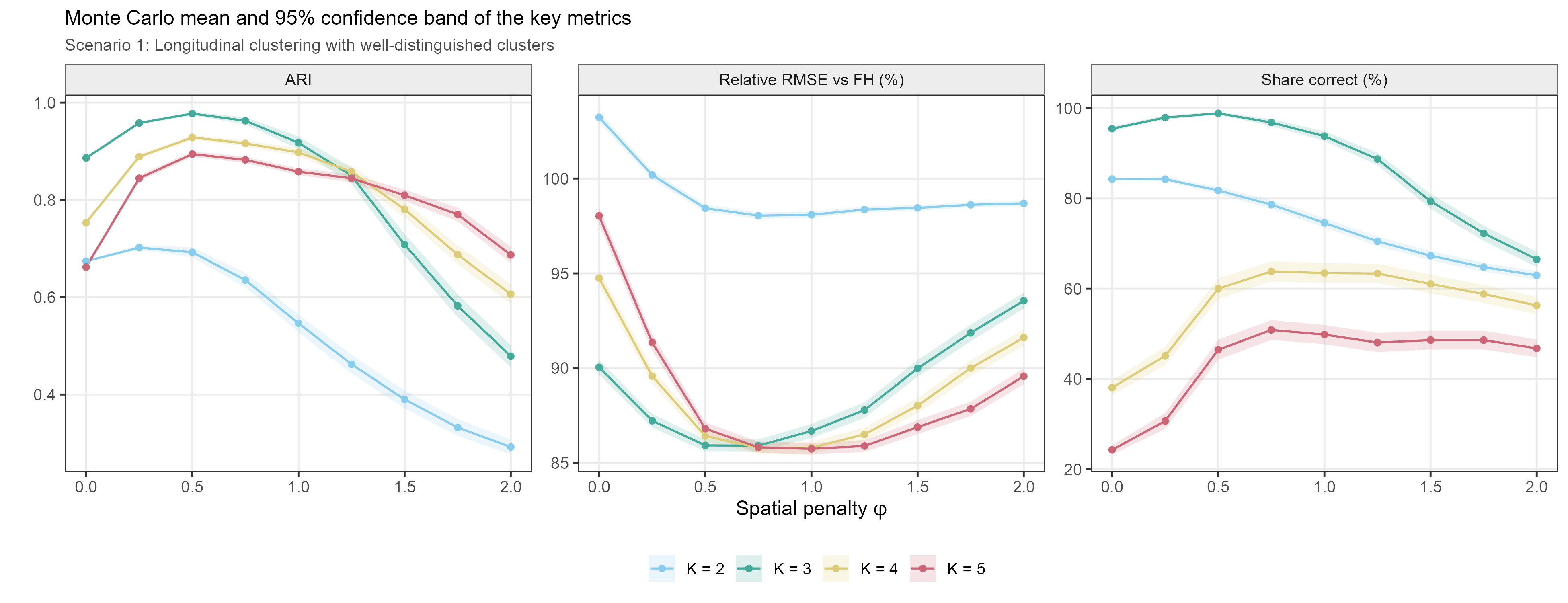}
  \caption{Scenario 1: Monte Carlo mean and 95\% band of ARI, share of correct assignments, and relative RMSE, as functions of $\phi$, by $K$.}
\end{figure}

\begin{figure}[!htb]
  \centering
  \includegraphics[width=0.8\linewidth]{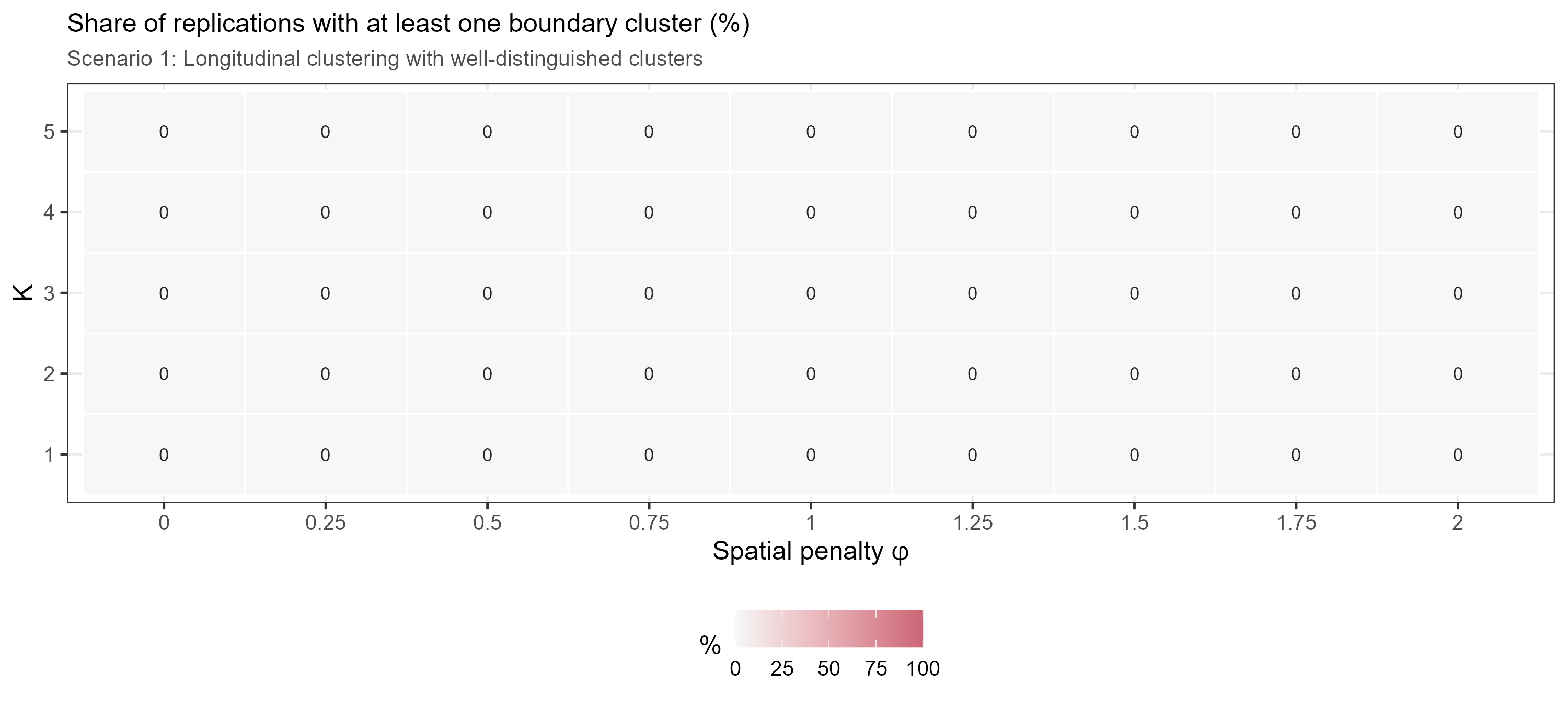}
  \caption{Scenario 1: share of fits with at least one boundary variance solution, by $K$ and $\phi$ (adjusted REML).}
\end{figure}

\begin{table}[!htb]
  \centering
  \caption{Scenario 1: Monte Carlo bias and standard deviation of the cluster-wise slopes and random-effect variances at the true $K=3$, by $\phi$ (auto-generated).}
  \adjustbox{max width=\linewidth}{
\begin{tabular}{lllll}
\toprule
Parameter & Group & Phi & Bias & SD \\
\midrule
beta1 & G1 & 0.000 & 0.085 & 0.609 \\
beta1 & G1 & 0.250 & 0.041 & 0.391 \\
beta1 & G1 & 0.500 & 0.018 & 0.301 \\
beta1 & G1 & 0.750 & 0.070 & 0.522 \\
beta1 & G1 & 1.000 & 0.230 & 1.029 \\
beta1 & G1 & 1.250 & 0.315 & 1.105 \\
beta1 & G1 & 1.500 & 0.742 & 1.679 \\
beta1 & G1 & 1.750 & 1.060 & 1.909 \\
beta1 & G1 & 2.000 & 1.372 & 2.040 \\
beta1 & G2 & 0.000 & -0.030 & 0.810 \\
beta1 & G2 & 0.250 & -0.052 & 0.596 \\
beta1 & G2 & 0.500 & -0.056 & 0.592 \\
beta1 & G2 & 0.750 & -0.186 & 1.088 \\
beta1 & G2 & 1.000 & -0.314 & 1.333 \\
beta1 & G2 & 1.250 & -0.639 & 1.753 \\
beta1 & G2 & 1.500 & -1.042 & 2.185 \\
beta1 & G2 & 1.750 & -1.211 & 2.316 \\
beta1 & G2 & 2.000 & -1.393 & 2.549 \\
beta1 & G3 & 0.000 & -0.297 & 1.226 \\
beta1 & G3 & 0.250 & -0.244 & 1.089 \\
beta1 & G3 & 0.500 & -0.164 & 1.144 \\
beta1 & G3 & 0.750 & -0.293 & 1.589 \\
beta1 & G3 & 1.000 & -0.578 & 2.124 \\
beta1 & G3 & 1.250 & -1.190 & 2.891 \\
beta1 & G3 & 1.500 & -2.243 & 3.594 \\
beta1 & G3 & 1.750 & -2.976 & 3.862 \\
beta1 & G3 & 2.000 & -3.614 & 4.050 \\
sigma2u & G1 & 0.000 & 4.779 & 60.840 \\
sigma2u & G1 & 0.250 & 3.308 & 48.029 \\
sigma2u & G1 & 0.500 & 2.851 & 49.362 \\
sigma2u & G1 & 0.750 & 9.180 & 70.500 \\
sigma2u & G1 & 1.000 & 25.448 & 143.037 \\
sigma2u & G1 & 1.250 & 41.686 & 168.517 \\
sigma2u & G1 & 1.500 & 91.225 & 244.949 \\
sigma2u & G1 & 1.750 & 127.041 & 275.580 \\
sigma2u & G1 & 2.000 & 158.902 & 288.140 \\
sigma2u & G2 & 0.000 & 3.052 & 47.815 \\
sigma2u & G2 & 0.250 & 0.755 & 22.623 \\
sigma2u & G2 & 0.500 & 2.063 & 41.541 \\
sigma2u & G2 & 0.750 & 6.157 & 51.729 \\
sigma2u & G2 & 1.000 & 25.948 & 139.917 \\
sigma2u & G2 & 1.250 & 50.471 & 150.596 \\
sigma2u & G2 & 1.500 & 86.932 & 216.796 \\
sigma2u & G2 & 1.750 & 125.152 & 263.805 \\
sigma2u & G2 & 2.000 & 150.827 & 284.518 \\
sigma2u & G3 & 0.000 & 7.184 & 40.621 \\
sigma2u & G3 & 0.250 & 1.684 & 16.302 \\
sigma2u & G3 & 0.500 & 4.932 & 44.485 \\
sigma2u & G3 & 0.750 & 7.342 & 40.108 \\
sigma2u & G3 & 1.000 & 17.477 & 89.234 \\
sigma2u & G3 & 1.250 & 30.230 & 131.519 \\
sigma2u & G3 & 1.500 & 68.586 & 205.667 \\
sigma2u & G3 & 1.750 & 121.033 & 372.848 \\
sigma2u & G3 & 2.000 & 136.401 & 277.512 \\
\bottomrule
\end{tabular}
}
\end{table}

\begin{table}[!htb]
  \centering
  \caption{Scenario 1: clustering and prediction performance summary (auto-generated).}
  \adjustbox{max width=\linewidth}{
\begin{tabular}{lllllllll}
\toprule
MC\_G & Phi & ARI\_mean & ARI\_sd & Share\_mean & Share\_sd & RelRMSE\_mean & RelRMSE\_sd & Boundary\_pct \\
\midrule
1.000 & 0.000 & 0.000 & 0.000 & 100.000 & 0.000 & 100.000 & 0.000 & 0.000 \\
1.000 & 0.250 & 0.000 & 0.000 & 100.000 & 0.000 & 100.000 & 0.000 & 0.000 \\
1.000 & 0.500 & 0.000 & 0.000 & 100.000 & 0.000 & 100.000 & 0.000 & 0.000 \\
1.000 & 0.750 & 0.000 & 0.000 & 100.000 & 0.000 & 100.000 & 0.000 & 0.000 \\
1.000 & 1.000 & 0.000 & 0.000 & 100.000 & 0.000 & 100.000 & 0.000 & 0.000 \\
1.000 & 1.250 & 0.000 & 0.000 & 100.000 & 0.000 & 100.000 & 0.000 & 0.000 \\
1.000 & 1.500 & 0.000 & 0.000 & 100.000 & 0.000 & 100.000 & 0.000 & 0.000 \\
1.000 & 1.750 & 0.000 & 0.000 & 100.000 & 0.000 & 100.000 & 0.000 & 0.000 \\
1.000 & 2.000 & 0.000 & 0.000 & 100.000 & 0.000 & 100.000 & 0.000 & 0.000 \\
2.000 & 0.000 & 0.674 & 0.034 & 84.301 & 1.126 & 103.235 & 4.642 & 0.000 \\
2.000 & 0.250 & 0.702 & 0.080 & 84.268 & 7.347 & 100.189 & 3.699 & 0.000 \\
2.000 & 0.500 & 0.692 & 0.166 & 81.796 & 13.242 & 98.436 & 3.121 & 0.000 \\
2.000 & 0.750 & 0.635 & 0.236 & 78.636 & 15.338 & 98.046 & 2.753 & 0.000 \\
2.000 & 1.000 & 0.546 & 0.295 & 74.598 & 16.828 & 98.090 & 2.430 & 0.000 \\
2.000 & 1.250 & 0.462 & 0.308 & 70.512 & 17.256 & 98.365 & 2.520 & 0.000 \\
2.000 & 1.500 & 0.390 & 0.297 & 67.307 & 16.650 & 98.454 & 2.438 & 0.000 \\
2.000 & 1.750 & 0.332 & 0.275 & 64.784 & 15.653 & 98.618 & 2.179 & 0.000 \\
2.000 & 2.000 & 0.292 & 0.253 & 62.968 & 14.726 & 98.693 & 2.160 & 0.000 \\
3.000 & 0.000 & 0.886 & 0.053 & 95.475 & 7.207 & 90.044 & 6.242 & 0.000 \\
3.000 & 0.250 & 0.958 & 0.047 & 97.936 & 6.991 & 87.227 & 5.768 & 0.000 \\
3.000 & 0.500 & 0.977 & 0.057 & 98.885 & 5.173 & 85.924 & 5.150 & 0.000 \\
3.000 & 0.750 & 0.963 & 0.119 & 96.845 & 12.324 & 85.910 & 5.342 & 0.000 \\
3.000 & 1.000 & 0.917 & 0.205 & 93.805 & 16.738 & 86.684 & 5.977 & 0.000 \\
3.000 & 1.250 & 0.850 & 0.276 & 88.738 & 21.870 & 87.785 & 6.615 & 0.000 \\
3.000 & 1.500 & 0.708 & 0.356 & 79.393 & 26.658 & 89.988 & 7.099 & 0.000 \\
3.000 & 1.750 & 0.582 & 0.379 & 72.296 & 27.500 & 91.860 & 7.061 & 0.000 \\
3.000 & 2.000 & 0.479 & 0.368 & 66.501 & 26.933 & 93.558 & 6.639 & 0.000 \\
4.000 & 0.000 & 0.753 & 0.075 & 38.084 & 25.844 & 94.752 & 6.838 & 0.000 \\
4.000 & 0.250 & 0.889 & 0.077 & 45.112 & 33.180 & 89.578 & 6.404 & 0.000 \\
4.000 & 0.500 & 0.928 & 0.074 & 60.004 & 36.603 & 86.421 & 5.158 & 0.000 \\
4.000 & 0.750 & 0.916 & 0.094 & 63.847 & 36.164 & 85.801 & 5.098 & 0.000 \\
4.000 & 1.000 & 0.898 & 0.136 & 63.471 & 35.666 & 85.799 & 5.187 & 0.000 \\
4.000 & 1.250 & 0.858 & 0.194 & 63.381 & 34.627 & 86.509 & 5.718 & 0.000 \\
4.000 & 1.500 & 0.781 & 0.264 & 61.053 & 33.542 & 88.021 & 6.449 & 0.000 \\
4.000 & 1.750 & 0.687 & 0.309 & 58.812 & 32.188 & 90.000 & 6.862 & 0.000 \\
4.000 & 2.000 & 0.606 & 0.323 & 56.285 & 30.523 & 91.614 & 6.955 & 0.000 \\
5.000 & 0.000 & 0.662 & 0.087 & 24.279 & 17.055 & 98.031 & 6.999 & 0.000 \\
5.000 & 0.250 & 0.844 & 0.094 & 30.687 & 28.265 & 91.355 & 7.043 & 0.000 \\
5.000 & 0.500 & 0.894 & 0.094 & 46.485 & 35.553 & 86.809 & 5.409 & 0.000 \\
5.000 & 0.750 & 0.882 & 0.100 & 50.855 & 35.281 & 85.823 & 4.880 & 0.000 \\
5.000 & 1.000 & 0.858 & 0.111 & 49.823 & 34.213 & 85.743 & 4.945 & 0.000 \\
5.000 & 1.250 & 0.844 & 0.132 & 48.066 & 34.117 & 85.892 & 5.051 & 0.000 \\
5.000 & 1.500 & 0.810 & 0.186 & 48.614 & 33.557 & 86.890 & 5.924 & 0.000 \\
5.000 & 1.750 & 0.770 & 0.226 & 48.616 & 33.538 & 87.849 & 6.185 & 0.000 \\
5.000 & 2.000 & 0.687 & 0.267 & 46.794 & 31.343 & 89.573 & 6.621 & 0.000 \\
\bottomrule
\end{tabular}
}
\end{table}

\clearpage

\subsection{Scenario 2: longitude-based clusters, poor separation}

\begin{figure}[!htb]
  \centering
  \includegraphics[width=0.95\linewidth]{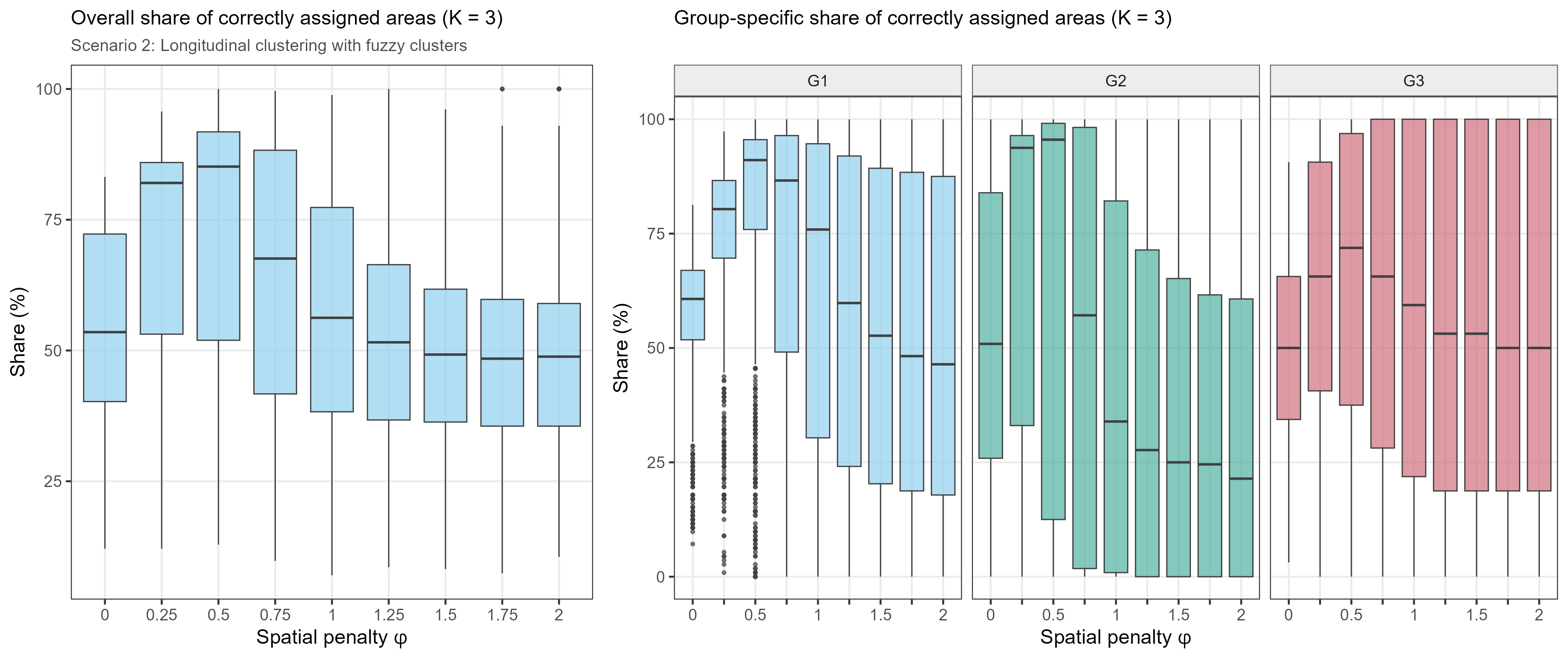}
  \caption{Scenario 2: overall (left) and group-specific (right) share of correctly assigned areas at the true $K=3$, by spatial penalty $\phi$.}
\end{figure}

\begin{figure}[!htb]
  \centering
  \includegraphics[width=0.9\linewidth]{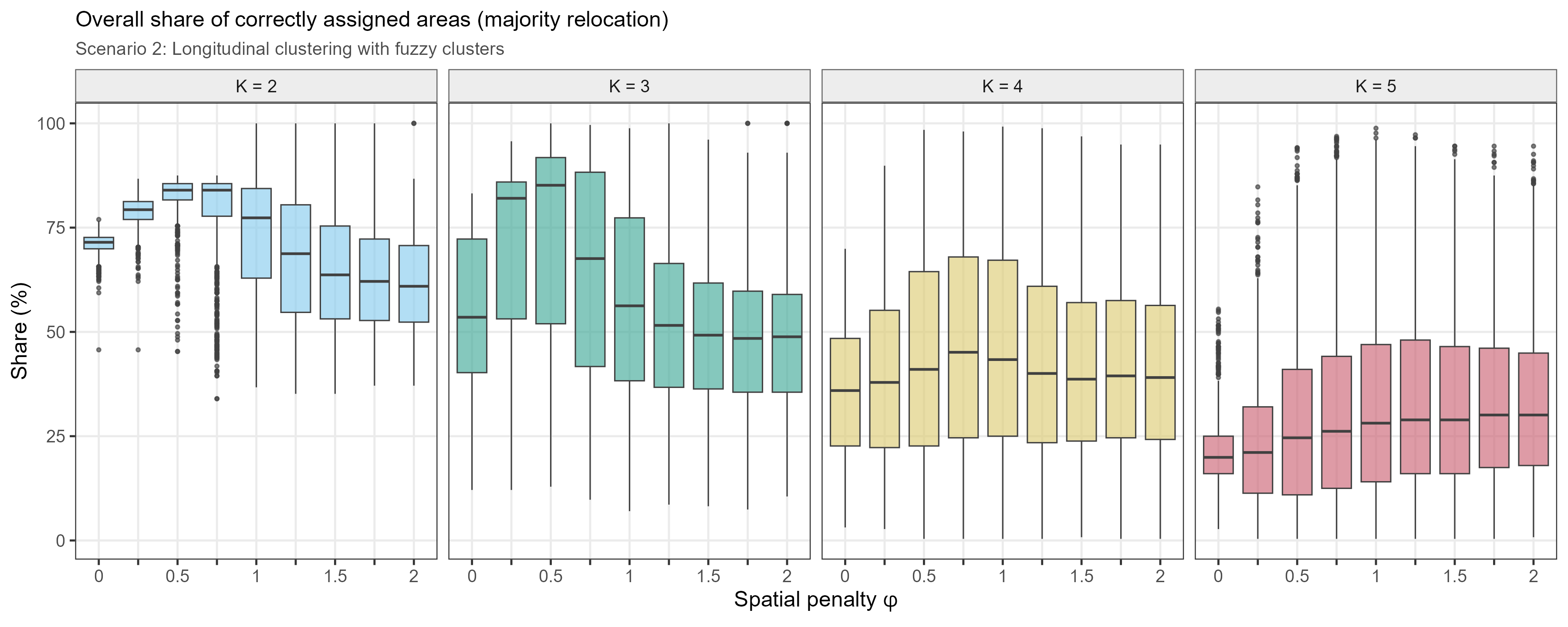}
  \caption{Scenario 2: overall share of correctly assigned areas by number of clusters $K$ and spatial penalty $\phi$.}
\end{figure}

\begin{figure}[!htb]
  \centering
  \includegraphics[width=0.9\linewidth]{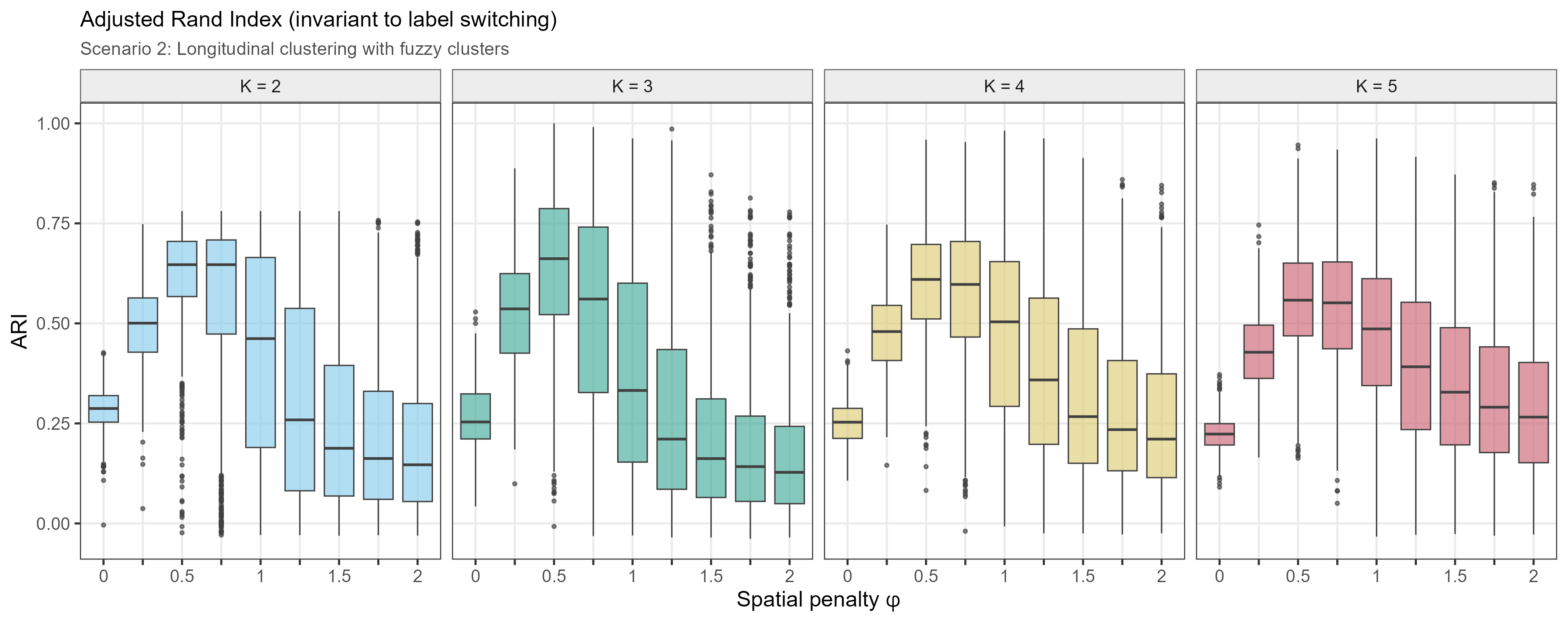}
  \caption{Scenario 2: Adjusted Rand Index between estimated and true partition, by $K$ and $\phi$.}
\end{figure}

\begin{figure}[!htb]
  \centering
  \includegraphics[width=0.95\linewidth]{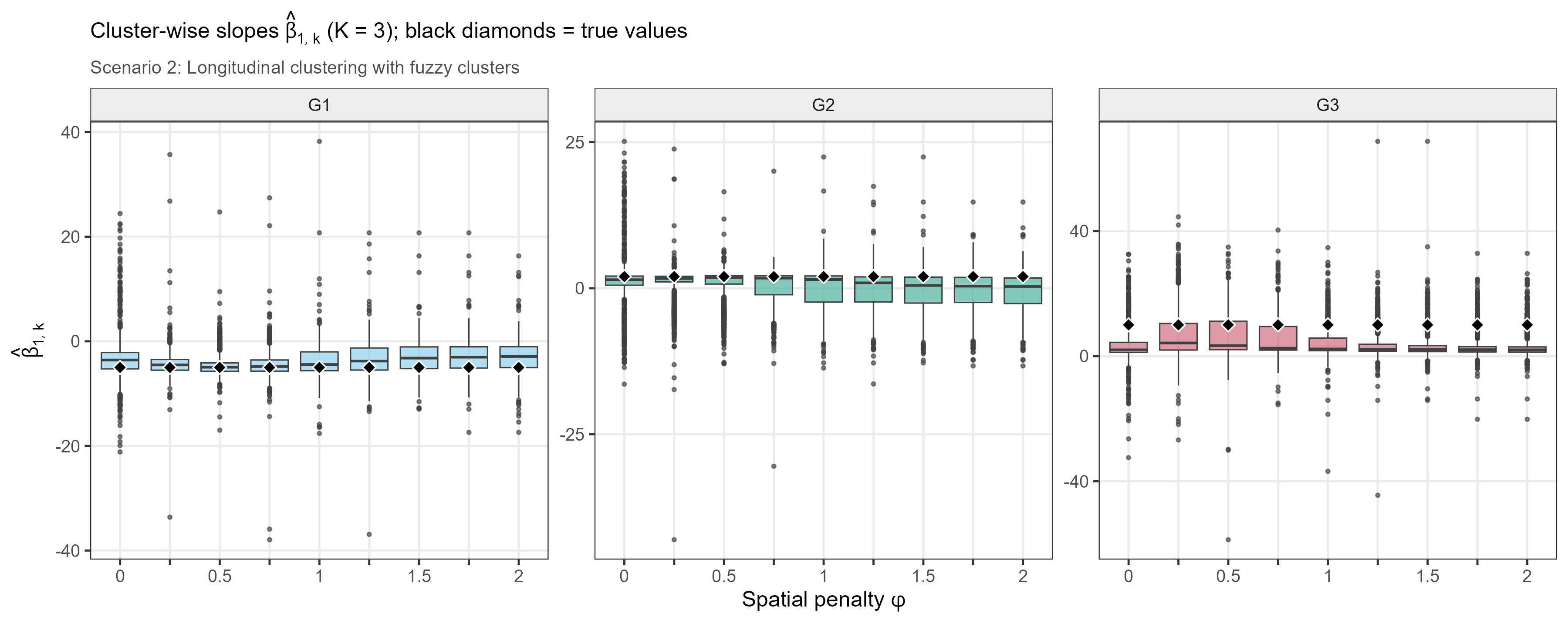}
  \caption{Scenario 2: cluster-specific slope estimates at the true $K=3$, by $\phi$. Black diamonds mark the true values.}
\end{figure}

\begin{figure}[!htb]
  \centering
  \includegraphics[width=0.95\linewidth]{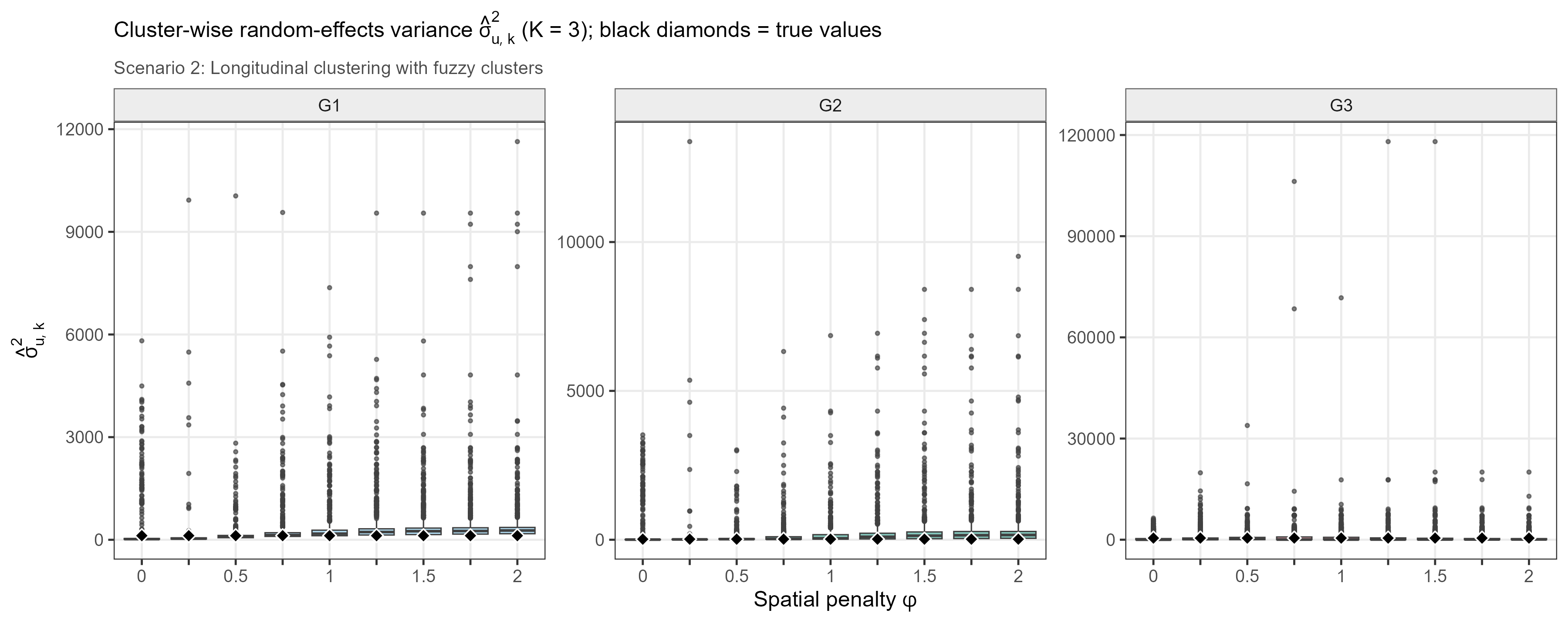}
  \caption{Scenario 2: cluster-specific random-effect variance estimates at the true $K=3$, by $\phi$. Black diamonds mark the true values.}
\end{figure}

\clearpage

\begin{figure}[!htb]
  \centering
  \includegraphics[width=0.95\linewidth]{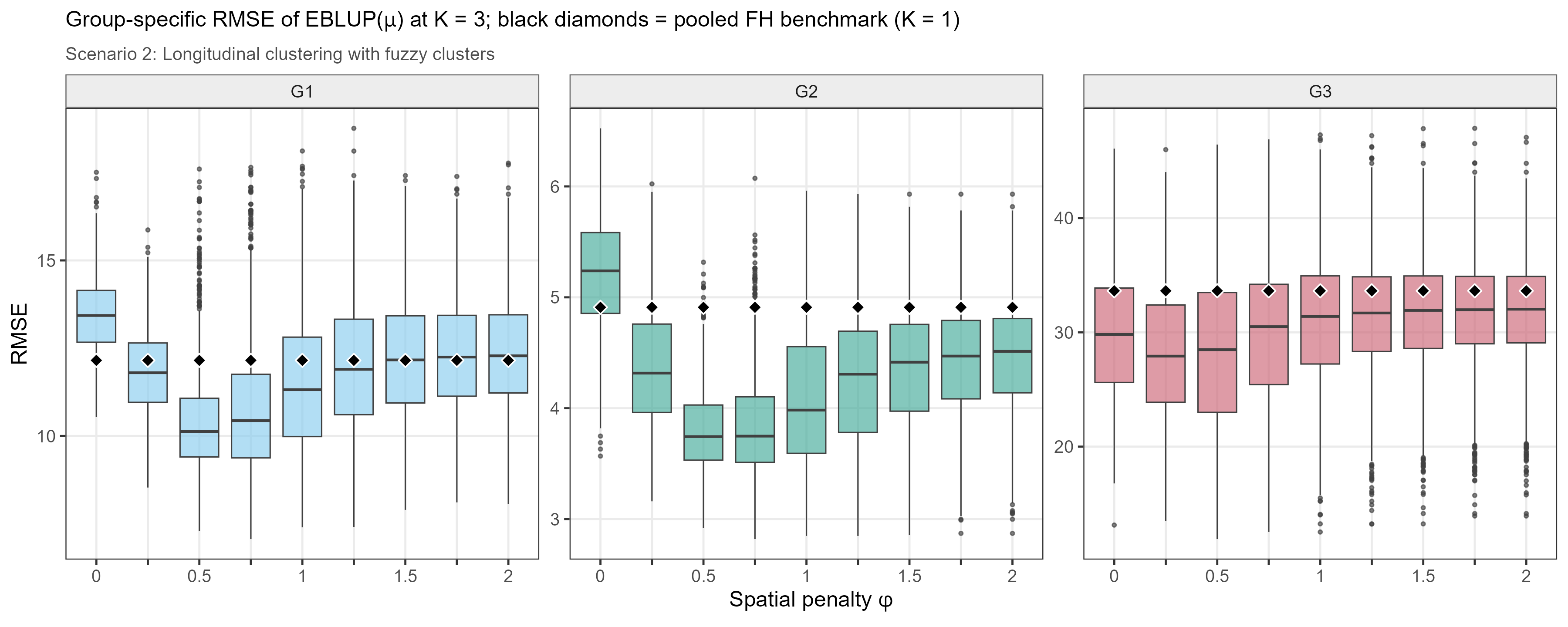}
  \caption{Scenario 2: RMSE of the EBLUP by true group at $K=3$. Black diamonds mark the mean RMSE of the pooled FH benchmark ($K=1$).}
\end{figure}

\begin{figure}[!htb]
  \centering
  \includegraphics[width=0.9\linewidth]{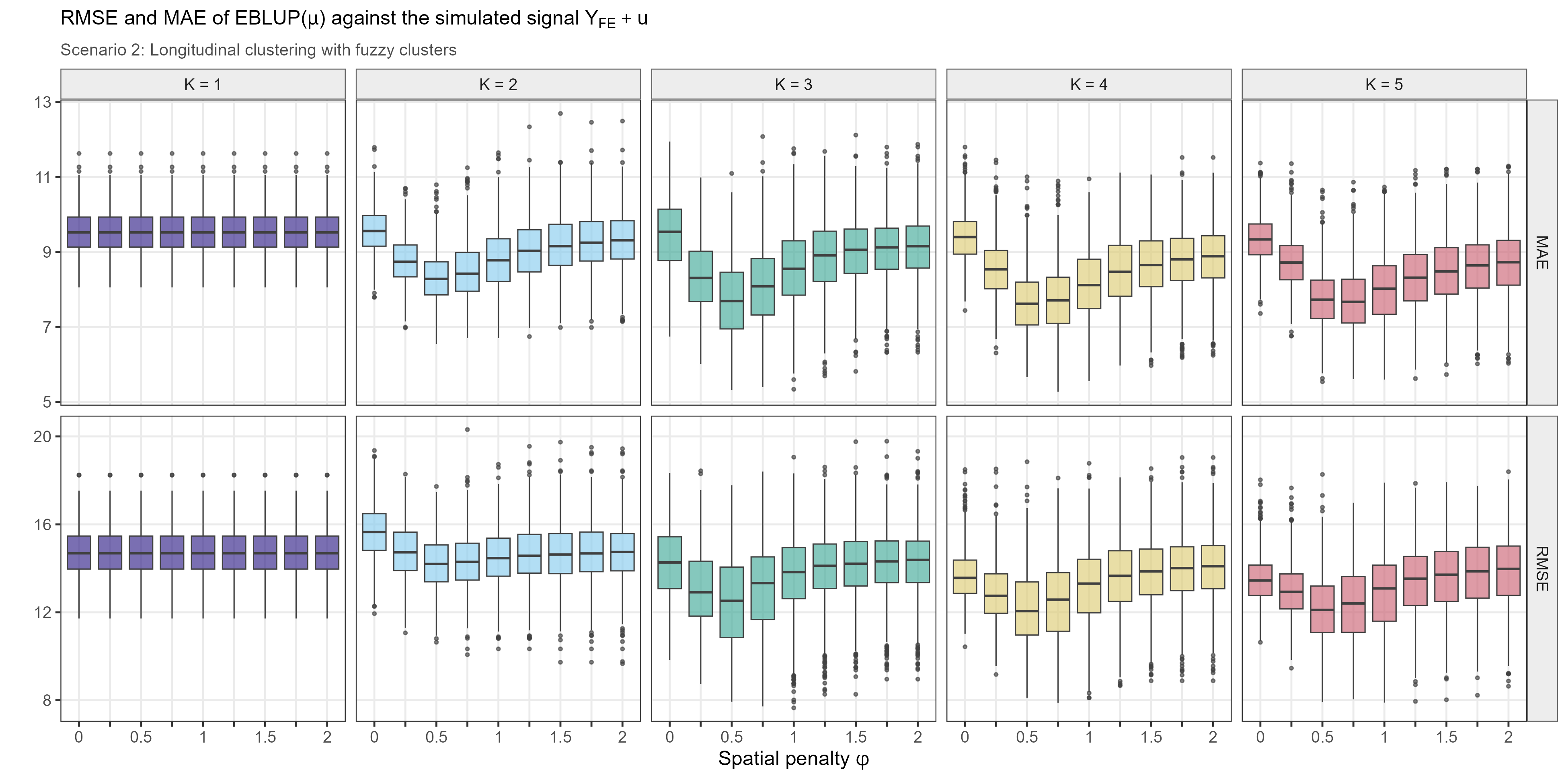}
  \caption{Scenario 2: RMSE of the EBLUP by $K$ and $\phi$.}
\end{figure}

\begin{figure}[!htb]
  \centering
  \includegraphics[width=0.9\linewidth]{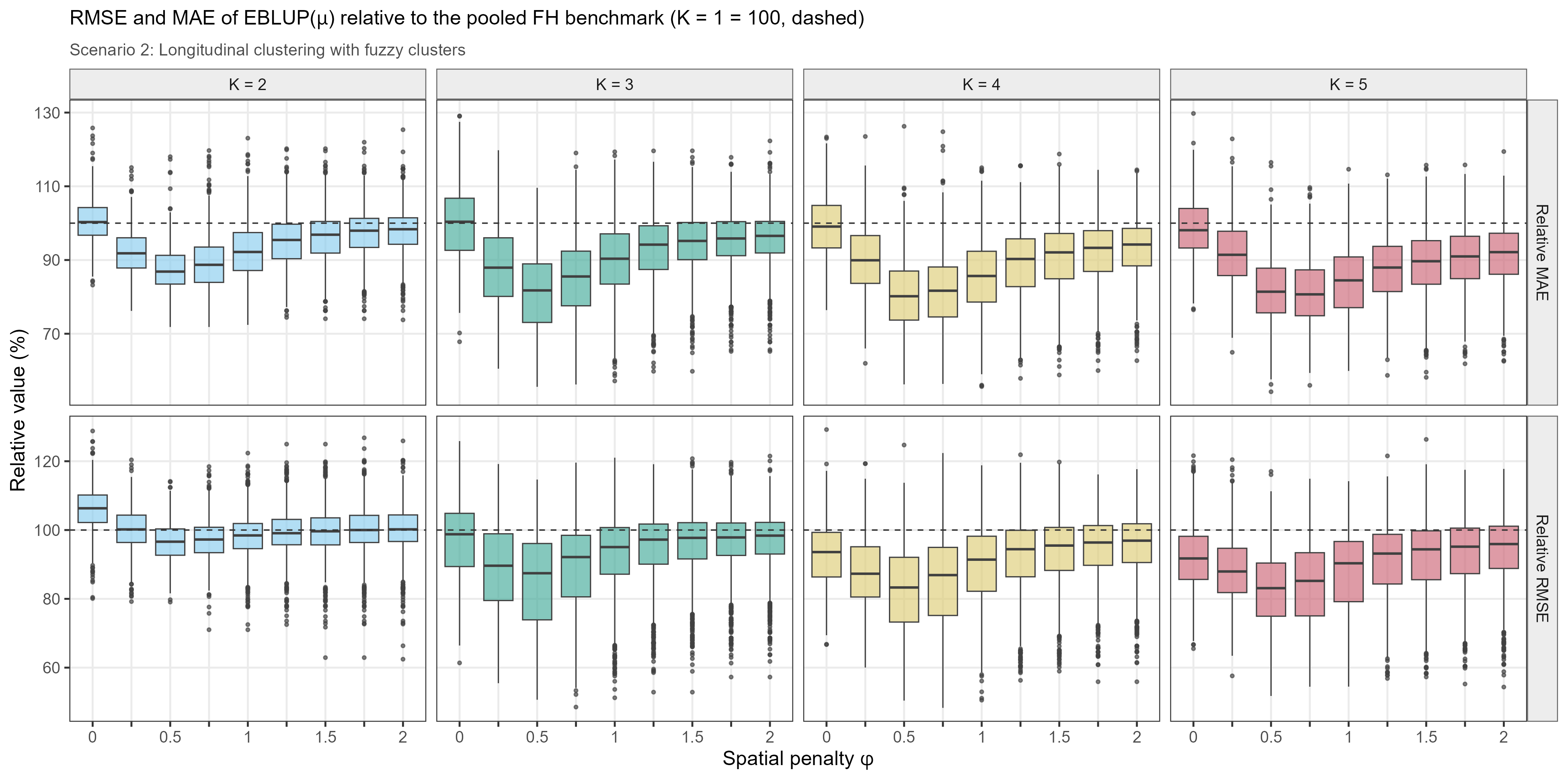}
  \caption{Scenario 2: RMSE of the EBLUP relative to the pooled FH benchmark (\%, dashed line at 100), by $K$ and $\phi$.}
\end{figure}

\begin{figure}[!htb]
  \centering
  \includegraphics[width=0.95\linewidth]{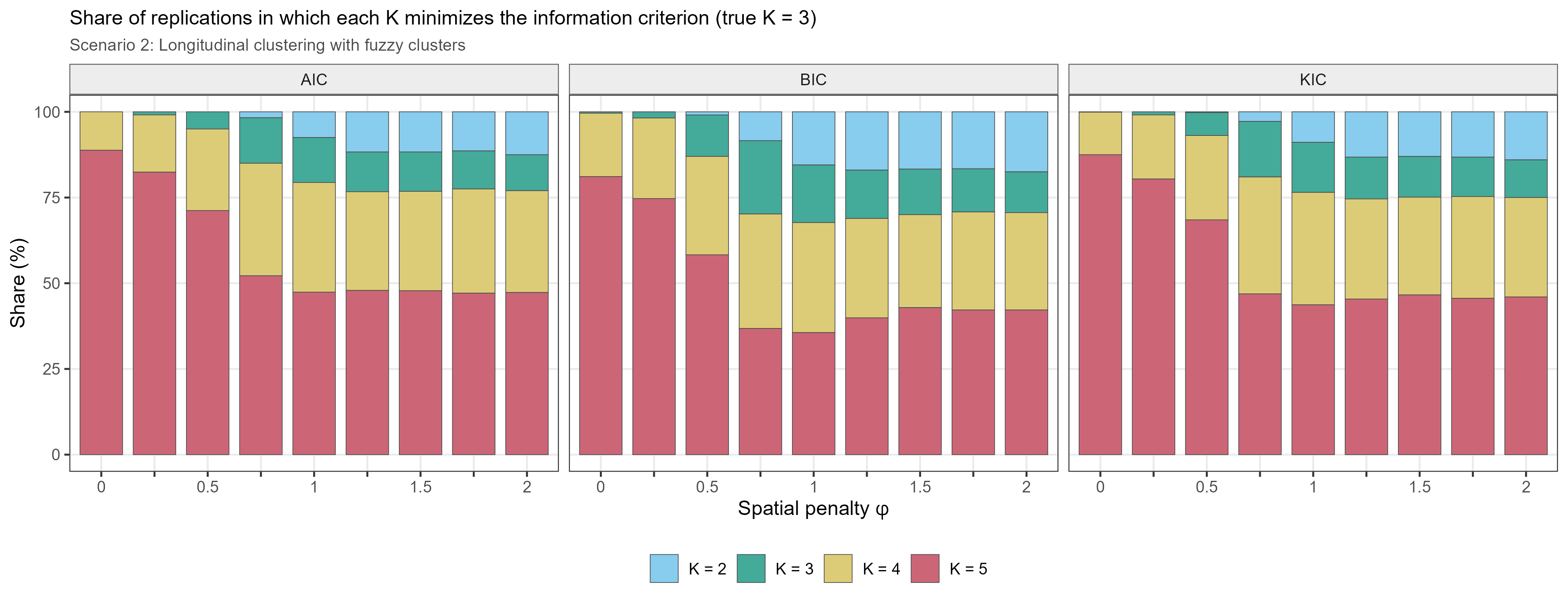}
  \caption{Scenario 2: share of replications in which each $K$ minimizes AIC, BIC, and KIC, by $\phi$.}
\end{figure}

\begin{figure}[!htb]
  \centering
  \includegraphics[width=0.95\linewidth]{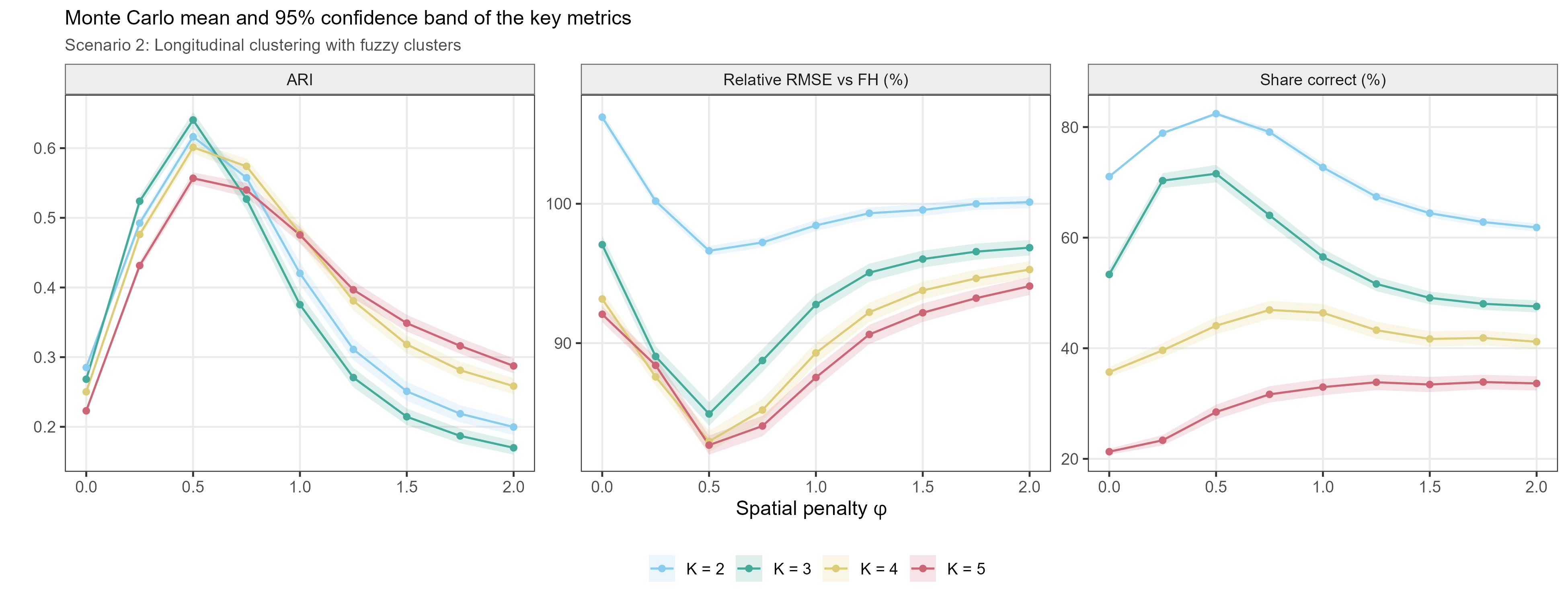}
  \caption{Scenario 2: Monte Carlo mean and 95\% band of ARI, share of correct assignments, and relative RMSE, as functions of $\phi$, by $K$.}
\end{figure}

\begin{figure}[!htb]
  \centering
  \includegraphics[width=0.8\linewidth]{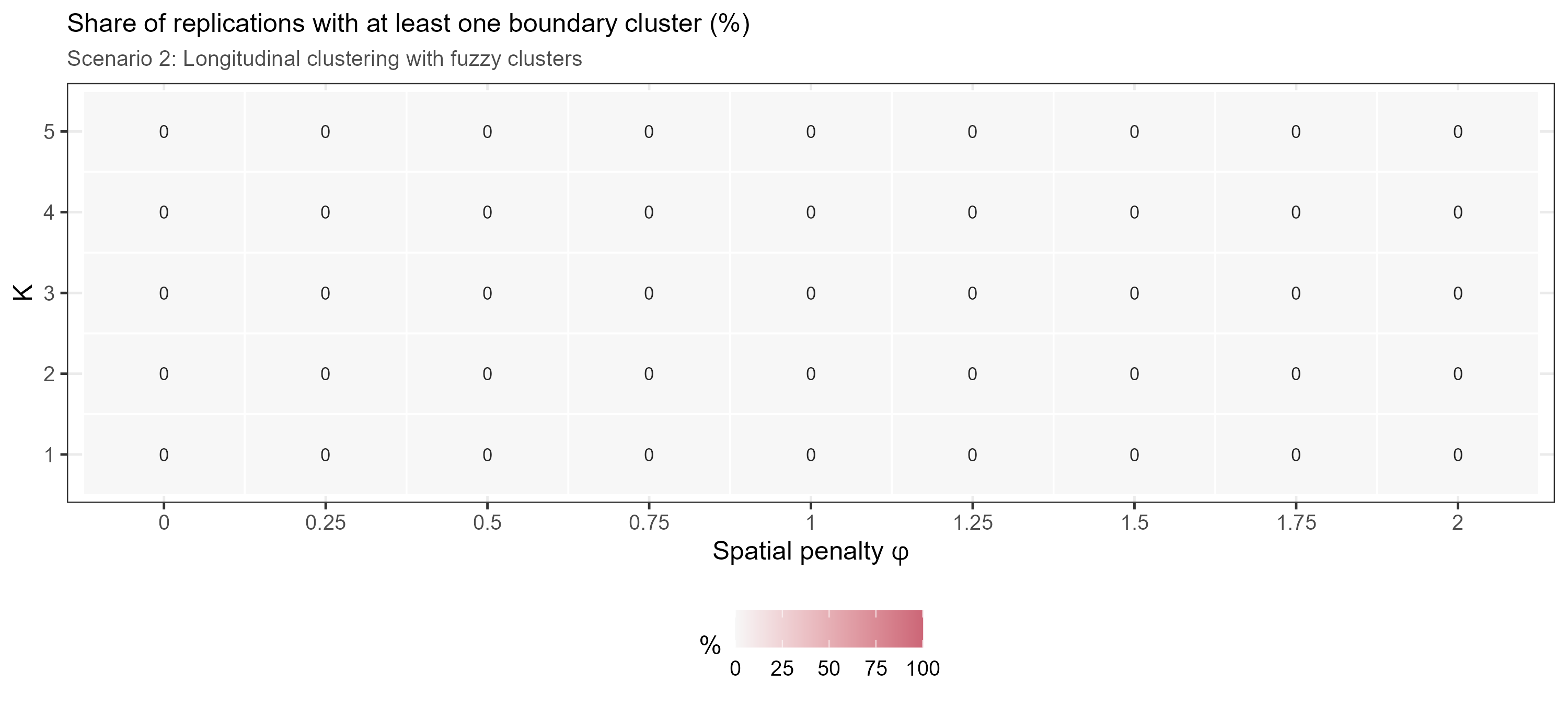}
  \caption{Scenario 2: share of fits with at least one boundary variance solution, by $K$ and $\phi$ (adjusted REML).}
\end{figure}

\begin{table}[!htb]
  \centering
  \caption{Scenario 2: Monte Carlo bias and standard deviation of the cluster-wise slopes and random-effect variances at the true $K=3$, by $\phi$ (auto-generated).}
  \adjustbox{max width=\linewidth}{
\begin{tabular}{lllll}
\toprule
Parameter & Group & Phi & Bias & SD \\
\midrule
beta1 & G1 & 0.000 & 1.714 & 4.808 \\
beta1 & G1 & 0.250 & 0.668 & 2.681 \\
beta1 & G1 & 0.500 & 0.363 & 2.198 \\
beta1 & G1 & 0.750 & 0.602 & 3.031 \\
beta1 & G1 & 1.000 & 1.195 & 3.200 \\
beta1 & G1 & 1.250 & 1.625 & 3.338 \\
beta1 & G1 & 1.500 & 1.857 & 3.003 \\
beta1 & G1 & 1.750 & 1.946 & 2.992 \\
beta1 & G1 & 2.000 & 1.929 & 3.041 \\
beta1 & G2 & 0.000 & -0.909 & 4.926 \\
beta1 & G2 & 0.250 & -1.339 & 3.301 \\
beta1 & G2 & 0.500 & -1.453 & 3.111 \\
beta1 & G2 & 0.750 & -1.771 & 3.253 \\
beta1 & G2 & 1.000 & -2.091 & 3.392 \\
beta1 & G2 & 1.250 & -2.359 & 3.448 \\
beta1 & G2 & 1.500 & -2.482 & 3.492 \\
beta1 & G2 & 1.750 & -2.519 & 3.353 \\
beta1 & G2 & 2.000 & -2.605 & 3.356 \\
beta1 & G3 & 0.000 & -6.343 & 6.265 \\
beta1 & G3 & 0.250 & -3.169 & 7.640 \\
beta1 & G3 & 0.500 & -3.396 & 6.807 \\
beta1 & G3 & 0.750 & -4.315 & 6.109 \\
beta1 & G3 & 1.000 & -5.475 & 5.649 \\
beta1 & G3 & 1.250 & -6.138 & 5.301 \\
beta1 & G3 & 1.500 & -6.338 & 5.032 \\
beta1 & G3 & 1.750 & -6.599 & 4.489 \\
beta1 & G3 & 2.000 & -6.730 & 4.461 \\
sigma2u & G1 & 0.000 & 58.178 & 616.730 \\
sigma2u & G1 & 0.250 & -47.077 & 421.592 \\
sigma2u & G1 & 0.500 & 28.179 & 400.826 \\
sigma2u & G1 & 0.750 & 143.175 & 577.929 \\
sigma2u & G1 & 1.000 & 198.210 & 575.795 \\
sigma2u & G1 & 1.250 & 257.246 & 634.415 \\
sigma2u & G1 & 1.500 & 258.735 & 586.765 \\
sigma2u & G1 & 1.750 & 279.621 & 719.262 \\
sigma2u & G1 & 2.000 & 291.661 & 796.405 \\
sigma2u & G2 & 0.000 & 121.202 & 523.861 \\
sigma2u & G2 & 0.250 & 26.283 & 497.366 \\
sigma2u & G2 & 0.500 & 51.447 & 230.143 \\
sigma2u & G2 & 0.750 & 99.236 & 370.962 \\
sigma2u & G2 & 1.000 & 145.742 & 423.958 \\
sigma2u & G2 & 1.250 & 210.075 & 568.344 \\
sigma2u & G2 & 1.500 & 264.190 & 714.664 \\
sigma2u & G2 & 1.750 & 271.382 & 681.623 \\
sigma2u & G2 & 2.000 & 283.654 & 720.319 \\
sigma2u & G3 & 0.000 & -32.409 & 996.276 \\
sigma2u & G3 & 0.250 & 205.293 & 1635.083 \\
sigma2u & G3 & 0.500 & 123.203 & 1573.638 \\
sigma2u & G3 & 0.750 & 338.010 & 4240.819 \\
sigma2u & G3 & 1.000 & 206.695 & 2657.997 \\
sigma2u & G3 & 1.250 & 226.418 & 4027.011 \\
sigma2u & G3 & 1.500 & 216.796 & 4071.079 \\
sigma2u & G3 & 1.750 & 45.153 & 1351.520 \\
sigma2u & G3 & 2.000 & -3.030 & 1212.869 \\
\bottomrule
\end{tabular}
}
\end{table}

\begin{table}[!htb]
  \centering
  \caption{Scenario 2: clustering and prediction performance summary (auto-generated).}
  \adjustbox{max width=\linewidth}{
\begin{tabular}{lllllllll}
\toprule
MC\_G & Phi & ARI\_mean & ARI\_sd & Share\_mean & Share\_sd & RelRMSE\_mean & RelRMSE\_sd & Boundary\_pct \\
\midrule
1.000 & 0.000 & 0.000 & 0.000 & 100.000 & 0.000 & 100.000 & 0.000 & 0.000 \\
1.000 & 0.250 & 0.000 & 0.000 & 100.000 & 0.000 & 100.000 & 0.000 & 0.000 \\
1.000 & 0.500 & 0.000 & 0.000 & 100.000 & 0.000 & 100.000 & 0.000 & 0.000 \\
1.000 & 0.750 & 0.000 & 0.000 & 100.000 & 0.000 & 100.000 & 0.000 & 0.000 \\
1.000 & 1.000 & 0.000 & 0.000 & 100.000 & 0.000 & 100.000 & 0.000 & 0.000 \\
1.000 & 1.250 & 0.000 & 0.000 & 100.000 & 0.000 & 100.000 & 0.000 & 0.000 \\
1.000 & 1.500 & 0.000 & 0.000 & 100.000 & 0.000 & 100.000 & 0.000 & 0.000 \\
1.000 & 1.750 & 0.000 & 0.000 & 100.000 & 0.000 & 100.000 & 0.000 & 0.000 \\
1.000 & 2.000 & 0.000 & 0.000 & 100.000 & 0.000 & 100.000 & 0.000 & 0.000 \\
2.000 & 0.000 & 0.285 & 0.051 & 71.048 & 2.570 & 106.212 & 6.174 & 0.000 \\
2.000 & 0.250 & 0.492 & 0.098 & 78.898 & 3.657 & 100.188 & 6.048 & 0.000 \\
2.000 & 0.500 & 0.616 & 0.126 & 82.420 & 5.384 & 96.626 & 5.630 & 0.000 \\
2.000 & 0.750 & 0.558 & 0.212 & 79.100 & 10.815 & 97.226 & 5.958 & 0.000 \\
2.000 & 1.000 & 0.420 & 0.256 & 72.698 & 13.533 & 98.447 & 6.466 & 0.000 \\
2.000 & 1.250 & 0.311 & 0.248 & 67.400 & 13.695 & 99.320 & 6.705 & 0.000 \\
2.000 & 1.500 & 0.251 & 0.221 & 64.425 & 12.868 & 99.557 & 6.971 & 0.000 \\
2.000 & 1.750 & 0.219 & 0.199 & 62.820 & 12.170 & 99.992 & 7.062 & 0.000 \\
2.000 & 2.000 & 0.200 & 0.186 & 61.843 & 11.661 & 100.117 & 7.003 & 0.000 \\
3.000 & 0.000 & 0.268 & 0.079 & 53.346 & 18.575 & 97.066 & 10.405 & 0.000 \\
3.000 & 0.250 & 0.524 & 0.131 & 70.313 & 21.806 & 89.049 & 12.070 & 0.000 \\
3.000 & 0.500 & 0.640 & 0.189 & 71.567 & 25.550 & 84.919 & 13.516 & 0.000 \\
3.000 & 0.750 & 0.527 & 0.252 & 64.036 & 26.023 & 88.752 & 13.298 & 0.000 \\
3.000 & 1.000 & 0.375 & 0.258 & 56.511 & 23.489 & 92.772 & 11.801 & 0.000 \\
3.000 & 1.250 & 0.271 & 0.226 & 51.627 & 20.852 & 95.056 & 10.539 & 0.000 \\
3.000 & 1.500 & 0.214 & 0.191 & 49.108 & 18.772 & 96.031 & 9.811 & 0.000 \\
3.000 & 1.750 & 0.187 & 0.171 & 48.039 & 17.988 & 96.566 & 9.294 & 0.000 \\
3.000 & 2.000 & 0.170 & 0.159 & 47.564 & 17.451 & 96.849 & 8.997 & 0.000 \\
4.000 & 0.000 & 0.250 & 0.056 & 35.690 & 14.234 & 93.167 & 9.132 & 0.000 \\
4.000 & 0.250 & 0.476 & 0.101 & 39.623 & 20.857 & 87.579 & 10.245 & 0.000 \\
4.000 & 0.500 & 0.601 & 0.140 & 44.078 & 25.294 & 82.953 & 12.182 & 0.000 \\
4.000 & 0.750 & 0.574 & 0.180 & 46.928 & 26.254 & 85.202 & 12.745 & 0.000 \\
4.000 & 1.000 & 0.478 & 0.223 & 46.402 & 25.865 & 89.297 & 12.259 & 0.000 \\
4.000 & 1.250 & 0.381 & 0.224 & 43.267 & 24.577 & 92.218 & 11.147 & 0.000 \\
4.000 & 1.500 & 0.318 & 0.208 & 41.684 & 22.599 & 93.780 & 10.154 & 0.000 \\
4.000 & 1.750 & 0.281 & 0.195 & 41.868 & 21.827 & 94.644 & 9.847 & 0.000 \\
4.000 & 2.000 & 0.258 & 0.185 & 41.178 & 21.030 & 95.278 & 9.679 & 0.000 \\
5.000 & 0.000 & 0.223 & 0.042 & 21.294 & 8.642 & 92.071 & 9.069 & 0.000 \\
5.000 & 0.250 & 0.431 & 0.099 & 23.357 & 15.182 & 88.406 & 9.441 & 0.000 \\
5.000 & 0.500 & 0.557 & 0.138 & 28.466 & 21.561 & 82.684 & 11.038 & 0.000 \\
5.000 & 0.750 & 0.540 & 0.164 & 31.663 & 23.832 & 84.057 & 11.842 & 0.000 \\
5.000 & 1.000 & 0.475 & 0.187 & 32.974 & 23.728 & 87.537 & 12.047 & 0.000 \\
5.000 & 1.250 & 0.397 & 0.198 & 33.847 & 22.906 & 90.625 & 11.108 & 0.000 \\
5.000 & 1.500 & 0.349 & 0.192 & 33.445 & 22.022 & 92.179 & 10.704 & 0.000 \\
5.000 & 1.750 & 0.316 & 0.184 & 33.889 & 21.233 & 93.228 & 10.487 & 0.000 \\
5.000 & 2.000 & 0.287 & 0.173 & 33.640 & 20.668 & 94.092 & 10.317 & 0.000 \\
\bottomrule
\end{tabular}
}
\end{table}

\clearpage

\subsection{Scenario 3: altitude-based clusters, clear separation}

\begin{figure}[!htb]
  \centering
  \includegraphics[width=0.95\linewidth]{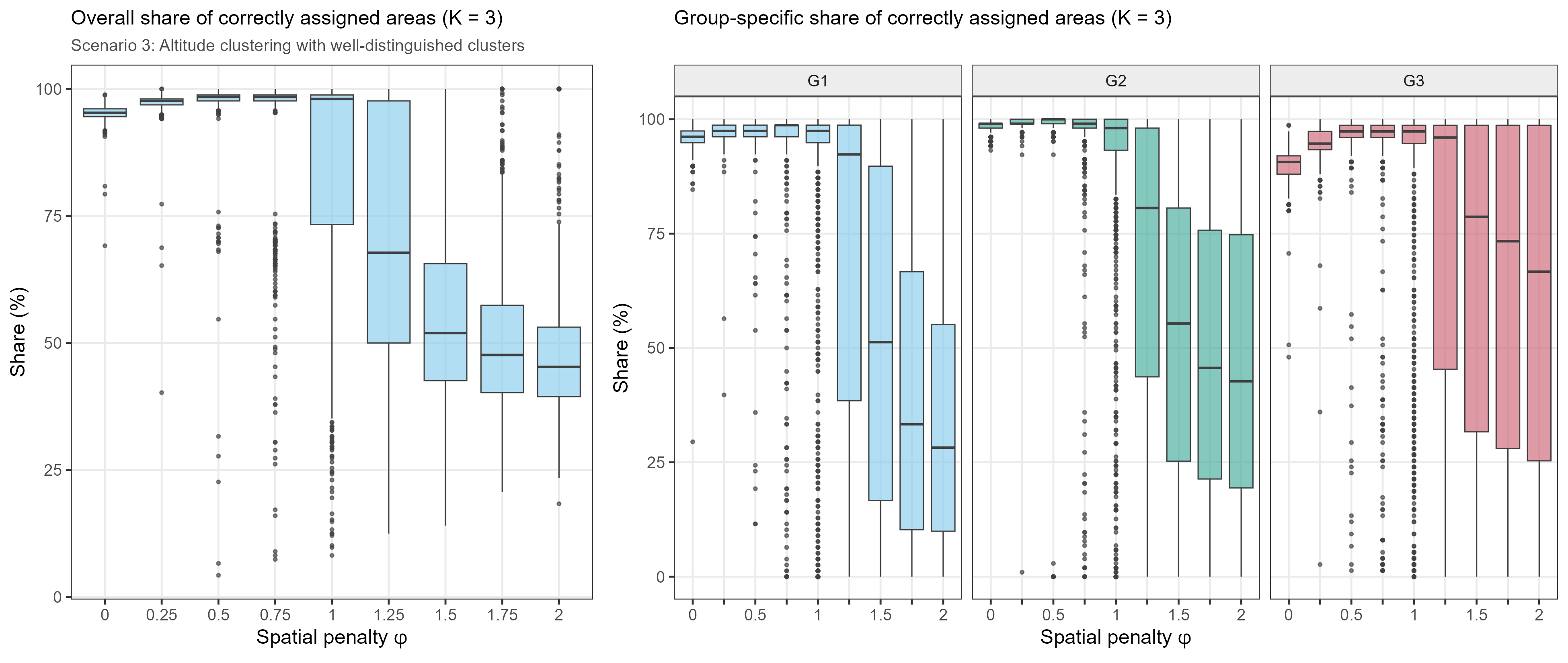}
  \caption{Scenario 3: overall (left) and group-specific (right) share of correctly assigned areas at the true $K=3$, by spatial penalty $\phi$.}
\end{figure}

\begin{figure}[!htb]
  \centering
  \includegraphics[width=0.9\linewidth]{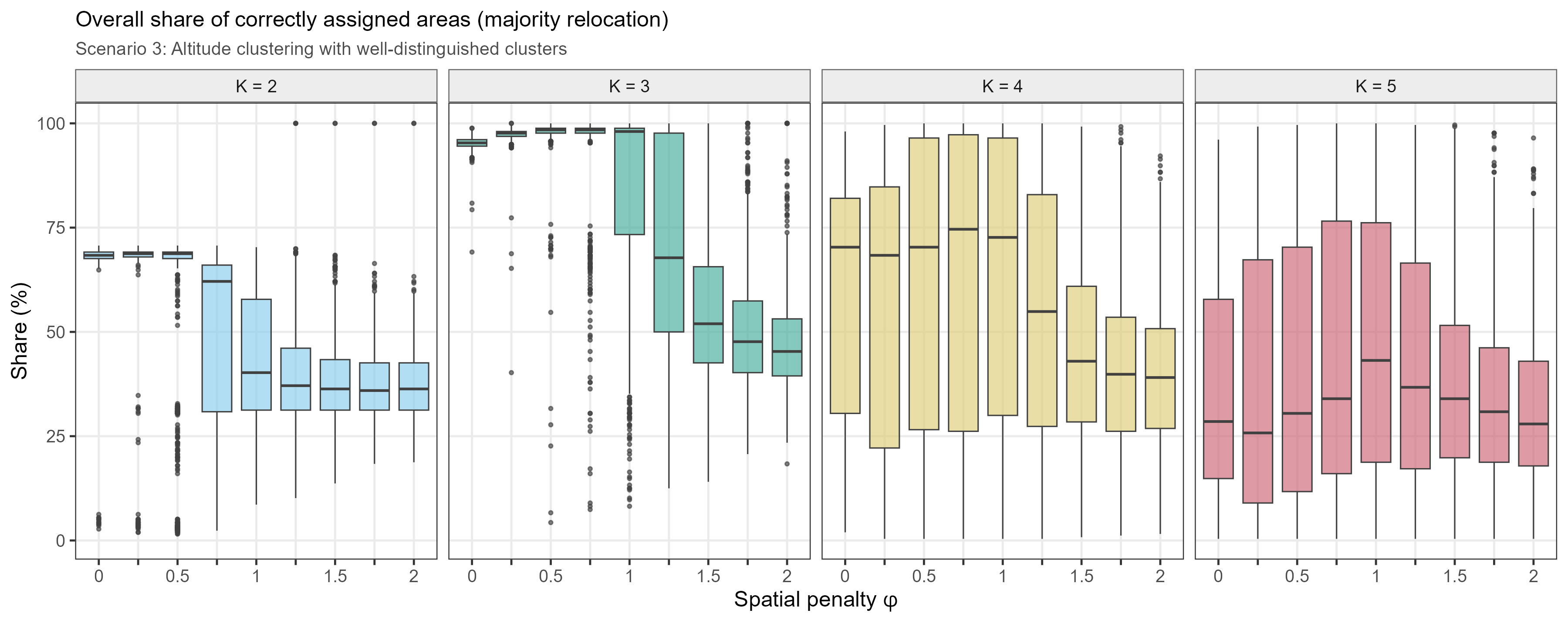}
  \caption{Scenario 3: overall share of correctly assigned areas by number of clusters $K$ and spatial penalty $\phi$.}
\end{figure}

\begin{figure}[!htb]
  \centering
  \includegraphics[width=0.9\linewidth]{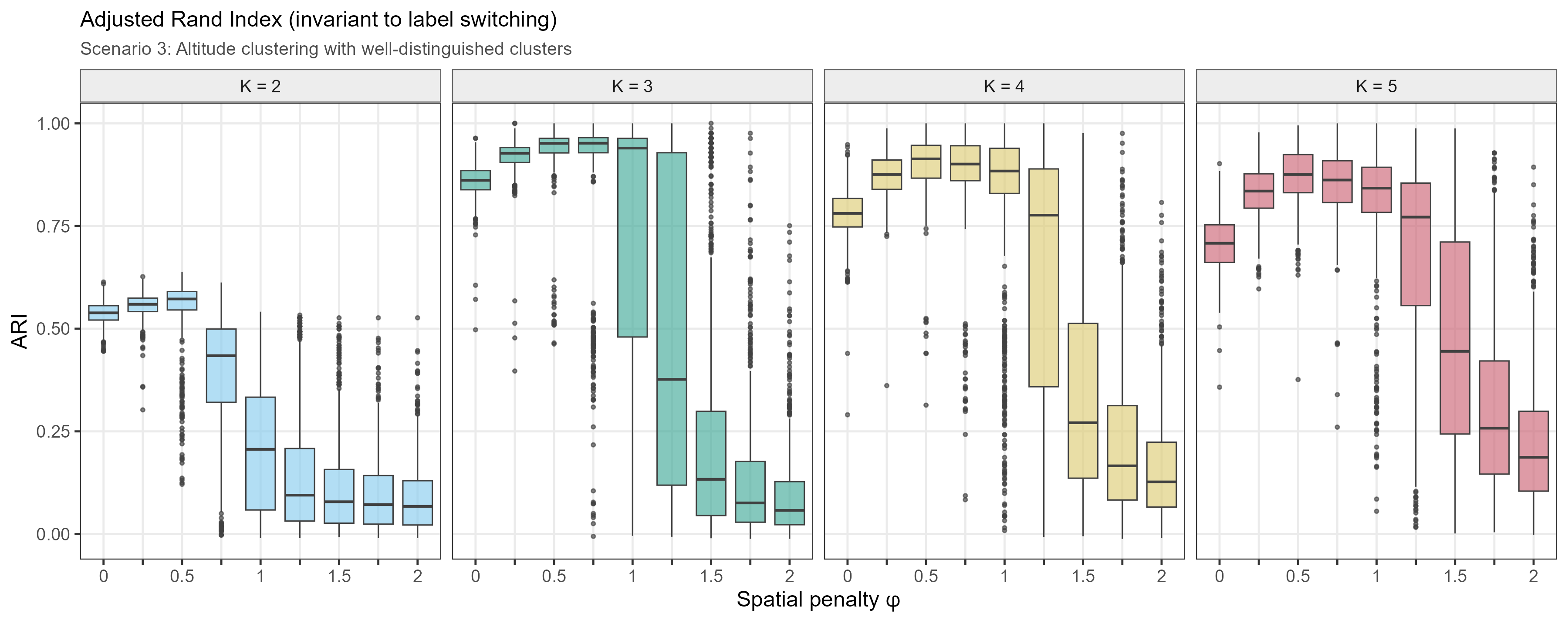}
  \caption{Scenario 3: Adjusted Rand Index between estimated and true partition, by $K$ and $\phi$.}
\end{figure}

\begin{figure}[!htb]
  \centering
  \includegraphics[width=0.95\linewidth]{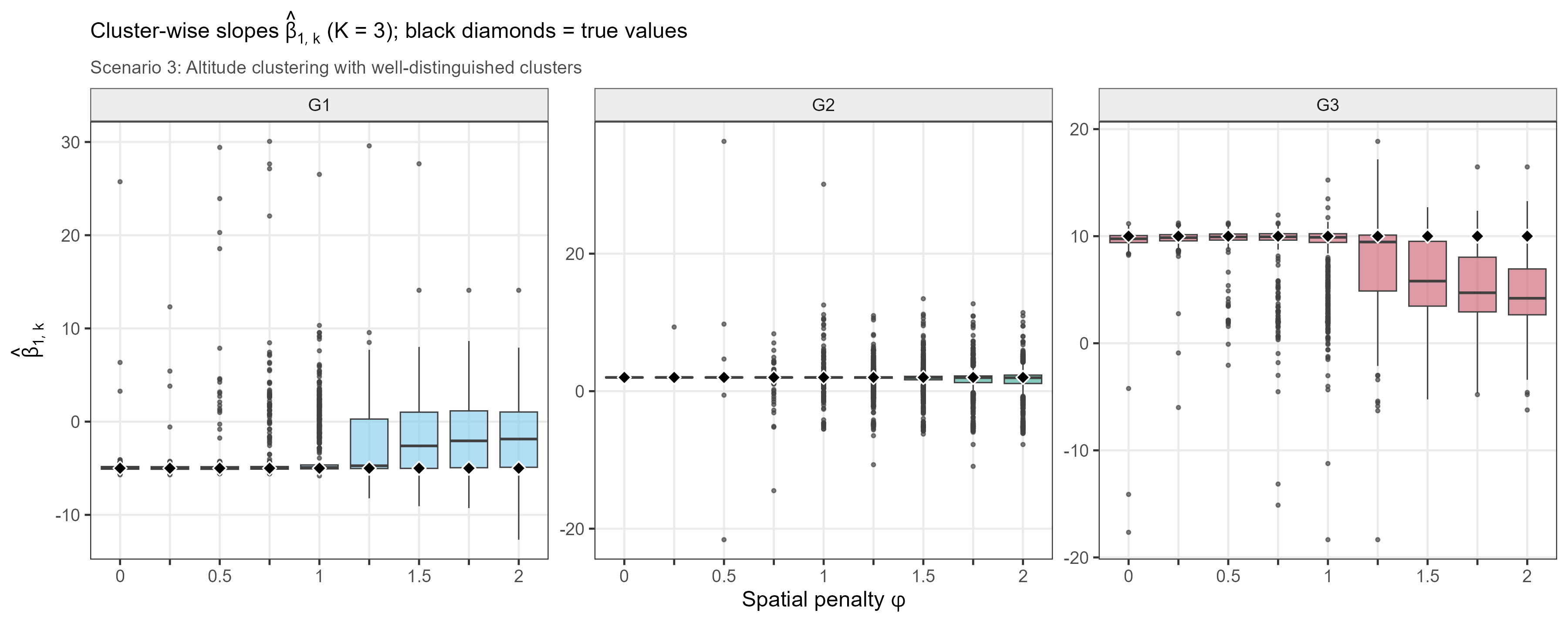}
  \caption{Scenario 3: cluster-specific slope estimates at the true $K=3$, by $\phi$. Black diamonds mark the true values.}
\end{figure}

\begin{figure}[!htb]
  \centering
  \includegraphics[width=0.95\linewidth]{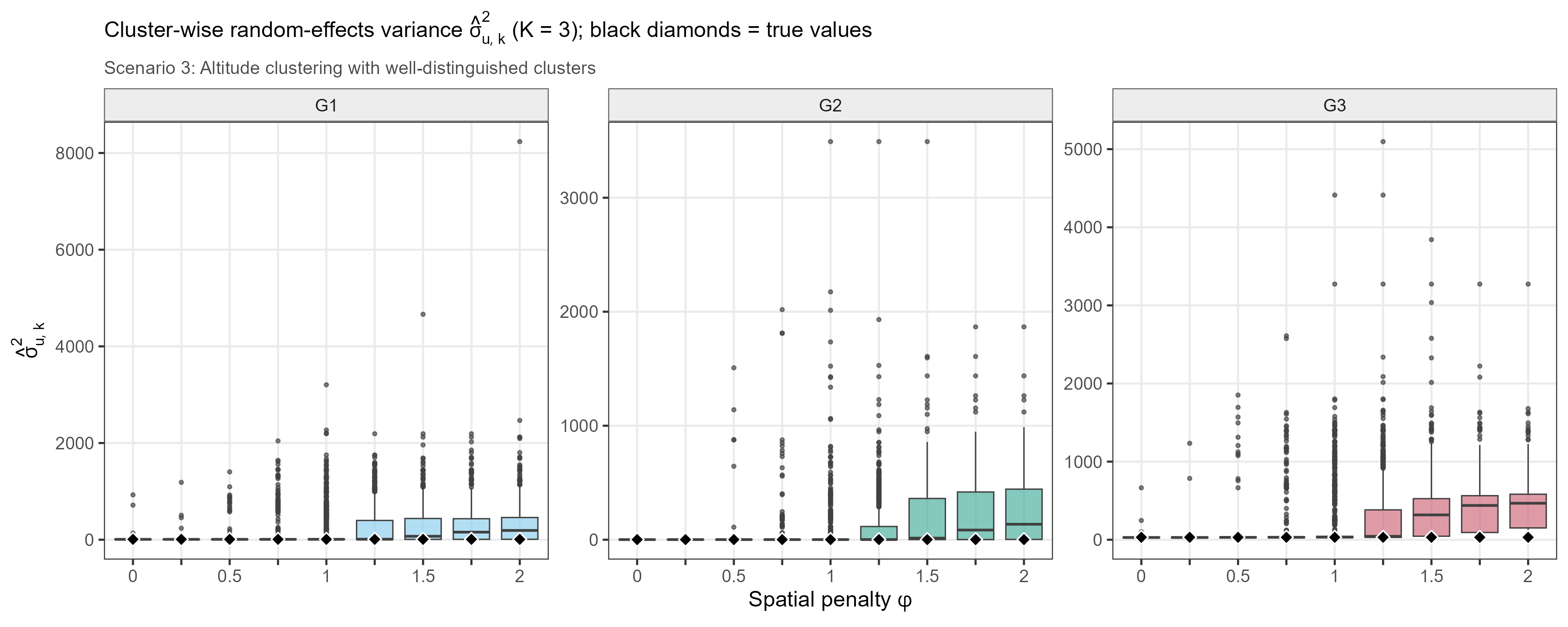}
  \caption{Scenario 3: cluster-specific random-effect variance estimates at the true $K=3$, by $\phi$. Black diamonds mark the true values.}
\end{figure}

\clearpage

\begin{figure}[!htb]
  \centering
  \includegraphics[width=0.95\linewidth]{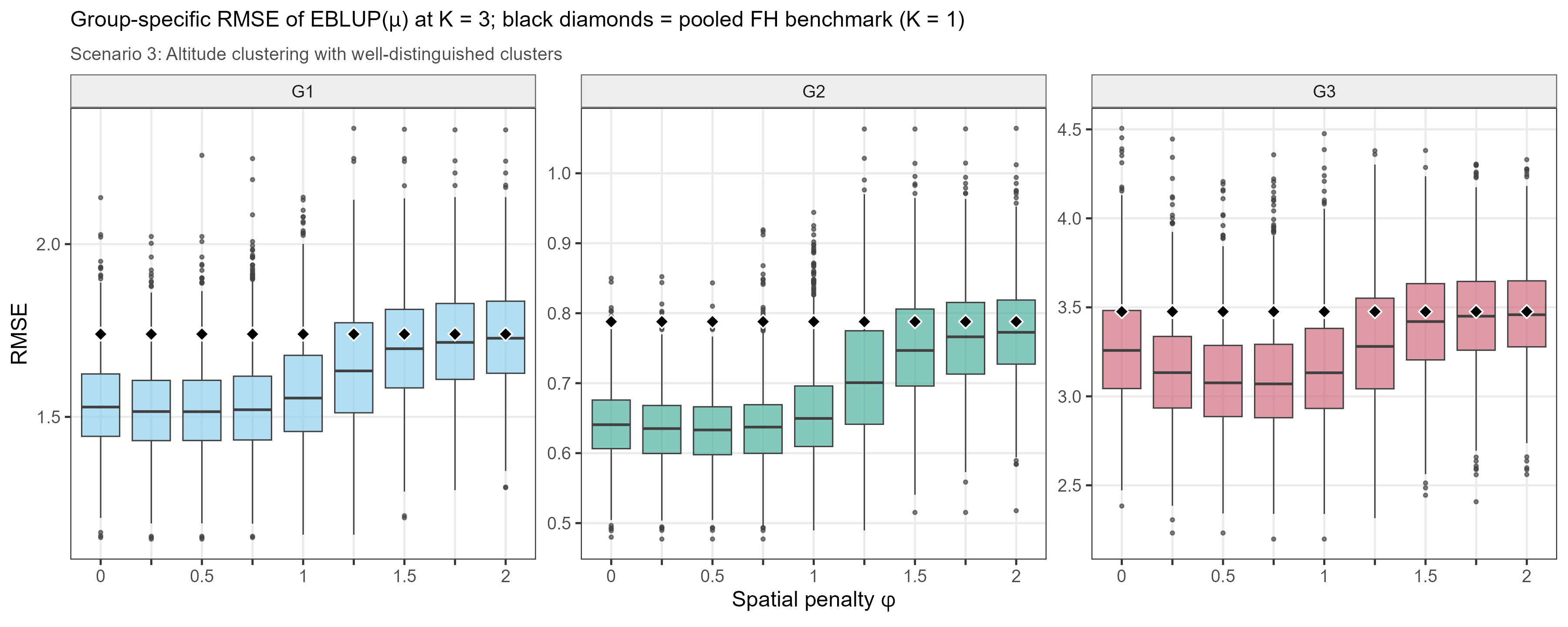}
  \caption{Scenario 3: RMSE of the EBLUP by true group at $K=3$. Black diamonds mark the mean RMSE of the pooled FH benchmark ($K=1$).}
\end{figure}

\begin{figure}[!htb]
  \centering
  \includegraphics[width=0.9\linewidth]{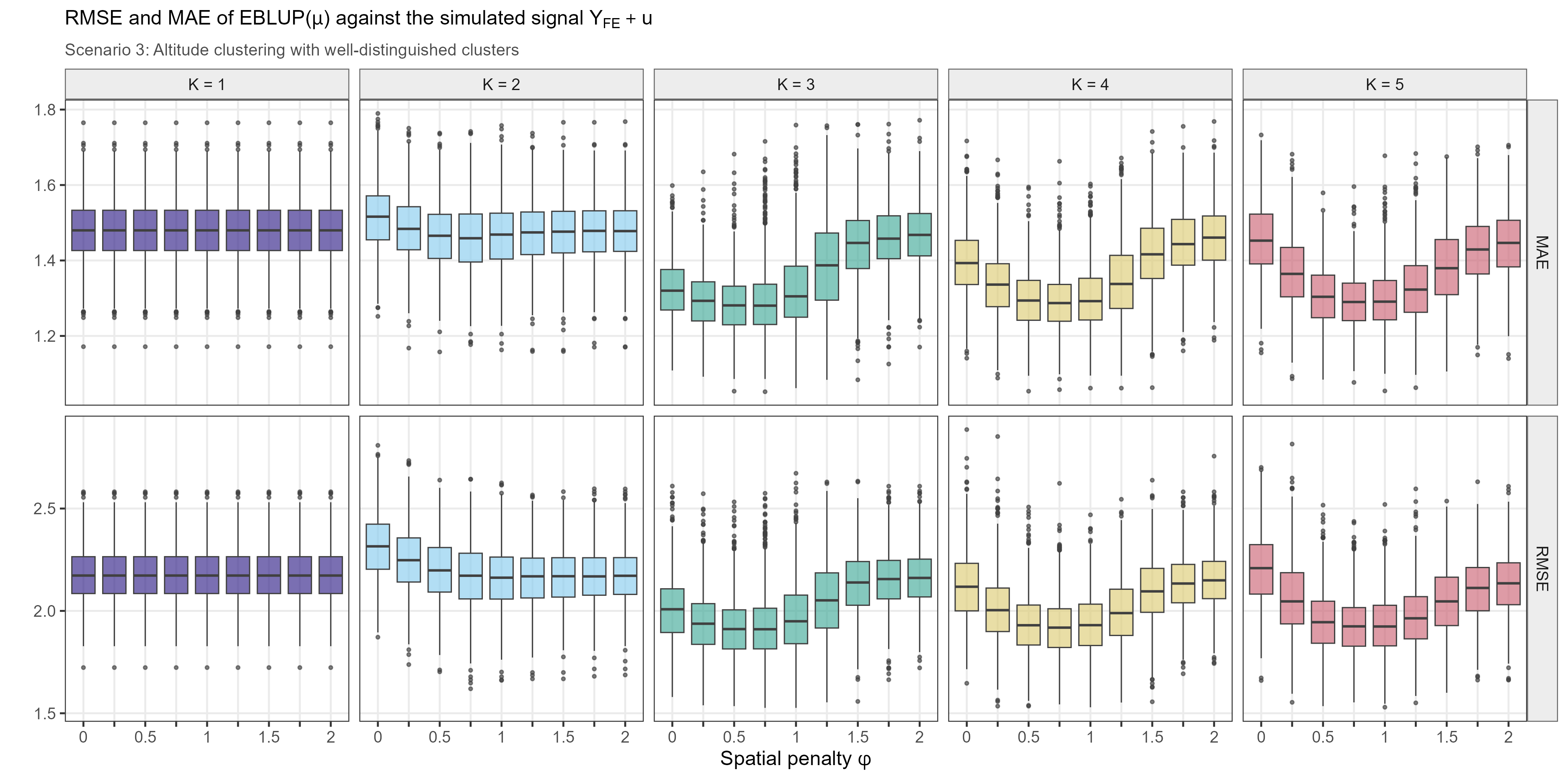}
  \caption{Scenario 3: RMSE of the EBLUP by $K$ and $\phi$.}
\end{figure}

\begin{figure}[!htb]
  \centering
  \includegraphics[width=0.9\linewidth]{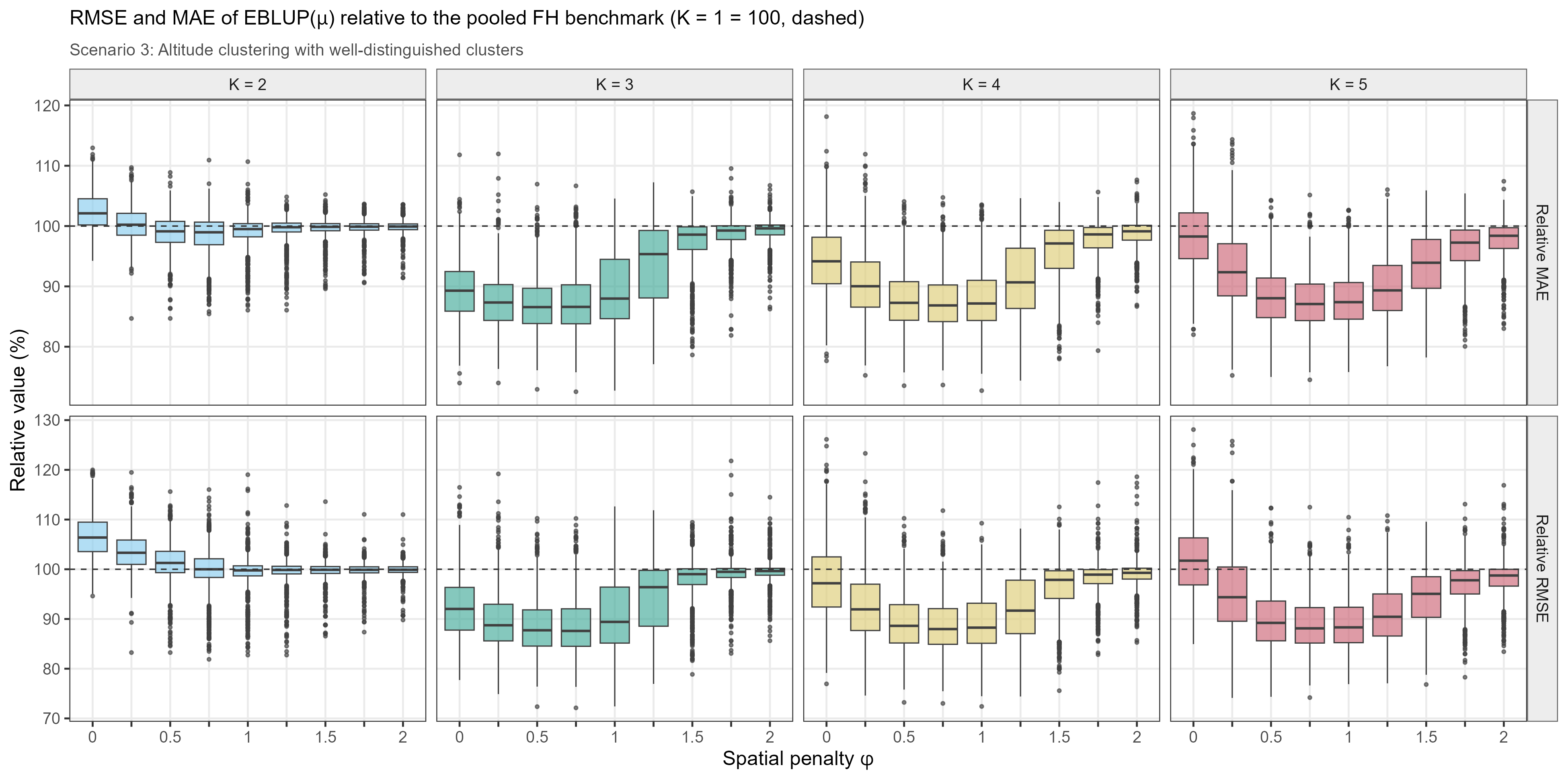}
  \caption{Scenario 3: RMSE of the EBLUP relative to the pooled FH benchmark (\%, dashed line at 100), by $K$ and $\phi$.}
\end{figure}

\begin{figure}[!htb]
  \centering
  \includegraphics[width=0.95\linewidth]{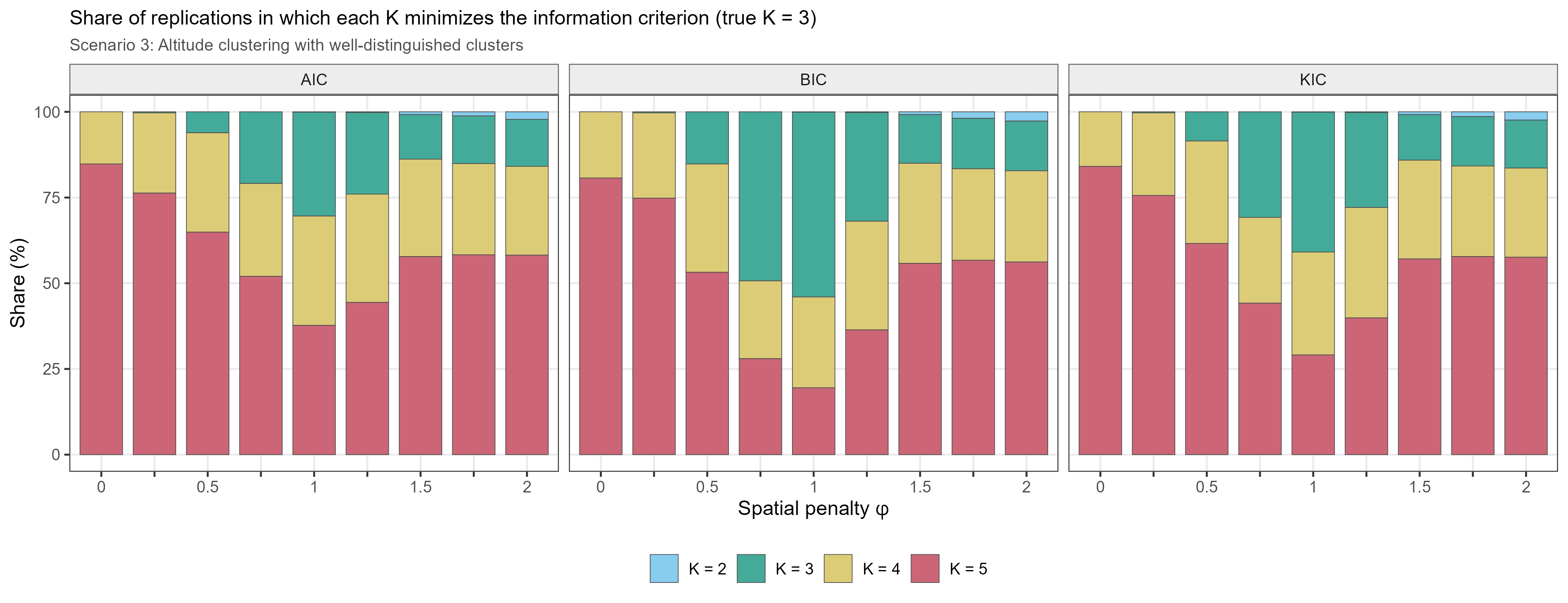}
  \caption{Scenario 3: share of replications in which each $K$ minimizes AIC, BIC, and KIC, by $\phi$.}
\end{figure}

\begin{figure}[!htb]
  \centering
  \includegraphics[width=0.95\linewidth]{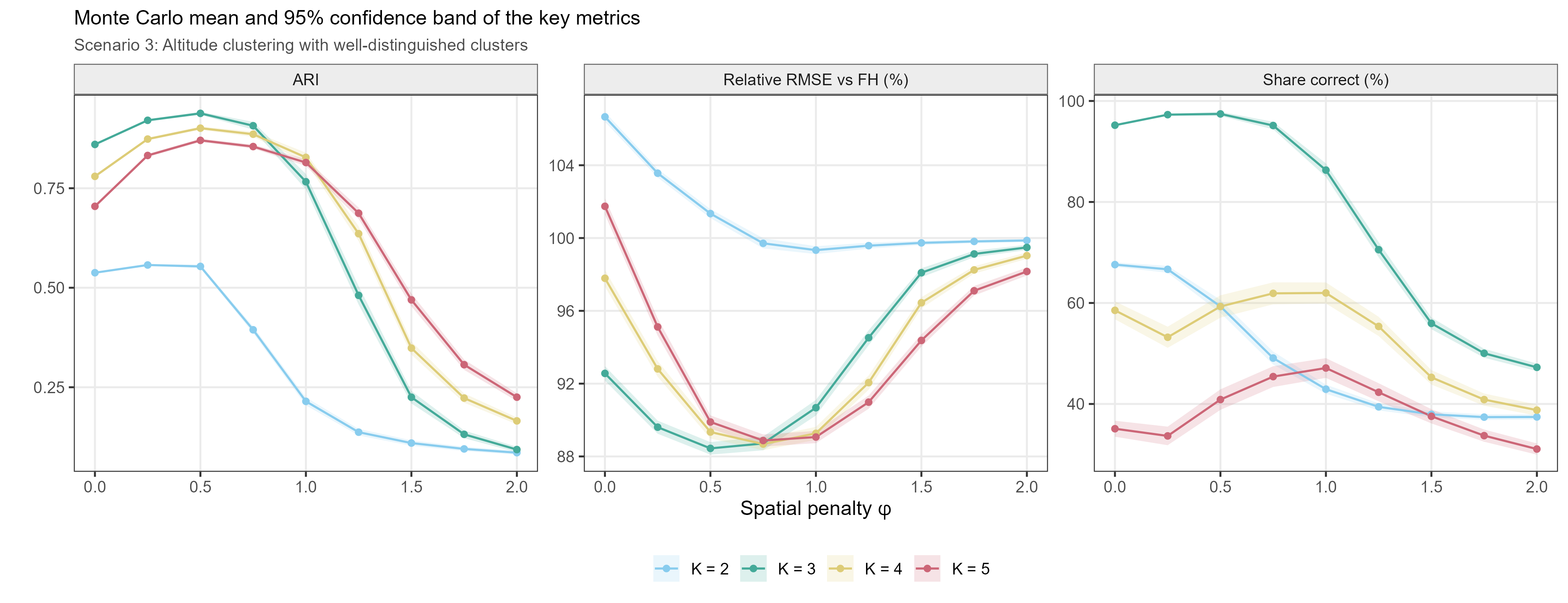}
  \caption{Scenario 3: Monte Carlo mean and 95\% band of ARI, share of correct assignments, and relative RMSE, as functions of $\phi$, by $K$.}
\end{figure}

\begin{figure}[!htb]
  \centering
  \includegraphics[width=0.8\linewidth]{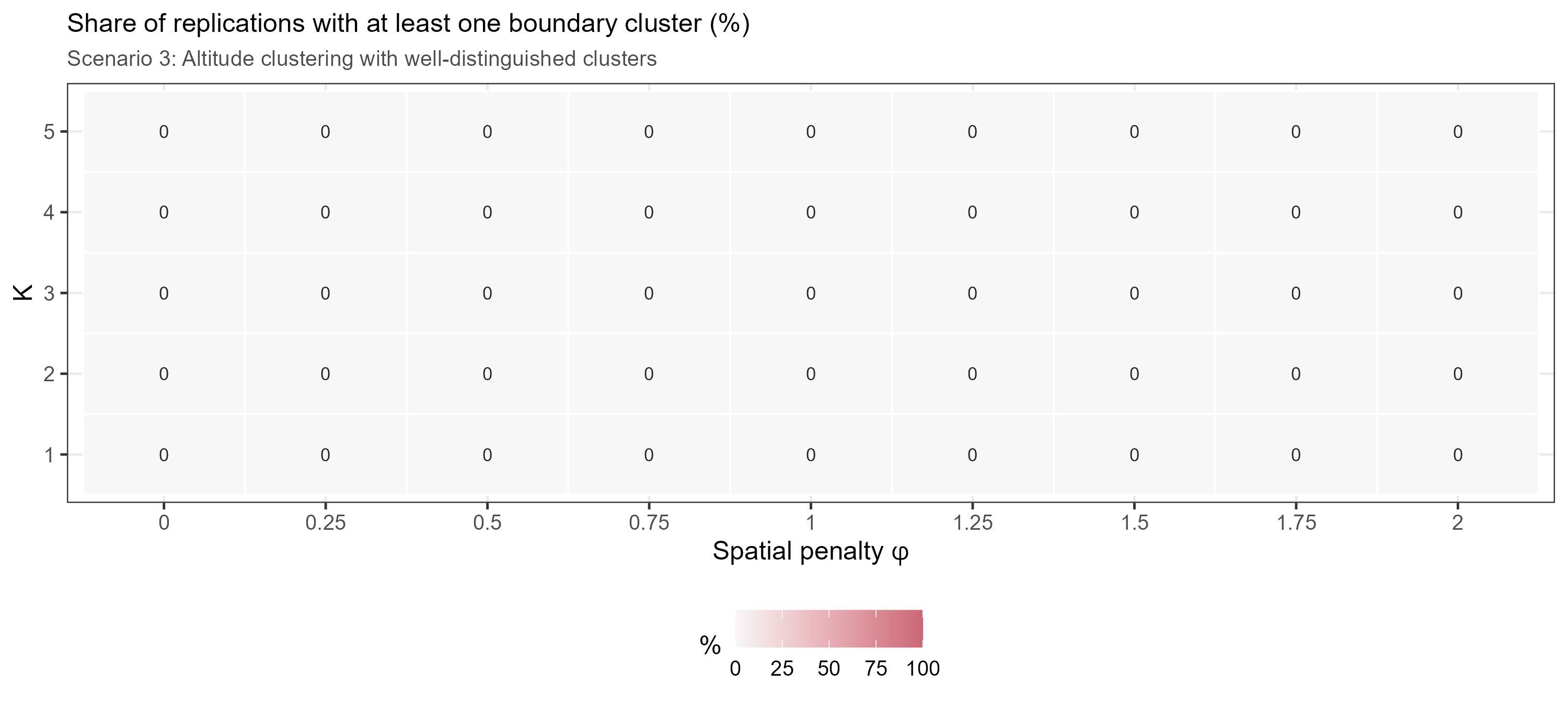}
  \caption{Scenario 3: share of fits with at least one boundary variance solution, by $K$ and $\phi$ (adjusted REML).}
\end{figure}

\begin{table}[!htb]
  \centering
  \caption{Scenario 3: Monte Carlo bias and standard deviation of the cluster-wise slopes and random-effect variances at the true $K=3$, by $\phi$ (auto-generated).}
  \adjustbox{max width=\linewidth}{
\begin{tabular}{lllll}
\toprule
Parameter & Group & Phi & Bias & SD \\
\midrule
beta1 & G1 & 0.000 & 0.088 & 1.089 \\
beta1 & G1 & 0.250 & 0.067 & 0.738 \\
beta1 & G1 & 0.500 & 0.242 & 2.022 \\
beta1 & G1 & 0.750 & 0.560 & 2.661 \\
beta1 & G1 & 1.000 & 1.325 & 2.890 \\
beta1 & G1 & 1.250 & 2.448 & 3.336 \\
beta1 & G1 & 1.500 & 3.020 & 3.382 \\
beta1 & G1 & 1.750 & 3.218 & 3.239 \\
beta1 & G1 & 2.000 & 3.256 & 3.226 \\
beta1 & G2 & 0.000 & -0.001 & 0.073 \\
beta1 & G2 & 0.250 & 0.008 & 0.242 \\
beta1 & G2 & 0.500 & 0.019 & 1.348 \\
beta1 & G2 & 0.750 & -0.030 & 0.784 \\
beta1 & G2 & 1.000 & 0.024 & 1.568 \\
beta1 & G2 & 1.250 & -0.200 & 1.594 \\
beta1 & G2 & 1.500 & -0.241 & 1.961 \\
beta1 & G2 & 1.750 & -0.352 & 2.110 \\
beta1 & G2 & 2.000 & -0.411 & 2.115 \\
beta1 & G3 & 0.000 & -0.325 & 1.314 \\
beta1 & G3 & 0.250 & -0.176 & 0.780 \\
beta1 & G3 & 0.500 & -0.192 & 1.074 \\
beta1 & G3 & 0.750 & -0.536 & 2.180 \\
beta1 & G3 & 1.000 & -1.225 & 2.937 \\
beta1 & G3 & 1.250 & -2.463 & 3.467 \\
beta1 & G3 & 1.500 & -3.884 & 3.243 \\
beta1 & G3 & 1.750 & -4.644 & 3.157 \\
beta1 & G3 & 2.000 & -5.088 & 3.161 \\
sigma2u & G1 & 0.000 & 1.714 & 36.960 \\
sigma2u & G1 & 0.250 & 2.383 & 43.454 \\
sigma2u & G1 & 0.500 & 12.840 & 102.765 \\
sigma2u & G1 & 0.750 & 51.579 & 222.820 \\
sigma2u & G1 & 1.000 & 141.836 & 344.473 \\
sigma2u & G1 & 1.250 & 220.369 & 356.340 \\
sigma2u & G1 & 1.500 & 249.048 & 371.903 \\
sigma2u & G1 & 1.750 & 262.723 & 347.336 \\
sigma2u & G1 & 2.000 & 287.548 & 440.665 \\
sigma2u & G2 & 0.000 & -0.053 & 0.242 \\
sigma2u & G2 & 0.250 & -0.000 & 0.408 \\
sigma2u & G2 & 0.500 & 5.163 & 74.226 \\
sigma2u & G2 & 0.750 & 16.713 & 127.473 \\
sigma2u & G2 & 1.000 & 49.851 & 215.609 \\
sigma2u & G2 & 1.250 & 108.398 & 234.312 \\
sigma2u & G2 & 1.500 & 170.249 & 255.527 \\
sigma2u & G2 & 1.750 & 205.545 & 242.308 \\
sigma2u & G2 & 2.000 & 226.428 & 235.705 \\
sigma2u & G3 & 0.000 & -0.729 & 22.740 \\
sigma2u & G3 & 0.250 & 0.218 & 45.589 \\
sigma2u & G3 & 0.500 & 13.770 & 135.715 \\
sigma2u & G3 & 0.750 & 50.118 & 240.026 \\
sigma2u & G3 & 1.000 & 128.489 & 361.922 \\
sigma2u & G3 & 1.250 & 231.794 & 425.267 \\
sigma2u & G3 & 1.500 & 334.895 & 374.344 \\
sigma2u & G3 & 1.750 & 380.112 & 331.616 \\
sigma2u & G3 & 2.000 & 404.486 & 317.888 \\
\bottomrule
\end{tabular}
}
\end{table}

\begin{table}[!htb]
  \centering
  \caption{Scenario 3: clustering and prediction performance summary (auto-generated).}
  \adjustbox{max width=\linewidth}{
\begin{tabular}{lllllllll}
\toprule
MC\_G & Phi & ARI\_mean & ARI\_sd & Share\_mean & Share\_sd & RelRMSE\_mean & RelRMSE\_sd & Boundary\_pct \\
\midrule
1.000 & 0.000 & 0.000 & 0.000 & 100.000 & 0.000 & 100.000 & 0.000 & 0.000 \\
1.000 & 0.250 & 0.000 & 0.000 & 100.000 & 0.000 & 100.000 & 0.000 & 0.000 \\
1.000 & 0.500 & 0.000 & 0.000 & 100.000 & 0.000 & 100.000 & 0.000 & 0.000 \\
1.000 & 0.750 & 0.000 & 0.000 & 100.000 & 0.000 & 100.000 & 0.000 & 0.000 \\
1.000 & 1.000 & 0.000 & 0.000 & 100.000 & 0.000 & 100.000 & 0.000 & 0.000 \\
1.000 & 1.250 & 0.000 & 0.000 & 100.000 & 0.000 & 100.000 & 0.000 & 0.000 \\
1.000 & 1.500 & 0.000 & 0.000 & 100.000 & 0.000 & 100.000 & 0.000 & 0.000 \\
1.000 & 1.750 & 0.000 & 0.000 & 100.000 & 0.000 & 100.000 & 0.000 & 0.000 \\
1.000 & 2.000 & 0.000 & 0.000 & 100.000 & 0.000 & 100.000 & 0.000 & 0.000 \\
2.000 & 0.000 & 0.538 & 0.027 & 67.596 & 7.015 & 106.660 & 4.424 & 0.000 \\
2.000 & 0.250 & 0.557 & 0.028 & 66.665 & 11.064 & 103.561 & 3.805 & 0.000 \\
2.000 & 0.500 & 0.554 & 0.073 & 59.326 & 20.937 & 101.344 & 3.999 & 0.000 \\
2.000 & 0.750 & 0.394 & 0.136 & 49.090 & 20.676 & 99.710 & 4.533 & 0.000 \\
2.000 & 1.000 & 0.215 & 0.161 & 42.927 & 14.531 & 99.336 & 3.501 & 0.000 \\
2.000 & 1.250 & 0.137 & 0.133 & 39.431 & 11.477 & 99.581 & 2.599 & 0.000 \\
2.000 & 1.500 & 0.110 & 0.110 & 37.949 & 9.742 & 99.734 & 1.945 & 0.000 \\
2.000 & 1.750 & 0.095 & 0.089 & 37.397 & 8.967 & 99.814 & 1.599 & 0.000 \\
2.000 & 2.000 & 0.086 & 0.080 & 37.430 & 8.451 & 99.867 & 1.387 & 0.000 \\
3.000 & 0.000 & 0.860 & 0.040 & 95.202 & 1.659 & 92.562 & 6.408 & 0.000 \\
3.000 & 0.250 & 0.921 & 0.041 & 97.289 & 2.567 & 89.607 & 5.814 & 0.000 \\
3.000 & 0.500 & 0.938 & 0.062 & 97.437 & 6.640 & 88.442 & 5.548 & 0.000 \\
3.000 & 0.750 & 0.907 & 0.159 & 95.140 & 12.118 & 88.714 & 6.057 & 0.000 \\
3.000 & 1.000 & 0.766 & 0.313 & 86.307 & 21.603 & 90.677 & 6.936 & 0.000 \\
3.000 & 1.250 & 0.481 & 0.375 & 70.554 & 24.347 & 94.522 & 6.653 & 0.000 \\
3.000 & 1.500 & 0.225 & 0.254 & 55.963 & 18.676 & 98.095 & 3.996 & 0.000 \\
3.000 & 1.750 & 0.132 & 0.156 & 50.039 & 14.317 & 99.126 & 2.918 & 0.000 \\
3.000 & 2.000 & 0.093 & 0.108 & 47.255 & 12.094 & 99.484 & 2.295 & 0.000 \\
4.000 & 0.000 & 0.780 & 0.057 & 58.540 & 28.527 & 97.789 & 7.357 & 0.000 \\
4.000 & 0.250 & 0.874 & 0.052 & 53.209 & 33.857 & 92.810 & 6.979 & 0.000 \\
4.000 & 0.500 & 0.901 & 0.067 & 59.307 & 35.778 & 89.342 & 5.777 & 0.000 \\
4.000 & 0.750 & 0.886 & 0.106 & 61.908 & 35.114 & 88.678 & 5.429 & 0.000 \\
4.000 & 1.000 & 0.827 & 0.194 & 61.984 & 33.658 & 89.268 & 5.793 & 0.000 \\
4.000 & 1.250 & 0.636 & 0.300 & 55.359 & 30.773 & 92.057 & 6.304 & 0.000 \\
4.000 & 1.500 & 0.348 & 0.266 & 45.273 & 23.197 & 96.445 & 4.941 & 0.000 \\
4.000 & 1.750 & 0.223 & 0.189 & 40.867 & 19.472 & 98.249 & 3.301 & 0.000 \\
4.000 & 2.000 & 0.165 & 0.139 & 38.782 & 17.576 & 99.034 & 2.851 & 0.000 \\
5.000 & 0.000 & 0.705 & 0.068 & 35.099 & 25.567 & 101.743 & 7.159 & 0.000 \\
5.000 & 0.250 & 0.832 & 0.064 & 33.674 & 29.685 & 95.116 & 7.727 & 0.000 \\
5.000 & 0.500 & 0.870 & 0.068 & 40.870 & 32.819 & 89.893 & 5.862 & 0.000 \\
5.000 & 0.750 & 0.855 & 0.080 & 45.420 & 33.154 & 88.874 & 5.373 & 0.000 \\
5.000 & 1.000 & 0.815 & 0.138 & 47.122 & 31.581 & 89.065 & 5.419 & 0.000 \\
5.000 & 1.250 & 0.687 & 0.229 & 42.318 & 28.925 & 90.983 & 5.855 & 0.000 \\
5.000 & 1.500 & 0.470 & 0.260 & 37.560 & 22.786 & 94.374 & 5.450 & 0.000 \\
5.000 & 1.750 & 0.307 & 0.209 & 33.730 & 19.549 & 97.104 & 4.230 & 0.000 \\
5.000 & 2.000 & 0.225 & 0.159 & 31.077 & 17.916 & 98.163 & 3.555 & 0.000 \\
\bottomrule
\end{tabular}
}
\end{table}

\clearpage

\subsection{Scenario 4: altitude-based clusters, poor separation}

\begin{figure}[!htb]
  \centering
  \includegraphics[width=0.95\linewidth]{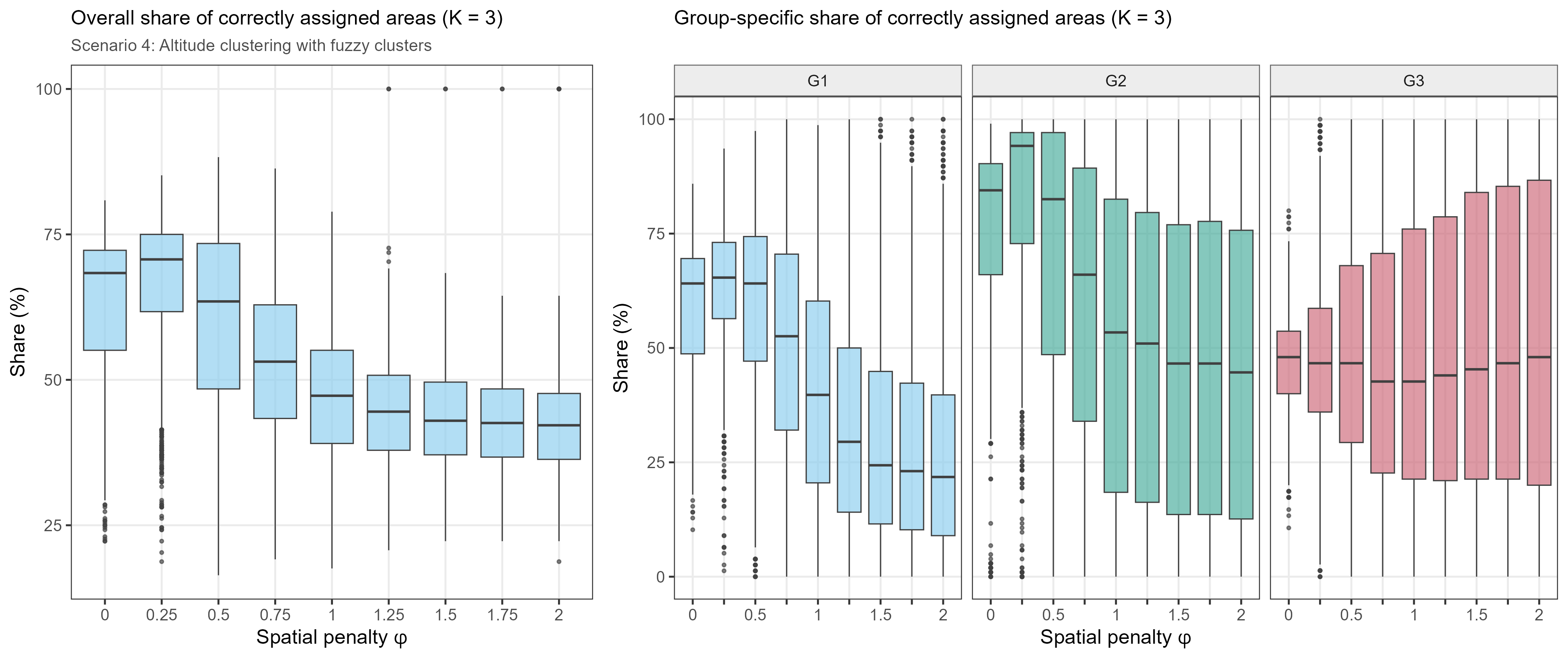}
  \caption{Scenario 4: overall (left) and group-specific (right) share of correctly assigned areas at the true $K=3$, by spatial penalty $\phi$.}
\end{figure}

\begin{figure}[!htb]
  \centering
  \includegraphics[width=0.9\linewidth]{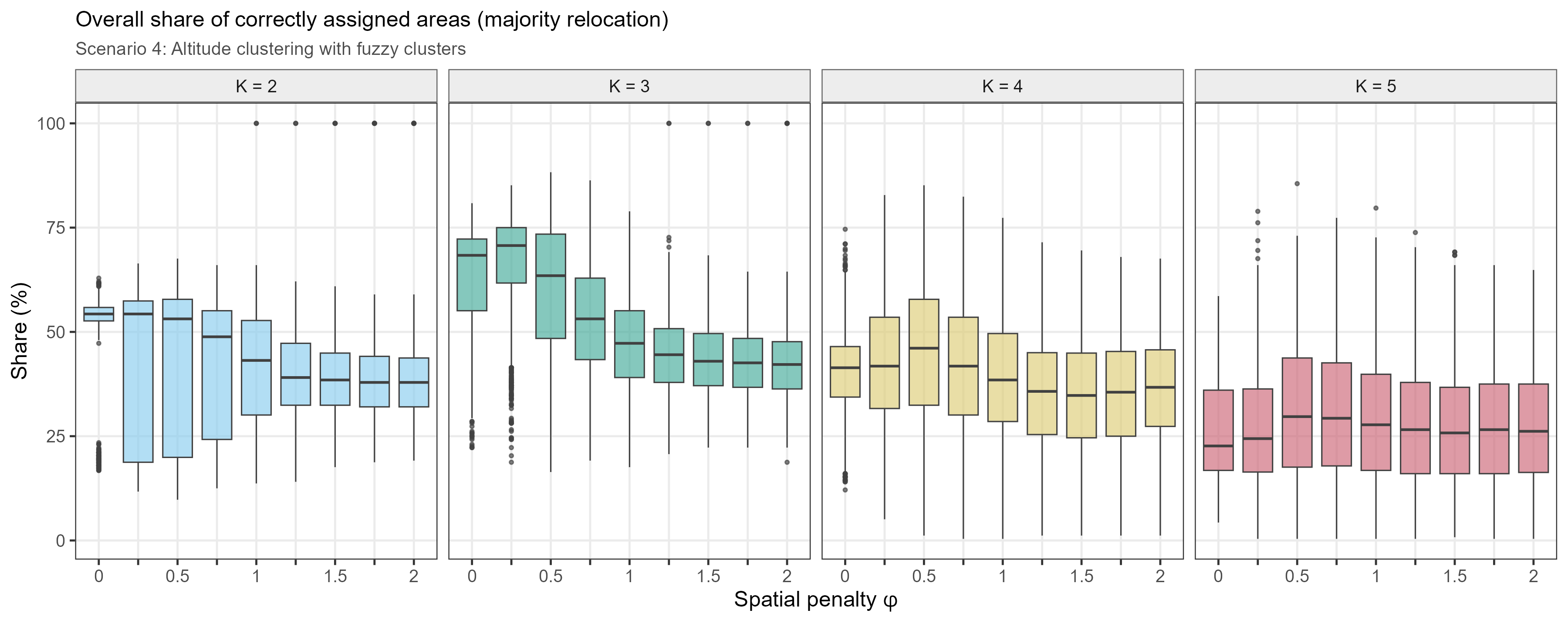}
  \caption{Scenario 4: overall share of correctly assigned areas by number of clusters $K$ and spatial penalty $\phi$.}
\end{figure}

\begin{figure}[!htb]
  \centering
  \includegraphics[width=0.9\linewidth]{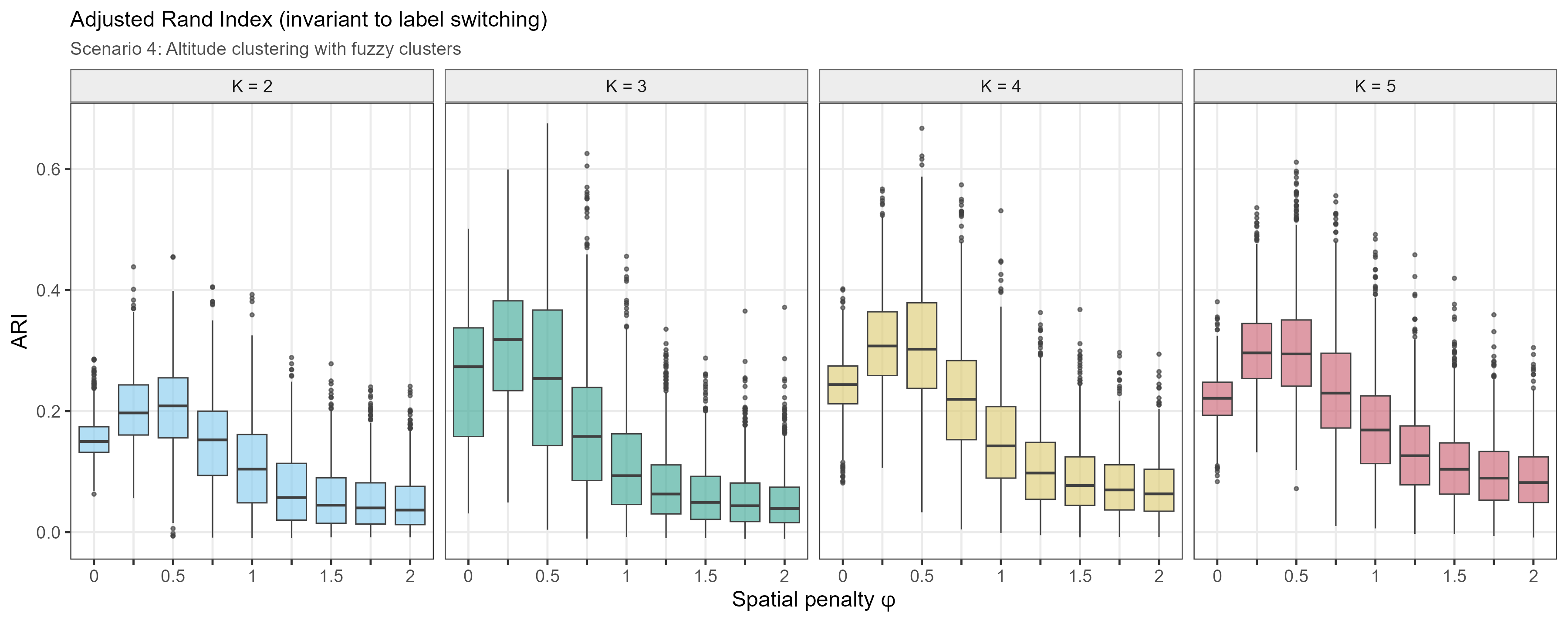}
  \caption{Scenario 4: Adjusted Rand Index between estimated and true partition, by $K$ and $\phi$.}
\end{figure}

\begin{figure}[!htb]
  \centering
  \includegraphics[width=0.95\linewidth]{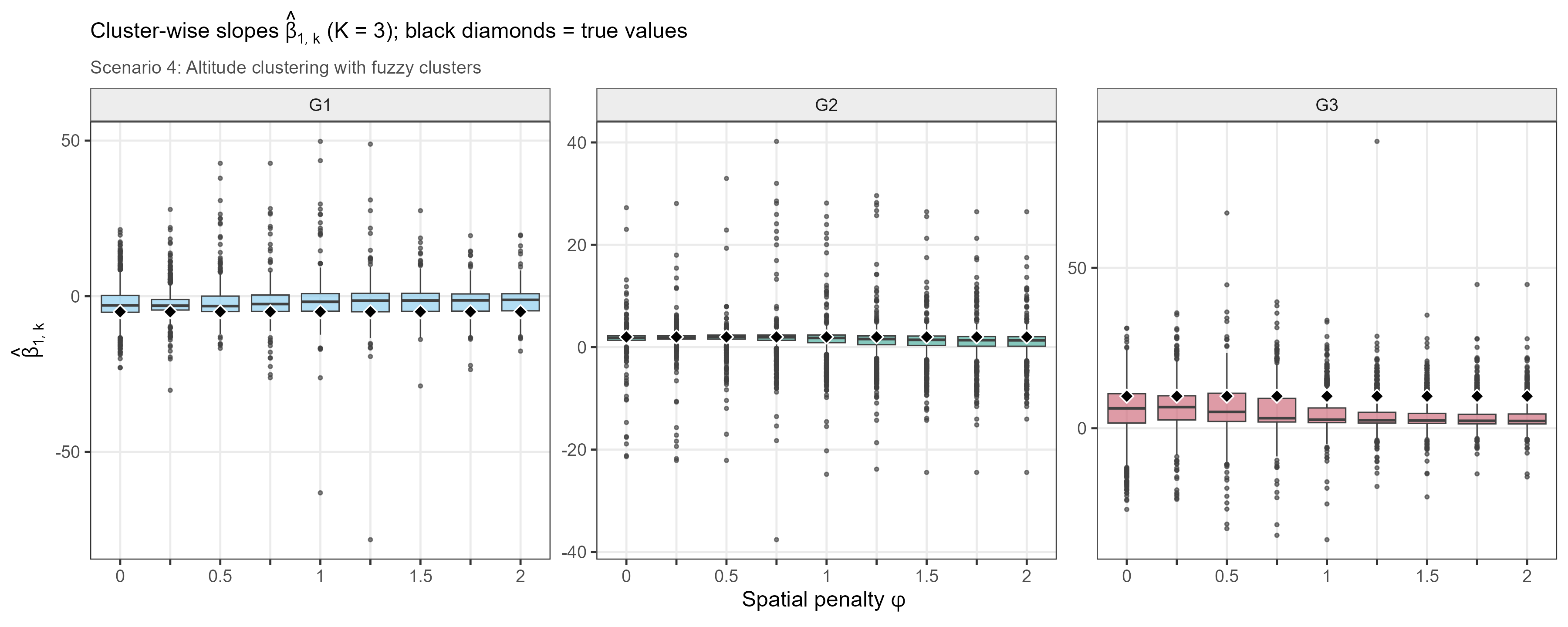}
  \caption{Scenario 4: cluster-specific slope estimates at the true $K=3$, by $\phi$. Black diamonds mark the true values.}
\end{figure}

\begin{figure}[!htb]
  \centering
  \includegraphics[width=0.95\linewidth]{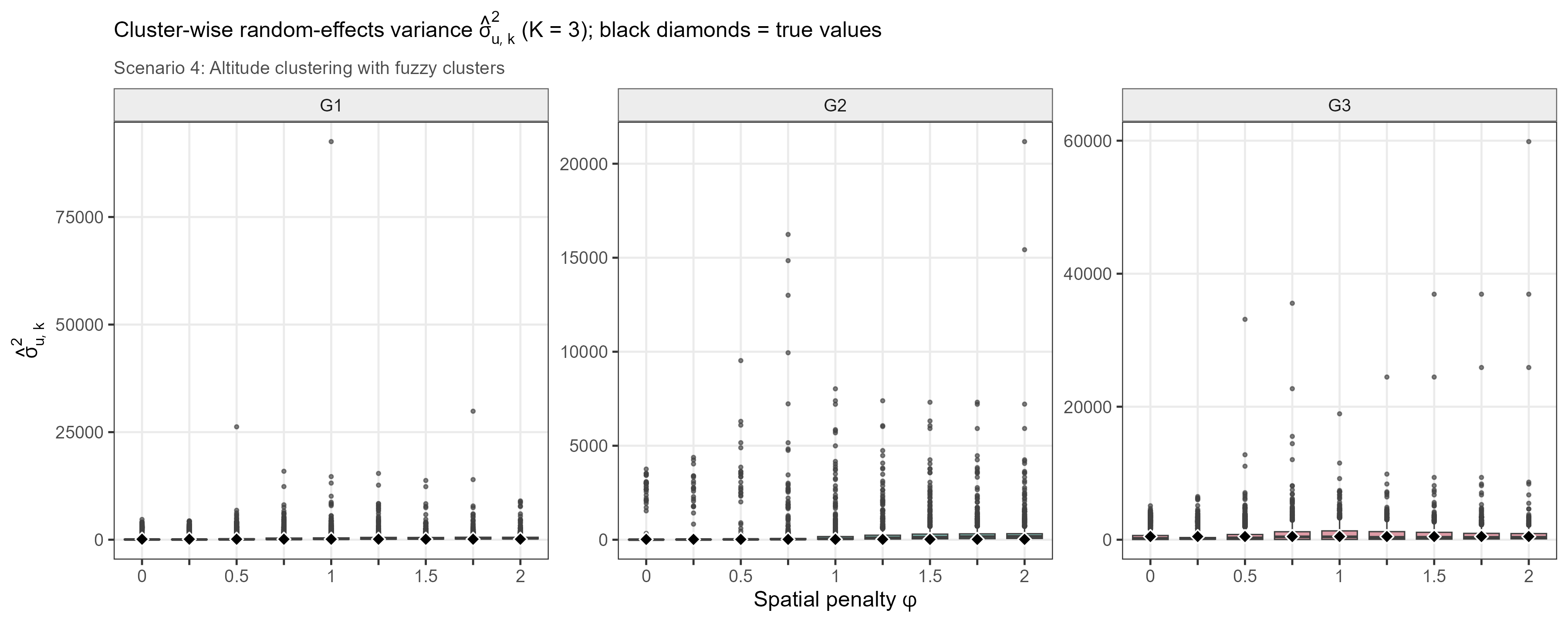}
  \caption{Scenario 4: cluster-specific random-effect variance estimates at the true $K=3$, by $\phi$. Black diamonds mark the true values.}
\end{figure}

\clearpage

\begin{figure}[!htb]
  \centering
  \includegraphics[width=0.95\linewidth]{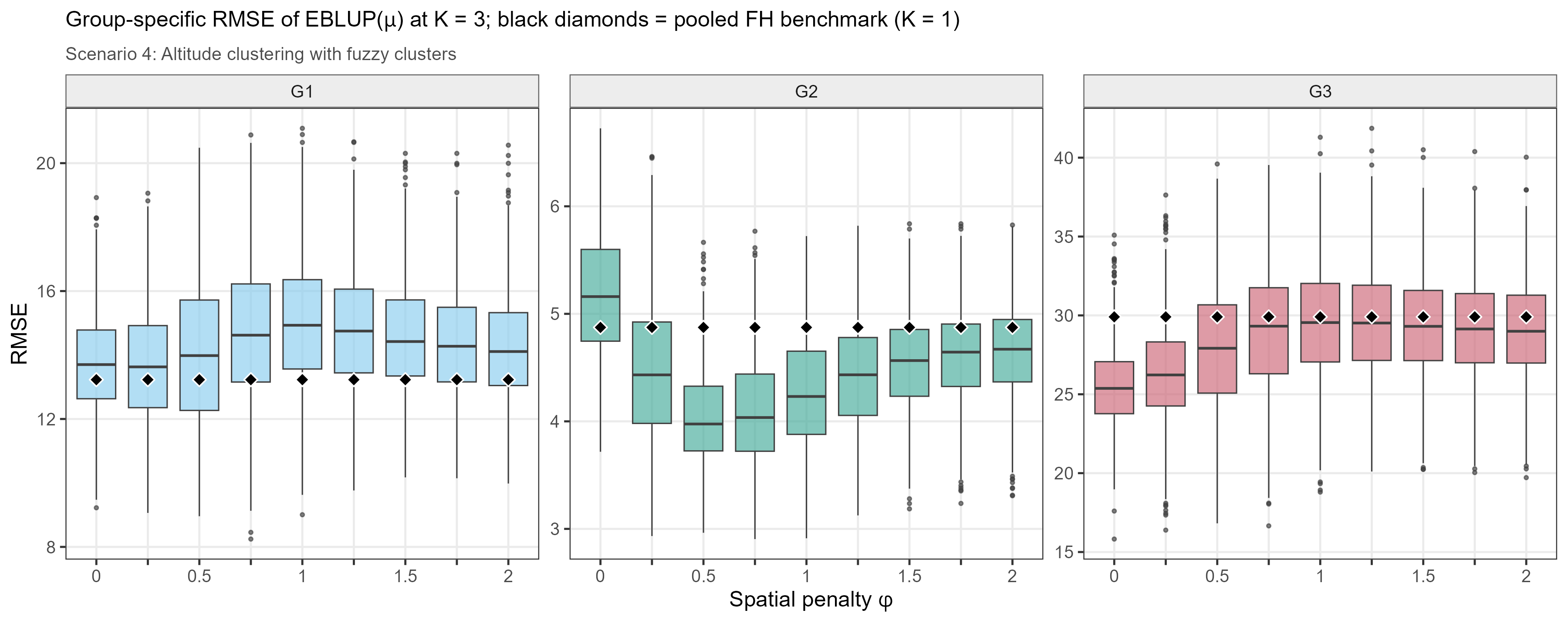}
  \caption{Scenario 4: RMSE of the EBLUP by true group at $K=3$. Black diamonds mark the mean RMSE of the pooled FH benchmark ($K=1$).}
\end{figure}

\begin{figure}[!htb]
  \centering
  \includegraphics[width=0.9\linewidth]{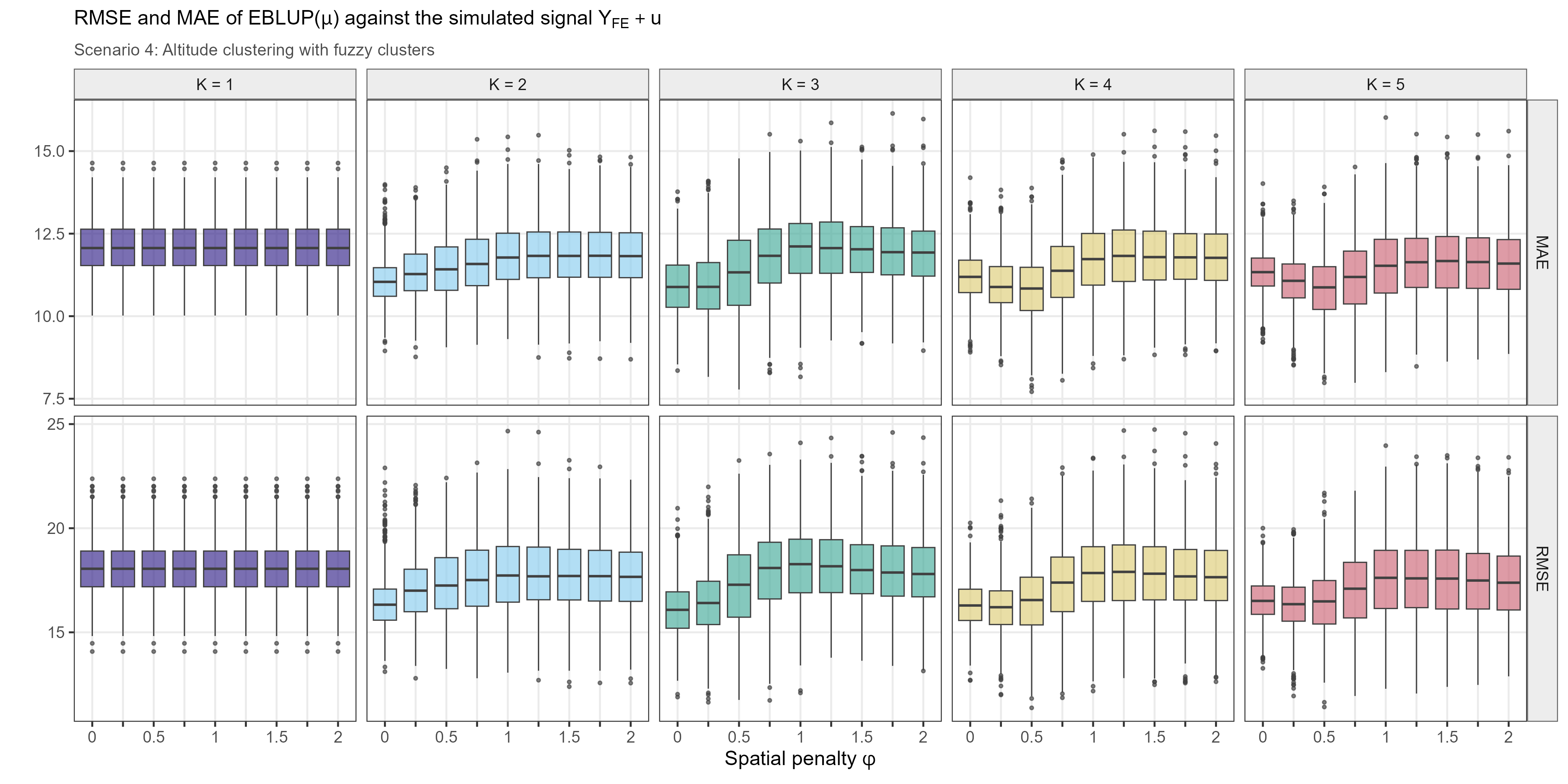}
  \caption{Scenario 4: RMSE of the EBLUP by $K$ and $\phi$.}
\end{figure}

\begin{figure}[!htb]
  \centering
  \includegraphics[width=0.9\linewidth]{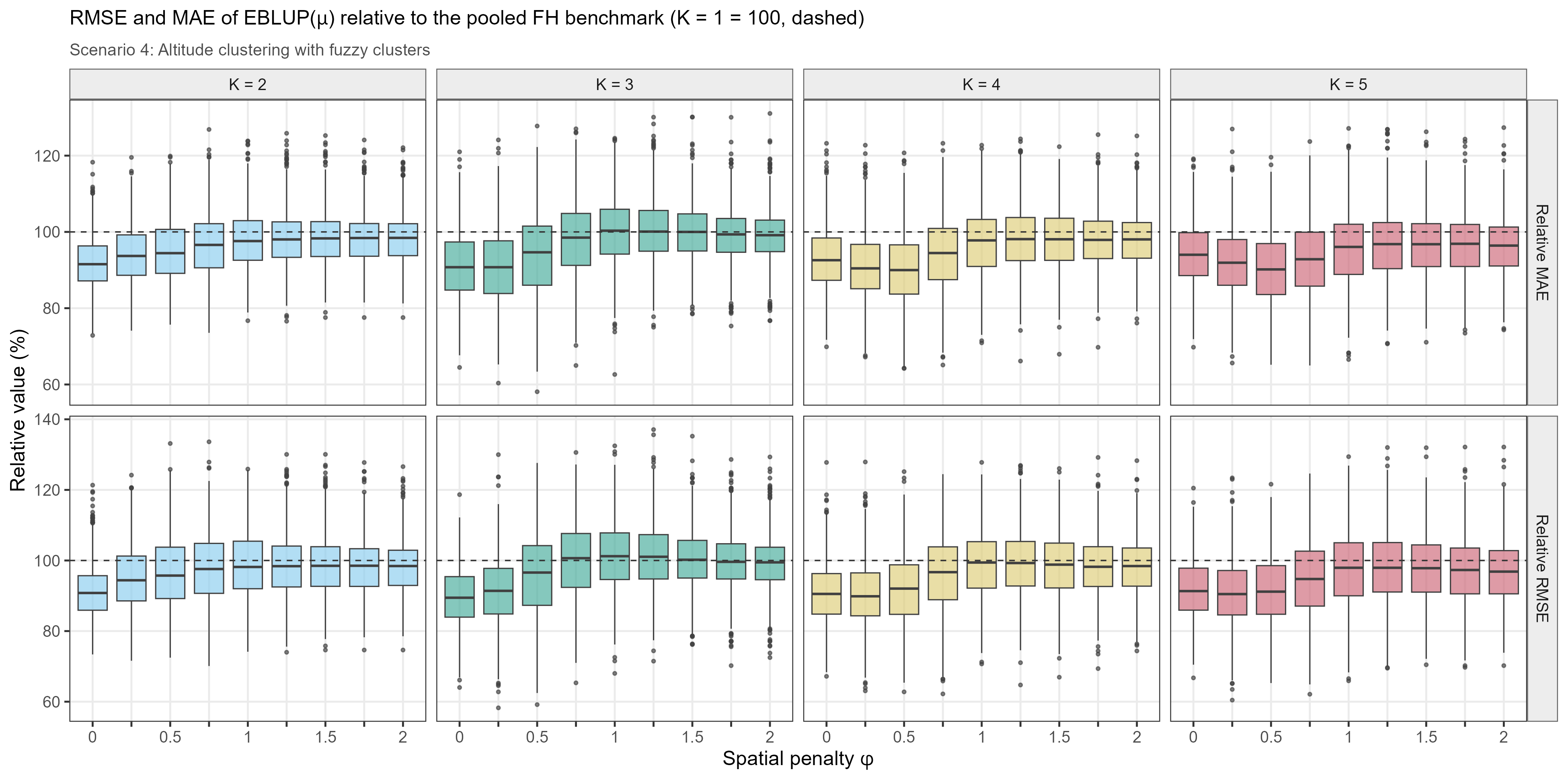}
  \caption{Scenario 4: RMSE of the EBLUP relative to the pooled FH benchmark (\%, dashed line at 100), by $K$ and $\phi$.}
\end{figure}

\begin{figure}[!htb]
  \centering
  \includegraphics[width=0.95\linewidth]{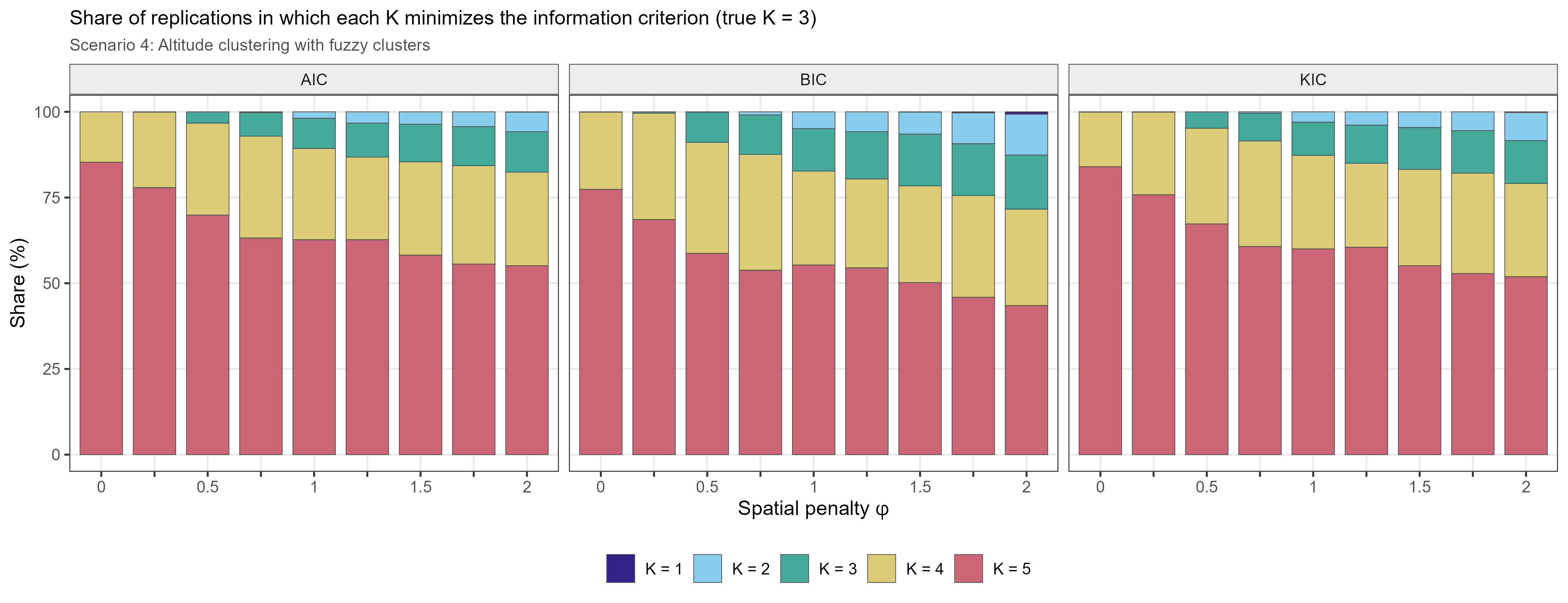}
  \caption{Scenario 4: share of replications in which each $K$ minimizes AIC, BIC, and KIC, by $\phi$.}
\end{figure}

\begin{figure}[!htb]
  \centering
  \includegraphics[width=0.95\linewidth]{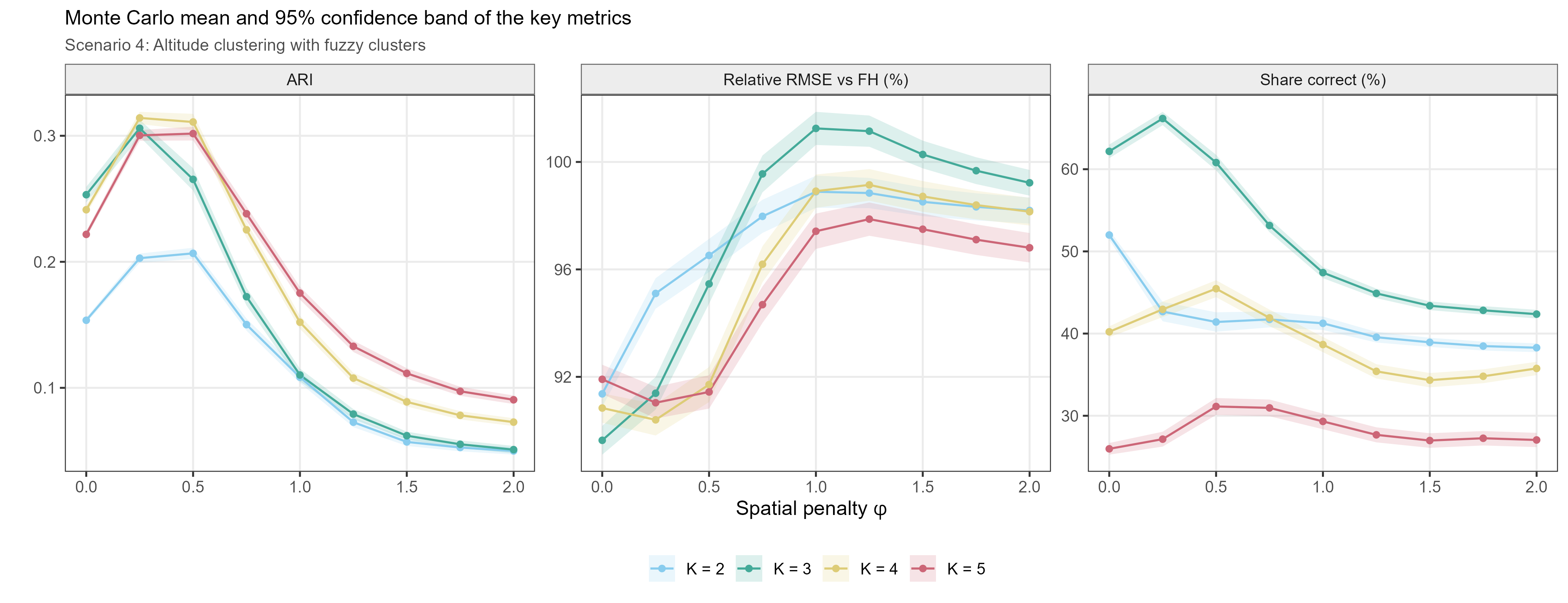}
  \caption{Scenario 4: Monte Carlo mean and 95\% band of ARI, share of correct assignments, and relative RMSE, as functions of $\phi$, by $K$.}
\end{figure}

\begin{figure}[!htb]
  \centering
  \includegraphics[width=0.8\linewidth]{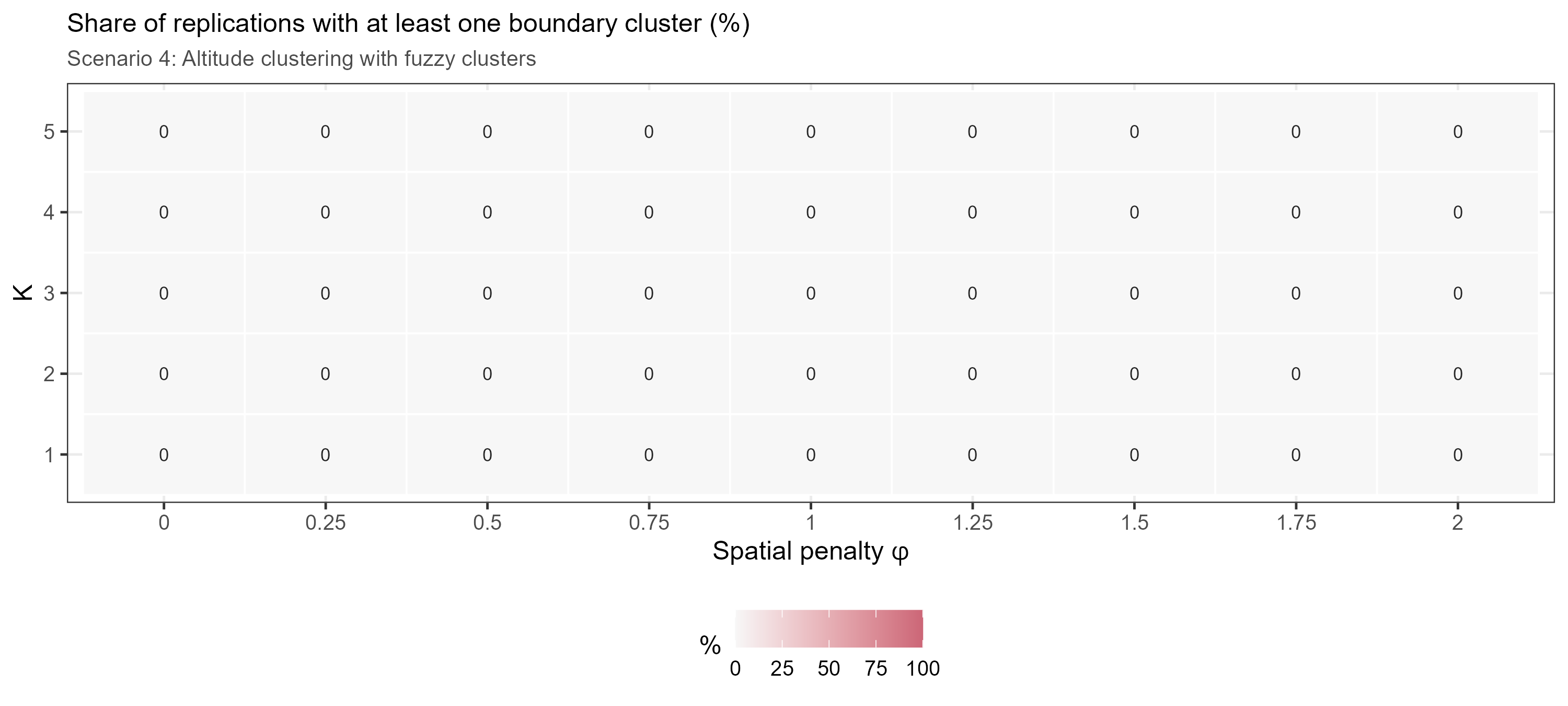}
  \caption{Scenario 4: share of fits with at least one boundary variance solution, by $K$ and $\phi$ (adjusted REML).}
\end{figure}

\begin{table}[!htb]
  \centering
  \caption{Scenario 4: Monte Carlo bias and standard deviation of the cluster-wise slopes and random-effect variances at the true $K=3$, by $\phi$ (auto-generated).}
  \adjustbox{max width=\linewidth}{
\begin{tabular}{lllll}
\toprule
Parameter & Group & Phi & Bias & SD \\
\midrule
beta1 & G1 & 0.000 & 2.581 & 5.706 \\
beta1 & G1 & 0.250 & 2.551 & 4.200 \\
beta1 & G1 & 0.500 & 2.804 & 4.855 \\
beta1 & G1 & 0.750 & 2.879 & 4.776 \\
beta1 & G1 & 1.000 & 3.268 & 5.606 \\
beta1 & G1 & 1.250 & 3.134 & 5.401 \\
beta1 & G1 & 1.500 & 3.150 & 4.220 \\
beta1 & G1 & 1.750 & 3.129 & 4.038 \\
beta1 & G1 & 2.000 & 3.246 & 4.003 \\
beta1 & G2 & 0.000 & -0.275 & 2.406 \\
beta1 & G2 & 0.250 & -0.181 & 2.341 \\
beta1 & G2 & 0.500 & -0.240 & 2.243 \\
beta1 & G2 & 0.750 & -0.327 & 3.556 \\
beta1 & G2 & 1.000 & -0.667 & 3.331 \\
beta1 & G2 & 1.250 & -0.845 & 3.437 \\
beta1 & G2 & 1.500 & -0.999 & 3.127 \\
beta1 & G2 & 1.750 & -1.064 & 3.212 \\
beta1 & G2 & 2.000 & -1.140 & 3.035 \\
beta1 & G3 & 0.000 & -3.943 & 8.357 \\
beta1 & G3 & 0.250 & -3.147 & 6.544 \\
beta1 & G3 & 0.500 & -3.673 & 7.146 \\
beta1 & G3 & 0.750 & -4.544 & 6.552 \\
beta1 & G3 & 1.000 & -5.564 & 5.529 \\
beta1 & G3 & 1.250 & -5.986 & 5.460 \\
beta1 & G3 & 1.500 & -6.281 & 4.563 \\
beta1 & G3 & 1.750 & -6.381 & 4.430 \\
beta1 & G3 & 2.000 & -6.423 & 4.406 \\
sigma2u & G1 & 0.000 & 365.230 & 1014.404 \\
sigma2u & G1 & 0.250 & 263.569 & 891.988 \\
sigma2u & G1 & 0.500 & 404.142 & 1294.993 \\
sigma2u & G1 & 0.750 & 496.487 & 1255.734 \\
sigma2u & G1 & 1.000 & 577.920 & 3233.101 \\
sigma2u & G1 & 1.250 & 473.607 & 1177.100 \\
sigma2u & G1 & 1.500 & 419.061 & 989.902 \\
sigma2u & G1 & 1.750 & 456.365 & 1334.474 \\
sigma2u & G1 & 2.000 & 433.794 & 876.265 \\
sigma2u & G2 & 0.000 & 60.333 & 454.223 \\
sigma2u & G2 & 0.250 & 46.086 & 397.079 \\
sigma2u & G2 & 0.500 & 88.374 & 599.215 \\
sigma2u & G2 & 0.750 & 171.580 & 999.696 \\
sigma2u & G2 & 1.000 & 206.173 & 696.632 \\
sigma2u & G2 & 1.250 & 231.389 & 585.527 \\
sigma2u & G2 & 1.500 & 280.825 & 620.431 \\
sigma2u & G2 & 1.750 & 285.566 & 589.741 \\
sigma2u & G2 & 2.000 & 331.512 & 1004.907 \\
sigma2u & G3 & 0.000 & 257.282 & 1144.594 \\
sigma2u & G3 & 0.250 & 3.813 & 940.225 \\
sigma2u & G3 & 0.500 & 313.104 & 1587.701 \\
sigma2u & G3 & 0.750 & 492.878 & 1925.712 \\
sigma2u & G3 & 1.000 & 479.535 & 1425.134 \\
sigma2u & G3 & 1.250 & 369.987 & 1345.651 \\
sigma2u & G3 & 1.500 & 344.128 & 1714.038 \\
sigma2u & G3 & 1.750 & 292.500 & 1705.478 \\
sigma2u & G3 & 2.000 & 306.158 & 2492.016 \\
\bottomrule
\end{tabular}
}
\end{table}

\begin{table}[!htb]
  \centering
  \caption{Scenario 4: clustering and prediction performance summary (auto-generated).}
  \adjustbox{max width=\linewidth}{
\begin{tabular}{lllllllll}
\toprule
MC\_G & Phi & ARI\_mean & ARI\_sd & Share\_mean & Share\_sd & RelRMSE\_mean & RelRMSE\_sd & Boundary\_pct \\
\midrule
1.000 & 0.000 & 0.000 & 0.000 & 100.000 & 0.000 & 100.000 & 0.000 & 0.000 \\
1.000 & 0.250 & 0.000 & 0.000 & 100.000 & 0.000 & 100.000 & 0.000 & 0.000 \\
1.000 & 0.500 & 0.000 & 0.000 & 100.000 & 0.000 & 100.000 & 0.000 & 0.000 \\
1.000 & 0.750 & 0.000 & 0.000 & 100.000 & 0.000 & 100.000 & 0.000 & 0.000 \\
1.000 & 1.000 & 0.000 & 0.000 & 100.000 & 0.000 & 100.000 & 0.000 & 0.000 \\
1.000 & 1.250 & 0.000 & 0.000 & 100.000 & 0.000 & 100.000 & 0.000 & 0.000 \\
1.000 & 1.500 & 0.000 & 0.000 & 100.000 & 0.000 & 100.000 & 0.000 & 0.000 \\
1.000 & 1.750 & 0.000 & 0.000 & 100.000 & 0.000 & 100.000 & 0.000 & 0.000 \\
1.000 & 2.000 & 0.000 & 0.000 & 100.000 & 0.000 & 100.000 & 0.000 & 0.000 \\
2.000 & 0.000 & 0.154 & 0.034 & 51.995 & 9.337 & 91.366 & 7.740 & 0.000 \\
2.000 & 0.250 & 0.203 & 0.059 & 42.675 & 19.066 & 95.105 & 9.151 & 0.000 \\
2.000 & 0.500 & 0.207 & 0.072 & 41.409 & 19.076 & 96.520 & 9.951 & 0.000 \\
2.000 & 0.750 & 0.150 & 0.078 & 41.723 & 15.670 & 97.975 & 9.953 & 0.000 \\
2.000 & 1.000 & 0.108 & 0.074 & 41.250 & 13.211 & 98.892 & 9.666 & 0.000 \\
2.000 & 1.250 & 0.073 & 0.063 & 39.559 & 10.445 & 98.843 & 9.081 & 0.000 \\
2.000 & 1.500 & 0.057 & 0.052 & 38.935 & 8.989 & 98.516 & 8.593 & 0.000 \\
2.000 & 1.750 & 0.053 & 0.049 & 38.478 & 8.778 & 98.329 & 8.133 & 0.000 \\
2.000 & 2.000 & 0.050 & 0.047 & 38.280 & 8.746 & 98.192 & 7.828 & 0.000 \\
3.000 & 0.000 & 0.253 & 0.100 & 62.172 & 13.376 & 89.637 & 8.611 & 0.000 \\
3.000 & 0.250 & 0.306 & 0.109 & 66.170 & 13.095 & 91.389 & 9.797 & 0.000 \\
3.000 & 0.500 & 0.265 & 0.143 & 60.811 & 15.246 & 95.460 & 11.834 & 0.000 \\
3.000 & 0.750 & 0.172 & 0.113 & 53.155 & 13.129 & 99.555 & 11.073 & 0.000 \\
3.000 & 1.000 & 0.110 & 0.082 & 47.419 & 10.868 & 101.249 & 9.954 & 0.000 \\
3.000 & 1.250 & 0.079 & 0.064 & 44.891 & 9.372 & 101.147 & 9.382 & 0.000 \\
3.000 & 1.500 & 0.062 & 0.052 & 43.385 & 8.596 & 100.278 & 8.316 & 0.000 \\
3.000 & 1.750 & 0.055 & 0.050 & 42.829 & 8.263 & 99.675 & 8.063 & 0.000 \\
3.000 & 2.000 & 0.051 & 0.048 & 42.364 & 8.481 & 99.225 & 7.715 & 0.000 \\
4.000 & 0.000 & 0.241 & 0.054 & 40.214 & 10.707 & 90.838 & 8.828 & 0.000 \\
4.000 & 0.250 & 0.314 & 0.078 & 42.948 & 15.266 & 90.399 & 9.425 & 0.000 \\
4.000 & 0.500 & 0.311 & 0.105 & 45.461 & 16.621 & 91.712 & 10.549 & 0.000 \\
4.000 & 0.750 & 0.225 & 0.101 & 41.910 & 16.245 & 96.192 & 10.817 & 0.000 \\
4.000 & 1.000 & 0.152 & 0.084 & 38.664 & 14.938 & 98.912 & 10.002 & 0.000 \\
4.000 & 1.250 & 0.108 & 0.068 & 35.402 & 14.034 & 99.146 & 9.503 & 0.000 \\
4.000 & 1.500 & 0.089 & 0.059 & 34.329 & 14.021 & 98.718 & 9.206 & 0.000 \\
4.000 & 1.750 & 0.078 & 0.053 & 34.803 & 13.597 & 98.395 & 8.580 & 0.000 \\
4.000 & 2.000 & 0.073 & 0.050 & 35.756 & 13.124 & 98.149 & 8.329 & 0.000 \\
5.000 & 0.000 & 0.222 & 0.042 & 25.976 & 11.436 & 91.908 & 8.783 & 0.000 \\
5.000 & 0.250 & 0.300 & 0.069 & 27.153 & 14.601 & 91.036 & 9.508 & 0.000 \\
5.000 & 0.500 & 0.302 & 0.088 & 31.127 & 16.612 & 91.438 & 10.006 & 0.000 \\
5.000 & 0.750 & 0.238 & 0.093 & 30.959 & 16.384 & 94.687 & 10.940 & 0.000 \\
5.000 & 1.000 & 0.175 & 0.083 & 29.312 & 15.391 & 97.420 & 10.619 & 0.000 \\
5.000 & 1.250 & 0.133 & 0.073 & 27.677 & 14.776 & 97.872 & 10.005 & 0.000 \\
5.000 & 1.500 & 0.112 & 0.066 & 26.984 & 14.325 & 97.496 & 9.445 & 0.000 \\
5.000 & 1.750 & 0.097 & 0.058 & 27.262 & 14.143 & 97.107 & 9.181 & 0.000 \\
5.000 & 2.000 & 0.091 & 0.054 & 27.052 & 13.983 & 96.806 & 8.858 & 0.000 \\
\bottomrule
\end{tabular}
}
\end{table}

\clearpage

\section{Extended results for the robustness designs}\label{sm:robust}

\subsection{Scenario 5: perfect separation (level-separated regimes)}

\begin{figure}[!htb]
  \centering
  \includegraphics[width=0.95\linewidth]{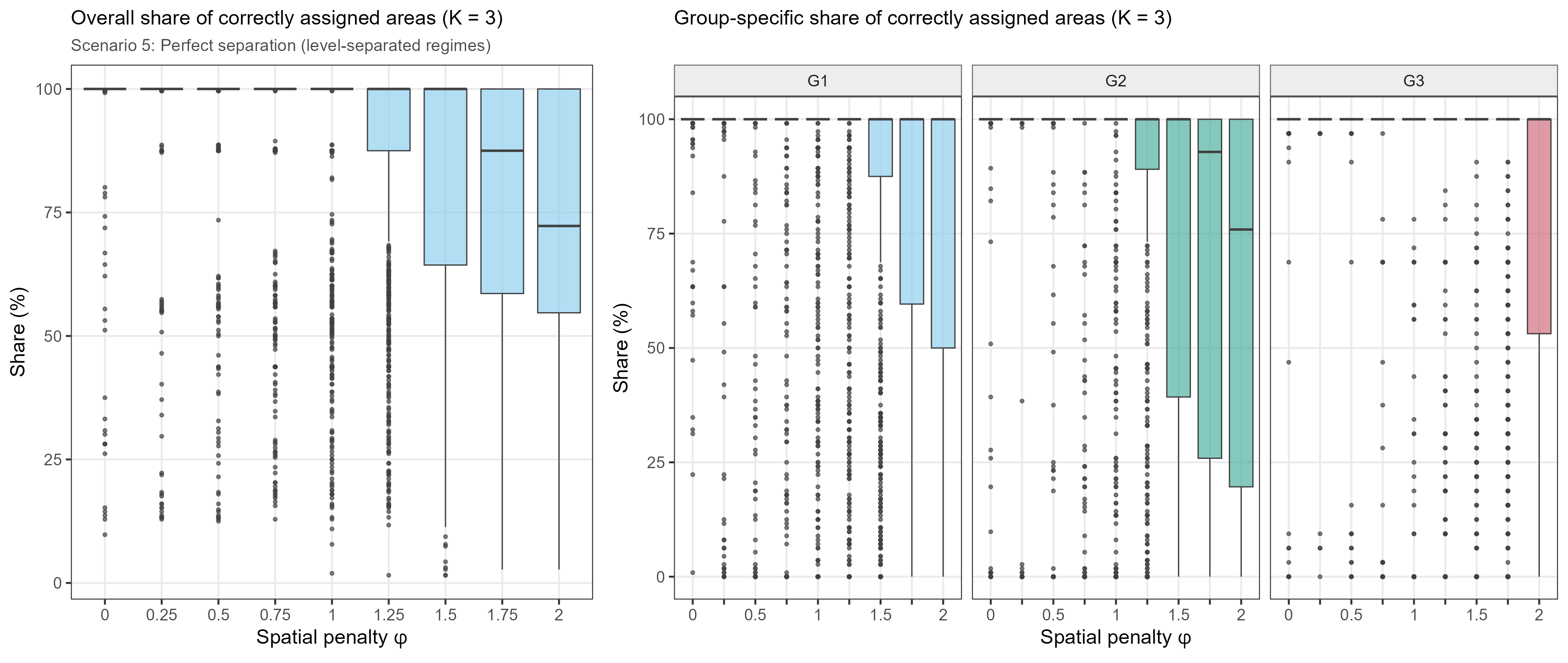}
  \caption{Scenario 5: overall (left) and group-specific (right) share of correctly assigned areas at the true $K=3$, by spatial penalty $\phi$.}
\end{figure}

\begin{figure}[!htb]
  \centering
  \includegraphics[width=0.9\linewidth]{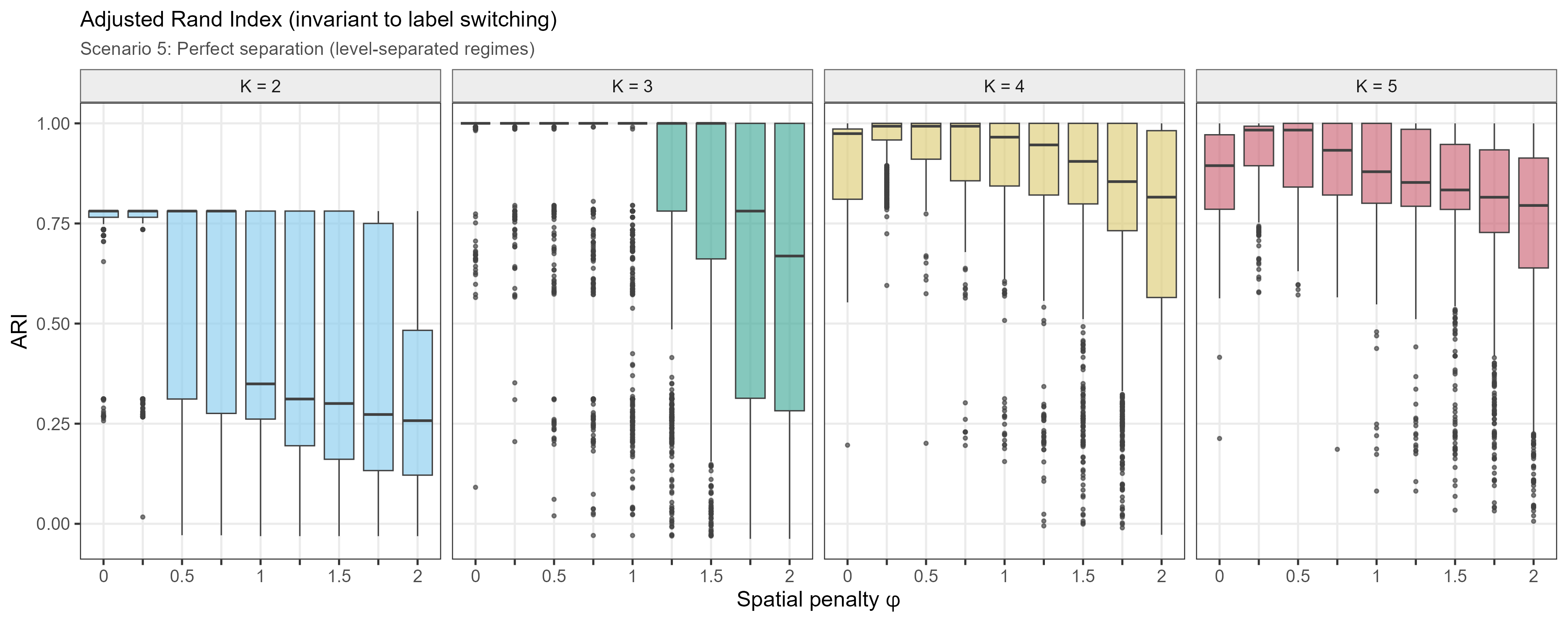}
  \caption{Scenario 5: Adjusted Rand Index between estimated and true partition, by $K$ and $\phi$.}
\end{figure}

\begin{figure}[!htb]
  \centering
  \includegraphics[width=0.95\linewidth]{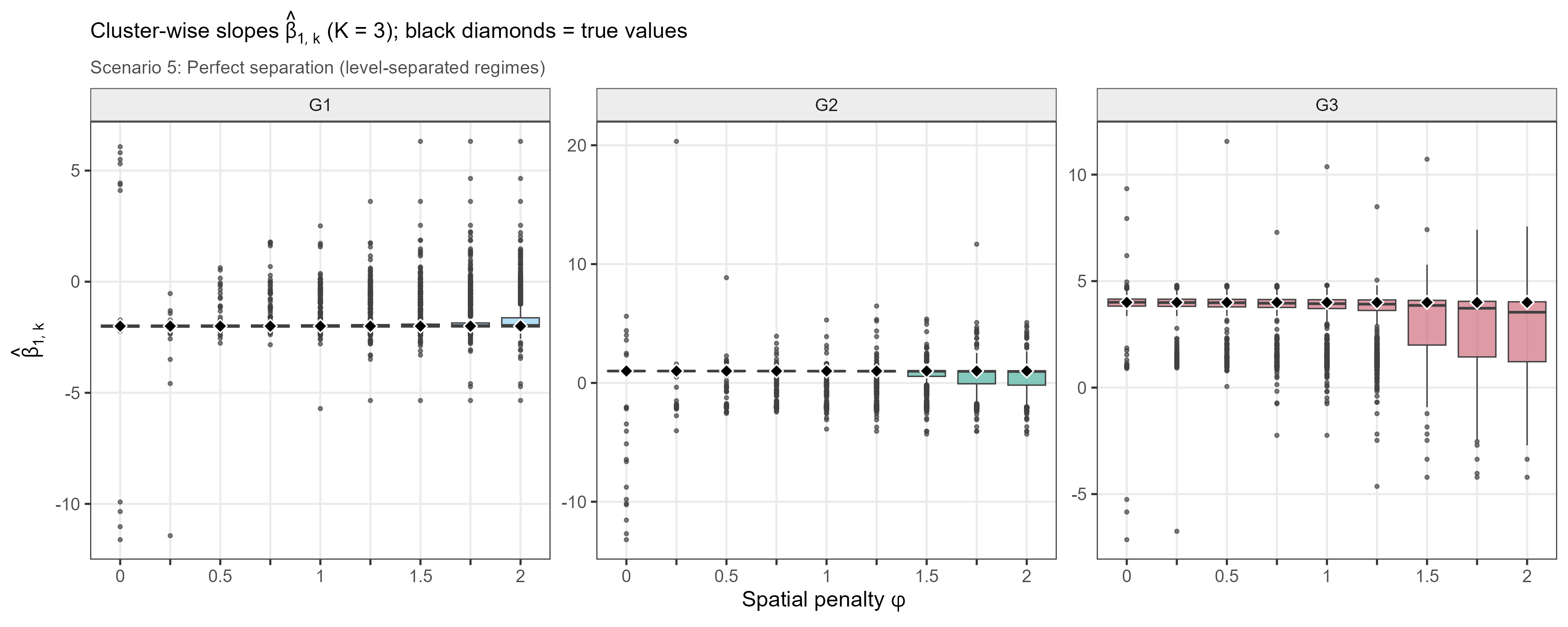}
  \caption{Scenario 5: cluster-specific slope estimates at the true $K=3$, by $\phi$. Black diamonds mark the true values.}
\end{figure}

\begin{figure}[!htb]
  \centering
  \includegraphics[width=0.9\linewidth]{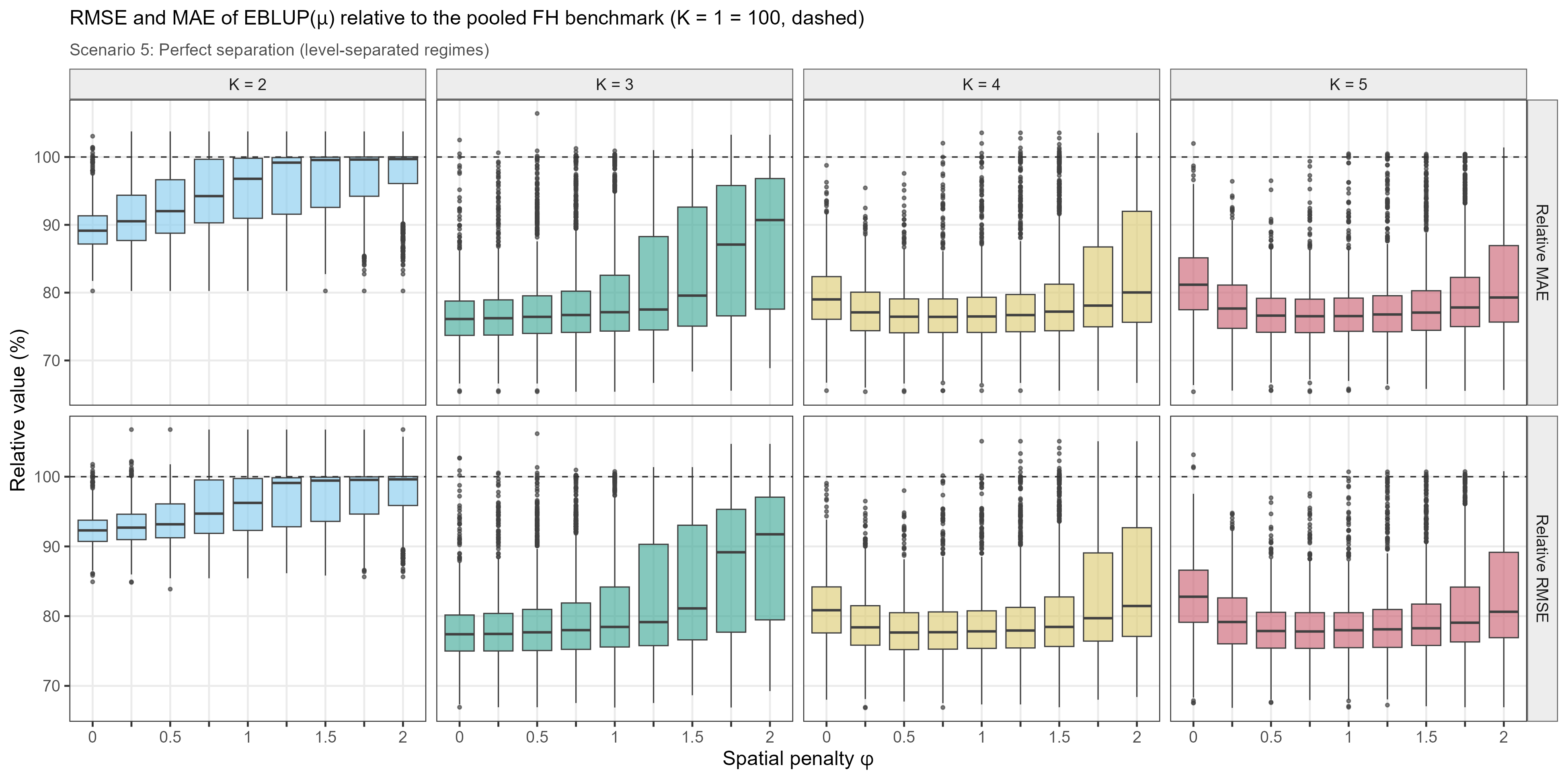}
  \caption{Scenario 5: RMSE of the EBLUP relative to the pooled FH benchmark (\%, dashed line at 100), by $K$ and $\phi$.}
\end{figure}

\begin{figure}[!htb]
  \centering
  \includegraphics[width=0.95\linewidth]{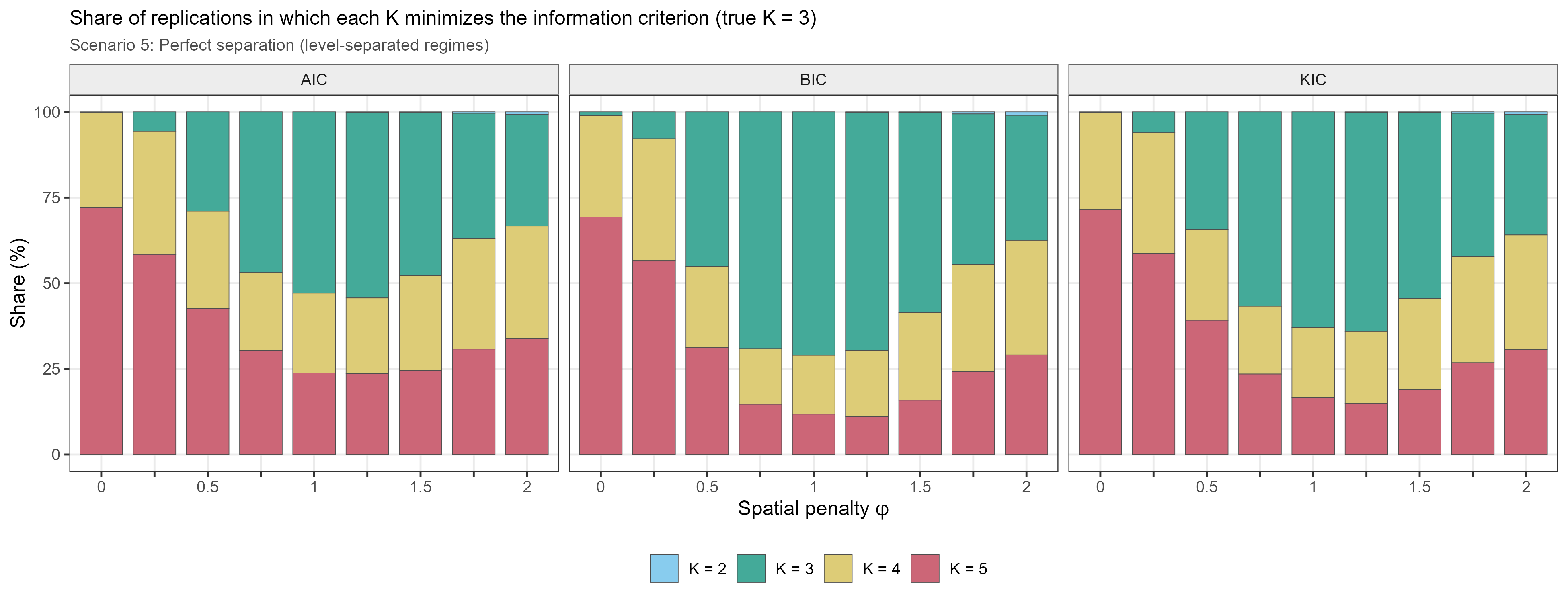}
  \caption{Scenario 5: share of replications in which each $K$ minimizes AIC, BIC, and KIC, by $\phi$.}
\end{figure}

\begin{figure}[!htb]
  \centering
  \includegraphics[width=0.95\linewidth]{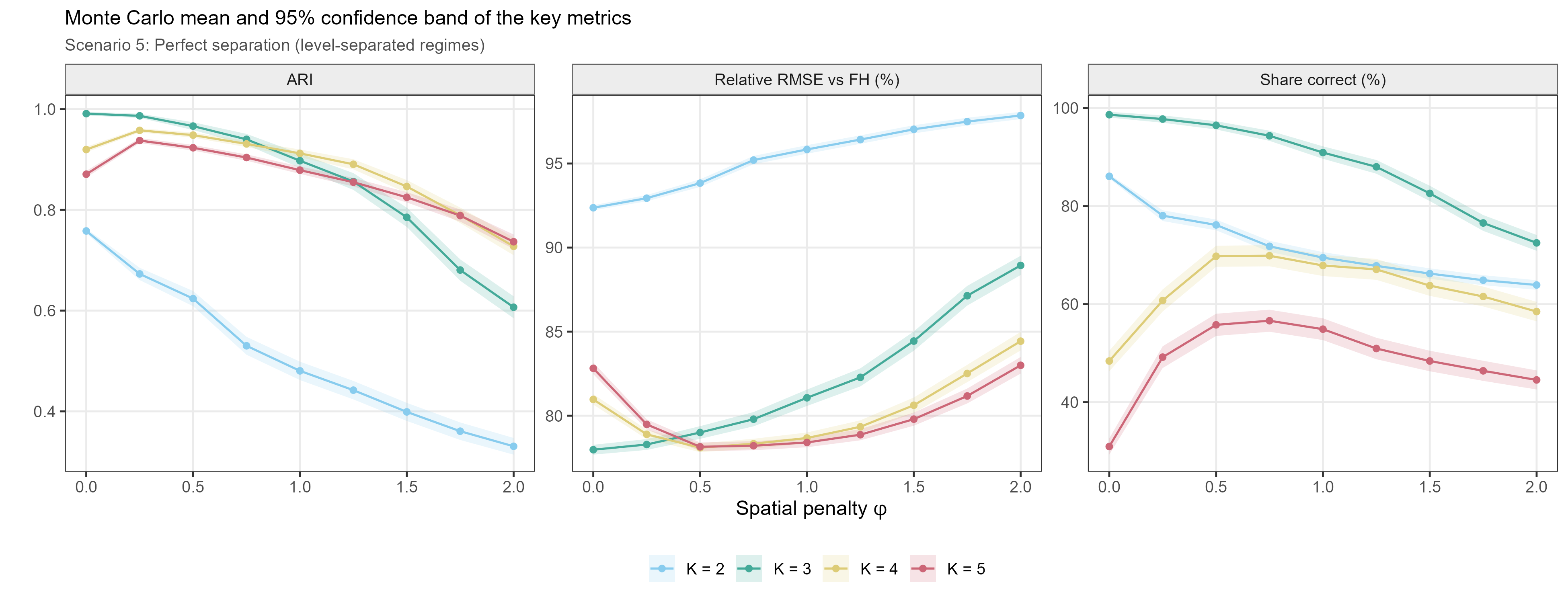}
  \caption{Scenario 5: Monte Carlo mean and 95\% band of ARI, share of correct assignments, and relative RMSE, as functions of $\phi$, by $K$.}
\end{figure}

\begin{table}[!htb]
  \centering
  \caption{Scenario 5: Monte Carlo bias and standard deviation of the cluster-wise slopes and random-effect variances at the true $K=3$, by $\phi$ (auto-generated).}
  \adjustbox{max width=\linewidth}{
\begin{tabular}{lllll}
\toprule
Parameter & Group & Phi & Bias & SD \\
\midrule
beta1 & G1 & 0.000 & 0.014 & 0.817 \\
beta1 & G1 & 0.250 & -0.011 & 0.327 \\
beta1 & G1 & 0.500 & 0.021 & 0.214 \\
beta1 & G1 & 0.750 & 0.056 & 0.356 \\
beta1 & G1 & 1.000 & 0.099 & 0.463 \\
beta1 & G1 & 1.250 & 0.146 & 0.569 \\
beta1 & G1 & 1.500 & 0.216 & 0.697 \\
beta1 & G1 & 1.750 & 0.303 & 0.827 \\
beta1 & G1 & 2.000 & 0.374 & 0.879 \\
beta1 & G2 & 0.000 & -0.108 & 1.132 \\
beta1 & G2 & 0.250 & -0.070 & 0.808 \\
beta1 & G2 & 0.500 & -0.146 & 0.713 \\
beta1 & G2 & 0.750 & -0.213 & 0.780 \\
beta1 & G2 & 1.000 & -0.337 & 0.947 \\
beta1 & G2 & 1.250 & -0.385 & 1.041 \\
beta1 & G2 & 1.500 & -0.486 & 1.141 \\
beta1 & G2 & 1.750 & -0.540 & 1.256 \\
beta1 & G2 & 2.000 & -0.608 & 1.257 \\
beta1 & G3 & 0.000 & -0.064 & 0.746 \\
beta1 & G3 & 0.250 & -0.104 & 0.641 \\
beta1 & G3 & 0.500 & -0.158 & 0.720 \\
beta1 & G3 & 0.750 & -0.264 & 0.862 \\
beta1 & G3 & 1.000 & -0.387 & 1.023 \\
beta1 & G3 & 1.250 & -0.534 & 1.177 \\
beta1 & G3 & 1.500 & -0.788 & 1.377 \\
beta1 & G3 & 1.750 & -1.127 & 1.517 \\
beta1 & G3 & 2.000 & -1.287 & 1.571 \\
sigma2u & G1 & 0.000 & 2.548 & 25.049 \\
sigma2u & G1 & 0.250 & 0.522 & 10.114 \\
sigma2u & G1 & 0.500 & 4.380 & 36.109 \\
sigma2u & G1 & 0.750 & 11.571 & 66.449 \\
sigma2u & G1 & 1.000 & 23.819 & 88.764 \\
sigma2u & G1 & 1.250 & 31.195 & 94.934 \\
sigma2u & G1 & 1.500 & 46.766 & 124.574 \\
sigma2u & G1 & 1.750 & 68.152 & 164.271 \\
sigma2u & G1 & 2.000 & 77.882 & 160.980 \\
sigma2u & G2 & 0.000 & 2.923 & 24.548 \\
sigma2u & G2 & 0.250 & 0.817 & 9.688 \\
sigma2u & G2 & 0.500 & 4.701 & 34.893 \\
sigma2u & G2 & 0.750 & 11.648 & 62.136 \\
sigma2u & G2 & 1.000 & 24.201 & 86.147 \\
sigma2u & G2 & 1.250 & 32.480 & 99.480 \\
sigma2u & G2 & 1.500 & 57.068 & 133.614 \\
sigma2u & G2 & 1.750 & 82.181 & 146.889 \\
sigma2u & G2 & 2.000 & 97.726 & 153.352 \\
sigma2u & G3 & 0.000 & 1.684 & 10.954 \\
sigma2u & G3 & 0.250 & 4.927 & 21.312 \\
sigma2u & G3 & 0.500 & 8.316 & 29.233 \\
sigma2u & G3 & 0.750 & 12.523 & 39.398 \\
sigma2u & G3 & 1.000 & 20.720 & 66.745 \\
sigma2u & G3 & 1.250 & 33.865 & 131.470 \\
sigma2u & G3 & 1.500 & 48.572 & 156.436 \\
sigma2u & G3 & 1.750 & 69.047 & 137.888 \\
sigma2u & G3 & 2.000 & 87.391 & 153.753 \\
\bottomrule
\end{tabular}
}
\end{table}

\begin{table}[!htb]
  \centering
  \caption{Scenario 5: clustering and prediction performance summary (auto-generated).}
  \adjustbox{max width=\linewidth}{
\begin{tabular}{lllllllll}
\toprule
MC\_G & Phi & ARI\_mean & ARI\_sd & Share\_mean & Share\_sd & RelRMSE\_mean & RelRMSE\_sd & Boundary\_pct \\
\midrule
1.000 & 0.000 & 0.000 & 0.000 & 100.000 & 0.000 & 100.000 & 0.000 & 0.000 \\
1.000 & 0.250 & 0.000 & 0.000 & 100.000 & 0.000 & 100.000 & 0.000 & 0.000 \\
1.000 & 0.500 & 0.000 & 0.000 & 100.000 & 0.000 & 100.000 & 0.000 & 0.000 \\
1.000 & 0.750 & 0.000 & 0.000 & 100.000 & 0.000 & 100.000 & 0.000 & 0.000 \\
1.000 & 1.000 & 0.000 & 0.000 & 100.000 & 0.000 & 100.000 & 0.000 & 0.000 \\
1.000 & 1.250 & 0.000 & 0.000 & 100.000 & 0.000 & 100.000 & 0.000 & 0.000 \\
1.000 & 1.500 & 0.000 & 0.000 & 100.000 & 0.000 & 100.000 & 0.000 & 0.000 \\
1.000 & 1.750 & 0.000 & 0.000 & 100.000 & 0.000 & 100.000 & 0.000 & 0.000 \\
1.000 & 2.000 & 0.000 & 0.000 & 100.000 & 0.000 & 100.000 & 0.000 & 0.000 \\
2.000 & 0.000 & 0.758 & 0.076 & 86.069 & 7.253 & 92.370 & 2.449 & 0.000 \\
2.000 & 0.250 & 0.673 & 0.197 & 78.043 & 17.679 & 92.940 & 3.000 & 0.000 \\
2.000 & 0.500 & 0.624 & 0.243 & 76.153 & 17.732 & 93.835 & 3.667 & 0.000 \\
2.000 & 0.750 & 0.530 & 0.287 & 71.786 & 18.235 & 95.205 & 4.012 & 0.000 \\
2.000 & 1.000 & 0.480 & 0.295 & 69.481 & 18.199 & 95.838 & 4.113 & 0.000 \\
2.000 & 1.250 & 0.442 & 0.296 & 67.812 & 17.971 & 96.425 & 4.051 & 0.000 \\
2.000 & 1.500 & 0.399 & 0.290 & 66.225 & 17.261 & 97.036 & 3.929 & 0.000 \\
2.000 & 1.750 & 0.361 & 0.281 & 64.883 & 16.501 & 97.494 & 3.717 & 0.000 \\
2.000 & 2.000 & 0.331 & 0.267 & 63.903 & 15.712 & 97.858 & 3.500 & 0.000 \\
3.000 & 0.000 & 0.991 & 0.058 & 98.632 & 9.383 & 77.978 & 4.719 & 0.000 \\
3.000 & 0.250 & 0.987 & 0.067 & 97.741 & 12.093 & 78.288 & 5.201 & 0.000 \\
3.000 & 0.500 & 0.966 & 0.124 & 96.466 & 13.662 & 79.003 & 6.075 & 0.000 \\
3.000 & 0.750 & 0.940 & 0.175 & 94.335 & 16.760 & 79.795 & 6.843 & 0.000 \\
3.000 & 1.000 & 0.898 & 0.229 & 90.912 & 20.500 & 81.067 & 7.861 & 0.000 \\
3.000 & 1.250 & 0.856 & 0.270 & 88.000 & 22.596 & 82.282 & 8.652 & 0.000 \\
3.000 & 1.500 & 0.785 & 0.308 & 82.587 & 25.202 & 84.441 & 9.244 & 0.000 \\
3.000 & 1.750 & 0.681 & 0.342 & 76.554 & 26.051 & 87.138 & 9.449 & 0.000 \\
3.000 & 2.000 & 0.607 & 0.351 & 72.491 & 26.018 & 88.942 & 9.236 & 0.000 \\
4.000 & 0.000 & 0.920 & 0.088 & 48.398 & 33.716 & 80.970 & 4.966 & 0.000 \\
4.000 & 0.250 & 0.958 & 0.066 & 60.745 & 36.350 & 78.894 & 4.573 & 0.000 \\
4.000 & 0.500 & 0.948 & 0.081 & 69.758 & 34.748 & 78.098 & 4.236 & 0.000 \\
4.000 & 0.750 & 0.931 & 0.103 & 69.870 & 34.808 & 78.336 & 4.726 & 0.000 \\
4.000 & 1.000 & 0.912 & 0.125 & 67.878 & 34.676 & 78.675 & 5.294 & 0.000 \\
4.000 & 1.250 & 0.891 & 0.163 & 67.086 & 34.389 & 79.342 & 6.225 & 0.000 \\
4.000 & 1.500 & 0.846 & 0.208 & 63.775 & 33.723 & 80.624 & 7.419 & 0.000 \\
4.000 & 1.750 & 0.789 & 0.250 & 61.557 & 32.711 & 82.514 & 8.314 & 0.000 \\
4.000 & 2.000 & 0.728 & 0.284 & 58.486 & 32.043 & 84.435 & 8.891 & 0.000 \\
5.000 & 0.000 & 0.871 & 0.105 & 30.964 & 26.743 & 82.820 & 5.558 & 0.000 \\
5.000 & 0.250 & 0.938 & 0.081 & 49.187 & 35.795 & 79.484 & 4.899 & 0.000 \\
5.000 & 0.500 & 0.923 & 0.090 & 55.780 & 36.560 & 78.154 & 4.170 & 0.000 \\
5.000 & 0.750 & 0.904 & 0.101 & 56.615 & 36.235 & 78.215 & 4.322 & 0.000 \\
5.000 & 1.000 & 0.879 & 0.120 & 54.893 & 35.852 & 78.411 & 4.600 & 0.000 \\
5.000 & 1.250 & 0.855 & 0.137 & 50.952 & 35.351 & 78.872 & 5.374 & 0.000 \\
5.000 & 1.500 & 0.825 & 0.165 & 48.403 & 33.838 & 79.797 & 6.295 & 0.000 \\
5.000 & 1.750 & 0.789 & 0.195 & 46.402 & 33.110 & 81.174 & 7.155 & 0.000 \\
5.000 & 2.000 & 0.737 & 0.231 & 44.540 & 31.353 & 83.000 & 7.975 & 0.000 \\
\bottomrule
\end{tabular}
}
\end{table}

\clearpage

\subsection{Scenario 6: high variability (widely spread coefficients)}

\begin{figure}[!htb]
  \centering
  \includegraphics[width=0.95\linewidth]{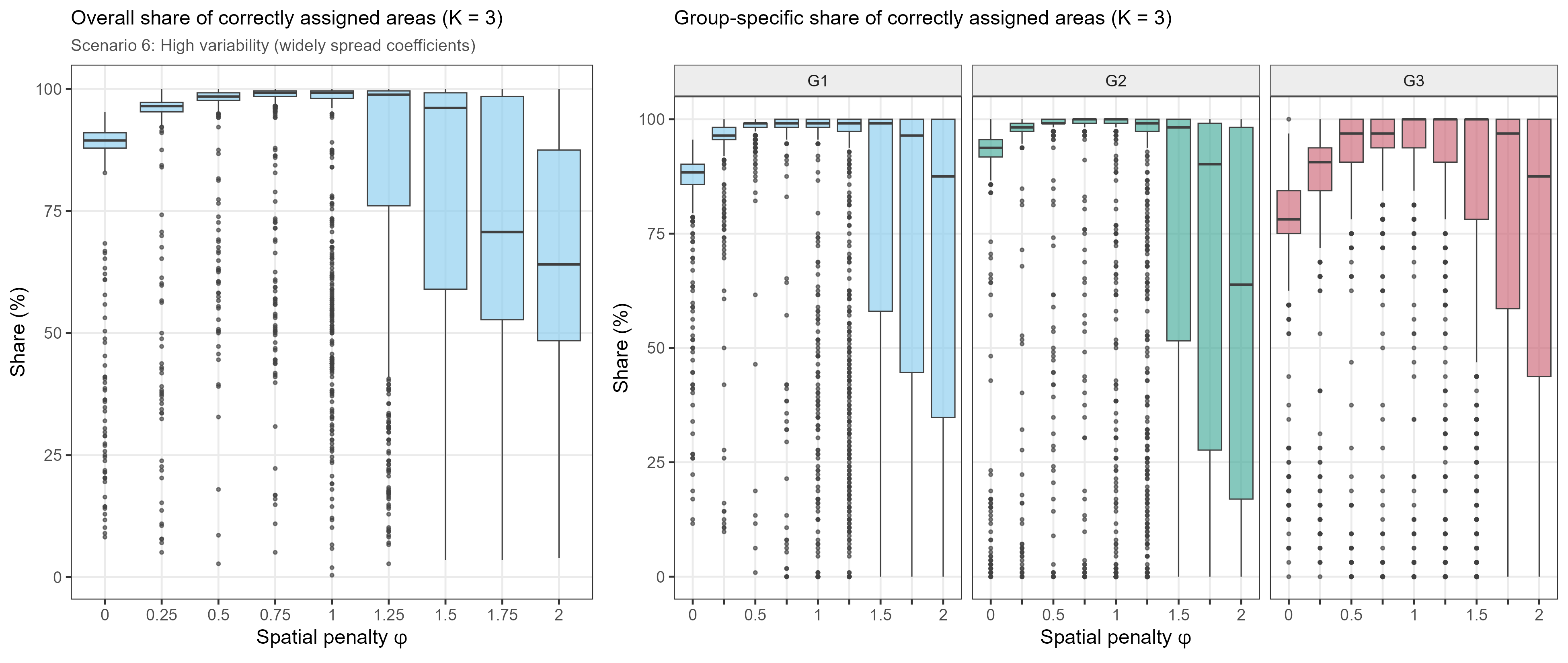}
  \caption{Scenario 6: overall (left) and group-specific (right) share of correctly assigned areas at the true $K=3$, by spatial penalty $\phi$.}
\end{figure}

\begin{figure}[!htb]
  \centering
  \includegraphics[width=0.9\linewidth]{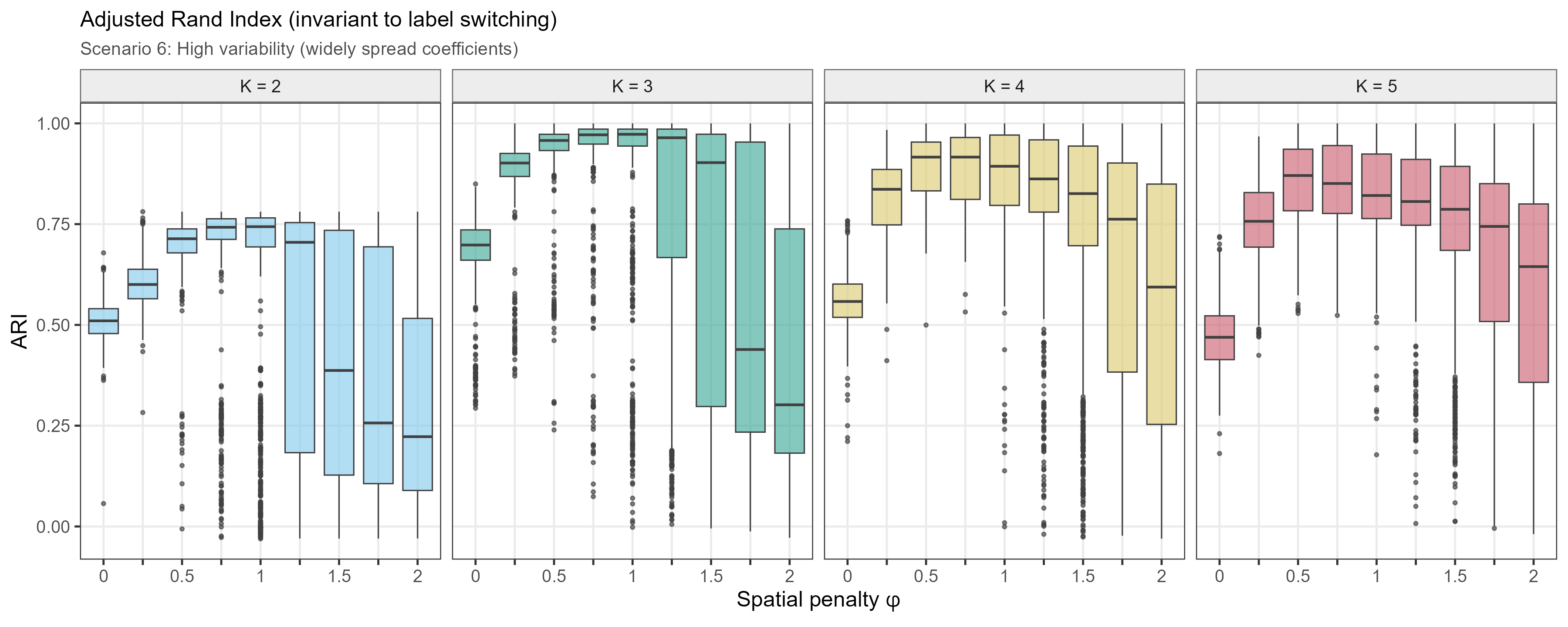}
  \caption{Scenario 6: Adjusted Rand Index between estimated and true partition, by $K$ and $\phi$.}
\end{figure}

\begin{figure}[!htb]
  \centering
  \includegraphics[width=0.95\linewidth]{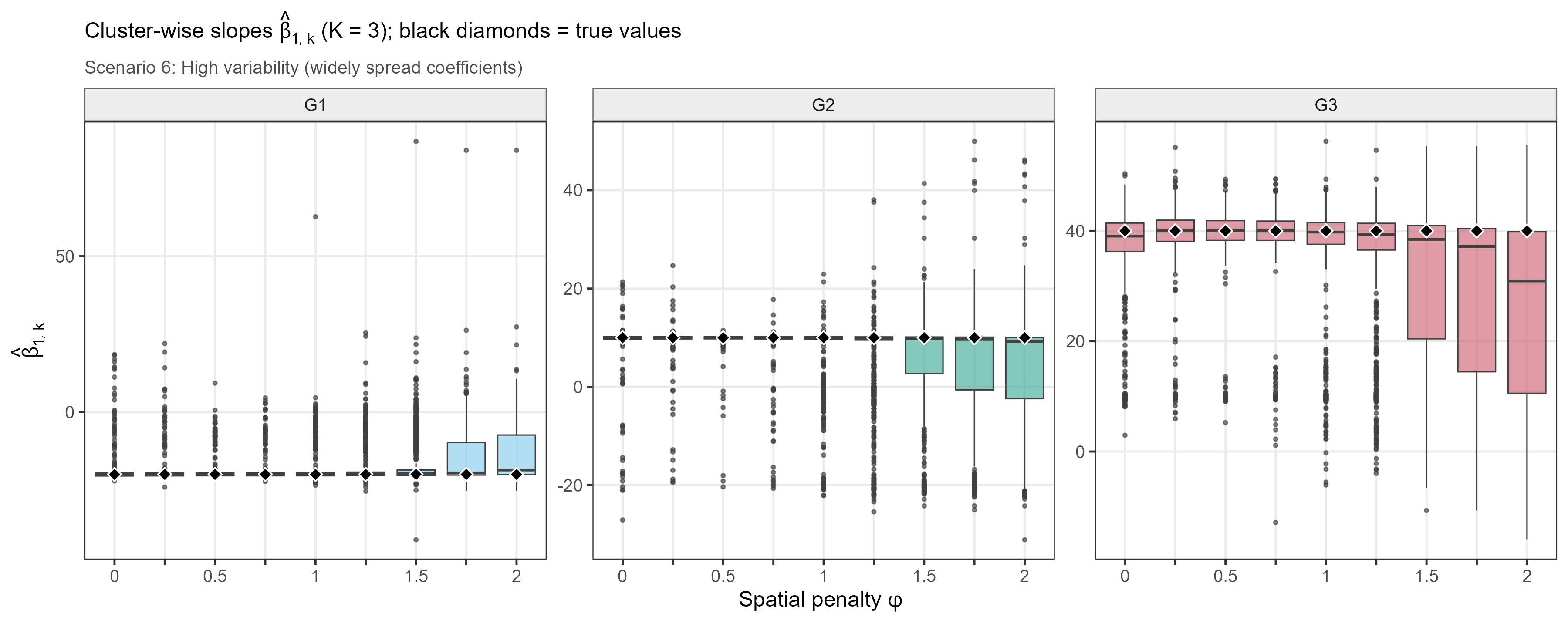}
  \caption{Scenario 6: cluster-specific slope estimates at the true $K=3$, by $\phi$. Black diamonds mark the true values.}
\end{figure}

\begin{figure}[!htb]
  \centering
  \includegraphics[width=0.9\linewidth]{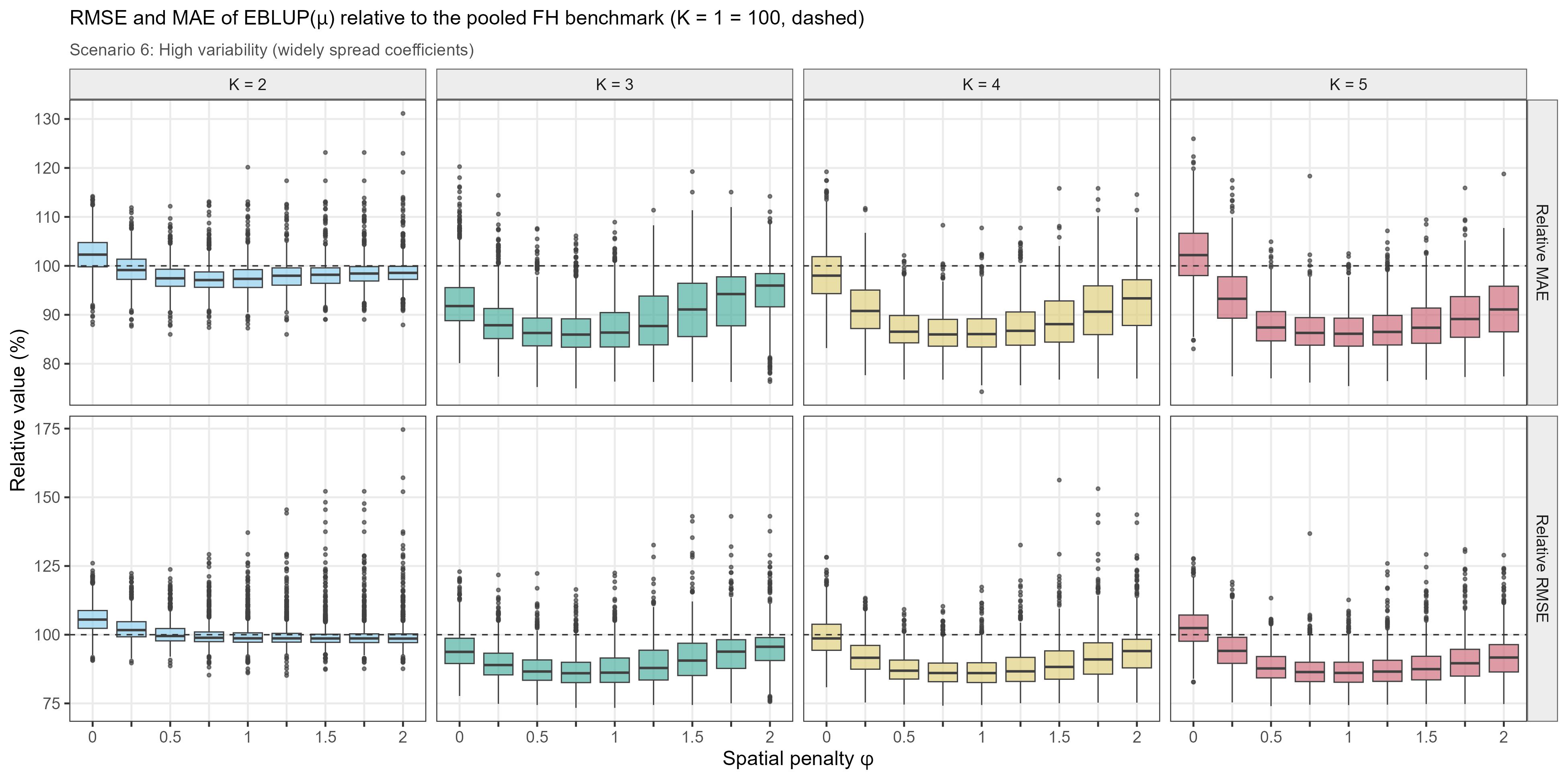}
  \caption{Scenario 6: RMSE of the EBLUP relative to the pooled FH benchmark (\%, dashed line at 100), by $K$ and $\phi$.}
\end{figure}

\begin{figure}[!htb]
  \centering
  \includegraphics[width=0.95\linewidth]{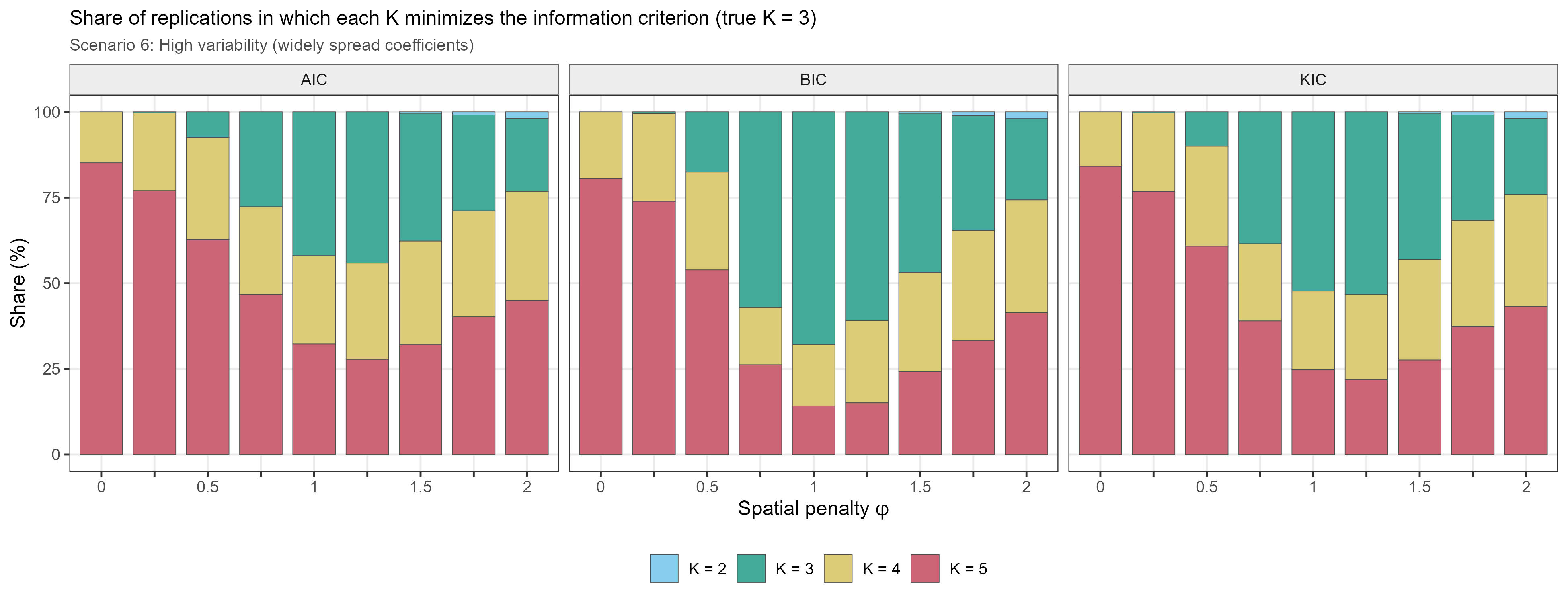}
  \caption{Scenario 6: share of replications in which each $K$ minimizes AIC, BIC, and KIC, by $\phi$.}
\end{figure}

\begin{figure}[!htb]
  \centering
  \includegraphics[width=0.95\linewidth]{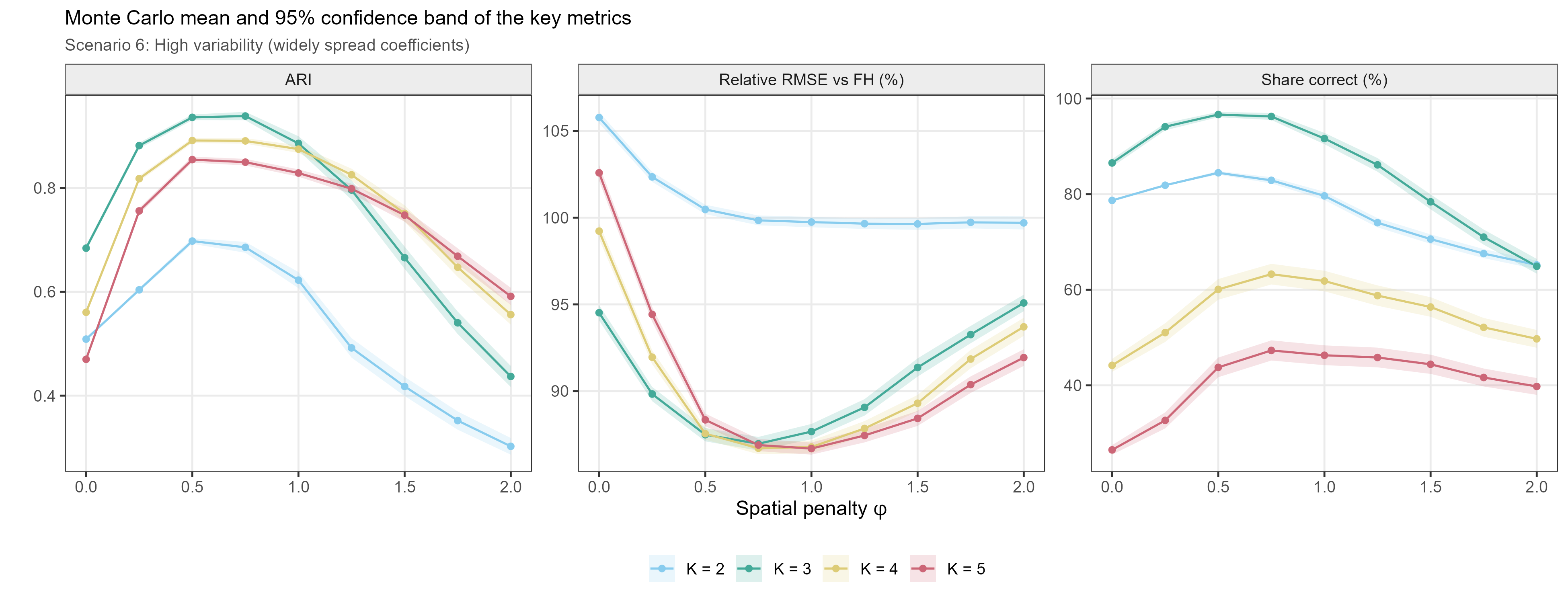}
  \caption{Scenario 6: Monte Carlo mean and 95\% band of ARI, share of correct assignments, and relative RMSE, as functions of $\phi$, by $K$.}
\end{figure}

\begin{table}[!htb]
  \centering
  \caption{Scenario 6: Monte Carlo bias and standard deviation of the cluster-wise slopes and random-effect variances at the true $K=3$, by $\phi$ (auto-generated).}
  \adjustbox{max width=\linewidth}{
\begin{tabular}{lllll}
\toprule
Parameter & Group & Phi & Bias & SD \\
\midrule
beta1 & G1 & 0.000 & 0.774 & 4.334 \\
beta1 & G1 & 0.250 & 0.601 & 3.715 \\
beta1 & G1 & 0.500 & 0.444 & 2.659 \\
beta1 & G1 & 0.750 & 0.508 & 2.972 \\
beta1 & G1 & 1.000 & 1.174 & 4.947 \\
beta1 & G1 & 1.250 & 2.058 & 5.716 \\
beta1 & G1 & 1.500 & 3.342 & 7.625 \\
beta1 & G1 & 1.750 & 4.920 & 8.097 \\
beta1 & G1 & 2.000 & 6.445 & 8.549 \\
beta1 & G2 & 0.000 & -0.450 & 3.610 \\
beta1 & G2 & 0.250 & -0.247 & 2.571 \\
beta1 & G2 & 0.500 & -0.173 & 1.891 \\
beta1 & G2 & 0.750 & -0.516 & 3.300 \\
beta1 & G2 & 1.000 & -1.562 & 5.588 \\
beta1 & G2 & 1.250 & -2.764 & 7.382 \\
beta1 & G2 & 1.500 & -4.016 & 8.508 \\
beta1 & G2 & 1.750 & -5.504 & 9.759 \\
beta1 & G2 & 2.000 & -6.621 & 10.683 \\
beta1 & G3 & 0.000 & -2.624 & 7.715 \\
beta1 & G3 & 0.250 & -1.086 & 6.751 \\
beta1 & G3 & 0.500 & -0.867 & 6.226 \\
beta1 & G3 & 0.750 & -1.128 & 6.757 \\
beta1 & G3 & 1.000 & -3.217 & 9.855 \\
beta1 & G3 & 1.250 & -5.089 & 11.557 \\
beta1 & G3 & 1.500 & -8.233 & 13.382 \\
beta1 & G3 & 1.750 & -10.847 & 14.260 \\
beta1 & G3 & 2.000 & -13.495 & 14.873 \\
sigma2u & G1 & 0.000 & 183.376 & 1134.001 \\
sigma2u & G1 & 0.250 & 160.539 & 1004.699 \\
sigma2u & G1 & 0.500 & 139.747 & 788.041 \\
sigma2u & G1 & 0.750 & 151.579 & 965.374 \\
sigma2u & G1 & 1.000 & 344.049 & 1497.386 \\
sigma2u & G1 & 1.250 & 508.428 & 1682.448 \\
sigma2u & G1 & 1.500 & 772.904 & 2305.309 \\
sigma2u & G1 & 1.750 & 1070.697 & 2535.862 \\
sigma2u & G1 & 2.000 & 1383.569 & 2618.806 \\
sigma2u & G2 & 0.000 & 94.623 & 848.134 \\
sigma2u & G2 & 0.250 & 48.190 & 574.826 \\
sigma2u & G2 & 0.500 & 15.747 & 276.884 \\
sigma2u & G2 & 0.750 & 74.535 & 730.393 \\
sigma2u & G2 & 1.000 & 213.526 & 985.037 \\
sigma2u & G2 & 1.250 & 408.682 & 1395.255 \\
sigma2u & G2 & 1.500 & 688.265 & 1748.783 \\
sigma2u & G2 & 1.750 & 857.614 & 1995.217 \\
sigma2u & G2 & 2.000 & 1075.937 & 2168.019 \\
sigma2u & G3 & 0.000 & 280.106 & 772.971 \\
sigma2u & G3 & 0.250 & -10.014 & 350.160 \\
sigma2u & G3 & 0.500 & -1.466 & 216.153 \\
sigma2u & G3 & 0.750 & 42.832 & 434.086 \\
sigma2u & G3 & 1.000 & 115.716 & 717.596 \\
sigma2u & G3 & 1.250 & 672.726 & 7574.497 \\
sigma2u & G3 & 1.500 & 750.546 & 2331.632 \\
sigma2u & G3 & 1.750 & 1118.483 & 3038.912 \\
sigma2u & G3 & 2.000 & 1643.119 & 7577.791 \\
\bottomrule
\end{tabular}
}
\end{table}

\begin{table}[!htb]
  \centering
  \caption{Scenario 6: clustering and prediction performance summary (auto-generated).}
  \adjustbox{max width=\linewidth}{
\begin{tabular}{lllllllll}
\toprule
MC\_G & Phi & ARI\_mean & ARI\_sd & Share\_mean & Share\_sd & RelRMSE\_mean & RelRMSE\_sd & Boundary\_pct \\
\midrule
1.000 & 0.000 & 0.000 & 0.000 & 100.000 & 0.000 & 100.000 & 0.000 & 0.000 \\
1.000 & 0.250 & 0.000 & 0.000 & 100.000 & 0.000 & 100.000 & 0.000 & 0.000 \\
1.000 & 0.500 & 0.000 & 0.000 & 100.000 & 0.000 & 100.000 & 0.000 & 0.000 \\
1.000 & 0.750 & 0.000 & 0.000 & 100.000 & 0.000 & 100.000 & 0.000 & 0.000 \\
1.000 & 1.000 & 0.000 & 0.000 & 100.000 & 0.000 & 100.000 & 0.000 & 0.000 \\
1.000 & 1.250 & 0.000 & 0.000 & 100.000 & 0.000 & 100.000 & 0.000 & 0.000 \\
1.000 & 1.500 & 0.000 & 0.000 & 100.000 & 0.000 & 100.000 & 0.000 & 0.000 \\
1.000 & 1.750 & 0.000 & 0.000 & 100.000 & 0.000 & 100.000 & 0.000 & 0.000 \\
1.000 & 2.000 & 0.000 & 0.000 & 100.000 & 0.000 & 100.000 & 0.000 & 0.000 \\
2.000 & 0.000 & 0.509 & 0.047 & 78.686 & 1.924 & 105.770 & 5.297 & 0.000 \\
2.000 & 0.250 & 0.604 & 0.058 & 81.859 & 2.237 & 102.344 & 4.615 & 0.000 \\
2.000 & 0.500 & 0.698 & 0.086 & 84.463 & 6.013 & 100.472 & 4.385 & 0.000 \\
2.000 & 0.750 & 0.686 & 0.169 & 82.886 & 10.602 & 99.840 & 4.566 & 0.000 \\
2.000 & 1.000 & 0.623 & 0.251 & 79.632 & 14.013 & 99.743 & 4.857 & 0.000 \\
2.000 & 1.250 & 0.492 & 0.303 & 74.025 & 15.726 & 99.652 & 4.905 & 0.000 \\
2.000 & 1.500 & 0.418 & 0.299 & 70.597 & 15.896 & 99.637 & 5.517 & 0.000 \\
2.000 & 1.750 & 0.352 & 0.283 & 67.565 & 15.279 & 99.728 & 5.691 & 0.000 \\
2.000 & 2.000 & 0.302 & 0.259 & 65.188 & 14.411 & 99.697 & 6.093 & 0.000 \\
3.000 & 0.000 & 0.684 & 0.091 & 86.521 & 13.089 & 94.517 & 7.186 & 0.000 \\
3.000 & 0.250 & 0.882 & 0.091 & 94.096 & 12.186 & 89.830 & 6.635 & 0.000 \\
3.000 & 0.500 & 0.936 & 0.090 & 96.649 & 9.053 & 87.487 & 6.208 & 0.000 \\
3.000 & 0.750 & 0.939 & 0.131 & 96.237 & 11.671 & 86.963 & 6.351 & 0.000 \\
3.000 & 1.000 & 0.886 & 0.223 & 91.625 & 19.128 & 87.666 & 7.087 & 0.000 \\
3.000 & 1.250 & 0.796 & 0.299 & 86.120 & 23.073 & 89.060 & 7.659 & 0.000 \\
3.000 & 1.500 & 0.666 & 0.346 & 78.393 & 25.058 & 91.361 & 8.465 & 0.000 \\
3.000 & 1.750 & 0.540 & 0.349 & 71.013 & 25.579 & 93.258 & 8.073 & 0.000 \\
3.000 & 2.000 & 0.437 & 0.327 & 64.911 & 24.880 & 95.081 & 7.667 & 0.000 \\
4.000 & 0.000 & 0.561 & 0.065 & 44.191 & 21.819 & 99.222 & 7.221 & 0.000 \\
4.000 & 0.250 & 0.818 & 0.084 & 51.042 & 31.586 & 91.952 & 6.546 & 0.000 \\
4.000 & 0.500 & 0.891 & 0.078 & 60.080 & 34.928 & 87.545 & 5.535 & 0.000 \\
4.000 & 0.750 & 0.891 & 0.085 & 63.267 & 34.833 & 86.706 & 5.542 & 0.000 \\
4.000 & 1.000 & 0.875 & 0.119 & 61.831 & 34.671 & 86.775 & 5.963 & 0.000 \\
4.000 & 1.250 & 0.825 & 0.190 & 58.782 & 34.366 & 87.844 & 6.893 & 0.000 \\
4.000 & 1.500 & 0.751 & 0.256 & 56.394 & 33.054 & 89.299 & 7.682 & 0.000 \\
4.000 & 1.750 & 0.647 & 0.299 & 52.138 & 31.627 & 91.846 & 8.583 & 0.000 \\
4.000 & 2.000 & 0.556 & 0.311 & 49.734 & 30.048 & 93.703 & 8.411 & 0.000 \\
5.000 & 0.000 & 0.470 & 0.076 & 26.510 & 16.055 & 102.582 & 7.284 & 0.000 \\
5.000 & 0.250 & 0.756 & 0.096 & 32.677 & 27.135 & 94.421 & 6.933 & 0.000 \\
5.000 & 0.500 & 0.855 & 0.093 & 43.766 & 33.161 & 88.342 & 5.844 & 0.000 \\
5.000 & 0.750 & 0.850 & 0.103 & 47.309 & 34.026 & 86.889 & 5.704 & 0.000 \\
5.000 & 1.000 & 0.829 & 0.114 & 46.300 & 33.406 & 86.685 & 5.645 & 0.000 \\
5.000 & 1.250 & 0.799 & 0.157 & 45.837 & 33.263 & 87.441 & 6.453 & 0.000 \\
5.000 & 1.500 & 0.748 & 0.203 & 44.416 & 32.450 & 88.425 & 6.983 & 0.000 \\
5.000 & 1.750 & 0.669 & 0.247 & 41.654 & 30.317 & 90.371 & 7.702 & 0.000 \\
5.000 & 2.000 & 0.591 & 0.265 & 39.780 & 28.352 & 91.932 & 7.732 & 0.000 \\
\bottomrule
\end{tabular}
}
\end{table}

\clearpage

\section{Misspecification study}\label{sm:misspec}
The misspecification experiments explore the behavior of the SC-FH algorithm when the data generating process departs from the assumed model. The reference design is Scenario~3 (altitude-based clusters, clear separation); each design is replicated $500$ times and fitted through adjusted REML with $K=3$, $\phi=0.5$, alongside the pooled FH benchmark. Six DGPs are considered:
\begin{itemize}
  \item \textbf{GAUSS}: the correctly specified control, identical to Scenario~3, against which every other design is read;
  \item \textbf{SKEW-E}: sampling errors drawn from a standardized shifted lognormal distribution (skewness $\approx 3.7$) rescaled to the stated variances $\sigma^2_{ed}$; the Gaussian working likelihood faces positively asymmetric direct estimates;
  \item \textbf{SKEW-U}: area random effects fdrawn from a standardized shifted lognormal distribution (skewness $\approx 3.7$), rescaled to $\sigma^2_{u k}$;
  \item \textbf{SMOOTH}: no discrete regimes --- intercept and slope vary continuously with longitude between the extreme cluster values; the fit selects $K \in \{1,\ldots,4\}$ by BIC at $\phi=0.5$, and the question is whether spurious clusters are created and at what predictive cost;
  \item \textbf{VARDIR-UNDER}: the true sampling variances are twice the stated ones, mimicking understated design-based variance estimates;
  \item \textbf{SAR-U}: the area random effects follow a simultaneous autoregressive process on the true contiguity graph ($\rho = 0.4$, marginal variances rescaled to $\sigma^2_{u k}$), so that the working assumption of independent random effects is violated by residual spatial correlation calibrated on the application data.
\end{itemize}

\begin{table}[!htb]
\centering
\caption{SC-FH under model misspecification (reference design: Scenario 3; $500$ replications; $K=3$, $\phi=0.5$). Classification accuracy (ARI and share of correct assignments, where a true partition exists), RMSE of the EBLUP for the true $\mu_d$, boundary rate, EBLUP RMSE ratio against the pooled FH benchmark, and modal BIC-selected number of clusters (SMOOTH only).}
\label{tab:misspec}
\adjustbox{max width=\linewidth}{
\begin{tabular}{lllllll}
\toprule
DGP & ARI & Share\_correct & RMSE\_Beta1 & Boundary\_rate & RMSE\_ratio\_vs\_FH & G\_BIC\_mode \\
\midrule
GAUSS & 0.938 & 0.975 & 0.424 & 0.000 & 0.884 & -- \\
SKEW-E & 0.936 & 0.971 & 0.413 & 0.000 & 0.910 & -- \\
SKEW-U & 0.926 & 0.968 & 0.420 & 0.000 & 0.861 & -- \\
SMOOTH & -- & -- & -- & 0.000 & 0.997 & 4 \\
VARDIR-UNDER & 0.926 & 0.968 & 0.475 & 0.000 & 0.859 & -- \\
SAR-U & 0.938 & 0.975 & 0.398 & 0.000 & 0.882 & -- \\
\bottomrule
\end{tabular}
}
\end{table}

The algorithm proves to be remarkably robust (Table~\ref{tab:misspec}). Against the correctly specified control (GAUSS: ARI $0.938$, EBLUP RMSE ratio $0.884$), positive skewness in the sampling errors or in the random effects barely moves the classification (SKEW-E $0.936$, SKEW-U $0.926$) and leaves the predictive gain over the pooled benchmark intact (ratios $0.910$ and $0.861$); understated sampling variances behave similarly (VARDIR-UNDER, ARI $0.926$, ratio $0.859$). The spatially correlated random effects of the SAR-U design --- the most relevant departure for spatial applications --- are the mildest of all: the classification is indistinguishable from the Gaussian control (ARI $0.938$) and the prediction gain is preserved (ratio $0.882$), so that ignoring a moderate residual spatial correlation in the random effects costs essentially nothing. No boundary solution occurs in any design. The single instructive failure is SMOOTH, where by construction there is no discrete regime: the clusterwise predictor neither gains nor loses against the pooled benchmark (ratio $0.997$) and BIC most frequently selects $\widehat{K}=4$, a reminder that the discrete-regime model should not be forced on a genuinely smooth surface --- exactly the situation the stability-based selection rule of Section~\ref{sec:tuning} is designed to expose.

\begin{table}[!htb]
  \centering
  \caption{Detailed breakdown of the misspecification study (reference design: Scenario~3; $K=3$, $\phi=0.5$; $500$ replications per design). Columns, left to right: the mean over replications of the relocated cluster-wise variance-component estimates $\widehat{\sigma}^2_{u,g}$ for $g=1,2,3$ (aligned to the true groups by a majority rule; true values $\sigma^2_{u,g}=(7.76,\,1.24,\,31.02)$); the first quartile, median, and third quartile across replications of the RMSE ratio $r=\mathrm{RMSE}(\widehat{\mu}^{\text{SC-FH}})/\mathrm{RMSE}(\widehat{\mu}^{\text{FH}})$, each RMSE taken over the $D$ areas of a replication, so that $r<1$ signals a gain over the pooled benchmark; the over-clustering share, i.e.\ the fraction of replications in which BIC selects more than one cluster ($\widehat{K}>1$), meaningful only for the regime-free SMOOTH design (``--'' for the others); the mean number of algorithm iterations; and the number of valid and failed replications. The classification and prediction summaries (ARI, share of correct assignments, EBLUP RMSE ratio, modal $\widehat{K}$) appear in the corresponding appendix table of the main paper.}
  \label{tab:sm_misspec}
  \adjustbox{max width=\linewidth}{
\begin{tabular}{lllllllllll}
\toprule
DGP & S2u\_G1 & S2u\_G2 & S2u\_G3 & RMSEratio\_Q1 & RMSEratio\_med & RMSEratio\_Q3 & ShareOverclust & Iter\_mean & nValid & nFailed \\
\midrule
GAUSS & 17.764 & 4.331 & 50.184 & 0.848 & 0.878 & 0.912 & -- & 7.1 & 500.000 & 0.000 \\
SKEW-E & 22.982 & 15.154 & 44.387 & 0.864 & 0.897 & 0.948 & -- & 7.1 & 500.000 & 0.000 \\
SKEW-U & 26.266 & 4.167 & 61.116 & 0.823 & 0.863 & 0.903 & -- & 7.4 & 500.000 & 0.000 \\
SMOOTH & NaN & NaN & NaN & 0.985 & 0.996 & 1.008 & 1.000 & 10.3 & 500.000 & 0.000 \\
VARDIR-UNDER & 28.674 & 6.168 & 57.385 & 0.836 & 0.852 & 0.875 & -- & 7.3 & 500.000 & 0.000 \\
SAR-U & 21.492 & 4.489 & 52.757 & 0.846 & 0.876 & 0.909 & -- & 7.0 & 500.000 & 0.000 \\
\bottomrule
\end{tabular}
}
\end{table}

\begin{figure}[!htb]
  \centering
  \includegraphics[width=0.98\linewidth]{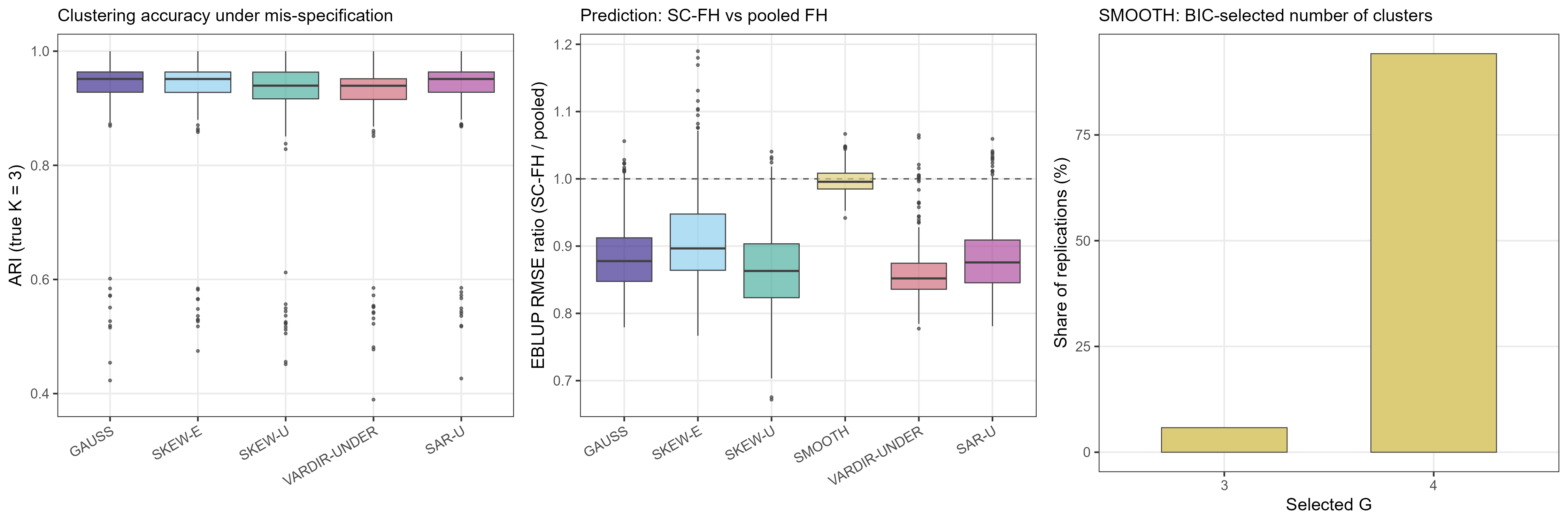}
  \caption{SC-FH under misspecification: clustering accuracy by design (left), EBLUP RMSE ratio against the pooled FH benchmark with the dashed line at $1$ (center), and distribution of the BIC-selected number of clusters under the SMOOTH design (right).}
  \label{fig:sm_misspec}
\end{figure}

Table~\ref{tab:sm_misspec} and Figure~\ref{fig:sm_misspec} refine the summary of the main text along three axes. First, \emph{variance-component recovery}: the mean relocated estimates lie above their true values $\sigma^2_{u,g}=(7.76,\,1.24,\,31.02)$ in every design (e.g.\ $17.8/4.3/50.2$ under GAUSS), an upward bias that combines the strict positivity of adjusted REML --- which trades a small positive bias for the elimination of boundary solutions --- with the residual misclassification, whose mis-assigned areas inflate the within-cluster dispersion; crucially, this bias is essentially the same across the misspecified designs as in the correctly specified control, and it is benign for prediction, since over-estimating $\sigma^2_{u}$ only makes the EBLUP shrink less toward the synthetic part. Second, \emph{dispersion of the predictive gain}: the interquartile range of the replication-level RMSE ratio is narrow and lies entirely below one in the regime designs (e.g.\ $[0.848,0.912]$ under GAUSS, $[0.823,0.903]$ under SKEW-U), so the advantage over the pooled benchmark is systematic rather than driven by a few favorable replications; only under SMOOTH does it straddle one ($[0.985,1.008]$), i.e.\ neither a systematic gain nor a loss. Third, \emph{over-clustering}: in the absence of discrete regimes BIC selects more than one cluster in every SMOOTH replication (share $1.000$), with a modal $\widehat{K}=4$, confirming that an information criterion evaluated on the endogenously optimized partition cannot by itself detect the lack of a regime structure. Across all designs the algorithm converges in $7$--$10$ iterations and every one of the $500$ replications completes without failures.

\clearpage
\section{Comparison of the REML updating schemes}\label{sm:reml}
On all six simulation designs, the full SC-FH procedure is run with each of the three REML updating schemes of \citet[Sect.~16.5.4]{MoralesSAE} --- Fisher scoring on the REML score (M1), its residual-based variant (M2), and the fixed-point iteration on the estimated random effects (M3) --- and with the adjusted REML estimator of \citet{LiLahiri2010}. Since M1--M3 maximize the same restricted likelihood, they must agree up to numerical tolerance wherever the solution is interior; the experiment verifies this equivalence within the full endogenous-clustering loop and quantifies the incidence of boundary solutions under unadjusted REML.

\begin{table}[!htb]
  \centering
  \caption{Agreement of the four variance-estimation routines inside the full SC-FH algorithm, by scenario and $\phi$ ($500$ replications): ARI against the truth and against the M1 partition, median and maximum absolute deviation of the cluster-wise variance components from M1, boundary rate, RMSE of the cluster-wise slopes, mean iterations and mean time per fit.}
  \label{tab:sm_reml}
  \adjustbox{max width=\linewidth}{
\begin{tabular}{lllllllllll}
\toprule
Scenario & Phi & Method & ARI\_truth & ARI\_vs\_M1 & MedDiffS2u & MaxDiffS2u & Boundary\_rate & RMSE\_Beta1 & Iter\_mean & Time\_sec \\
\midrule
Scenario 1 & 0.000 & REML-M1 & 0.883 & 1.000 & 0.0e+00 & 0.0e+00 & 0.002 & 1.182 & 7.5 & 0.68 \\
Scenario 1 & 0.000 & REML-M2 & 0.883 & 1.000 & 2.8e-14 & 3.4e-13 & 0.002 & 1.182 & 7.5 & 0.65 \\
Scenario 1 & 0.000 & REML-M3 & 0.883 & 1.000 & 6.3e-05 & 1.3e-04 & 0.002 & 1.182 & 7.5 & 0.71 \\
Scenario 1 & 0.000 & AdjREML & 0.886 & 0.991 & 4.4e+00 & 9.3e+02 & 0.000 & 0.846 & 7.4 & 0.63 \\
Scenario 1 & 0.500 & REML-M1 & 0.976 & 1.000 & 0.0e+00 & 0.0e+00 & 0.002 & 0.914 & 6.1 & 0.59 \\
Scenario 1 & 0.500 & REML-M2 & 0.976 & 1.000 & 2.5e-14 & 3.1e-13 & 0.002 & 0.914 & 6.1 & 0.59 \\
Scenario 1 & 0.500 & REML-M3 & 0.976 & 1.000 & 7.1e-05 & 1.3e-04 & 0.002 & 0.914 & 6.1 & 0.64 \\
Scenario 1 & 0.500 & AdjREML & 0.975 & 0.992 & 4.3e+00 & 1.3e+03 & 0.000 & 0.853 & 6.0 & 0.58 \\
Scenario 2 & 0.000 & REML-M1 & 0.250 & 1.000 & 0.0e+00 & 0.0e+00 & 0.960 & 6.419 & 12.5 & 1.78 \\
Scenario 2 & 0.000 & REML-M2 & 0.250 & 1.000 & 2.1e-14 & 2.1e-12 & 0.960 & 6.419 & 12.5 & 1.75 \\
Scenario 2 & 0.000 & REML-M3 & 0.250 & 1.000 & 3.9e-04 & 5.8e-03 & 0.960 & 6.419 & 12.5 & 5.42 \\
Scenario 2 & 0.000 & AdjREML & 0.266 & 0.803 & 5.8e+01 & 5.1e+03 & 0.000 & 6.622 & 12.2 & 0.88 \\
Scenario 2 & 0.500 & REML-M1 & 0.627 & 1.000 & 0.0e+00 & 0.0e+00 & 0.440 & 5.366 & 9.5 & 0.82 \\
Scenario 2 & 0.500 & REML-M2 & 0.627 & 1.000 & 7.1e-14 & 4.3e-12 & 0.440 & 5.366 & 9.5 & 0.80 \\
Scenario 2 & 0.500 & REML-M3 & 0.627 & 1.000 & 1.7e-03 & 3.9e-02 & 0.440 & 5.366 & 9.5 & 2.23 \\
Scenario 2 & 0.500 & AdjREML & 0.646 & 0.859 & 2.0e+02 & 9.2e+03 & 0.000 & 4.930 & 9.5 & 0.74 \\
Scenario 3 & 0.000 & REML-M1 & 0.859 & 1.000 & 0.0e+00 & 0.0e+00 & 0.000 & 1.084 & 6.9 & 0.64 \\
Scenario 3 & 0.000 & REML-M2 & 0.859 & 1.000 & 2.1e-14 & 1.7e-13 & 0.000 & 1.084 & 6.9 & 0.63 \\
Scenario 3 & 0.000 & REML-M3 & 0.859 & 1.000 & 6.7e-05 & 1.2e-04 & 0.000 & 1.084 & 6.9 & 0.68 \\
Scenario 3 & 0.000 & AdjREML & 0.859 & 0.996 & 1.7e+00 & 9.6e+02 & 0.000 & 1.197 & 6.9 & 0.61 \\
Scenario 3 & 0.500 & REML-M1 & 0.934 & 1.000 & 0.0e+00 & 0.0e+00 & 0.000 & 1.757 & 7.0 & 0.64 \\
Scenario 3 & 0.500 & REML-M2 & 0.934 & 1.000 & 2.1e-14 & 3.4e-13 & 0.000 & 1.757 & 7.0 & 0.63 \\
Scenario 3 & 0.500 & REML-M3 & 0.934 & 1.000 & 6.7e-05 & 1.2e-04 & 0.000 & 1.757 & 7.0 & 0.68 \\
Scenario 3 & 0.500 & AdjREML & 0.938 & 0.991 & 1.7e+00 & 1.4e+03 & 0.000 & 1.390 & 7.0 & 0.61 \\
Scenario 4 & 0.000 & REML-M1 & 0.224 & 1.000 & 0.0e+00 & 0.0e+00 & 0.892 & 7.088 & 11.4 & 1.10 \\
Scenario 4 & 0.000 & REML-M2 & 0.224 & 1.000 & 2.8e-13 & 5.4e-12 & 0.892 & 7.088 & 11.4 & 1.08 \\
Scenario 4 & 0.000 & REML-M3 & 0.224 & 1.000 & 1.2e-03 & 7.3e-03 & 0.892 & 7.088 & 11.4 & 4.33 \\
Scenario 4 & 0.000 & AdjREML & 0.252 & 0.802 & 1.7e+02 & 3.8e+03 & 0.000 & 6.657 & 11.4 & 0.83 \\
Scenario 4 & 0.500 & REML-M1 & 0.255 & 1.000 & 0.0e+00 & 0.0e+00 & 0.604 & 6.037 & 10.2 & 1.01 \\
Scenario 4 & 0.500 & REML-M2 & 0.255 & 1.000 & 1.7e-13 & 6.8e-12 & 0.604 & 6.037 & 10.2 & 1.00 \\
Scenario 4 & 0.500 & REML-M3 & 0.255 & 1.000 & 1.7e-03 & 1.1e-02 & 0.604 & 6.037 & 10.2 & 2.65 \\
Scenario 4 & 0.500 & AdjREML & 0.263 & 0.754 & 2.4e+02 & 1.7e+04 & 0.000 & 5.825 & 10.2 & 0.77 \\
Scenario 5 & 0.000 & REML-M1 & 0.991 & 1.000 & 0.0e+00 & 0.0e+00 & 0.000 & 0.814 & 6.1 & 0.61 \\
Scenario 5 & 0.000 & REML-M2 & 0.991 & 1.000 & 1.2e-14 & 1.4e-13 & 0.000 & 0.814 & 6.1 & 0.59 \\
Scenario 5 & 0.000 & REML-M3 & 0.991 & 1.000 & 9.8e-06 & 2.3e-05 & 0.000 & 0.814 & 6.1 & 0.64 \\
Scenario 5 & 0.000 & AdjREML & 0.992 & 0.995 & 6.7e-01 & 1.4e+02 & 0.000 & 0.724 & 6.0 & 0.57 \\
Scenario 5 & 0.500 & REML-M1 & 0.962 & 1.000 & 0.0e+00 & 0.0e+00 & 0.000 & 0.666 & 6.0 & 0.59 \\
Scenario 5 & 0.500 & REML-M2 & 0.962 & 1.000 & 1.2e-14 & 2.8e-13 & 0.000 & 0.666 & 6.0 & 0.59 \\
Scenario 5 & 0.500 & REML-M3 & 0.962 & 1.000 & 9.6e-06 & 2.3e-05 & 0.000 & 0.666 & 6.0 & 0.63 \\
Scenario 5 & 0.500 & AdjREML & 0.964 & 0.987 & 6.9e-01 & 1.9e+02 & 0.000 & 0.632 & 6.1 & 0.58 \\
Scenario 6 & 0.000 & REML-M1 & 0.672 & 1.000 & 0.0e+00 & 0.0e+00 & 0.040 & 6.340 & 9.4 & 0.76 \\
Scenario 6 & 0.000 & REML-M2 & 0.672 & 1.000 & 2.3e-13 & 3.6e-12 & 0.040 & 6.340 & 9.4 & 0.74 \\
Scenario 6 & 0.000 & REML-M3 & 0.672 & 1.000 & 6.3e-04 & 2.2e-03 & 0.040 & 6.340 & 9.4 & 0.92 \\
Scenario 6 & 0.000 & AdjREML & 0.683 & 0.964 & 8.6e+01 & 9.3e+03 & 0.000 & 5.719 & 9.0 & 0.71 \\
Scenario 6 & 0.500 & REML-M1 & 0.930 & 1.000 & 0.0e+00 & 0.0e+00 & 0.014 & 4.826 & 6.9 & 0.63 \\
Scenario 6 & 0.500 & REML-M2 & 0.930 & 1.000 & 2.3e-13 & 2.7e-12 & 0.014 & 4.826 & 6.9 & 0.63 \\
Scenario 6 & 0.500 & REML-M3 & 0.930 & 1.000 & 1.2e-03 & 2.2e-03 & 0.014 & 4.826 & 6.9 & 0.69 \\
Scenario 6 & 0.500 & AdjREML & 0.935 & 0.985 & 7.0e+01 & 8.7e+03 & 0.000 & 4.311 & 6.8 & 0.61 \\
\bottomrule
\end{tabular}
}
\end{table}

\begin{figure}[!htb]
  \centering
  \includegraphics[width=0.98\linewidth]{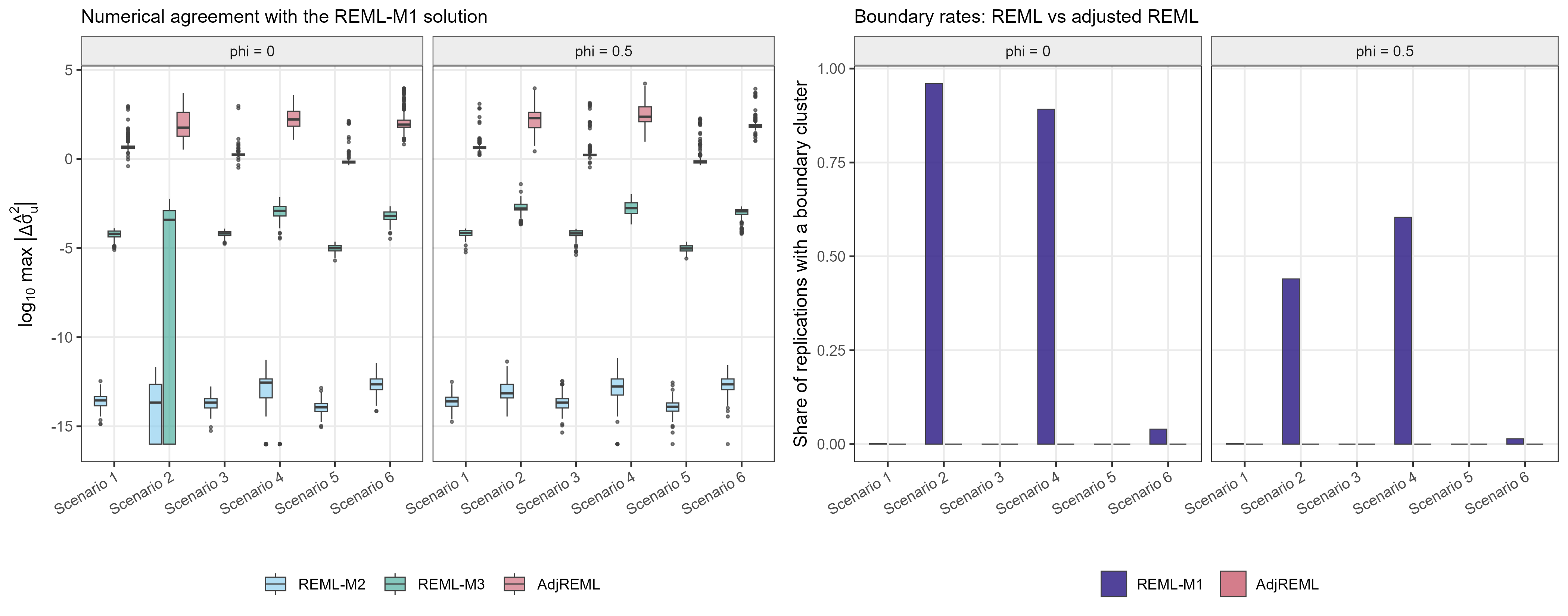}
  \caption{Left: distribution across replications of the maximum absolute difference of the cluster-wise variance components from the REML-M1 solution ($\log_{10}$ scale), by scenario and penalty --- M2 sits at machine precision, M3 at the tolerance of its fixed-point recursion, adjusted REML visibly apart because of its strictly positive estimates. Right: share of replications with a boundary cluster under REML-M1 and under adjusted REML.}
  \label{fig:sm_reml}
\end{figure}

\clearpage

\section{Validation of the parametric bootstrap}\label{sm:boot}
The refit-with-clustering parametric bootstrap (Algorithm~2 of the main paper) is validated on the two altitude-based designs, fitted at the true $K=3$ and $\phi=0.5$: each of $500$ Monte Carlo replications is paired with a full bootstrap of $B=200$ refits, and three properties are measured: (i) the empirical coverage of the percentile confidence intervals for the cluster-wise regression coefficients at the $90\%$ and $95\%$ nominal levels; (ii) the ratio between the average bootstrap standard error and the Monte Carlo standard deviation of the same estimators; (iii) the agreement between the bootstrap MSE of the EBLUPs and their Monte Carlo mean squared error, area by area. Because the bootstrap refits the entire clustering procedure at every draw, the experiment explicitly checks that label uncertainty is propagated. A handful of replications failed at the initial fit and are excluded ($499$ valid in Scenario~3 and $495$ in Scenario~4), and the number of valid bootstrap refits per replication is close to the full $B=200$ in both designs.

\begin{figure}[!htb]
  \centering
  \includegraphics[width=0.95\linewidth]{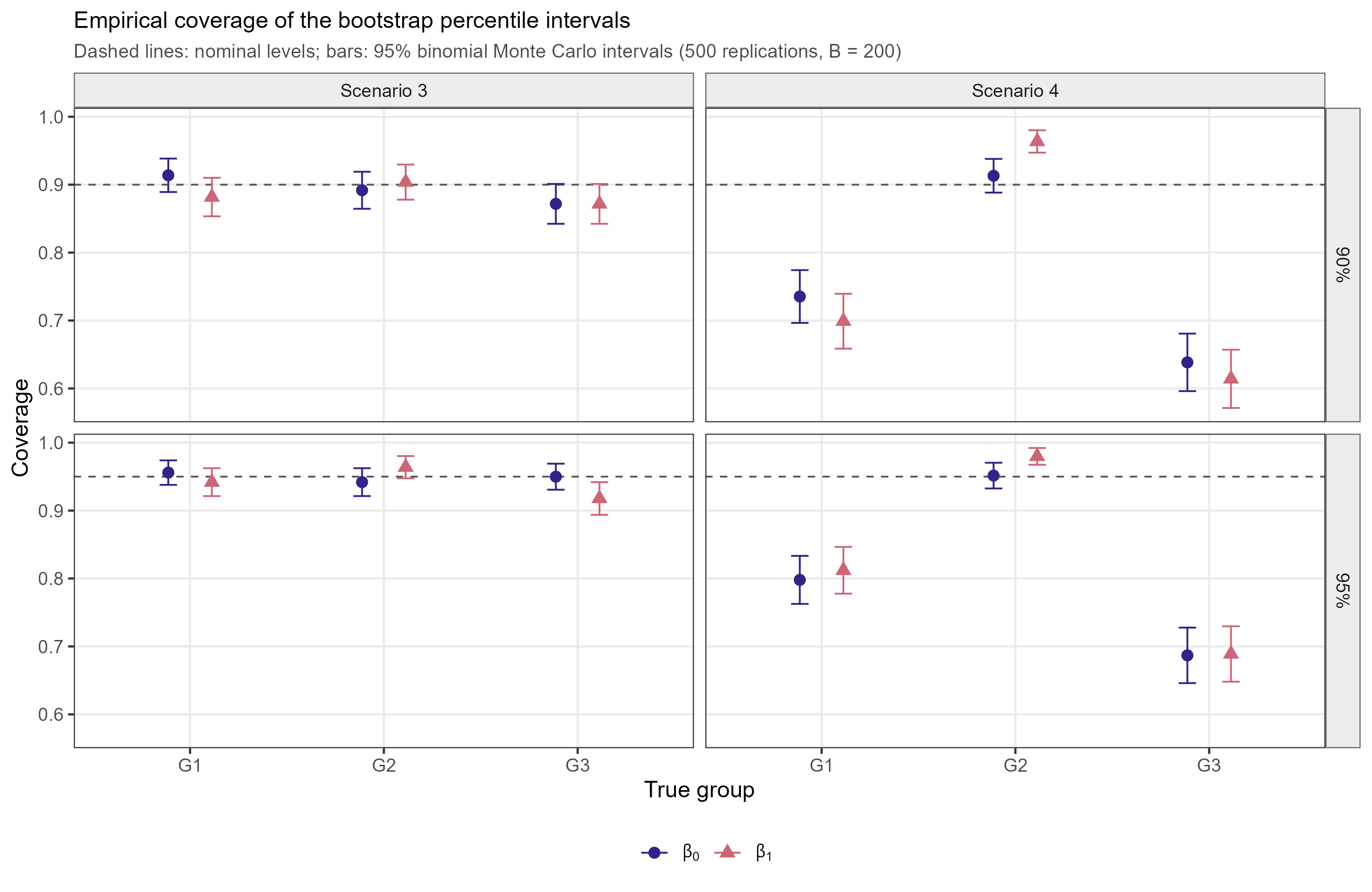}
  \caption{Empirical coverage of the bootstrap percentile intervals for the cluster-wise coefficients, by true group, coefficient, and nominal level (dashed lines). Error bars are $95\%$ binomial Monte Carlo intervals.}
  \label{fig:sm_boot_cov}
\end{figure}

\paragraph{What the coverage experiment measures.}
A confidence interval is a procedure, not a single object, and its coverage is a property of the procedure: if the entire pipeline --- generate the data, fit the SC-FH model, run the bootstrap, build the interval at nominal level $1-\alpha$ --- were repeated on fresh data, how often would the resulting interval contain the true parameter? The experiment answers this question literally. For each Monte Carlo replication the true cluster-wise coefficients are known by design; the percentile interval is built from the $B$ aligned draws of that replication alone, and a $0/1$ indicator records whether the true value falls inside it. The average of the indicators across replications estimates the true coverage of the procedure. Since each indicator is a Bernoulli draw, the empirical rate carries a binomial Monte Carlo error of $\sqrt{p(1-p)/\mathrm{NRep}}$ --- about $\pm 0.013$ at $p = 0.9$ with $500$ replications --- which is displayed as the $95\%$ error bars of Figure~\ref{fig:sm_boot_cov}: an empirical rate whose bar crosses the nominal line is compatible with exact calibration.

\paragraph{Label alignment.}
In a clustered model the coverage question is well posed only after the labels are matched. Two relocations are composed: the clusters of the original fit are relocated onto the true groups by the majority rule on the confusion matrix (so that ``the slope of cluster $k$'' refers to the regime it actually captured), and the clusters of every bootstrap refit are relocated onto those of the original fit, so that all bootstrap draws are expressed in the labeling of the true groups. Draws in which the second matching is not one-to-one (a collapsed refit cluster) are flagged by the analysis code and can be excluded; in the present runs they are a negligible fraction.

\paragraph{Why the Monte Carlo distribution is the benchmark.}
The bootstrap is an estimator of the sampling distribution of the SC-FH estimates: its promise is that the dispersion of the bootstrap refits around the original fit --- in the bootstrap world, the fitted model and partition play the role of the truth --- reproduces the dispersion of the estimates around the true parameters. In a simulation the latter is directly observable: it is the Monte Carlo distribution of the estimates across replications, which therefore provides the gold standard against which the bootstrap output is judged. The Monte Carlo results are therefore the gold standard of exactly the quantity the bootstrap is trying to estimate: a well-calibrated bootstrap yields empirical coverage close to the nominal levels and bootstrap-to-Monte-Carlo ratios close to one; for a perfect bootstrap, the coverage would match the nominal levels, and the bootstrap-SE--to--Monte-Carlo-SD and bootstrap-MSE--to--Monte-Carlo-MSE ratios would equal one. However, exact agreement is not expected, for three reasons. First, simulation noise: both $B$ and the number of replications are finite. Second, and substantively most important, the \emph{plug-in error}: the bootstrap world is generated from the \emph{estimated} model rather than from the true one, so any systematic bias of the point estimates is transmitted to the bootstrap distribution. When the point estimates are nearly unbiased --- Scenario~3, where the partition is recovered almost exactly --- the two worlds nearly coincide and the bootstrap is calibrated. When the estimates are systematically attenuated by misclassification --- Scenario~4 --- the bootstrap world is generated by a flatter model than the true one: the intervals are centered on the attenuated values and undercover precisely for the extreme regimes, and the bootstrap prediction errors are milder than the real ones, so the bootstrap MSE understates its Monte Carlo counterpart. The validation experiment is, in essence, a measurement of how much the plug-in error matters in practice: it is negligible where the clustering is reliable and it reproduces the known biases of the point estimator where the clustering is not.

\paragraph{Bootstrap standard errors versus percentile intervals.}
The results display an apparent paradox: the ratio between the average bootstrap standard error and the Monte Carlo standard deviation is highly dispersed across coefficients (from about $0.2$ to $1.3$, well below one on average in Scenario~3), yet the percentile intervals are essentially calibrated. The resolution is that the two quantities live at different levels. The Monte Carlo standard deviation is a single global summary computed across \emph{all} replications, and it is inflated by the minority of replications with degraded partitions, whose relocated estimates are far from the truth. The percentile intervals, instead, are computed replication by replication: each interval only needs to cover the truth for its own dataset, and the outlying replications affect their own indicator without contaminating the others. A Wald-type interval built as ``estimate $\pm z \cdot$ average bootstrap SE'' would inherit the global understatement and undercover; the percentile intervals do not. This is the operational reason why the percentile intervals, and not standard-error-based intervals, are the recommended inferential output of the procedure.

\begin{figure}[!htb]
  \centering
  \includegraphics[width=0.9\linewidth]{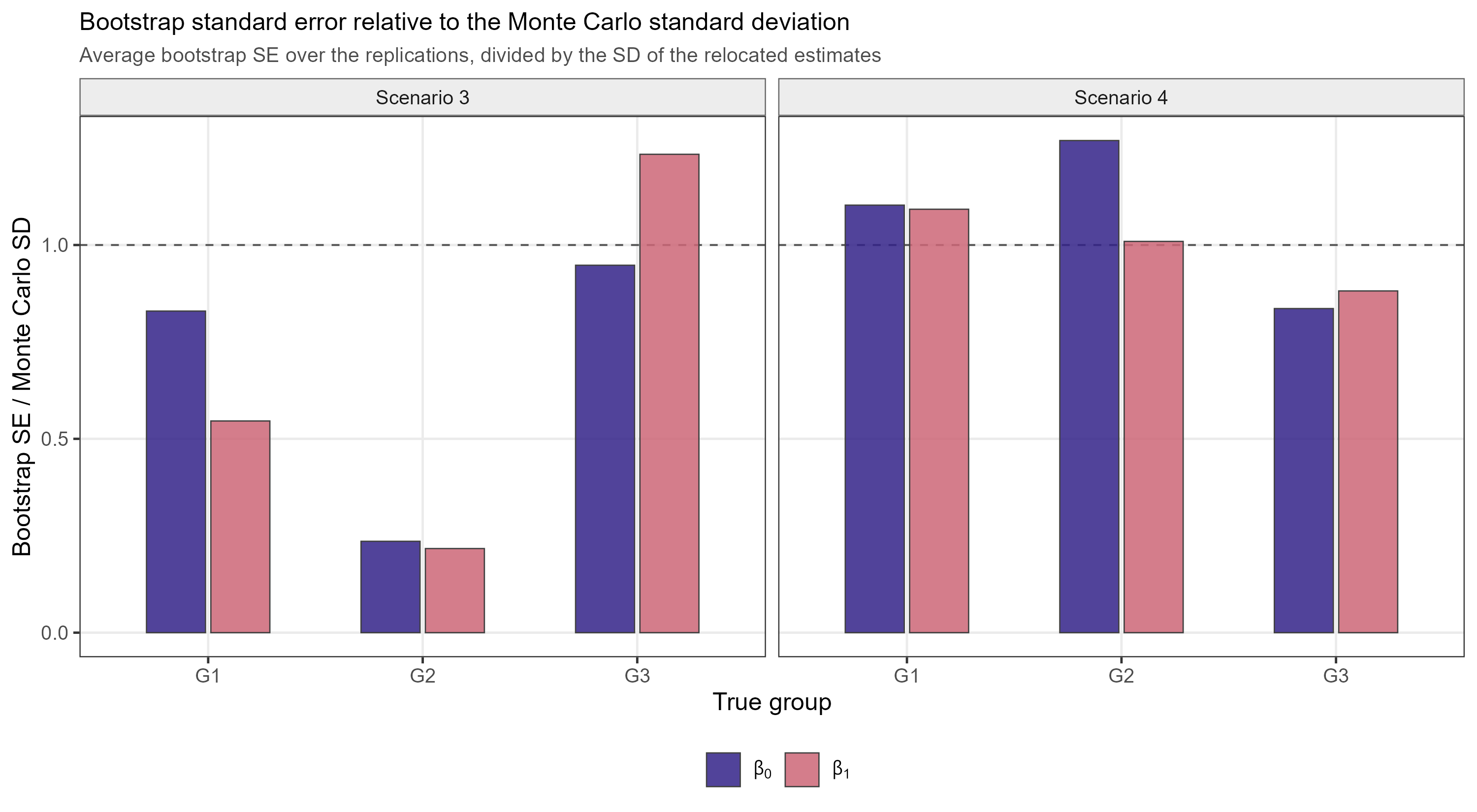}
  \caption{Average bootstrap standard error relative to the Monte Carlo standard deviation of the cluster-wise coefficient estimates, by true group and coefficient.}
  \label{fig:sm_boot_se}
\end{figure}

\begin{figure}[!htb]
  \centering
  \includegraphics[width=0.75\linewidth]{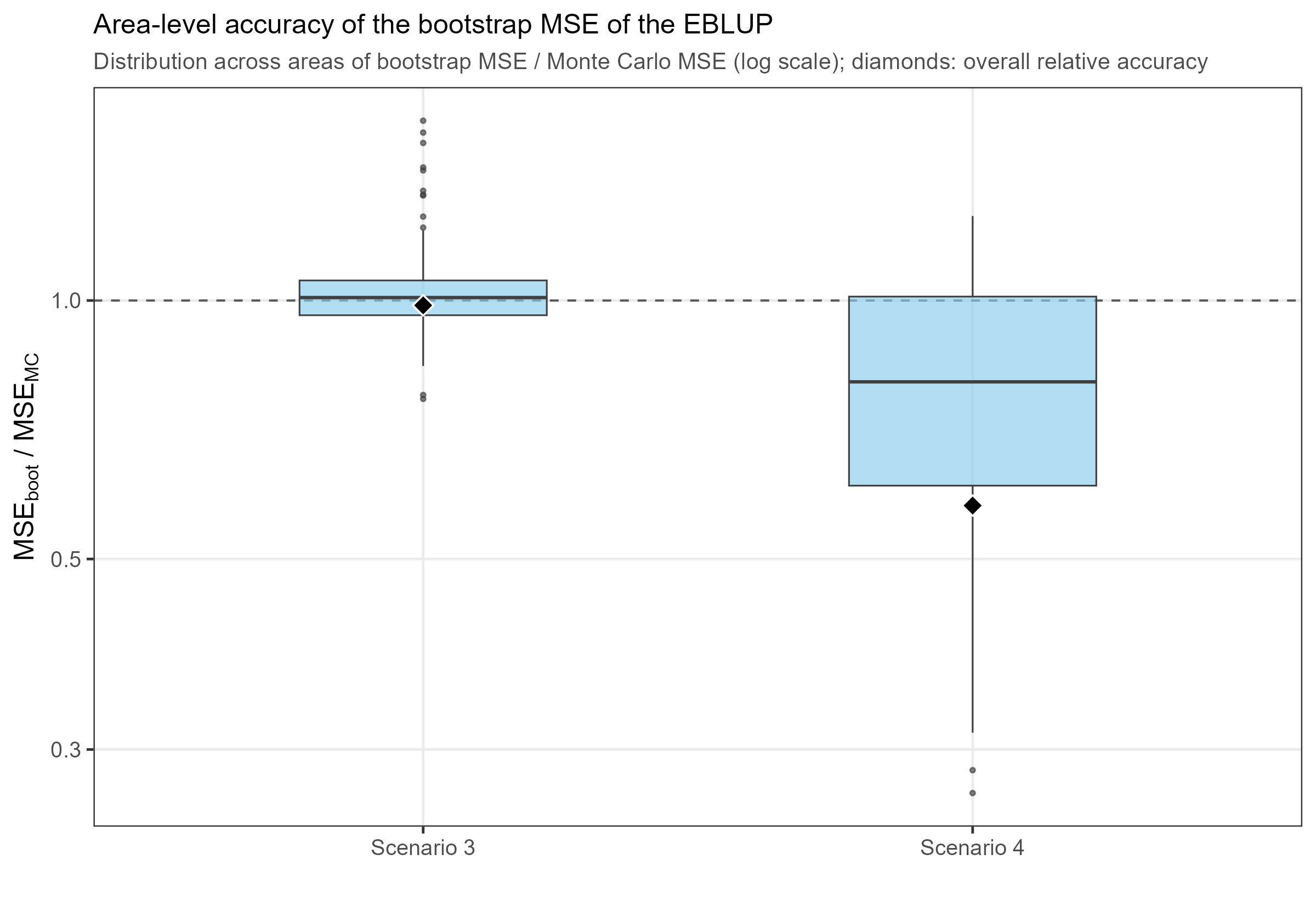}
  \caption{Distribution across areas of the ratio between the bootstrap MSE and the Monte Carlo MSE of the EBLUP (log scale). Black diamonds mark the overall relative accuracy (ratio of study-wide means).}
  \label{fig:sm_boot_mse}
\end{figure}

Table~\ref{tab:sm_boot} reports the scenario-level summary and Table~\ref{tab:sm_boot_detail} the full breakdown by coefficient and group; Figures~\ref{fig:sm_boot_cov}--\ref{fig:sm_boot_mse} visualize the three properties. In Scenario~3 the percentile intervals are essentially calibrated at the $95\%$ level (empirical coverage $0.95$ on average across the six coefficient--group combinations) with a mild undercoverage at the $90\%$ level ($0.89$ on average), and the bootstrap MSE tracks the Monte Carlo MSE almost exactly (median area-level ratio $1.01$, overall relative accuracy $0.99$). In Scenario~4 the bootstrap inherits the attenuation of the point estimates documented in the main paper: the intervals for the well-identified central regime remain calibrated ($0.95$--$0.98$ at the $95\%$ level), those for the two extreme regimes undercover (around $0.69$--$0.81$), and the bootstrap MSE understates the Monte Carlo MSE (overall relative accuracy $0.58$). The dispersion of the bootstrap-SE--to--Monte-Carlo-SD ratio (Figure~\ref{fig:sm_boot_se}) reflects the same mechanism: the ratio between the average bootstrap standard error and the Monte Carlo standard deviation is highly dispersed across coefficients (from about $0.2$ to $1.3$), because the Monte Carlo standard deviation is inflated by the minority of replications with degraded partitions, to which the replication-wise percentile intervals are instead robust --- which is why the percentile intervals, and not Wald-type intervals based on the bootstrap standard error, are the recommended inferential output.

Eventually, we remark that bootstrap inference should be read together with the classification diagnostics: it is trustworthy precisely in the regimes that the model identifies well.

\begin{table}[!htb]
  \centering
  \caption{Bootstrap validation, scenario-level summary: valid replications, empirical coverage of the percentile intervals (average over the three groups, by coefficient and nominal level), average bootstrap-SE to Monte-Carlo-SD ratio, median area-level bootstrap-MSE to Monte-Carlo-MSE ratio, and overall relative accuracy of the bootstrap MSE. Auto-generated by \texttt{Sims\_Bootstrap\_Analysis\_20260710.R}.}
  \label{tab:sm_boot}
  \adjustbox{max width=\linewidth}{
\begin{tabular}{lllllllll}
\toprule
Scenario & NRep\_valid & beta0\_90\% & beta1\_90\% & beta0\_95\% & beta1\_95\% & SE\_ratio & MSE\_ratio\_med & Rel\_accuracy \\
\midrule
Scenario 3 & 499.000 & 0.892 & 0.886 & 0.949 & 0.941 & 0.668 & 1.008 & 0.987 \\
Scenario 4 & 495.000 & 0.762 & 0.759 & 0.812 & 0.827 & 1.032 & 0.804 & 0.577 \\
\bottomrule
\end{tabular}
}
\end{table}

\begin{table}[!htb]
  \centering
  \caption{Bootstrap validation, full breakdown by coefficient and true group: empirical coverage at the $90\%$ and $95\%$ nominal levels and bootstrap-SE to Monte-Carlo-SD ratio. Auto-generated by \texttt{Sims\_Bootstrap\_Analysis\_20260710.R}.}
  \label{tab:sm_boot_detail}
  \adjustbox{max width=0.85\linewidth}{
\begin{tabular}{llllll}
\toprule
Scenario & Coef & Group & Cov\_90\% & Cov\_95\% & SE\_ratio \\
\midrule
Scenario 3 & beta0 & G1 & 0.914 & 0.956 & 0.830 \\
Scenario 3 & beta0 & G2 & 0.892 & 0.942 & 0.236 \\
Scenario 3 & beta0 & G3 & 0.872 & 0.950 & 0.948 \\
Scenario 3 & beta1 & G1 & 0.882 & 0.942 & 0.546 \\
Scenario 3 & beta1 & G2 & 0.904 & 0.964 & 0.217 \\
Scenario 3 & beta1 & G3 & 0.872 & 0.918 & 1.234 \\
Scenario 4 & beta0 & G1 & 0.735 & 0.798 & 1.103 \\
Scenario 4 & beta0 & G2 & 0.913 & 0.952 & 1.269 \\
Scenario 4 & beta0 & G3 & 0.638 & 0.687 & 0.836 \\
Scenario 4 & beta1 & G1 & 0.699 & 0.812 & 1.092 \\
Scenario 4 & beta1 & G2 & 0.964 & 0.980 & 1.009 \\
Scenario 4 & beta1 & G3 & 0.614 & 0.689 & 0.882 \\
\bottomrule
\end{tabular}
}
\end{table}

\clearpage

\section{Additional material for the application}\label{sm:app}
\begin{figure}[!htb]
  \centering
  \includegraphics[width=0.95\linewidth]{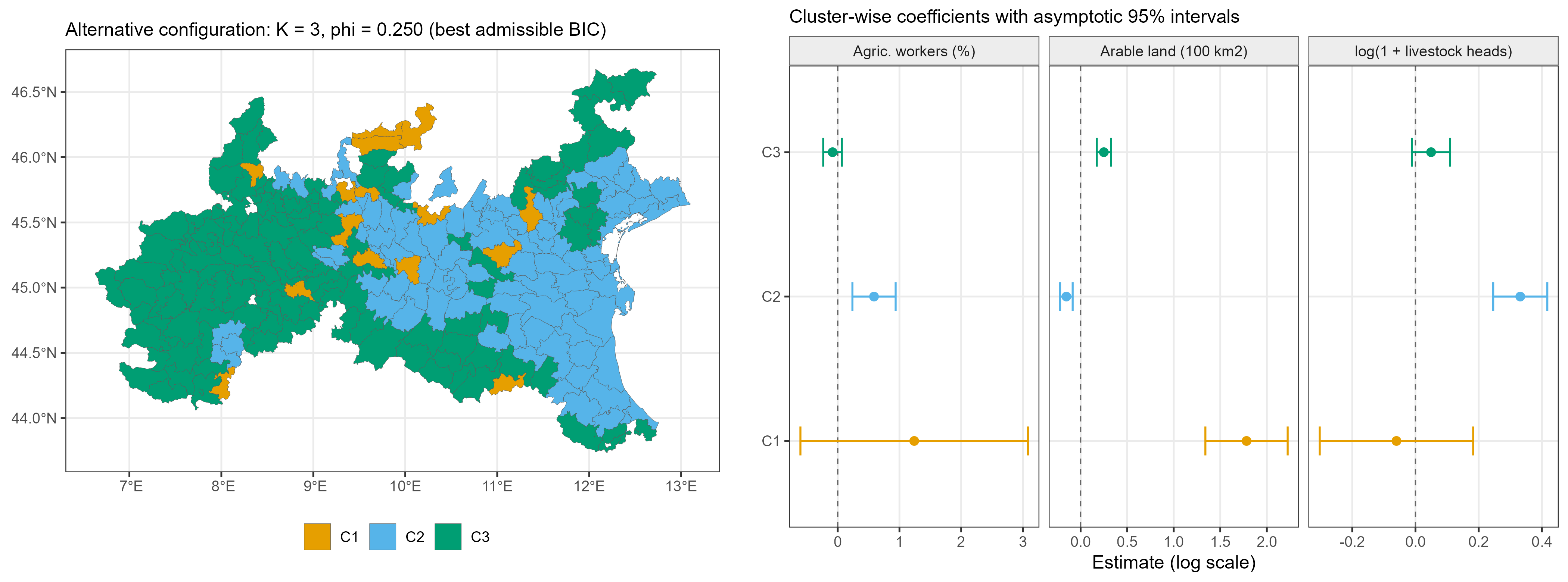}
  \caption{Alternative configuration with $K=3$ at $\phi=0.25$ (best admissible BIC within the moderate band): map of the regimes and cluster-wise coefficients with asymptotic $95\%$ intervals. The third cluster is small, spatially scattered, and poorly identified.}
  \label{fig:sm_altk}
\end{figure}

The material below complements the application of the main paper (average standard output per farm, Po Valley, 2020). Table~\ref{tab:sm_screen} reports the univariate screening of the candidate covariates (pooled FH fits on the log scale, one candidate at a time): the CAMS ammonia concentration is the strongest single predictor, but it is strongly collinear with the livestock endowment (correlation $0.79$) and is an outcome of the production regimes rather than an input, which motivates its ex-post profiling role in the main analysis. Table~\ref{tab:sm_tests} reports the bootstrap percentile tests for the between-regime differences of the cluster-wise coefficients at the selected configuration. Figure~\ref{fig:sm_altk} shows the best admissible alternative configuration with $K=3$ ($\phi=0.25$, the strongest admissible BIC for that $K$ inside the moderate band): the two main regimes are essentially preserved, while the third cluster is a small, spatially scattered group with poorly identified coefficients, which corroborates the selected $K=2$.

\begin{table}[!htb]
  \centering
  \caption{Univariate screening of the candidate covariates: slope, $t$-value and BIC of the pooled FH fit on the log scale, one candidate at a time. Auto-generated by the application script.}
  \label{tab:sm_screen}
  \adjustbox{max width=0.7\linewidth}{
\begin{tabular}{llll}
\toprule
Variable & Beta & tvalue & BIC \\
\midrule
AQcams\_nh3\_Mean\_BKm & 0.121 & 12.000 & 512.876 \\
lAnimals & 0.304 & 10.538 & 532.440 \\
AQeea\_PM25\_BKm & 0.092 & 10.248 & 535.517 \\
EM\_NH3\_MNM\_BKm & 0.001 & 9.700 & 545.242 \\
AltMean & -0.001 & -8.280 & 562.234 \\
WE\_t2m\_Mean\_BKm & 0.166 & 7.735 & 569.045 \\
ArableLand100 & 0.263 & 7.498 & 571.618 \\
SE\_ShareWorkers\_Agriculture & -0.271 & -2.523 & 617.176 \\
\bottomrule
\end{tabular}
}
\end{table}

\begin{table}[!htb]
  \centering
  \caption{Bootstrap percentile tests for the between-regime differences of the cluster-wise coefficients at the selected configuration ($K=2$, $\phi=0.625$; $B=200$). Auto-generated by the application script.}
  \label{tab:sm_tests}
  \adjustbox{max width=0.85\linewidth}{
\begin{tabular}{llllll}
\toprule
Coefficient & Pair & Difference & Lower & Upper & Pvalue \\
\midrule
Intercept & C1 - C2 & -1.096 & -2.820 & 0.532 & 0.240 \\
lAnimals & C1 - C2 & 0.215 & 0.052 & 0.371 & 0.020 \\
SE\_ShareWorkers\_Agriculture & C1 - C2 & 0.507 & -0.116 & 1.311 & 0.090 \\
ArableLand100 & C1 - C2 & -0.391 & -0.582 & -0.283 & 0.005 \\
\bottomrule
\end{tabular}
}
\end{table}

\end{document}